\documentclass[
11pt, 
singlespacing, 
headsepline, 
]{MastersDoctoralThesis} 
\usepackage{amsfonts}

\usepackage[utf8]{inputenc}
\usepackage[russian,english]{babel}

\usepackage{newtxtext} 

\usepackage[upint]{newtxmath}
\usepackage{tikz-cd}
\usepackage{mathtools}
\usepackage{multirow}
\usepackage{diagbox}

\usepackage{slashed}
\usepackage{comment}
\usepackage{changepage}
\usepackage[utf8]{inputenc} 
\usepackage{psl-cover}
\usepackage{caption}
\usepackage{subcaption}
\usepackage{verse}
\usepackage{graphicx}
\usepackage{tikz}
\usepackage{tikz-cd}
\usetikzlibrary{arrows,arrows.meta}
\tikzcdset{arrow style=tikz, diagrams={>= stealth }}

\usepackage{setspace}
\usepackage[all]{xy}
\usepackage[utf8]{inputenc}

\usepackage[autostyle=true]{csquotes} 

\thesistitle{Topological strings and Higgsing trees} 
\supervisor{Dr. Amir-Kian Kashani-Poor} 
\examiner{} 
\degree{} 
\author{David Jaramillo Duque} 
\addresses{} 

\subject{} 
\university{\href{https://www.lpens.ens.psl.eu}{Laboratoire de Physique de l’Ecole Normale Supérieure}} 
\department{} 
\group{{}} 
\faculty{{}} 

\AtBeginDocument{
\hypersetup{pdftitle=\ttitle} 
\hypersetup{pdfauthor=\authorname} 
\hypersetup{pdfkeywords=\keywordnames} 
}

\newcommand{\cW}{{\cal W}}
\newcommand{\cN}{{\cal N}}
\newcommand{\cM}{{\cal M}}

\newcommand{\cH}{{\cal H}}

\newcommand{\cO}{{\cal O}}

\newcommand{\cF}{{\cal F}}
\newcommand{\cG}{{\cal G}}

\newcommand{\cD }{{\cal D}}

\newcommand{\cK }{{\cal K}}
\newcommand{\cZ }{{\cal Z}}

\newcommand{\mg}{\mathfrak{g}}

\newcommand{\LambdaCR}{\Lambda_{cr}}
\newcommand{\LambdaR}{\Lambda_{r}}
\newcommand{\LambdaCW}{\Lambda_{cw}}
\newcommand{\LambdaW}{\Lambda_{w}}

\newcommand{\weyl}{W}
\newcommand{\dynkinsym}{\text{DynkinSym}}
\newcommand{\Eucl}{\IE}

\newcommand{\Sc}{\mathcal{S}}

\renewcommand{\P}{\mathbb{P}}
\renewcommand{\C}{\mathbb{C}}
\newcommand{\R}{\mathbb{R}}

\newcommand{\Z}{\mathbb{Z}}

\newcommand{\Hom}{\text{Hom}}
\newcommand{\acts}{\circlearrowright}

\newcommand{\gs}{g_{top}}

\newcommand{\E}{{\mathbb E_\tau}}
\newcommand{\vol}{\text{vol}\,}
\newcommand{\M}{\text{M}}
\newcommand{\Oc}{\mathcal O}
\newcommand{\Lc}{\mathcal L}

\newcommand{\g}{\mathfrak g}
\renewcommand{\P}{\mathbb P}

\def\Z{\mathbb Z}

\def\abSolLong{t_{k,L}}
\def\abSolShort{t_{k,S}}
\def\shortNorm{\kappa}
\def\interMatrix{\Omega}
\newcommand{\be}{\begin{equation}}
\newcommand{\ee}{\end{equation}}
\def\beq{\begin{eqnarray}}
\def\eeq{\end{eqnarray}}
\def\ba{\begin{eqnarray}}
\def\ea{\end{eqnarray}}
\def\a{\alpha}
\def\b{\beta}
\def\ep1{\epsilon_1}
\def\eps2{\epsilon_2}
\def\t1{\theta_1}

\newcommand{\zero}{{\tiny 0}}
\def\beq{\begin{eqnarray}}
\def\eeq{\end{eqnarray}}
\def\ba{\begin{eqnarray}}
\def\ea{\end{eqnarray}}
\def\a{\alpha}
\def\b{\beta}
\def\ep1{\epsilon_1}
\def\eps2{\epsilon_2}

\newcommand{\IZ}{\mathbb{Z}}
\newcommand{\IC}{\mathbb{C}}
\newcommand{\IP}{\mathbb{P}}
\newcommand{\IN}{\mathbb{N}}
\newcommand{\IR}{\mathbb{R}}
\newcommand{\IQ}{\mathbb{Q}}
\newcommand{\IH}{\mathbb{H}}

\newcommand{\IF}{\mathbb{F}}
\newcommand{\IE}{\mathbb{E}}
\usepackage{marginnote}
\newcommand{\Bn}{C} 

\newcommand{\tr}{\mathrm{Tr\, }}

\newcommand{\rk}{\mathrm{rk\, }}

\newcommand{\IIB}{\mathrm{IIB}}

\newcommand{\nn}{\nonumber}

\newcommand{\ER}{E_{\mathrm{R}}}

\newcommand{\ENS}{E_{\mathrm{NS}}}
\newcommand{\fpq}{x} 
\newcommand{\lpq}{\fpq_0}
\newcommand{\tM}{\widetilde \cM}
\newcommand{\casimirNS}{E_{\mathrm{NS}}^0}
\newcommand{\casimirR}{E_{\mathrm{R}}^0}
\newcommand{\Fff}{(F_4)_4}
\newcommand{\dominant}{\Lambda^+}
\newcommand{\affinedominant}{\hat \Lambda^+}
\newcommand{\dominantflavor}{\affinedominant(F)_{k_F}}
\newcommand{\dominantgauge}{\dominant(G)_{-n}}
\newcommand{\affineg}{\hat{\mathfrak{g}}}
\newcommand{\affineWeyl}{\hat{\cW}}
\newcommand{\finiteWeyl}{\cW}

\newcommand{\cha}{\mathrm{ch}\,}

\tikzcdset{scale cd/.style={every label/.append style={scale=#1},
    cells={nodes={scale=#1}}}}
\usepackage[utf8]{inputenc}

\begin{document}
\renewcommand{\C}{\mathbb{C}}

\frontmatter 

\pagestyle{plain} 

\title{Cordes topologiques et arbres d'Higgsing}

\institute{Laboratoire de physique de l'Ecole Normale Supérieure}
\doctoralschool{École Doctorale Physique en Île-de-France}{564}
\specialty{Physique}

\date{22/09/23}
\jurymember{1}{Kashani-Poor, Amir-Kian}{LPENS}{Directeur de thèse}
\jurymember{2}{Klemm, Albrecht}{Universität Bonn}{Examinateur}
\jurymember{3}{Petrini, Michela}{LPTHE}{Examinatrice}
\jurymember{4}{Troost, Jan}{LPENS}{Examinateur}
\jurymember{5}{Weigand, Timo}{Universität Hamburg}{Raporteur}
\jurymember{6}{Zaffaroni, Alberto}{Universita' di Milano-Bicocca}{Raporteur}

\frabstract{
Les théories des champs superconformes à 6 dimensions sont exotiques et fascinantes. Elles émergent des compactifications de la théorie F sur des fibrations elliptiques de Calabi-Yau, ce qui leur confère une riche gamme de dualités avec diverses autres formulations de la théorie des cordes et de la théorie M. Dans cette thèse, nous considérons des familles étendues de fibrations elliptiques, donnant lieu à des théories 6D reliées par des transitions de Higgs. Ces familles englobent non seulement l'espace des modules d'une variété spécifique, mais comprennent également d'autres variétés avec des topologies différentes.

Notre investigation se concentre sur les théories des champs superconformes 6D de rang 1, selon deux angles distincts. Dans le chapitre \ref{ch:paperJacobi}, nous utilisons la modularité, qui découle des équations d'anomalie holomorphes, pour calculer la fonction de partition de la corde topologique en termes de formes modulaires de Jacobi. Nous proposons également une méthode pour obtenir la fonction de partition topologique d'une théorie Higgsée à partir de sa théorie parente. Grâce à cette approche, nous pouvons expliquer de nombreuses extensions de symetrie que nous avons observées dans notre étude.

D'autre part, dans le chapitre \ref{ch:paperWorldsheet}, nous explorons le soliton à 2 dimensions de la théorie 6D, la corde non-critique. Le genre elliptique de cette corde non-critique coïncide avec une partie de la fonction de partition de la corde topologique. En étudiant attentivement cette corde non-critique, nous proposons un ansatz pour les genres elliptiques exprimés en termes de caractères des algèbres de courant associées. Nous présentons des preuves convaincantes soutenant la validité de cette hypothèse et révélons de nouvelles expressions sous forme fermée pour les genres elliptiques de ces cordes non-critiques.

Grâce à ces investigations, nous espérons éclairer le monde intrigant des théories des champs superconformes à 6 dimensions et découvrir de nouvelles perspectives sur leurs propriétés et connexions remarquables.
}
\enabstract{
 6-dimensional superconformal field theories are exotic and fascinating. They emerge from compactifications of F-theory on Calabi-Yau elliptic fibrations, which grants them a rich array of dualities with various other formulations of string and M-theory. In this thesis, we consider extended families of elliptic fibrations, giving rise 6d theories connected by Higgs transitions. These families not only encompass the moduli space of a specific manifold but also include other manifolds with different topologies.

Our investigation focuses on rank 1 6D superconformal field theories from two distinct angles. In Chapter \ref{ch:paperJacobi}, we employ modularity, which arises from the holomorphic anomaly equations, to compute the topological string partition function in terms of Jacobi modular forms. We also provide a prescription for obtaining the topological partition function of a Higgsed theory from its parent. Through this approach, we can explain numerous symmetry enhancements that we observed in our study.

On the other hand, in Chapter \ref{ch:paperWorldsheet}, we explore the 2D soliton of the 6D theory, the non-critical string. The elliptic genus of this non-critical string coincides with a part of the topological string partition function. By carefully studying this non-critical string, we propose an ansatz for the elliptic genera expressed in terms of characters of the associated current algebras. We present compelling evidence supporting the validity of this ansatz and unveil novel closed form expressions for the elliptic genera of these non-critical strings.

Through these investigations, we hope to shed light on the intriguing world of 6D superconformal field theories and uncover new insights into their remarkable properties and connections. 
}
\frkeywords {Corde topologique, SCFT à 6 dimensions, Théorie F}
\enkeywords{Topological string, 6D SCFTs, F-theory}

\pslassetspath{chemin/relatif/vers/le/dossier/contenant/le/template}

\maketitle{}


\renewenvironment{verse}
  {\itshape\begin{center}\setlength{\parskip}{1em}}
  {\end{center}}

\begin{verse}
\textbf{F-theory's Cosmic Tapestry: Unraveling Dimensions' Dance}

In realms of physics, F-theory sings,\\
A symphony of strings, what wonder it brings!\\
In spaces vast, where dimensions dance,\\
A tapestry of secrets, they entrance.

Elegant and bold, this theory soars,\\
Melding forces, unifying cores.\\
Through dualities, worlds intertwine,\\
In hidden folds, a grand design.

A tapestry of strings, they weave,\\
A cosmic quilt, where dreams believe.\\
Beyond the veil, where wonders gleam,\\
F-theory reigns, a potent scheme.

From Calabi-Yau's curves to fluxes' sway,\\
This theory guides us on our way.\\
With branes and flux, a dance profound,\\
In higher realms, new truths are found.

With duality's embrace, we soar,\\
In holographic realms, we explore.\\
Through duality's lens, we find,\\
A glimpse of truth, the ties that bind.

In holographic dreams, dimensions blend,\\
Emerging holograms, they transcend.\\
In shadows cast, black holes conceal,\\
The essence of reality's appeal.

So let this theory's wisdom show,\\
A quest for truth, a path to grow.\\
In F-theory's grasp, we learn to see,\\
The hidden fabric of reality.

As stars above, in cosmic sway,\\
Unravel truths, by night and day.\\
In this thesis, F-theory gleams,\\
A beacon bright, in science's streams.

So may these words, a tribute be,\\
To F-theory's grand tapestry.\\
With wonder and awe, our minds set free,\\
In quest of knowledge's ecstasy.
\end{verse}

\vspace{1em}

\begin{flushright}
\textit{ChatGPT}
\end{flushright}

\cleardoublepage


\title{Topological strings and Higgsing trees}

\begin{abstract}
\addchaptertocentry{\abstractname} 
6-dimensional superconformal field theories are exotic and fascinating. They emerge from compactifications of F-theory on Calabi-Yau elliptic fibrations, which grants them a rich array of dualities with various other formulations of string and M-theory. In this thesis, we consider extended families of elliptic fibrations, giving rise to 6d theories connected by Higgs transitions. These families not only encompass the moduli space of a specific manifold but also include other manifolds with different topologies.

Our investigation focuses on rank 1 6D superconformal field theories from two distinct angles. In Chapter \ref{ch:paperJacobi}, we employ modularity, which arises from the holomorphic anomaly equations, to compute the topological string partition function in terms of Jacobi modular forms. We also provide a prescription for obtaining the topological partition function of a Higgsed theory from its parent. Through this approach, we can explain numerous symmetry enhancements that we observed in our study.

On the other hand, in Chapter \ref{ch:paperWorldsheet}, we explore the 2D soliton of the 6D theory, the non-critical string. The elliptic genus of this non-critical string coincides with a part of the topological string partition function. By carefully studying this non-critical string, we propose an ansatz for the elliptic genera expressed in terms of characters of the associated current algebras. We present compelling evidence supporting the validity of this ansatz and unveil novel closed form expressions for the elliptic genera of these non-critical strings.

Through these investigations, we hope to shed light on the intriguing world of 6D superconformal field theories and uncover new insights into their remarkable properties and connections. 
\end{abstract}

\cleardoublepage


\title{Cordes topologiques et arbres d'Higgsing}

\begin{frabstractmdt}
 \addchaptertocentry{\frabstractmdtname} 
Les théories des champs superconformes à 6 dimensions sont exotiques et fascinantes. Elles émergent des compactifications de la théorie F sur des fibrations elliptiques de Calabi-Yau, ce qui leur confère une riche gamme de dualités avec diverses autres formulations de la théorie des cordes et de la théorie M. Dans cette thèse, nous considérons des familles étendues de fibrations elliptiques, donnant lieu à des théories 6D reliées par des transitions de Higgs. Ces familles englobent non seulement l'espace des modules d'une variété spécifique, mais comprennent également d'autres variétés avec des topologies différentes.

Notre investigation se concentre sur les théories des champs superconformes 6D de rang 1, selon deux angles distincts. Dans le chapitre \ref{ch:paperJacobi}, nous utilisons la modularité, qui découle des équations d'anomalie holomorphes, pour calculer la fonction de partition de la corde topologique en termes de formes modulaires de Jacobi. Nous proposons également une méthode pour obtenir la fonction de partition topologique d'une théorie Higgsée à partir de sa théorie parente. Grâce à cette approche, nous pouvons expliquer de nombreuses extensions de symetrie que nous avons observées dans notre étude.

D'autre part, dans le chapitre \ref{ch:paperWorldsheet}, nous explorons le soliton à 2 dimensions de la théorie 6D, la corde non-critique. Le genre elliptique de cette corde non-critique coïncide avec une partie de la fonction de partition de la corde topologique. En étudiant attentivement cette corde non-critique, nous proposons un ansatz pour les genres elliptiques exprimés en termes de caractères des algèbres de courant associées. Nous présentons des preuves convaincantes soutenant la validité de cet ansatz et révélons de nouvelles expressions sous forme fermée pour les genres elliptiques de ces cordes non-critiques.

Grâce à ces investigations, nous espérons éclairer le monde intrigant des théories des champs superconformes à 6 dimensions et découvrir de nouvelles perspectives sur leurs propriétés et connexions remarquables.
\end{frabstractmdt}

\cleardoublepage


\begin{acknowledgements}
\addchaptertocentry{\acknowledgementname} 
Researching this fascinating subject and writing about it has been one of the most incredible experiences of my life. Not only because of the beautiful mathematics I have had the opportunity to delve into but also because of the amazing people it has brought me into contact with. In these pages, I would like to extend my gratitude to some of these people. If, inadvertently, I have overlooked anyone's name, please know that it was not intentional, and I apologize for any omissions. I have taken care to properly scramble every list of names using a quantum number generator.\footnote{The whole universe is governed by quantum mechanics, so any random number generator is quantum. However, I am using Mathematica's RandomSample, which is pseudo-random.}

First and foremost, I would like to express my gratitude to my family. Permit me to do so in Spanish:

Quiero expresar mi profundo agradecimiento a mis padres, a quienes dedico esta tesis. Su apoyo incondicional ha sido el cimiento de mi camino académico y personal, y sin ellos, este logro no habría sido posible. No solo me respaldaron financieramente, sino que también estuvieron ahí para mí en cada momento de mi vida, brindándome apoyo y compañía.

Mi madre, Liliana, ha sido mi guía y mi inspiración. Siempre orgullosa de mis logros, sus consejos sutiles pero poderosos han sido mi faro en los momentos de duda. Cada palabra que me ha transmitido ha sido un tesoro de sabiduría que atesoro en mi corazón.

Mi padre, Juan, ha sido mi roca y mi mentor. Su constante apoyo y aliento me han dado la fuerza para superar obstáculos y perseguir mis sueños con determinación. Siempre dispuesto a escuchar y brindar consejos, su sabiduría y experiencia han sido invaluables en mi desarrollo como persona y profesional.

Gracias a ambos por creer en mí y por ser mis pilares inquebrantables. Su amor y confianza me han dado la fortaleza para enfrentar cualquier desafío. Esta tesis es un tributo a su amor y sacrificio, y espero que estén tan orgullosos de mí como yo lo estoy de ser su hijo.

Quiero también agradecer a Yulissa, quien ha sido un pilar fundamental en mi desarrollo como persona y ha sabido apoyarme desde mi infancia. Mauricio, con su generosidad inigualable. Mi querido hermano, Daniel, quien siempre está, y espero estará, presente escuchándome divagar sobre pequeñas cuerdas mientras intenta recordarme que no tienen nada que ver con el mundo real, lo que sea que ``el mundo real" signifique.

Quiero expresar mi sincero agradecimiento a mis abuelos, quienes han dejado una huella imborrable en mi vida y han sido verdaderos ejemplos a seguir.

Mi amada abuela Margarita, quien me enseñó el valor inmenso de servir a los demás. Su amor desinteresado y dedicación hacia los demás han dejado una marca indeleble en mi corazón, motivándome a buscar siempre formas de contribuir positivamente a la vida de los demás.

Mi entrañable abuelo Eduardo (que en paz descanse), con quien compartí preciosos momentos y quien, a través de su ejemplo, inculcó en mí un profundo amor por el conocimiento y el pensamiento crítico. Sus conversaciones y sabios consejos me animaron a explorar el mundo que nos rodea con curiosidad y mente abierta. 

Mi amado abuelo Ariel (que en paz descanse), que con su ejemplo me mostró la valiosa lección del trabajo arduo y la dedicación. Su incansable esfuerzo y pasión fueron un modelo para mí, impulsándome a esforzarme al máximo en todo lo que emprendo. 

Y por último, mi querida abuela Judith (que en paz descanse), aunque no tuve el privilegio de conocerla en persona, su legado perdura en mi corazón. Su enfoque positivo de la vida y su eterna sonrisa han sido una inspiración constante para mí, recordándome la importancia de enfrentar cada día con optimismo y gratitud.

Cada uno de ellos ha dejado una impronta única en mi formación como persona, y a todos ellos les debo mi más profundo agradecimiento. Su amor, sabiduría y bondad han sido un regalo invaluable en mi vida, y siempre llevaré con orgullo su legado a medida que continúo mi camino.

Quiero tomar un momento para agradecer también al resto de mi familia: mis tías, tíos, primas y primos. Su presencia en mi vida ha sido significativa, y aprecio sinceramente el apoyo y cariño que me han brindado a lo largo de los años. Siempre es reconfortante saber que puedo contar con ustedes, y valoro mucho los momentos compartidos juntos. Gracias por formar parte de mi vida y por enriquecerla con su presencia.

I would also like to thank my new family in France. Moving 10,000 kilometers away from home was one of the most challenging experiences I have ever gone through. However, I found the most amazing group of people I could have never dreamt of. This incredible group of people became not only my friends but also a new family. To give a witty comment about each of them would make them feel too important, so I won't do it. However, I'd like to mention some names.

During my first years in France, I met Pietro and \foreignlanguage{russian}{Алла} with whom I had many interesting conversations about the origin and reality of mathematics. Throughout the bulk of my PhD, I shared many conversations and dinners with incredibly talented individuals: Arthur, \includegraphics[]{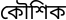}, Gabriel, Manuel, Augustin, Lucija,  \includegraphics[]{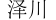},  \includegraphics[]{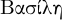},  \includegraphics[]{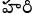}, Gauthier, Nat, Ludwig, and my dear friend Marko, may he rest in peace. The scientific discussions I had with them not only shaped this thesis but also deepened my understanding of physics. I would also like to thank Pauline and  \includegraphics[]{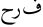}, whose support and encouragement were invaluable. I am proud to call all of them close friends. This last year has brought many changes in the lab and consequently a lot of new friends, with whom I discussed everything from physics to politics. Some of them include Iason, \includegraphics[]{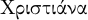}, \foreignlanguage{russian}{Михаил}, Olalla, Pavel, Lior, Bastian, and Paul. These friendships have enriched my academic journey, and I am grateful for the meaningful connections we've formed. 

To my dear Colombian friends, who have influenced not only this thesis but also my entire worldview, I am incredibly grateful. Special thanks to Daniel, Paula, Sebastian, Angie, Hugo, Arturo, Liliana, Jose, Victor, Diego, Juan, and Santiago for their unwavering support. During the Covid crisis, Jaime's generous lending of his house and Lily's warm hospitality, along with Nestor, Ivan, and Paula's kindness, made a significant difference. Your friendship has been invaluable, shaping both my academic journey and personal growth. I am honored to have you in my life.

I want to extend my sincere appreciation to Professor Florent Schaffhauser, whose guidance and mentorship during my undergraduate years were instrumental in shaping my research interests and instilling in me a passion for learning. Additionally, I am deeply thankful for his invaluable assistance in helping me prepare for the scholarship opportunity that ultimately led me to France and paved the way for the completion of this thesis.

I would like to express my heartfelt gratitude to my supervisor, Amir, who has been instrumental in shaping the ideas conveyed in these pages. Any success of this thesis is owed to his sharp insights, while any mistakes are solely a result of my own lack of understanding. Amir is not only a vast repository of knowledge with seemingly endless references, but also one of the few people I have had the pleasure of meeting, who takes the time to contemplate before providing an answer. Working with him has been an immense treasure, and as our time together comes to an end, I can't help but feel a tinge of sorrow. I would also like to extend my thanks to my colleagues, Zhihao and Thorsten, with whom I have closely worked over the past few years. The discussions with them have greatly contributed to shaping this thesis.

Finally, I would like to thank: Professor Jan Troost, Professor Michela Petrini, Professor Timo Weigand, Professor Albrecht Klemm, and Professor Alberto Zaffaroni for accepting the invitation to be part of the jury.

\end{acknowledgements}



\tableofcontents 




\dedicatory{A mis padres.} 


\mainmatter 

\pagestyle{thesis} 



\chapter{Introduction} 
\label{cha:Chapter1}

There are four known fundamental interactions in nature. Three of them—the electromagnetic, weak, and strong interactions—are very well described by the formalism of quantum field theory and are fabulously summarized by the Standard Model. However, as successful as the Standard Model is, every attempt to put the fourth force, gravity, on the same footing as the other three has been unsatisfactory. It is thus believed that some major reformulation is required to include gravity in the framework of quantum mechanics. String theory is a leading contender for a quantum theory of gravity.

The topological string is a sub-sector of string theory that shares many interesting features with the full theory. There is a structural similarity between both theories, as they have a worldsheet description that gives rise to a perturbative expansion in the surface genus for observables, and a target space description that provides information on non-perturbative effects. These similarities suggest that a better understanding of the topological string will lead to a better understanding of the full theory. More concretely, the topological string also computes observables of the full theory, particularly part of the BPS spectrum of string theory compactified on a Calabi-Yau manifold. Additionally, the topological string has proven to have a place not only in physics but also in mathematics, where it has been a source of inspiration in algebraic geometry. It has been particularly successful in enumerative geometry, where conjectures from the topological string have become mathematical theorems, and calculations that were notably difficult with previously available tools have become much simpler in the topological string framework. A significant focus of this thesis has been on computing Gromov-Witten and Gopakumar-Vafa invariants of Calabi-Yau threefolds using the topological string.

The data needed to define the topological string is a Calabi-Yau manifold of 3 complex dimensions (6 real dimensions). In the physical theory, these so-called internal dimensions get extended by 4, 5, or 6 space-time dimensions to obtain the 10 dimensions of type II or heterotic string theory, the 11 dimensions of M-theory, or the 12 dimensions of F-theory. In each case, the topological string yields information about the BPS spectrum of the 4, 5, or 6-dimensional effective theory.

In this thesis, we focused on elliptically fibered Calabi-Yau manifolds where there has recently been much progress in computing the topological string partition function. String theory on an elliptically fibered Calabi-Yau manifold is intriguing as it gives rise to various dualities among F, M, type II, and heterotic theories. In particular, elliptically fibered Calabi-Yau manifolds serve as the starting point for F-theory. 

The goal of the thesis was to study the relations among extended families of elliptically fibered Calabi-Yau manifolds called Higgsing trees. These families include not only modifications along the moduli space of a single Calabi-Yau manifold but also topologically different Calabi-Yau manifolds. From a physical point of view, these families are related by Higgsing the gauge symmetries of the 6-dimensional effective theory. From a mathematical point of view, these manifolds are obtained by reaching a singular point in the moduli space and resolving the geometry. This process actually corresponds to un-Higgsing, which is more natural from the geometrical point of view. In general, this procedure changes the topology of the manifold. Briefly, the link between un-Higgsing and enhancing singularities is as follows: Using M-theory/F-theory duality, the gauge bosons come from M2 branes wrapping the curves that resolve the singularity above the singular locus. After blowing up the singularities, the structure of the fiber is a collection of rational curves whose intersection structure is encoded by a Dynkin diagram. Up to subtleties due to non-trivial monodromies, the gauge symmetry of the effective 6-dimensional theory is the Lie algebra associated with the given diagram. In this picture, taking the most generic fibration—the one that only has the singularities enforced by topology—corresponds to the maximally higgsed theory, while un-Higgsing corresponds to specializing the fiber so that it has an enhanced singularity above the singular locus. This requires more blow-ups and thus gives rise to a larger gauge group in the effective theory and a different topology for the resolution manifold.

During this thesis, we computed the topological partition function for several elliptically fibered Calabi-Yau threefolds with a single compact curve in the base of the fibration. The effective 6D theories obtained from the compactification of F-theory on these spaces have one self-dual tensor multiplet coming from the expansion of the type IIB 4-form along the base curve. These theories are said to have rank 1 and have been classified (see Appendix \ref{sec:HiggsingTrees}). We computed the topological string partition function through two different approaches in \cite{Duan:2020imo} and \cite{Duque:2022tub}.

In Chapter \ref{ch:paperJacobi}, which reproduces \cite{Duan:2020imo}, we use the modular properties derived from the holomorphic anomaly equations to compute the topological string partition function in terms of Jacobi modular forms. The process is relatively straightforward. We consider an ansatz for $Z_{top}$ given as a fraction:
\begin{equation}
    Z_{top}\propto\frac{\cN}{\cD}\,,
    \label{eq:modAnsatz-intro}
\end{equation}
where $\cD$ is a known function, and $\cN$ is a function in a finite-dimensional vector space of Jacobi forms. By using low-order Gromov-Witten invariants, we can determine the exact value of $\cN$. This computation allowed us to (1) conjecture a very simple rule for obtaining the partition function of a daughter theory from its parent. We extensively tested this conjecture. Additionally, the computation also allowed us to (2) discover many symmetries that are not immediately apparent when examining a single theory but become clear when considering the neighboring theories in the Higgsing trees. These new symmetries helped us refine the ansätze for $\cN$. In some cases, these refined ansätze reduced the number of coefficients to fix by several orders of magnitude (see Appendix \ref{app:explicit}). Within this new framework, we were able to explain the extra symmetry pointed out in \cite{DelZotto:2017mee}. The authors of \cite{DelZotto:2017mee} observed that by using a particular map between the $D_4$ and $F_4$ fugacities, the elliptic genus of the $(D_4)_4$ theory exhibits the Weyl symmetry of $F_4$ and an enhanced translational symmetry. As we explain in Chapter \ref{ch:paperJacobi}, this is one of many places where we observe these symmetry enhancements. Furthermore, we described a symmetry that, to the best of our knowledge, had not been previously recognized, wherein the parent theory inherits a symmetry of its daughter. We coined this type of symmetry \emph{reverse inheritance}.

In Chapter \ref{ch:paperWorldsheet}, which reproduces \cite{Duque:2022tub}, we approached the problem from a
different direction. The 6D effective theory has a self-dual tensor field that sources a 2D soliton, the \emph{non-critical} string. It turns out that the topological string partition can be written as:
\begin{equation}
    Z_{top}\propto\sum_k Q_b^kZ_k\,
\end{equation}
where $Q_b$ is the fugacity associated to the unique curve in the base, and $Z_k$ is the elliptic genus of the non-critical string. In \cite{Duque:2022tub}, we extended the work in \cite{DelZotto:2018tcj} and gave an ansatz for the elliptic genus at degree 1 in terms of characters of the affine current algebras in the worldsheet. Schematically, our ansatz reads (see Chapter \ref{ch:paperWorldsheet} for the precise statement):
\begin{equation}
    Z_1\propto \sum_{(\omega,\lambda)}p^\lambda_{\omega,k}\hat \chi_\omega \hat \chi_\lambda
    \label{eq:affAnsatzIntro}
\end{equation}
where the sum over $\lambda$ is over the finitely many dominant weights of the flavor current algebra, $\omega$ are weights with dominant finite part of the current algebra coming from the 6D gauge algebra, and $p$ is a (finite) polynomial that doesn't depend on the fugacities of the aforementioned currents. Combining the modular ansatz \eqref{eq:modAnsatz-intro} with the affine ansatz \eqref{eq:affAnsatzIntro} and information coming from the low-lying modes of the 2D theory, we were able to compute the elliptic genera of several theories.

The organization of the thesis is as follows: Chapter \ref{ch:stringOnCY} describes the geometry of Calabi-Yau spaces and introduces topological string theory and the topological string partition function. Particular attention is given to the description of the moduli spaces. Chapter \ref{ch:6D} provides a high-level introduction to the engineering of 6D superconformal field theories from F-theory, explains the many dualities that these theories enjoy, and introduces the non-critical string. A more detailed discussion about these theories is carried out in Section \ref{s:theories_considered}. Chapter \ref{ch:paperJacobi} reproduces \cite{Duan:2020imo} and deals with computations using the modular ansatz and the 5D/6D effective theory picture. 
The biggest contribution from this paper comes from the discovery of new symmetries of the topological string partition function. Chapter \ref{ch:paperWorldsheet}, which reproduces \cite{Duque:2022tub}, introduces the ansatz \eqref{eq:affAnsatzIntro} and deals with the worldsheet theory of the non-critical string. The thesis ends with several technical appendices. Appendix \ref{ap:ToricGeometry} gives a short introduction to Toric geometry, which was the main tool used to describe the manifolds used in the thesis. Appendices \ref{app:explicit} to \ref{app:Higgsing_trees} come from \cite{Duan:2020imo}, and they include several tables of Gromov-Witten invariant that provide evidence that the Higgsing formulas we give are correct, our conventions for the rings of Jacobi forms, and the different relations between different rings of Jacobi rings. Appendices \ref{a:elliptic_genera} to \ref{sec:HiggsingTrees} come from \cite{Duque:2022tub}. Particularly interesting are Appendix \ref{a:elliptic_genera}, where we describe very briefly what the elliptic genus is, and Appendix \ref{a:Kazhdan-Lusztig}, where we explain how to compute affine characters at negative level using Kazhdan-Lusztig polynomials.

\chapter{String theory on Calabi-Yau manifolds}
\label{ch:stringOnCY}

Calabi-Yau spaces are the all time favorite spaces to compactify type IIA and type IIB string theory on, because they preserve an interesting amount of supersymmetry in the effective low dimensional theory. These spaces and the behaviour of strings on them can be analyzed in great detail see, for instance, \cite{Greene:1997} for a review and an extensive list of references. In this chapter, we describe very briefly the main tools used in this thesis. In particular, we give an explicit description of the moduli space of Calabi-Yau spaces. 

\section{The Geometry of Calabi-Yau Manifolds}
Calabi-Yau spaces are particularly well suited to study compactifications of type IIA and type IIB string because they preserve a large amount symmetry making it possible to analyse the effective theories in detail while leaving ample room for interesting phenomena. See \cite{Candelas:1985en} for the detail argument on how the geometry of Calabi-Yau spaces leads to low dimensional supersymmetry. In this section, we will describe the classical moduli spaces of Calabi-Yau manifolds.

Let us begin by considering a complex manifold $M$ with a Hermitian metric. Having a Hermitian metric implies that, given the complex structure $I:TM\to TM$ of $M$, the metric $g$ satisfies:
\begin{equation*}
g(I x, I y)=g(x,y)
\end{equation*}
for any vectors $x$ and $y$ in $T_p M$ at any point $p\in M$. Equivalently, in coordinate notation, this means that the only non-vanishing entries in $g$ are the mixed entries $g_{\bar{i} j}$ (and $g_{i\bar{j}}$). We also impose the condition that the complex structure $I$ is covariantly constant throughout the manifold. This condition is equivalent\footnote{See, for instance, \cite{Nakahara:2003nw}.} to requiring the Kähler form:
\begin{equation}
J(x,y):=g(Ix,y), \Leftrightarrow , J:=ig_{i\bar{j}}dz^idz^{\bar{j}}\label{eq:kahlerForm}
\end{equation}
to be closed: $dJ=0$. The final condition we want to impose is the existence of a covariantly constant spinor on the manifold. This is a necesary requisit to obtain low dimensional supersymmetry \cite{Greene:1997}. This condition is equivalent to having a vanishing Ricci curvature \cite{nakahara_geometry_2003,griffiths_principles_2014}.

Finding a metric on a manifold that yields an everywhere vanishing curvature is an almost impossible task. The only manifolds with an explicit metric where we know the curvature vanishes are flat spacetime and the products of $S^1$. Fortunately, Yau's theorem provides a solution \cite{Yau:1977ms}. The theorem states that given a Kähler manifold $M$ with a vanishing first Chern class (vanishing curvature implies vanishing first Chern class), there exists a unique metric on $M$ whose associated Kähler form lies in the same cohomology class as the original Kähler form and has vanishing curvature.

This completes the definition of the type of manifolds we are interested in: Kähler manifolds with vanishing curvature, also known as \emph{Calabi-Yau manifolds}. Following standard practice, we will also impose the constraint that the holonomy must be maximal. This ensures that the space $H^{3,0}(M)$ has dimension 1. In practice, obtaining a Calabi-Yau manifold is relatively straightforward: We begin with a complex manifold with a vanishing first Chern class and choose an element $[J] \in H^{1,1}(M)$\footnote{The actual elements we can choose lie in a smaller subset called the Kähler cone. See the discussion in section \ref{sec:ModSpacesCY}.}. Then, according to Yau's theorem, we know that there exists a metric that yields vanishing curvature and whose Kähler class lies in $[J]$. The manifold $M$ equipped with this metric is a Calabi-Yau space. In the generic case, we do not have an explicit metric, but we will see that this is not important as we can work with the Kähler class or, more precisely, its cohomology class.

\section{Classical Moduli Spaces}
\label{sec:ModSpacesCY}

In this section, we will describe the classical moduli space of Calabi-Yau manifolds by considering the possible deformations of the metric that satisfy the Calabi-Yau constraints.

One of these deformations arises from changing the cohomology class to which the Kähler form belongs, resulting in a different Calabi-Yau space. Equation \ref{eq:kahlerForm} imposes a constraint on the admissible cohomology classes: the integral of $J$ or products of $J$ must be positive:
\begin{equation}
\int J\geq 0\,,\int J^2\geq 0\,, \int J^3 \geq 0\,,
\end{equation}
where the integrals are taken over cycles of the appropriate dimension. The space of cohomology classes in $H^{1,1}(M)$ that satisfy this condition is known as the \emph{Kähler cone} of the Calabi-Yau manifold.

To explore the other possible deformation of the metric, let us consider the general metric deformation:
\begin{equation}
\delta g = \delta g_{ij}dx^i \otimes dx^j+\delta g_{i\bar j}dx^i\otimes dx^{\bar j}+ \text{c.c.},.
\end{equation}
The mixed part simply modifies the Kähler form and corresponds to infinitesimal movement within the Kähler cone. However, the first part introduces a new direction in moduli space. At first glance, it may appear that this spoils the Hermiticity of $g$. However, this issue can be resolved by a non-holomorphic change of variables, which eliminates the non-Hermitian contribution. In summary, besides the change $\delta g$, we need to make a non-holomorphic change in the variables $x'=x'(x,\bar{x})$ so that the metric $g'$ in the new $x'$ variables is Hermitian. This modification affects the complex structure of the manifold and gives rise to the \emph{Complex Structure} moduli space.

It can be shown \cite{Candelas:1987se,Greene:1997} that $\Omega_{ijk}g^{k\bar k}\delta g_{\bar k\bar l}dx^idx^jdx^{\bar l}$ is a harmonic form, where $\Omega$ is a generator of $H^{3,0}(M)$. Consequently, the space of complex structure deformations corresponds (locally) to the cohomology group $H^{2,1}(M)$.

This completes our description of the moduli space of classical Calabi-Yau manifolds. The term "classical" is used because, as we will discuss in detail later, we are missing an ingredient in the Kähler moduli space: the B-field. As mentioned earlier, the Kähler cone is a real space due to the positivity of the metric. However, when studying the behavior of strings in these spaces (as we will see in section \ref{subsec:Amod}), the B-field combines with the Kähler form to form a complex $(1,1)$ form. This new space is sometimes referred to as the \emph{Stringy Kähler Moduli Space}. In this work, we will use the term ``Kähler Moduli Space" and specify when necessary to avoid confusion.

\section{The world-sheet description, the A \& B models}
We now turn our attention to topological string theory on the Calabi-Yau manifolds we have constructed. This theory, first introduced in the 1980s by Witten \cite{witten:1988xj}, has seen remarkable progress in various directions since then. For our treatment, we will follow the approach presented in \cite{Witten:1991zz}, for more recent reviews see \cite{Hori:2003ic,Neitzke:2004ni,Marino:lectNotes,Aspinwall:2004jr,Vonk:2005yv}.

The Lagrangian of the $N=2$ sigma model with a Calabi-Yau manifold as the target space is given by:
\begin{equation}
L=2t\int d^2z\, \left(g_{IJ}\partial x^I \bar \partial x^J+i g_{\bar i i}\psi^{\bar i}_-D\psi-^i+ig_{\bar i i}\psi^{\bar i}+\bar D\psi_+^i+R_{i\bar i j\bar j}\psi_+^i\psi_+^{\bar i}\psi_-^j\psi_-^{\bar j}\right)\,,
\label{eq:untwistedLagrangian}
\end{equation}
where $x:\Sigma\to M$ represents the worldsheet embedding, $\psi_+$ and $\psi_-$ are sections of $K^{1/2}\otimes x^*(TM)$ and $\bar K^{1/2}\otimes x^*(TM)$, respectively, $R$ denotes the Riemann curvature, and $t$ is the string coupling constant. The $A$-model and $B$-model of topological string theory are obtained by performing two types of twists.

We can classify these twists as positive and negative. The positive twist is achieved by considering $\psi^i_+$ and $\psi_+^{\bar i }$ as sections of $x^*(T^{1,0}M)$ and $K\otimes x^*(T^{0,1})$, respectively. On the other hand, the negative twist is obtained by considering them as sections of $K\otimes x^*(T^{1,0}M)$ and $x^*(T^{0,1})$, respectively. The same twists can be applied to $\psi_-$ by exchanging $K$ with $\bar K$. If we choose different twists for $\psi_+$ and $\psi_-$, we arrive at the \emph{$A$-model}. However, if we employ the same twist on both sides, we obtain the \emph{$B$-model}.

\subsection{The $A$-model}
\label{subsec:Amod}
For the $A$-model, let us consider a $+$ twist for $\psi_+$ and a $-$ twist for $\psi_-$. We can then rename the fields according to their transformation properties:
\begin{align*}
    \psi_+^i\in \Gamma\left(K^{1/2}\otimes x^*(T^{1,0}M)\right)&\to \chi^i\in \Gamma\left( x^*(T^{1,0}M)\right)\,,\\
    \psi_+^{\bar i}\in \Gamma\left(K^{1/2}\otimes x^*(T^{0,1}M)\right)&\to \psi^{\bar i}_z\in \Gamma\left(K\otimes x^*(T^{0,1}M)\right)\,,\\
    \psi_-^i\in \Gamma\left(\bar K^{1/2}\otimes x^*(T^{1,0}M)\right)&\to \psi^{i}_{\bar z}\in \Gamma\left( \bar K \otimes x^*(T^{1,0}M)\right)\,,\\
    \psi_-^{\bar i}\in \Gamma\left(\bar K^{1/2}\otimes x^*(T^{0,1}M)\right)&\to \chi^{\bar i}\in \Gamma\left(x^*(T^{0,1}M)\right)\,.
\end{align*}
One of the supersymmetries of \eqref{eq:untwistedLagrangian} becomes very simple:
\begin{equation}
    \delta x^I = i \epsilon \chi^I\,,\quad \delta \chi^I =0 \,,
    \label{eq:susyA}
\end{equation}
where $\epsilon$ is the infinitesimal fermionic parameter. There are additional transformations for $\psi$ as well, but they are not relevant for our purposes.
In terms of the BRST operator $Q$, defined by $\delta_\epsilon \phi=-i\epsilon\{Q,\phi\}$ for any field $\phi$, the Lagrangian can be written as:
\begin{equation}
    L_A=it\int  d^2z\, \left\{Q,V\right\}+t\int x^*(J)\,,
    \label{eq:A-modelLagrangian}
\end{equation}
where $J$ is the Kähler form and 
\begin{equation}
    V=g_{i\bar j}(\psi_z^{\bar i}\bar \partial x^j +\psi_{\bar z}^i\partial x^{\bar i})\,.
\end{equation}
The second term involving the Kähler form is topological in the sense that it only depends on the cohomology class of $J$ and the homotopy type of $x$:
\begin{equation}
    \int_\Sigma x^*(J)=\int_{x(\Sigma)} J\,.
    \label{eq:A-modelJpart}
\end{equation}

The first term in the Lagrangian, which involves the BRST operator $Q$, is BRST-exact and therefore irrelevant. However, instead of simply removing it, we can multiply it by an infinite constant so that the behavior is completely dominated by the classical solutions of this term. The classical equations of motion derived from this part of the Lagrangian are given by:
\begin{equation}
    \partial x^{\bar i}=0\,, \bar \partial x^i=0\,.
    \label{eq:A-modelEOM}
\end{equation}
The solutions are then holomorphic maps $x:\Sigma \to M$.

Due to the presence of the BRST operator $Q$, the correlation functions remain unchanged if we replace any operator $\mathcal{O}$ by $\mathcal{O}+\{Q,\mathcal{O}'\}$. The operators of interest in the theory therefore correspond to BRST cohomology classes. By examining the transformations in \eqref{eq:susyA}, we see that if we identify $\chi^I$ with $dx^I$, the BRST operator $Q$ simply becomes the de Rham operator $d$ and the observables of the theory correspond to the cohomology classes of the manifold $M$:
\begin{align}
    \omega = \phi_{I_1\dots I_k}(x)dx^{I_1}\cdots dx^{I_k}&\leftrightarrow \mathcal O_\omega = \phi_{I_1\dots I_k}(x)\chi^{I_1}\cdots \chi^{I_k}\,,\\
    \{Q,\mathcal O_\omega\}=- O_{d\omega}\,. 
    \label{eq:A-modelOperators}
\end{align}
This is only a plausibility argument. It is not immediately clear that there are no operators that depend on $\psi$. However, a careful analysis of the symmetry structure of the Lagrangian reveals that this is indeed true \cite{witten:1988xj}.

If we assign ghost numbers $-1$, $0$, and $1$ to $\psi$, $x$, and $\chi$, respectively, the Lagrangian \eqref{eq:A-modelLagrangian} exhibits classical ghost number conservation. In the quantum theory, we must take into account the zero modes of the fields $\psi$ and $\chi$. As it often happens for anomalies, the zero modes and the ghost number anomaly are captured by a topological invariant. The difference between the $\chi$ and $\psi$ zero modes is given by:
\begin{equation}
    0_\chi-0_\psi=6(1-g), 
\end{equation}
where $g$ is the genus of $x(\Sigma)$. This anomaly implies that the only non-vanishing correlation functions come from operators $\mathcal{O}$ whose ghost numbers add up to $6(1-g)$. Consequently, since all the operators have positive ghost numbers, the theory is trivial for any $g>1$.

We can now compute the correlation functions of an arbitrary set of operators. A convenient basis of operators is obtained by considering the Poincaré duals of submanifolds $H\subset X$. The Poincaré dual of a submanifold $H$ can be thought of as a differential form with delta function support along the manifold $H$. In general, the intersection of $H$ with any other manifold $H'$ is given by the integral of the Poincaré dual form of $H$ along the manifold $H'$. For a submanifold $H$ of codimension $h$, its Poincaré form has dimension $h$. Thus, we can compute correlators obtained from operators whose codimensions add up to $6(1-g)$:
\begin{equation}
    \left<\prod_{i}O_{H_i}(p_i)\right>\propto \int Dx\, D\psi \, D \chi\, e^{-it\{Q,V\}} e^{-t\int_{x(\Sigma)}J}\prod_i O_{H_i}(p_i)\,. 
\end{equation}

This correlator can be greatly simplified by dividing the $Dx$ integral according to the different values that \eqref{eq:A-modelJpart} can take. We also notice that the integral vanishes unless the conditions
\begin{equation}
    x(p_i)\in H_i\,
    \label{eq:A-modelCorrelatorConstraint}
\end{equation}
are satisfied. Finally, in the generic case where there are no $\psi$ zero modes, we observe that (1) the dimension of the moduli space of holomorphic maps $x:\Sigma\to X$ coincides with the ghost number, since the holomorphic condition \eqref{eq:A-modelEOM} is the linearized version of the $\chi$ zero mode equation, and (2) the number of conditions imposed by \eqref{eq:A-modelCorrelatorConstraint} is also given by the ghost number. Therefore, the integral reduces to an integration over a set of points. As a result, the correlator reads:
\begin{equation}
    \left<\prod_{i}O_{H_i}(p_i)\right>=\sum_{\beta \in H_2(X)}e^{-t \int_\beta J}\times \# \tilde \cM_\beta(H)
\end{equation}
where $\tilde \cM_\beta(H)$ is the set of points in the moduli space $\cM_\beta$ of holomorphic maps $x:\Sigma \to X$ with $x(p_i) \in H_i$.\footnote{In the case where we actually find $\psi$ 0-modes, $\#\tilde \cM$ must be replaced by an integral of the Euler class of the line bundle over $\cM$ associated to the different jumps of $\psi$ 0-modes. For more details see \cite{witten:1988xj}.}

The most interesting case, and arguably the only interesting one, is when the operators are associated with 2-forms and therefore with hypersurfaces $H$. In this case, the only non-vanishing correlators are 3-point functions. The contribution from the $\beta=0$ class in $H_2(M)$ can be computed straightforwardly by realizing that if $[x(\Sigma)]=0\in H_2(M)$, then $\tilde{\mathcal{M}}_0(H)$ consists of points $x\in M$ such that $x\in H_i$ for every $i$. The correlator then reads:
\begin{equation}
    \left<\cO_{H_a}\cO_{H_b}\cO_{H_c}\right>=H_a\cap H_b \cap H_c +\sum_{\beta} N_{abc}e^{-t\int_\beta J}\,.
    \label{eq:A-model3pointFunction}
\end{equation}
Here, $N_{abc}$ are numbers related to the spaces $\tilde {\mathcal{M}}_\beta(H)$. Computing and properly defining these numbers can be a challenge, but in Section \ref{sec:mirrorSymmetry}, we will see that we can obtain them in a completely algorithmic fashion using the calculations in the $B$-model.

Finally, we note that the $A$-model correlation functions depend only on the Kähler class $J$ and the string coupling constant $t$. Their dependence on these variables occurs solely through the combination $tJ$. This combination then parametrizes the moduli space of the theory. It is important to note that $t$ can be a complex quantity, which makes the moduli space of the $A$-model a complexification of the Kähler moduli space. Furthermore, there exists a translational symmetry in the imaginary part of $tJ$: when we replace $tJ$ with $tJ+2\pi i \omega$, where $\omega$ is a form with integral periods over any $\beta \in H_2(X)$, the correlation functions remain unchanged. Consequently, the moduli space is the complexification of the Kähler moduli space modulo this symmetry.

\subsection{The $B$-model}
For concreteness, let us consider a $-$ twist for both $\psi_\pm$. The field redefinitions can be written as follows:
\begin{align}
     \psi_+^i\in \Gamma\left(K^{1/2}\otimes x^*(T^{1,0}M)\right)&\to  \Gamma\left( K\otimes x^*(T^{1,0}M)\right)\,,\\
    \psi_-^i\in \Gamma\left(\bar K^{1/2}\otimes x^*(T^{1,0}M)\right)&\to  \Gamma\left( \bar K \otimes x^*(T^{1,0}M)\right)\,,\\
    \psi_+^{\bar i}\in \Gamma\left(K^{1/2}\otimes x^*(T^{0,1}M)\right)&\to  \Gamma\left(x^*(T^{0,1}M)\right)\,,\\
    \psi_-^{\bar i}\in \Gamma\left(\bar K^{1/2}\otimes x^*(T^{0,1}M)\right)&\to \Gamma\left(x^*(T^{0,1}M)\right)\,.
\end{align}
We relabel the fields as follows:
\begin{align}
    \rho \in \Gamma\left(T^*\Sigma \otimes x^*(T^{1,0}M)\right)&\quad  \text{ with }\quad \rho^i=\psi^i_+dz+ \psi ^i_-d\bar z\,, \\
   \eta \in  \Gamma\left(x^*(T^{0,1}M)\right) & \quad \text{ with }\quad \eta^{\bar i} = \psi_+^{\bar i} +\psi_-^{\bar i} \,, \\
\theta \in  \Gamma\left(x^*(T^{*1,0}M)\right)& \quad\text{ with }\quad \theta_i = g_{i\bar i}(\psi_+^{\bar i} -\psi_-^{\bar i} ) \,.
\end{align}
The relevant supersymmetry transformation becomes 
\begin{equation}
    \delta x^i = 0\,\quad \delta x^{\bar i}=i\epsilon \eta^{\bar i}, \quad \delta \eta^{\bar i}=\delta \theta_i=0\,,\quad \delta \rho^i=-\alpha d\phi^i. 
\end{equation}
In terms of the associated BRST operator $Q$, the Lagrangian takes the form: 
\begin{equation}
    L=it\int \{Q,V\} +t W\,,
    \label{eq:B-modelLagragian}
\end{equation}
where
\begin{equation}
    V=g_{i\bar j}\left(\rho_z^i\bar \partial x^{\bar j}+ \rho_{\bar z}\partial x^{\bar j}\right)\,, \quad W=\int \left(-\theta_iD\rho^i-\frac{i}{2}R_{i\bar ij\bar j}\rho^i \rho ^j\eta^{\bar i}\theta_kg^{k\bar j}\right)\,. 
\end{equation}
Similarly to the $A$-model, the first term in the Lagrangian is BRST-exact and we can focus on its classical solutions. The classical solutions of the first part are constant maps $x:\Sigma \to M$, and the moduli space of solutions is isomorphic to $M$, a much simpler space than the spaces $\mathcal \cM_\beta$ we were dealing with for the A-model. 

By identifying $\eta^{\bar i}$ with $d x^{\bar i}$ and $\theta_i$ with $\partial_i$, we can identify $Q$ with the Dolbeault operator $\bar \partial$ on $M$. An operator $\mathcal{O}$ can then be associated with a $\Omega^p(\bigwedge^qT^{1,0}M)$ form as follows:
\begin{equation}
    \omega =dx^{\bar i_1}\cdots dx^{\bar i_p} \phi^{j_1\cdots j_q }_{\bar i_1\cdots \bar i_p}(x)\partial_{j_1}\cdots \partial_{j_q} \leftrightarrow \mathcal O_\omega= \eta^{\bar i_1}\cdots \eta^{\bar i_p} \phi^{j_1\cdots j_q }_{\bar i_1\cdots \bar i_p}(x)\theta_{j_1}\cdots \theta_{j_q}\,.\\
\end{equation}
The BRST cohomology is then identified with the Dolbeault cohomology:
\begin{equation}
    \{Q,\Oc_\omega\}=-\Oc_{\bar \partial \omega }
\end{equation}

The ghost charges in the $B$-model are $1$ for $\eta$ and $\theta$, $-1$ for $\rho$, and $0$ for $x$. From anomaly considerations, the correlation functions $\left<\prod \mathcal{O}_i\right>$ vanish unless the sum of ghost numbers adds up to $6(1-g)$. Furthermore, by using a refined $\mathbb{Z}\oplus \mathbb{Z}$ ghost cohomology, we can show that $\theta$ and $\eta$ must have exactly the same number of fields. Therefore, for a non-vanishing correlator $\left<\prod \mathcal{O}_i\right>$, the differential form associated with the product $\prod \mathcal{O}i$ must belong to $H^3_{\bar \partial}(M, \bigwedge^3 T^{1,0}M)$.

To analyze the role of the string coupling constant $t$ in the action \eqref{eq:B-modelLagragian}, we observe that the first term does not introduce any $t$ dependence since it is BRST-exact. For the second term, we can remove the $t$ dependence by the replacement $\theta\to \theta/t$ due to its linear dependence on $\theta$. Therefore, if we take a correlator $\left<\prod \mathcal{O}_i\right>$ and perform the $\theta\to \theta/t$ replacement, the only $t$ dependence arises from the $\theta$ dependence in the operators. Since each correlator requires three $\theta$'s to be non-vanishing, the overall $t$ dependence is a factor of $t^{-3}$.

The relevant fields in the $B$-model have one $\eta$ and one $\theta$, so the non-vanishing correlators involve three observables. We have already argued that the $Dx$ integral reduces to an integral $dx$ over $M$, and the fermionic integrals have been dealt with using anomaly considerations. The product of observables must correspond to a form in $H^3_{\bar \partial}(M, \bigwedge^3 T^{1,0}M)$. In order for the form to be integrated over $M$, we can uniquely and simply obtain a full-dimensional form as follows:
\begin{equation}
    \omega =dx^{\bar i_1}dx^{\bar i_2}dx^{\bar i_3} \omega^{j_1j_2j_3 }_{\bar i_1\bar i_2 \bar i_3}\partial_{j_1}\partial_{j_2} \partial_{j_3}\quad\to \quad \Omega_{j_1j_2j_3}\omega^{j_1j_2j_3 }_{\bar i_1\bar i_2 \bar i_3} \Omega_{k_1k_2k_3}dx^{\bar i_1}dx^{\bar i_2}dx^{\bar i_3}dx^{k_1}dx^{k_2}dx^{k_3}\,,
\end{equation}
where $\Omega$ is the unique $(3,0)$ form up to rescaling. Consequently, for three operators associated with the forms $\omega_\alpha=\omega_\alpha^i \partial_i\in H_{\bar \partial}^1(T^{1,0}M)$, labeled by $\alpha=a,b,c$, the correlator is given by:\footnote{We have omitted a constant factor coming from the string coupling $t$ because $\Omega$ is actually only defined up to a multiplicative constant. We would need to fix some normalization to ensure the correctness of the correlator below. However, this is a minor technical issue that will not cause any problems in what follows.}
\begin{equation}
    \left<\cO_{\omega_a}\cO_{\omega_b}\cO_{\omega_c}\right>= \int_M \Omega_{ijk}\omega_a^i\omega_b^j\omega_c^k\Omega\,. 
    \label{eq:B-correlator}
\end{equation}

This result shows that the $B$-model does not depend on the Kähler class of the Calabi-Yau manifold. The only relevant data is the holomorphic 3-form $\Omega$, which depends solely on the complex structure. The complex structure moduli space is already complex, and since there is no $t$ dependence anywhere, the moduli space of the quantum theory coincides with the classical moduli space of the Calabi-Yau manifold. Moreover, by comparing this result with the $A$-model result \eqref{eq:A-model3pointFunction}, we can see that calculations in the $B$-model are much simpler. The $B$-model reduces to a classical geometric problem where we only need to integrate differential forms on the Calabi-Yau manifold, while the $A$-model involves calculus over the moduli space of holomorphic curves. The simplicity of the $B$-model ultimately allows us to obtain explicit results.

\section{Mirror Symmetry - The Basic Picture}
\label{sec:mirrorSymmetry}

Mirror symmetry, in its simplest form, states that for any Calabi-Yau manifold $M$, there exists another Calabi-Yau manifold $\tilde{M}$ called the \emph{mirror} of $M$, such that the A-model on $M$ is equivalent to the B-model on $\tilde{M}$:
\begin{equation}
    \boxed{A(M)= B(\tilde M)}
\end{equation}
There are various ways to argue for the existence of a mirror manifold, but a complete description of mirror symmetry is beyond the scope of this thesis. Here, we will provide a physical argument for mirror symmetry and explain how one can obtain mirror pairs and describe their correlation functions in an algorithmic manner (see section \ref{sec:MirrorSymToricHyper}). For a comprehensive treatment of mirror symmetry, we refer to the textbook by Cox and Katz \cite{CoxKatz} and the early literature on the subject, such as \cite{Greene:1990ud, Candelas:1990rm, Batyrev:1994hm, Hosono:1993qy}.

Consider the untwisted Lagrangian \eqref{eq:untwistedLagrangian} and focus on the zero modes of the fermions with Ramond boundary conditions (we choose this for concreteness). The anticommutation relations are given by:
\begin{equation}
    \{\psi^i_\alpha,\psi^j_\alpha\}=\{\psi^{ \bar i}_\alpha,\psi^{\bar j}_\alpha\}=0,\, \{\psi^i_\alpha,\psi^{\bar j}_\alpha\}=g^{i\bar j}\,,\quad \alpha=+,-\,.
\end{equation}
To understand the excitations of the theory, we choose a Fock vacuum $\left|0\right>$ that is annihilated by either $\psi_+^i$ or $\psi_+^{\bar i}$, and $\psi_-^i$ or $\psi_-^{\bar i}$. If we had only one fermion, the choice of vacuum would be immaterial. However, with two fermions, the choice for one fermion affects the choice for the other. For concreteness, let us say that the vacuum is annihilated by $\psi_+^i$. Then, we have two choices, denoted as $A$ and $B$:
\begin{align}
    A:&\quad \psi_+^i\left|0\right>=\psi_-^{\bar i}\left|0\right>=0\,,\\
    B:&\quad \psi_+^i\left|0\right>=\psi_-^{i}\left|0\right>=0\,.
\end{align}
These two choices lead to different massless excitations:
\begin{align}
    A:&\quad \phi_{i_1\cdots i_r j_1 \cdots j_r} \psi_-^{i_1}\cdots \psi_-^{i_k}\psi_+^{\bar j_1}\cdots \psi_+^{\bar j_l}\left|0\right>\,,\\
    B:&\quad \phi^{i_1\cdots i_r}_{j_1 \cdots j_r} \psi_{-,i_1}\cdots \psi_{-,i_k}\psi_+^{\bar j_1}\cdots \psi_+^{\bar j_l}\left|0\right>\,.
\end{align}
where we have written $\psi_{-,i}=g_{i\bar i}\psi_-^{\bar i}$. The $A$ and $B$ excitations have the same charge on the right and opposite charge on the left. Assigning charges $+1$ to $\psi^{\bar i}\pm$ and $-1$ to $\psi^i\pm$, the states have charges $(-k,l)$ and $(k,l)$, respectively. Furthermore, by analyzing how the supersymmetry operators act on these states, similar to what we did for the topological models, we can identify the set of supersymmetry-preserving operators of type $A$ with harmonic forms of degree $H^{k,l}(M)$ and the set of operators of type $B$ with Dolbeault cohomology $H^l_{\bar \partial}(M,\bigwedge^k T^{1,0}M)$. Thus, we conclude that states of charge $(k,l)$ and $(-k,l)$ have completely different geometric origins.

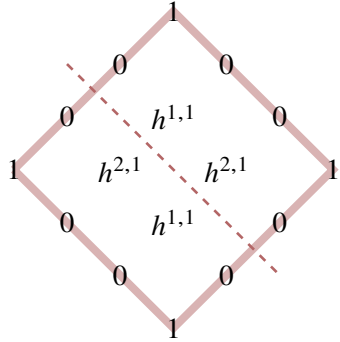
\begin{figure}
    \centering
    \begin{tikzpicture}[scale=0.7]
  
    \draw[line width=3pt,mdtRed!30] (-3,0)--(0,3)--(3,0)--(0,-3)--(-3,0);
    \draw[line width=1pt,mdtRed!60,dashed] (-2,2)--(2,-2);
    
    \node at (0,3){$1$};

\node at (-1,2){$0$};
\node at (1,2){$0$};

\node at (0,1){$h^{1,1}$};
\node at (2,1){$0$};
\node at (-2,1){$0$};

\node at (-3,0){$1$};
\node at (-1,0){$h^{2,1}$};
\node at (1,0){$h^{2,1}$};
\node at (3,0){$1$};

\node at (-2,-1){$0$};
\node at (0,-1){$h^{1,1}$};
\node at (2,-1){$0$};

\node at (-1,-2){$0$};
\node at (1,-2){$0$};

\node at (0,-3){$1$};

    \end{tikzpicture}
    \caption{\emph{Hodge diamond} for a threefold. reflection through the diagonal gives the Hodge diamond of the mirror threefold. $h^{p,q}=\dim H^{p,q}$}
    \label{fig:HodgeDiamond}
\end{figure}

This conclusion is rather surprising because, in the absence of any geometrical interpretation, for instance by considering the CFT as a set of operators and correlators, the sign of the R-charges would be irrelevant. We could simply change the sign of the left R-charge without altering the other. This led to the conjecture that there exist another manifold $\tilde{M}$ for which the roles of $H^{k,l}(M)$ and $H^l_{\bar \partial}(M,\bigwedge^k T^{1,0}M)$ are reversed. In particular, we have argued that the Kähler moduli are associated with $H^{1,1}(M)$ and the complex structure moduli space is associated with $H^{1}_{\bar \partial}(M, T^{*1,0}M)$ (note that $H^{2,1}(M)$ and $H^{1}_{\bar \partial}(M, T^{*1,0}M)$ are isomorphic through multiplication with $\Omega \in H^{3,0}(M)$). Hence, the complex moduli space and the Kähler moduli space are reversed. The name ``mirror symmetry" comes from the fact that the Hodge diamonds of each of the manifolds look like mirror images of each other (see Figure \ref{fig:HodgeDiamond}).

\section{The Quantum Moduli Space}
We now have a description of the $A$ and $B$ topological strings, but we have not performed any concrete calculations yet. In this and the next section, we will outline the general strategy that allows us to obtain the invariants in equation \eqref{eq:A-model3pointFunction} using mirror symmetry. We will provide a high-level picture and provide references for more detailed information. For a more comprehensive treatment, see \cite{CoxKatz, Hosono:1993qy}. In \cite{Candelas:1990rm}, the program we are about to describe is realized in great detail for the quintic hypersurface in $\mathbb{P}^4$.

Both the Kähler and complex structure moduli spaces are special Kähler manifolds, as explained in \cite{Candelas:1990pi}. A special Kähler manifold is a Kähler manifold with its Kähler form given by
\begin{equation}
    J\propto \partial \bar \partial \cK\,,
\end{equation}
where the Kähler potential $\cK$ is entirely determined by the holomorphic prepotential $\cG$ through
\begin{equation}
    e^{-\cK}=-i\left(\zeta^a \frac{\partial}{\partial \zeta^{\bar a}}\bar \cG-\zeta^{\bar a} \frac{\partial}{\partial \zeta^{a}} \cG\right)\,.
\end{equation}
Here $\zeta^a$ is any set of holomorphic coordinates. The prepotential is of crucial importance as it serves as the primary tool for describing these moduli spaces.

For the complex structure moduli space, we obtain the prepotential as follows: Consider the group $H_3$. From Figure \ref{fig:HodgeDiamond}, it has dimension $2h^{2,1}+2$. By taking the intersection of two cycles, $H_3$ acquires a symplectic structure. Fix a symplectic basis $A^a,\, B_b$ with $a,b=0,\dots,h^{2,1}$ and
\begin{equation}
    A^a \cdot B_b=\delta_b^a, \quad A^a \cdot A^b =0, \quad B_a \cdot B_b=0
\end{equation}
and consider the dual forms $\alpha_a,\,\beta^b$. As they form a basis of 3-forms, the holomorphic 3-form $\Omega$ can be expanded as 
\begin{equation}
    \Omega= \zeta^a \alpha_a -\cG_a \beta^a\,,
    \label{eq:OmegaExpansion}
\end{equation}
where 
\begin{equation}
    \zeta^a = \int _{A^a}\Omega\,,\quad \cG_a=\int_{B_a}\Omega\,, 
\end{equation}
are the periods of $\Omega$.

$\zeta$ and $\mathcal{G}$ are not independent because there are $2h^{2,1}+2$ of them, and the space of complex structures is parameterized by $\Omega$ and has dimension $h^{2,1}$. Therefore, we can solve for $\mathcal{G}$ in terms of $\zeta$. There is still one relation among the $\zeta$'s, but it is convenient to keep all the $\zeta$'s and use them as projective variables in the moduli space. This is in line with the fact that $\Omega$ is unique only up to a multiplicative constant. To simplify the notation, we define the period vector as:
\begin{equation}
    \Pi=\left(
    \begin{array}{c}
         \cG_a \\ \zeta^a
    \end{array}
\right)
\end{equation}
There is a relatively simple way to find these periods. If we go back to equation \eqref{eq:OmegaExpansion}, we notice that by differentiating $\Omega$ with respect to the $\zeta$'s, we are bound to find a system of differential equations satisfied by the periods due to $H_3$ being finite-dimensional. The set of differential equations satisfied by the periods is called the \emph{Picard-Fuchs system}, and solving it will give us all the periods.
    
Finally, the prepotential is given by
\begin{equation}
    \cG(\zeta)=\frac{1}{2}\zeta^a \cG_a
\end{equation}
and it is quadratic in $\zeta^a$.

For the Kähler side, classical geometry alone cannot fully capture the situation because, as we saw in equation \eqref{eq:A-model3pointFunction}, there are instanton corrections that are of purely quantum origin. Therefore, we expect classical geometry to fail us. We will describe the prepotential that comes from classical geometry and leave open the possibility of quantum corrections. Taking a basis $e_a \in H^2 = H^{1,1}$, we can expand the Kähler form as $J = w^a e_a$, and the $w^a$ give us coordinates in the Kähler moduli space. The classical prepotential is then given by
\begin{equation}
    \cF_\text{classical}=\frac{1}{3!}\kappa_{abc}w^a w^b w^c\,,
\end{equation}
where 
\begin{equation}
    \kappa_{abc}=\int_M e_ae_be_c
\end{equation}
are the intersection numbers\footnote{We could have deduced this for instance by looking at the 3 point functions in equation \eqref{eq:A-model3pointFunction}}. To match the structure of the complex structure moduli space coordinates, we introduce a new coordinate $w^0$ (which we think was set to 1 above) and consider the $w^a$ (including $a=0$) as projective coordinates. We also divide by $w^0$ to obtain a quadratic prepotential, similar to the one for the complex structure moduli space. The corrected prepotential is then given by
\begin{equation}
    \cF = \frac{1}{3!}\frac{\kappa_{abc}w^aw^bw^c}{w^0}+\dots
    \label{eq:A-prep}
\end{equation}
where the sum on top is only for non-zero indices and $\dots$ stands for eventual exponential corrections. Similarly, we can define $\mathcal{F}_a = \partial_{w^a} \mathcal{F}$ and write a period vector \footnote{note that these periods do not correspond to periods of any form; however, we use the same terminology as for the complex moduli space, as they will ultimately be related to each other}:
\begin{equation}
    \amalg=\left(
    \begin{array}{c}
         \cF_a \\w^a
    \end{array}\right)\,. 
\end{equation}

The general expectation from mirror symmetry is that the periods $\amalg$ of $M$ should be related to the periods $\Pi$ of $\tilde{M}$. In the next section, we will provide an explicit mapping for toric hypersurfaces, which are Calabi-Yau threefolds obtained as the zero locus of a single function on a toric ambient space. These methods, with some modifications, also work for complete intersections in toric spaces, but the notation becomes more cumbersome. For more information, see \cite{Hosono:1994ax}.

\section{Mirror symmetry for hypersurfaces in toric spaces}
\label{sec:MirrorSymToricHyper}
We will borrow the notation from Appendix \ref{ap:ToricGeometry}. Consider the Calabi-Yau threefold $X_\Delta$ obtained from the polytope $\Delta$. The periods of the mirror are given by \cite{Hosono:1993qy}:
\begin{equation}
    \Pi_i=\int_{\gamma_i}\frac{1}{f_{\Delta^\circ}}\prod_{j=0}^n\frac{dx_j}{x_j}\,,
\end{equation}
where $f_{\Delta^\circ}$ is the function analogous to the one in equation \eqref{eq:ToricFunction}:
\begin{equation}
    f_{\Delta^\circ}=\sum_{\rho \in \Sigma(1)} a_\rho x^{v_\rho+1}+a_0 x. 
\end{equation}

This explicit description of the periods allows us to obtain the Picard-Fuchs system in two steps: First, for every Mori cone vector $l$, consider $\bar{l} = \left(-\sum_\rho l_\rho, l\right) \in \mathbb{Z}^{\Sigma(1)+1}$. This $\bar{l}$ can be used to construct a differential operator that annihilates $f_{\Delta^\circ}$:
\begin{align*}
&\mathcal D_l=\prod_{\bar l_\rho>0}\left(\frac{\partial}{\partial a_\rho}\right)^{\bar l_\rho}-\prod_{\bar l_\rho<0}\left(\frac{\partial}{\partial a_\rho}\right)^{-\bar l_\rho },\\
&\mathcal D_l\frac{1}{f_{\Delta^\circ}}\propto \prod_{\bar l_\rho>0}x^{v_\rho \bar l_\rho}-\prod_{\bar l_\rho<0}x^{-v_\rho \bar l_\rho}=\prod_{\bar l_\rho<0}x^{-v_\rho \bar l_\rho}(x^{\sum_\rho \bar l_\rho v_\rho}-1)=0
\end{align*}
The last equality follows from the definition of the Mori cone. The system of equations can be solved trivially by setting $z_k = (-1)^{\bar{l}_0^{(k)}} a^{\bar{l}^{(k)}}$, where $k$ runs over all the vectors in the Mori cone. This means that the periods depend on the variables $a_\rho$ only through the $z$'s. 

The second set of differential equations can be obtained as:
\begin{equation*}
\mathcal Z_j=\sum_mv_{\rho,j}\frac{\partial}{\partial \ln a_\rho}-\delta_{0,j},\quad \mathcal Z_j\frac{1}{f_{\Delta^\circ}}=0
\end{equation*}
where $v_{\rho,j}$ is the $j$-th entry of $v_\rho$, and $v_{\rho,0} = 1$ for all $\rho$ (these are simply the vectors in $\Delta^\circ \cap N \times \{1\}$), these equations are nontrivial relations between the monomials $x^{v_\rho}$.

This set of differential equations can be solved around the point $z=0$ using the \emph{Frobenius method}. The Frobenius method consists of writing an ansatz
\begin{equation}
    \sum_{n\in \mathbb N^k}c(n,\eta)z^{n+\eta}
\end{equation}
and plugging it into the differential equation to solve for $\eta$ and $c$. In general, there is a unique solution for $\eta$: $\eta=0$. The unique holomorphic solution is given by
\begin{equation}
    \varpi_0=\sum_{n}c(n)z^n\,,
\end{equation}
and the remaining solutions are obtained by making ansätze with logarithmic singularities. Specifically, there are logarithmic, double logarithmic, and triple logarithmic singularities. Particularly important are the logarithmic solutions given by: 
\begin{equation}
    \varpi_i=\ln (z_i)\varpi_0 + \text{ Analytic in }z\,.
\end{equation}
In mathematical jargon, we would say that the indicial equation (the equation satisfied by $\eta$) is completely degenerate, as it has only a single solution. The point $z=0$ is therefore a point of maximally unipotent monodromy. This remark allows us to relate the periods to the Kähler moduli space of $X_\Delta$. 

The large-radius point, i.e., the point where $\int_{e_a} J \to \infty$ for every $e_a \in H_2(X_{\Delta})$, corresponds to a point where the period vector $\amalg$ has maximal unipotent monodromy. This can be shown explicitly in precise examples, such as the quintic hypersurface \cite{Candelas:1990rm}, and in general by a detailed study of the geometry \cite{morrison1992mirror}. 

We can intuitively understand these monodromies around the large-radius point as follows: Consider the expansion of the K\"ahler form given by $J = t^i e_i$ with $e_i$ having integrals that are multiples of $2\pi i$ over any cycle. The fact that the K\"ahler form only appears in the exponent implies that the points $t$ and $t+n$ for any integer $n$ represent the same point in moduli space. We then have the Kähler moduli space $\cM_{k}$ and a natural cover $\overline \cM$ such that $\cM_k=\overline \cM/\Z$. Consider a closed path $t:[0,1]\to \cM_{k}$. If the path $t$ is contractible, lifts $\tilde t:[0,1]\to \overline \cM$ give closed paths in $\overline \cM$. However, if $t$ encircles a singularity, the lifts are not closed. Hence, when going around singularities, $t$, or more precisely $\tilde t$, changes by an integer. Noting that $\ln$ changes by $2\pi i$ monodromies, a natural guess for the \emph{mirror map}, i.e., the map $t^i(z) = \frac{w^i}{w^0}(z)$ from the complex moduli space of $X_{\Delta^\circ}$ to the Kähler moduli space of $X_\Delta$, is
\begin{equation}
t^i=\frac{1}{2\pi i} \frac{\varpi_1^{(i)}}{\varpi_0}=\ln z_i + \text{analytic}\,.
\end{equation}
From this equality, we can determine the remaining periods in $\amalg$ by looking at the leading contributions in equation \eqref{eq:A-prep} and replacing each product of $t_i$'s with the period in $\Pi$ that has the appropriate logarithms.

\section[Topological string partition function, Gromov-Witten invariants, and Gopakumar-\\Vafa invariants]{Topological string partition function, Gromov-Witten invariants, and Gopakumar-Vafa invariants}
So far, we have focused on understanding the CFT obtained by twisting the Lagrangian \eqref{eq:untwistedLagrangian}. However, to reach the topological string, we need to couple our theory to gravity. This requires integrating over all possible Riemann surfaces $\Sigma$. Conformal transformations on $\Sigma$ correspond to holomorphic maps; therefore, integration over the moduli space of $\Sigma$ then reduces to integration over its complex structure moduli space. The moduli space can be decomposed according to the genus $g$. We denote the moduli space of genus $g$ curves by $\mathcal{M}_g$. The standard coordinates on $\mathcal{M}_g$ are the Beltrami differentials, which locally can be written as:
\begin{equation}
    \delta_{i} dz=dm^i \mu^{i\,\bar z}_zd\bar z\,,
\end{equation}
where $dm^i$ represents an infinitesimal movement in $\mathcal{M}_g$. To build a quantity we can integrate over $\mathcal{M}_g$ to obtain an interesting object, we contract the Beltrami differentials with the operator $G$ associated with the energy-momentum operator $T$ via $T = \{Q_{BRST}, G\}$, where $Q_{BRST}$ is the ghost operator associated with reparametrizations of the worldsheet. Thus, we define the topological string partition function as:
\begin{equation}
    F_g=\int_{\cM_g}dm^idm^{\bar i}\prod_{i=1}^{3g-3}\int_\Sigma dzd\bar z\, G_{zz}\mu^{iz}_{\bar z}\int_\Sigma dzd\bar z\, G_{\bar z\bar z}\mu^{\bar i \bar z}_{z}
\end{equation}

All these functions are combined to form the topological free energy $F$ and the topological string partition function $Z$:
 \begin{equation}
     F=\sum_{g\geq 0}g_s^{2g-2}F_g,\quad Z=\exp(F)\,
         \label{eq:Fg's}
 \end{equation}
 where $F_0$ is the prepotential $\mathcal F$.

Similarly to what happened for the prepotential, the contributions to $F$ can be separated with respect to the class of $x(\Sigma)$. \footnote{Here, we suppress the finite part of $F$, which is simply a polynomial in $t$, and focus on the part that depends on exponentials of the Kähler form.}
\begin{equation}
    F=\sum_g\sum_{\beta \in H_2M} g_s^{2g-2} r_{g,\beta}e^{-t \int_{\beta}J}
\end{equation}
The numbers $r_{g,\beta}$ are the \emph{Gromov-Witten invariants} of the manifold $M$. In general, they are rational numbers. By using duality to $M$-theory, we can give an alternative expansion of $F$ in terms of integer coefficients:
\begin{equation}
    F=\sum_{m\geq 1}\sum_g\sum_{\beta \in H_2M} I_g^\beta x_m^{2g-2} \frac{Q^{m\beta}}{m}\,,
\end{equation}
where
\begin{equation}
    x_m=2\sin\left(\frac{g_sm}{2}\right)\,,\quad Q^\beta=e^{-t\int_\beta J}\,,
\end{equation}
and $I_g^\beta$ are the integer \emph{Gopakumar-Vafa invariants} that correspond to the number of BPS particles in $M$ theory compactified in $X$. This calculation gives a physical meaning to the formal expressions in equation \eqref{eq:Fg's}. For the original derivation, see \cite{Gopakumar:1998ii, Gopakumar:1998jq}, or refer to \cite{Dedushenko:2014nya} for a detailed calculation.



\chapter{6D Superconformal theories}
\label{ch:6D}

\section{F-theory}
The 10-dimensional Type IIB supergravity action is invariant under $\mathrm{SL}(2,\mathbb{Z})$ transformations, and there is extensive evidence to believe that this symmetry is preserved even non-perturbatively in Type IIB superstrings. See \cite{Vafa:1996xn,Schwarz:1995dk} or \cite{Weigand:2018rez} for a review. The way this symmetry acts on the axio-dilaton $\tau$ led to the conjecture that this theory can be understood through a torus whose complex modulus is given by $\tau$. This description of the type IIB string theory in terms of a torus is known as \emph{F-theory } Sometimes, it is said that $F$-theory is a 12 dimensional theory because the 10 dimensions of type IIB and the 2 dimensions of the torus add up to 12; however, F-theory is not purely geometric in nature. The volume of the torus is not relevant. As we will explain in section \ref{sec:Mtheory}, after compactification on a circle, we arrive at M-theory, where the full moduli space of this torus becomes relevant.

The axio-dilaton, or more precisely $C_0$, is magnetically sourced by D7-branes. Therefore, allowing singularities in the behavior of $\tau$ describes a vacuum with D7-branes. In this sense, F-theory provides a description of Type IIB string theory in the presence of D7-branes. In this work, we will generally think of F-theory as a non-perturbative description of Type IIB string theory in the presence of D7-branes.

\section{F-theory/M-theory duality}
\label{sec:Mtheory}
M-theory compactified on a torus is dual to Type IIA string theory compactified on a circle. By T-dualizing the Type IIA circle, we also establish a duality with Type IIB string theory \cite{Schwarz:1995dk}. It is important to note that to transition from M-theory to Type IIB, we require two circles: one for the reduction from 11 dimensions to 10 dimensions and another for the T-duality transformation. When we take the volume of the M-theory torus to zero and recall that T-duality inverses the radius of the circle, the Type IIB circle decompactifies and we reach the duality:

\begin{equation}
    \M(\E)_{\vol \E\to 0} \text{ in } \mathbb R^{1,8} \sim \IIB \text{ in }\R^{1,9}\,.
\end{equation}

Using this duality fiber-wise, we can consider an elliptic fibration $Y_3$ with base $B_2$ and obtain the duality relation\footnote{We are labelling the spaces by their complex dimension and we are specializing to these particular dimensions because it is the case of importance for this work, but the relation holds for any torus fibration $Y_{n+1}\to B_n$}:
\begin{equation}
    \boxed{\M(Y_3)_{\vol \E \to 0}\text{ in } \mathbb R^{1,4}\sim \IIB(B_2) \text{ in }\mathbb \R^{1,5}}\,.
\end{equation}
This duality is at the core of the present understanding of $F$-theory. 

On the right-hand side, we have an effective 6D theory that only knows about the complex structure moduli of the torus (through the axio-dilaton). In contrast, on the left-hand side, before taking the limit, we have a theory that fully incorporates the structure of the torus fiber. We can establish a relationship between these two effective theories without resorting to higher-dimensional constructions. Here's how: Starting from the theory on the left and transitioning to the theory on the right, we consider the two circles $S_{R_M}\times S_{R_A}$. By compactifying the $M$-theory circle with $R_M\to 0$, we obtain type IIA theory with an inner circle $S_{R_A}$. T-dualizing this last circle, we arrive at type IIB theory with an inner circle $S_{R_B}$, where the radius is $R_B=1/R_A$. Finally, taking the limit $R_A\to 0$, which corresponds to taking $R_B\to \infty$, we reach the theory on the right. In this process, we observe that one of the flat directions on the right originates from the torus shrinking to zero size on the left. To reverse this process, we need to shrink one of the flat directions of the theory on the right to regrow the fiber on the left. Consequently, by compactifying the effective 6D theory on a circle, we arrive at the 5D theory.



\section{Elliptic fibrations}

In this section, we briefly review the relevant aspects of elliptic curves and elliptic fibrations, which will be important for our discussion. For a comprehensive introduction to elliptic curves, we refer the reader to \cite{husemoller_elliptic_1987}. In the context of F-theory, the arithmetic structure of elliptic curves also plays a crucial role, and interested readers can consult \cite{silverman_arithmetic_1986,sutherlad:2013} for more details.

Any elliptic curve can be represented in what is known as the \emph{Weierstrass form}, given by (see for instance \cite{Deligne:1975} or \cite{Braun:2014oya} for a more modern derivation):
\begin{equation}
    y^2=x^3+fxz^4+gz^6\,,
    \label{eq:Wform}
\end{equation}
where $[x:y:z]$ are homogeneous coordinates in $\mathbb{P}_{231}$, and $f$ and $g$ are constants. The discriminant of this equation is given by: 
\begin{equation}
    \Delta = 4f^3+27g^2 
\end{equation}
and it is well-known that the \emph{$j$-function}:
\begin{equation}
    j=4\frac{24^3f^3}{\Delta}
\end{equation}
completely characterizes the complex structure of the elliptic curve. In other words, two elliptic curves are isomorphic if and only if their Weierstrass forms lead to the same $j$-invariant. The value of $j$ is related to the complex structure modulus $\tau$ through the invertible relationship:
\begin{equation}
    j(\tau)=1728 \frac{E_4(\tau)^3}{E_4(\tau)^3-E_6(\tau)^2}\,,
\end{equation}
where $E_4$ and $E_6$ are the Eisenstein series, see \eqref{eq:EisensteinSeries}.

We are not only interested in individual elliptic curves but in fibrations of them. In the case of fibrations, the Weierstrass form is only obtained globally if there exists a global section. In all the theories studied in this thesis there is a section so this does not present a problem for us. In the absence of section we talk of ``genus-one fibration" instead of elliptic fibrations. This genus one fibration lead to discrete symmetries in the effective theories. For an introduction on discrete symmetries in F-theory see \cite{Cvetic:2018bni}.

In the case of fibrations, the variables $[x:y:z]$ become sections of a $\P_{231}$ fibration over the base manifold, and $f$ and $g$ are promoted to sections over the base as well. By analyzing the transformation properties under $\mathrm{SL}(2,\mathbb{Z})$ transformations, we can determine the bundles to which each of these elements belong:
\begin{equation}
\begin{gathered}
     f\in \Gamma(\Lc^4)\,, \quad g \in \Gamma(\Lc^6)\\  \label{eq:fgsections}
     x \in \Gamma(\Lc^2)\,, \quad y \in \Gamma(\Lc^3)\,,\quad z \in \Gamma(\Oc)\,.\,
\end{gathered}
\end{equation}
Here $\mathcal{L}$ is the holomorphic line bundle associated with holomorphic variations of the axio-dilaton $\tau$ (see \cite{Weigand:2018rez}, section 2.2). Using the adjunction formula, we can further conclude that in order for the fibration to be a Calabi-Yau manifold, the bundle $\mathcal{L}$ must be the anti-canonical bundle of the base manifold:

\begin{equation}
   \boxed{ \Lc=\bar K}\,.
\end{equation}

Hence, the process of constructing the fibration becomes entirely algorithmic: given a base manifold $B_2$, we can determine the generic sections of the different powers of the anti-canonical bundle, providing us with an explicit description of the Calabi-Yau manifold $Y_3$. In the next section, we will explore how specializing the sections $f$ and $g$ leads to more singular geometries, corresponding to effective theories with enhanced gauge symmetry. This will demonstrate how complex structure deformations correspond to un-Higgsing in the effective theory.

\section{Blowing up the singularities and the Kodaira classification}

The elliptic fibrations we have constructed are singular along the discriminant locus given by $\Sigma = \{\Delta = 0\}$. The nature of the singularity depends on the specific vanishing orders of $\Delta$, $f$, and $g$. These singularities for elliptic curves were classified by Kodaira \cite{MR0184257} long before physicists became interested in them. The effect of monodromies on the global structure of the resolution for fibrations, rather than individual fibers, was explained by Tate \cite{Tate:1975}. The relation to F-theory was later uncovered \cite{Bershadsky:1996nh,Grassi:2011hq}.

In general, the result is that at the discriminant locus, the singular fiber has to be blown up, and the generic fiber is given by the sum of several $\P^1$:
\begin{equation}
    [\E]=\sum_{i=0}^{\tilde n} a_i[\P^1_i]
\end{equation}
where the $\mathbb{P}^1$'s intersect according to the Dynkin diagram of a simply-laced Lie algebra. By considering one of these $\P^1$ and traversing a closed path around the discriminant locus, we may end up on a different $\mathbb{P}^1$. Therefore, we consider invariant orbits of curves, which are linear combinations of the resolution curves:
\begin{equation}
    C_i=\sum_{j\in I_i} [ \P^1_j]\,,
\end{equation}
where $I_i$ is a set of indices, and $i = 1, \ldots, n \leq \tilde{n}$. We also denote the divisors obtained by fibering each irreducible curve $C_i$ along one component of the discriminant locus as $E_i$. The Kodaira-Tate classification sates that: $n$ is the rank of a simple Lie group, and the intersection of the $E$'s and $C$'s is given by:
\begin{equation}
E_i \cdot C_j = -C_{ij},
\end{equation}
where $C_{ij}$ is the (affine) Cartan matrix of the same Lie algebra. In particular, we not only obtain simply-laced gauge algebras because non-trivial monodromies act as foldings of the Dynkin diagrams. Here, we reproduce the table from \cite{Grassi:2011hq} that gives the Lie algebra for every possible combination of vanishing orders of $\Delta$, $f$, and $g$ (see Table \ref{tab:KodairaClassification}).

\begin{table}[]
    \centering
    \begin{tabular}{c|c|c}
         Singularity type & Vanishing orders of $(f,g,\Delta)$ & Gauge algebra $\g$  \\\hline
         $I_0$& $(\geq 0,\geq 0,0)$ & $-$\\
         $I_1$ & $(0,0,1)$ & $-$\\
         $I_2$ & $(0,0,2)$ & $A_1$\\
         $I_m, \, m\geq 3$ &$(0,0,m)$& $C_m$ or $A_m$\\
         $II$ &$(\geq 1,1,2)$& $-$\\
         $III$ &$(1,\geq 2,3)$& $A_1$\\
         $IV$ &$(\geq 2,2,4)$& $C_2$ or $A_2$\\
         $I_0^*$ &$(\geq 2,\geq 3,6)$& $G_2$, $B_3$, or $D_4$\\
         $I_{2n-5}^*,\, n\geq 3$& $(2,3,2n+1)$ &$B_{2n-2}$ or $D_{2n-1}$\\
         $I_{2n-4}^*,\, n\geq 3$& $(2,3,2n+2)$ &$B_{2n-1}$ or $D_{2n}$\\
         $IV^*$& $(\geq 3,4,8)$ &  $F_4$ or $E_6$\\
         $III^*$ & $(3,\geq 5,9)$ & $E_7$\\
         $II^*$ & $(\geq 4,5,10)$ & $E_8$
    \end{tabular}
    \caption{Table of possible singularities of an elliptic fibration. When there is more than one possible gauge algebra, the choice comes from monodromy. For singulrities with higher vanishing orders: $(\geq 4,\geq 6, \geq 12)$ there are no crepant resolutions. Table taken from \cite{Grassi:2011hq}}
    \label{tab:KodairaClassification}
\end{table}

So far, we have discussed only a single component of the discriminant locus $\Sigma$. The discriminant locus may have several components, and each component gives rise to a set of divisors, curves, and the corresponding gauge algebra. At the points where two components of the discriminant locus intersect, the singularities are generally enhanced and need to be blown up further.

\section{The M-theory picture / the 5D theory}

The vector fields of the 5D theory arise from the M-theory 3-form $C_3$. This form can be expanded in a basis of $H^{1,1}(\hat Y_3)$.
\begin{equation}
    C_3=A^0[\tilde S_0]+\sum A_{U(1)}^i [\sigma(S_i)]+\sum A^i[E_i]+ \sum A^a_{U(1)} \pi^*(D_a)\,. 
    \label{eq:C3form}
\end{equation}
We will discuss the origin and roll of each of this terms in this section. 
 
We have already encountered the divisors $E_i$ in the previous section. In \eqref{eq:C3form}, the sum covers all components of the discriminant locus. The associated fields $A^i$ are the gauge fields of the effective 5D theory. The charge of an effective particle obtained by wrapping an M2 brane on a holomorphic or anti-holomorphic curve $C$ is given by:
\begin{equation}
    \int_{M2=C\times \R}C_3=\int_\R A^i \int_C [E_i]= q_i\int A^i\,, 
\end{equation}
where 
\begin{equation}
    q_i=\int_C [E_i]=[C]\cdot [E_i]\,. 
\end{equation}
In particular, the curves $C_i$ discussed in the previous section have charges given by the simple roots of the corresponding Lie algebra, and M2 branes wrapping these curves are associated with the corresponding gauge bosons. This provides our effective 5D theories with the gauge algebras listed in Table \ref{tab:KodairaClassification}.

The fields $A^0,\, A_{U(1)}^i$ correspond to sections of the elliptic fibration. A section is an embedding $s: B_2 \rightarrow Y_3$ of the base into the fibration such that $\pi \circ s = \text{id}_{B_2}$. We have explicitly separated the contribution of the zero section $A^0$ from the rest of the sections and have shifted the sections to satisfy necessary transversality conditions using the Shioda map $\sigma$. We do not provide the explicit details here, but refer to \cite{Morrison:1996na,Morrison:1996pp,schuett2010elliptic} or the review \cite{Weigand:2018rez} for more information. The gauge field corresponding to the zero section can be identified with the Kaluza-Klein (KK) gauge field, so it does not come from a 6D $U(1)$ gauge group. The other fields correspond to 6D $U(1)$ gauge groups. The group of sections on an elliptic fibration has been extensively studied in mathematical literature and is known as the \emph{Mordell-Weil group} of the fibration. Similar to before, the charges of M2 branes wrapping a given curve are obtained by taking the intersection of the curves with the corresponding divisors:
\begin{equation}
    q_{KK}=[C]\cdot [\tilde S_0], \quad q_{U(1)}^i=[C]\cdot[\sigma (S_i) ]\,. 
\end{equation}

Lastly, the fields $A_{U(1)}^a$ are associated to divisors obtained by pulling back curves in $B_2$. They correspond to 5D $U(1)$ gauge fields, but they do not come from 6D gauge fields. Instead, they arise from the dimensional reduction of 2-form fields in the 6D theory. The type IIB 4-form $C_4$ can be expanded as:
\begin{equation}
    C_4=\sum b_a [D_a]\,,
\end{equation}
where $D_a$ are the components of the discriminant locus. The expansion of the 2-form $b_a$ along the $6D \rightarrow 5D$ compactification circle gives the 5D fields $b_a = A_a d\theta + \dots$.


As we will argue shortly, we are interested in decompactifying the base to decouple gravity. This decompactification causes some of the gauge couplings to become infinte: $g_a \sim \text{vol}_{D_a} \rightarrow \infty$. Consequently, the algebras associated to these divisors become flavor algebras. Matter fields arise from M2 branes wrapping the resolution curves at the intersection of two components of the discriminant locus. These matter fields are usually referred to as \emph{localized matter}. By computing Gromov-Witten invariants corresponding to different curves, we can determine the matter content of the effective theory \cite{Kashani-Poor:2019jyo}.

\section{6D Anomaly and the non-critical string}

Among the 6D fields, there is one type that stands out and deserves special attention: the 2-form fields $b_a$. These fields arise from the self-dual $C_4$ form and, therefore, are also self-dual fields. The self-duality condition is given by:
\begin{equation}
    db_a=*db_a\,, 
\end{equation}
where $*$ is the Hodge star operator. It is well-known that self-dual fields are anomalous \cite{Alvarez-Gaume:1983ihn}. However, dismissing the theory as inconsistent would be premature. The way this 6D theory avoids anomalies is quite subtle. This subtle mechanism and the wrong belief that 6D theories were trivial in the infrared --corrected by Seiberg and Witten \cite{Seiberg:1996vs}-- were probably the reason why these theories took so long to be studied. In this section, we discuss how the presence of non-critical strings makes these theories actually consistent. Non-critical strings lie at the heart of this thesis and are extensively discussed in chapter \ref{ch:paperWorldsheet}. Here, we provide a general overview and refer to this chapter for explicit calculations and references. A detailed introduction can be found in Section \ref{s:theories_considered}.

The idea is relatively simple. The $b_a$ fields appear in the 6D action accompanied by the corresponding gauge field:
\begin{equation}
    \int b_a \tr F_a^2\,. 
\end{equation}
Hence, the $b$-field is sourced by gauge instantons, and quantization around a 6D instanton solution gives rise to a 2D soliton in the 6D theory spectrum: the \emph{non-critical string}. The effective description of this non-critical string is as a sigma model with the target space being the moduli space of 4D gauge instantons. This explains the terminology ``non-critical," as the target space is not of critical dimension. Therefore, the sigma model is actually anomalous. Surprisingly, this anomaly turns out to be what saves the full theory. If one adds up the anomaly coming from the self-dual $b$-field and the anomaly coming from the sigma model, they cancel each other out, resulting in a consistent theory.

It is important to note that describing the worldsheet theory of the non-critical string is challenging, and the way we approach it is somewhat indirect. We believe that the full 6D theory with its soliton should be consistent, and for this to be the case, the two anomalies must cancel each other. We therefore determine the anomaly that the non-critical string should have in order to cancel the anomaly of the self-dual $b$-field. Then, with some guesswork, we look for a sigma model that gives rise to precisely this anomaly. For extensive discussions on this subject, refer to Chapter \ref{ch:paperWorldsheet} and the references therein.

\section{The Coulomb branch and Higgsing trees}
\label{sec:CoulombBranchAndHiggsingTrees}

In this section, we specialize our discussion to the case of rank 1 6D SCFTs, focusing on the decompactification limit of \emph{Hirzebruch surfaces}. Rank 1 theories are engineered as elliptic fibrations over 2-dimensional manifolds with a single compact curve. The compact curve corresponds to the base divisor, and higher rank SCFTs can be constructed by gluing together different rank 1 theories. Each tensor multiplet corresponds to a base divisor, so higher rank theories are obtained by adding more curves in the base. For more information on the construction of higher rank 6D SCFTs, refer to \cite{Morrison:2012np}.
Hirzebruch surfaces are themselves fibrations. To decouple gravity, we consider the non-compact space obtained by taking the limit of infinite size of the fiber, denoted by $Y_3$ below. The Hirzebruch surface and its decompactification limit can be visualized as follows:

\[\begin{tikzcd}
	{\mathbb E_\tau} & {\tilde Y_3} && {\mathbb E_\tau} & {Y_3} \\
	{\mathbb P_1} & { \mathbb F_k} && {O(-k)} & {B_2} \\
	& {C_k} &&& {C_k}
	\arrow[from=1-2, to=2-2]
	\arrow[from=2-1, to=2-2]
    \arrow[from=1-5, to=2-5]
	\arrow[from=2-2, to=3-2]
	\arrow[from=1-1, to=1-2]
	\arrow[from=1-4, to=1-5]
	\arrow[from=2-4, to=2-5]
	\arrow[from=2-5, to=3-5]
\end{tikzcd}\]
The compact curve $C_k$ at the bottom of the diagram is the only compact curve in the base and has self-intersection $C_k \cdot C_k = -k$. 

By considering the most generic sections of the corresponding bundles for $f$ and $g$, we arrive at the \emph{maximally Higgsed} theories. These theories are the elliptic fibrations over the given base that have the least amount of singularity and, consequently, require the minimum number of blow-ups. Each blow-up divisor gives rise to a simple root of the gauge algebra, so maximally Higgsed theories have gauge groups that cannot be reduced further. This explains the terminology maximally Higgsed. 

On the other hand, if we specialize $f$ and $g$ in a way that enhances the singularities and requires more blow-ups for resolution, we arrive at theories with higher rank gauge groups. Physically, the inverse process is easier to understand: starting from a theory with more singular behavior and moving towards a theory with less singular behavior. We shrink the resolution divisors and perform a complex structure deformation by choosing more generic $f$ and $g$. This corresponds to giving a hypermultiplet a non-zero vacuum expectation value. The resulting theory has a less singular fibration that requires fewer blow-ups and, therefore, has a lower rank gauge group. This entire process is a Higgs transition.

The mathematical framework for dealing with singular elliptic fibrations was described in \cite{Tate:1975} and further developed in the context of F-theory in \cite{Grassi:2011hq,Katz:2011qp,Bershadsky:1996nh}. A systematic review of the different Hirzebruch surfaces that give rise to rank 1 6D SCFTs is presented in Appendix \ref{sec:HiggsingTrees}, which provides a collection of Higgsing trees. Each labeled node in the tree represents the gauge and flavor groups, with a sub-index indicating the level of the corresponding current algebra in the worldsheet of the non-critical string. This appendix is taken from \cite{Duque:2022tub}. The detailed study of these theories and their corresponding Higgs transitions from both the 6D/5D effective theory and the critical string worldsheet perspective is the focus of this thesis. For more information, refer to Chapters \ref{ch:paperJacobi} and \ref{ch:paperWorldsheet}.


\label{sec:HiggsingTrees}

\chapter{Jacobi forms along Higgsing trees}
\label{ch:paperJacobi}

\centerline{Zhihao Duan${}^{a}$, David Jaramillo Duque${}^{b}$,} 
		
		\centerline{Amir-Kian Kashani-Poor${}^{b}$}

		\vskip 0.2in
		\begin{center}{\footnotesize
				\begin{tabular}{c}
					${}^{\, a}${\em School of Physics, Korea Institute for Advanced Study, Seoul 02455, Korea}\\[0ex]
				\end{tabular}
		}\end{center}
		\begin{center}{\footnotesize
				\begin{tabular}{c}
					${}^{\, b}${\em LPENS, CNRS, PSL Research University, Sorbonne Universit\'{e}s, UPMC, 75005 Paris, France}\\[0ex]
				\end{tabular}
		}\end{center}

\begin{adjustwidth}{80pt}{80pt}
			{}{\bf Abstract:} Using topological string techniques, we compute BPS counting functions of 5d gauge theories which descend from 6d superconformal field theories upon circle compactification. Such theories are naturally organized in terms of nodes of Higgsing trees. We demonstrate that the specialization of the partition function as we move from the crown to the root of a tree is determined by homomorphisms between rings of Weyl invariant Jacobi forms. Our computations are made feasible by the fact that symmetry enhancements of the gauge theory which are manifest on the massless spectrum are inherited by the entire tower of BPS particles. In some cases, these symmetry enhancements have a nice relation to the 1-form symmetry of the associated gauge theory. 
\end{adjustwidth}	
\newpage

 \section{Introduction}
	
	Theories with extended supersymmetry have interesting massive spectra protected against decay by the BPS constraint (see e.g. \cite{Seiberg:1994rs}). Many techniques have been developed for computing exact counting functions on the BPS spectrum, notably topological string methods and localization, see \cite{Hori:2003ic} and \cite{Pestun:2016zxk} for reviews. Rather than starting with a set of fundamental fields and their interactions and deriving the resulting spectrum of the theory, the starting point of many of these approaches is a wholesale construction of the theory from the outset, in terms of its embedding in string theory. Having access to a subsector of the complete spectrum, we can ask how the latter is constrained by consideration of just the low lying spectrum. Ultimately, we would like to understand, if and if yes then in how many ways a given low lying spectrum can be completed to a consistent spectrum of a UV complete theory. The swampland program, as initiated in \cite{Vafa:2005ui}, is an overarching term for work attempting to address this question, with the emphasis that the UV complete theory encompass gravity; see \cite{Brennan:2017rbf, Palti:2019pca} for recent reviews.
	
	In this work, we will be concerned with the BPS spectra of 5d gauge theories with 8 supercharges obtained via compactification on a circle of rank 1 6d theories with (1,0) supersymmetry. These theories are dubbed 5d KK theories in \cite{Jefferson:2017ahm}. Much work has been expended in studying their superconformal limits \cite{DelZotto:2017pti,Bhardwaj:2018yhy, Bhardwaj:2018vuu,Bhardwaj:2019fzv}. Here, we will be interested in the theories at a generic point on their Coulomb branch and the associated spectrum of BPS particles. 
	
	Rank 1 theories are readily constructed via F-theory compactifications \cite{Morrison:1996na,Morrison:1996pp} on elliptically fibered Calabi-Yau manifolds $X$ \cite{Candelas:1996su,Bershadsky:1996nh,Perevalov:1997vw}. The rank in the context of 6d theories indicates the dimension of the tensor branch, which geometrically corresponds to the number of compact homology 2-cycles in the base of the fibration. Requiring that an elliptic fibration $X$ over a given base be a Calabi-Yau manifold imposes a minimal singularity on the elliptic fiber, which maps to the gauge algebra $\mathfrak{g}$ of the 6d theory obtained by F-theory compactification on $X$. Resolving the singularity via blow-ups does not change the 6d physics, but moves the associated 5d KK theory away from the origin of its Coulomb branch. Upon specializing the complex structure of $X$, the singularity can be enhanced, leading to a larger gauge algebra $\mathfrak{g}'$ of the compactified theory. Physically, the transition $\mathfrak{g}' \rightarrow \mathfrak{g}$ corresponds to Higgsing. The graph with nodes consisting of elliptically fibered Calabi-Yau manifolds and links describing the process of specializing complex structure to a singularity, then resolving it by blowing up, is referred to as a Higgsing tree. Rank 1 Higgsing trees, constructed over the base surfaces $\cO(-n) \rightarrow \IP^1$, for $n = 0, \ldots , 8, 12$, provide the setting for our computations.
	
	A counting function for 5d BPS states is famously given by the topological string partition function $Z_{top}$ \cite{Gopakumar:1998ii,Gopakumar:1998jq,Iqbal:2012xm}. In \cite{Candelas:1993dm,Katz:1996ht}, examples of the type of transformation encoded in the links of Higgsing trees were dubbed geometric transitions and studied in detail for several 2 and 3 parameter Calabi-Yau manifolds. The genus 0 Gromov-Witten invariants of the 1 parameter Calabi-Yau models obtained after the transition were shown to be related to those of the parent theory by summing over the K\"ahler parameters associated to the blown-down curves. In this work, we will obtain a stronger form of this result for the nodes and links of rank 1 Higgsing trees. By invoking modularity results \cite{Huang:2015sta,Gu:2017ccq,DelZotto:2016pvm,DelZotto:2017mee}, the computation of $Z_{top}$ at a given order in an expansion in the base curve exponentiated K\"ahler parameter $Q_B$, schematically $Z_{top} \sim \sum_k Z_k Q_B^k$, reduces to the determination of a holomorphic weak Weyl invariant Jacobi form \cite{EZ,Wirthmuller:Jacobi,Bertola}\footnote{This being the only class of Jacobi forms we will be concerned with in this paper, we will drop the adjective {\it weak} in the following.} of determined index and weight, with the coarsest choice of Weyl group possible being that of the gauge group of the engineered theory, $\weyl(\mathfrak{g})$. We conjecture that the topological string partition functions of nodes of the Higgsing tree specialize upon moving towards the root of the tree according to maps $J(\mathfrak{g}') \rightarrow J(\mathfrak{g})$ relating the ring of Jacobi forms of the associated Weyl groups.\footnote{The connection between different nodes of rank 1 Higgsing trees was studied for some examples in a different presentation of $Z_{top}$ in \cite{Kim:2018gjo,Gu:2020fem}.} We provide ample evidence for this conjecture by computing $Z_k$ at $k=1$ for a host of examples and demonstrating the specialization explicitly.
	
	Note that the topological string is insensitive to complex structure deformations\footnote{Here and in the following, we speak from the perspective of the so-called $A$-model \cite{Witten:1991zz}.\label{A-model}}. From its vantage point, all nodes of a Higgsing tree can hence be seen as singular geometries lying on subslices of the K\"ahler moduli space. One parent theory (which on infinite length Higgsing trees would depend on infinitely many parameters) should hence capture all geometries subsumed in a Higgsing tree. This perspective could offer a path towards the proof of our conjecture; the point to be addressed is that the topological string partition function exhibits singularities on the subslices of moduli space corresponding to the singular geometries. For the models that we consider, we demonstrate by computation that these singularities are only apparent.
	
	The elements of $J(\mathfrak{g})$ at given weight and index span a finite dimensional vector space over $\IQ$.
	The computation of $Z_k$ in the class of models we are considering is reduced to obtaining the expansion coefficients in an appropriate basis of this space. The reduction of the computation to a finite dimensional problem is conceptually important. Practically, the number of coefficients grows rapidly with the rank of $\mathfrak{g}$ and the base degree $k$. Luckily, $Z_k$ for many of the geometries we consider turns out to exhibit a higher symmetry than merely the Weyl group $\weyl(\mathfrak{g})$. These enhanced symmetries have various origins. The most straight-forward cases are 1-form symmetries as well as constraints arising from the fact that a theory arises via Higgsing. Somewhat more surprisingly, the fact that some maps $J(\mathfrak{g}') \rightarrow J(\mathfrak{g})$ are injective, hence invertible on their image, can lead to enhanced symmetries motivated by moving opposite the Higgsing arrow. We call this phenomenon ``reverse inheritance." This latter class of enhanced symmetries can also be explained intrinsically, without reference to Higgsing, via a cancellation mechanism described in \cite{Kashani-Poor:2019jyo}. Notably, we argue for symmetry enhancement at the level of the massless 6d spectrum of the parent theory of the 5d theory, yet find computationally that the symmetry extends to the full BPS spectrum, begging the question (which we leave for future study) whether this had to be the case. The elliptic genus in 6d or 4d theories as expressed in terms of Jacobi forms or related structures is enlisted to study various conjectures relating to the swampland program in the works \cite{Lee:2018spm,Lee:2018urn,Lee:2019tst,Lee:2020blx,Lee:2020gvu,1837146}. 
	
	We conclude this introduction with a summary of the ensuing sections. We begin with a rapid review of the physical setting, and discuss the general structure of $Z_k$ in section \ref{sec:how_to}, completing the discussion in the literature to encompass all rank 1 theories. In section \ref{sec:moving}, we take a first look at the constraints arising from being a node of a Higgsing tree, and discuss in detail the specialization of $Z_k$ upon descending a Higgsing tree towards its root. We also derive how our specialization results for $Z_k^{\mg}$ manifest themselves at the level of Gromov-Witten invariants. Section \ref{sec:enhanced} is dedicated to the discussion of symmetry enhancements of $Z_k$ derived both from the perspective of neighboring nodes in the Higgsing tree and intrinsically. Several technical appendices conclude the paper: in appendix \ref{app:explicit}, we explicitly give $Z_1$ for one among the many models for which we have computed it, to convey the general flavor of our results. Appendix \ref{app:GWinv} contains tables of Gromov-Witten data which exemplify the specialization results derived in section \ref{sec:moving}. Appendix \ref{app:root_systems} summarizes data on simple Lie algebras which is relevant for the discussion in the main text. Appendix \ref{app:Jacobi_forms} gives explicit formulae for the generators of the ring of Jacobi forms for all simple groups other than $E_n$, $n=6,7,8$. Appendix \ref{app:specializationformulas} provides further details regarding the specialization maps between these. Appendix \ref{app:Weyl_invariant_polynomials} provides a brief introduction to Weyl invariant polynomials, and explains how these enter in our calculations. Finally, appendix \ref{app:Higgsing_trees} provides a very brief review of elliptically fibered Calabi-Yau manifolds, and reproduces some rank 1 Higgsing trees for the reader's convenience.

    \section{How to capture BPS degeneracies via Weyl invariant Jacobi forms} \label{sec:how_to}
    \subsection{The topological string and BPS states of the 5d theory}
    The topological string was born as a worldsheet theory \cite{Witten:1991zz,Bershadsky:1993cx}. To a Calabi-Yau manifold $X$ and to each worldsheet genus $g$, it assigns a power series $F_g$ in K\"ahler parameters $Q_i$ associated to the homology 2-cycles of $X$ (see footnote \ref{A-model}). The coefficients are the celebrated Gromov-Witten invariants of $X$. Upon introducing a formal parameter $\gs$, a putative theory tying together all $F_g$ is assigned the partition function
    \be
    Z_{top}(X) = \exp\left( \sum_g F_g\, \gs^{2g-2} \right) \,.
    \ee
    In \cite{Gopakumar:1998ii,Gopakumar:1998jq}, it was shown that $Z_{top}(X)$ is not merely a formal construct; it captures the non-perturbative BPS spectrum of M-theory compactified on $X$, arising from M2 branes wrapping holomorphic curves in $X$ (see \cite{Iqbal:2012xm} for a succinct summary of these matters).
    
    The 5d theories of interest in this paper arise upon compactifying M-theory on a certain class of elliptically fibered, interlinked Calabi-Yau manifolds $X$ constituting the nodes of so-called Higgsing trees, as we review in appendix \ref{app:Higgsing_trees}. We will call the theory obtained upon compactification on $X$ $\mathrm{M}[X]$ for the purposes of this section. The massless perturbative spectrum of $\mathrm{M}[X]$ is captured by the compactification of 11 dimensional supergravity on $X$. For non-compact $X$, this spectrum consists of $h^{1,1}(X)$ massless vector fields and $h^{2,1}(X)+1$ uncharged massless hypermultiplets. 
    
    The symmetries constraining $Z_{top}(X)$ become manifest at special points in the K\"ahler moduli space of $X$ in which this massless spectrum is enhanced. The description of these points of enhancement is most natural from a 6d perspective, arising from compactifying F-theory on $X$. In this description, only the complex structure of the elliptic fiber of $X$ is physical. Considering a representative of the geometry in which all exceptional curves in the fiber are blown down brings the gauge symmetry $\mathfrak{g}$ associated to the resulting singular fiber to light, together with hypermultiplets charged under $\mathfrak{g}$. This theory yields $\mathrm{M}[X]$ upon circle compactification, as follows from M-theory/F-theory duality \cite{Vafa:1996xn}. In contrast to the F-theory compactification, the elliptic fiber now is fully physical, with its size inversely proportional to the size of the compactification circle \cite{Vafa:1996xn}. Away from the singular limit of the fiber, we recover the perturbative massless spectrum described above.

    \subsection{The topological string and Jacobi forms}
    On an elliptically fibered Calabi-Yau manifold $X$, an astute rewriting of the holomorphic anomaly equations \cite{Bershadsky:1993ta,Bershadsky:1993cx} can be used to demonstrate that the topological string partition function inherits modular properties of the elliptic fiber \cite{Hosono:1999qc,Klemm:2012,Huang:2015sta,Cota:2019cjx}: upon extracting a universal fiber independent contribution $Z_0$ and expanding in the base class $Q_B$,
	\be \label{eq:Zk_from_Ztop}
	Z_{top} = Z_0 \left( 1 + \sum_{k > 0} c_k(\mathbf{Q}) Q_B^k Z_k \right) \,,
	\ee
    with $c_k(\mathbf{Q})$ a coefficient on which we shall comment momentarily, the expansion coefficients $Z_k$ can be shown to be meromorphic Jacobi forms (with a simple multiplier due to a contribution from the Dedekind $\eta$ function). An alternative route towards unearthing this modular structure proceeds by identifying $Z_k$ with the elliptic genera of $k$ non-critical strings in the spectrum of the 6d theory describing F-theory compactified on $X$ \cite{Benini:2013a,Benini:2013xpa,Haghighat:2014vxa,Haghighat:2015ega,DelZotto:2016pvm,DelZotto:2017mee,Kim:2018gjo}. From either route, the following ansatz for $Z_k$ can be motivated:
    \be
	Z_k = \frac{1}{\eta^{n(k)}(q)} \frac{\cN}{\cD}(q,\mathbf{Q},\gs) \,,
	\label{eq:ZkAnsatz}
	\ee
    where $\cN$ and $\cD$ are holomorphic Jacobi forms with modular parameter the K\"ahler parameter $q$ of the generic fiber,
    \be \label{eq:generic_fiber}
	q = Q_0 \prod_i (Q_i)^{a_i} \,.
	\ee
	Here, $Q_0$ and $Q_i$, $i=1, \ldots, \rk(\mathfrak{g})$, denote exponentiated K\"ahler parameters of fibral curves ($Q_0$ being associated to the only curve among these which intersects the base $B$ of the fibration).\footnote{Note that in addition to these parameters, dependence on flavor fugacities can be introduced in the elliptic genus \cite{Kim:2018gjo,Gu:2020fem}. At the level of the geometry and the topological string partition function, this requires including additional divisors in $X$. For further discussion of flavor symmetry in the SCFT limit of 5d theories, see \cite{Bhardwaj:2020ruf,Bhardwaj:2020avz}.} The $a_i$ coincide with the marks\footnote{Note that this equation appears with the $a_i$ identified as comarks in several previous works. The distinction is of course irrelevant for simply laced groups.} of $\mathfrak{g}$. The elliptic parameters of the Jacobi forms are given by $Q_i$ as well as $g_{top}$, with the Weyl group $\weyl(\mathfrak{g})$ acting on the former.
    
    The denominator $\cD$ has a universal contribution present for all rank 1 models which depends only on $\tau$ and $\gs$. Its form is largely fixed by comparison with the Gopakumar-Vafa expansion \cite{Gopakumar:1998ii,Gopakumar:1998jq} of the topological string:
	\be \label{eq:denom_uni}
	\cD_{univ} = \prod_{m=1}^k \phi_{-2,1}(m \gs) \,.
	\ee
	For all $X$ leading to gauge symmetry, the denominator also depends on the K\"ahler parameters of the resolved curves in the fiber. The expression for this contribution that we will use is derived in \cite{Kim:2018gak} by lifting the result for $Z_{top}$ in the gauge theory limit \cite{Nekrasov:2002qd} to 6d \cite{Hollowood:2003cv}. Recall \cite{Bernard:1977nr} that an instanton solution for the gauge group $SU(2)$ \cite{Belavin:1975fg} can be embedded into the gauge group $G$  via the embedding of the gauge algebra $\mathfrak{a}_1$ into $\mathfrak{g}$, with image a generator $T^{\alpha}$ of the Cartan subalgebra associated to a given root $\alpha$ and the corresponding lowering and raising operators. The bilinear form on $\mathfrak{g}$ takes the form
    \begin{equation}
        \text{tr}(T_{a}^{\alpha} T_{b}^{\alpha}) = c_\alpha \delta_{ab}
    \end{equation}
    on these three generators. Choosing the normalization of the bilinear form such that $c_\alpha$ is equal to 1 for all long roots, the constant takes the following values for short roots:
    \begin{equation} \label{eq:def calpha}
    \begin{aligned}
        c_\alpha &= 2, \quad G = B_n, C_n\ \text{and}\ F_4,\\
        c_\alpha &= 3, \quad G = G_2.
    \end{aligned}
    \end{equation}
    In the following, we will drop the index $\alpha$: $c$ will refer to the appropriate value given in \eqref{eq:def calpha}.
    An $SU(2)$ instanton with instanton number $k_{SU(2)}$ maps under this embedding to a $G$ instanton with instanton number $k_G = c_\alpha k_{SU(2)}$.
    
	Following \cite{Kim:2018gak}, we identify the contribution of an $SU(2)$ instanton of instanton number $k$ to the denominator $\cD$ as
	\be \label{eq:denom_SU2}
	\cD^{A_1}_{k,\alpha} =	 \prod_{ab \le k, a,b >0} \phi_{-1,\frac{1}{2}} ((a-b) \gs + m_\alpha)  \phi_{-1,\frac{1}{2}} ((a-b) \gs - m_\alpha)
	\ee 
    $m_\alpha$ is the contribution of the gauge fugacity or K\"ahler parameter associated to the positive root $\alpha$ of $A_1$ to which we will return below. 
  
    The contribution for a general gauge group $G$ can then be expressed as
    \begin{equation}  \label{eq:denom_G}
        \cD^{\mathfrak{g}}_k=\cD_{k,L}^G\cD_{k/c,S}^\mathfrak{g} \,,
    \end{equation}
    with
    \begin{equation}
        \cD^{\mathfrak{g}}_{k,L}=\prod_{\alpha\in\Delta_L^+}\cD^{A_1}_{k,\alpha}, \quad \cD^{\mathfrak{g}}_{k,S}=\prod_{\alpha\in\Delta_S^+}\cD^{A_1}_{k,\alpha} \,,
    \end{equation}
    where we have indicated the set of positive long and short roots as $\Delta_L^+$ and $\Delta_S^+$ respectively.
   Note that \eqref{eq:denom_G} is invariant under permutations on the sets of long and short positive roots, and $\cD^{A_1}_{k,\alpha}$ is invariant under $\alpha \rightarrow -\alpha$. As the Weyl group $\weyl(\mathfrak{g})$ is a subgroup of the permutation group on all roots that does not mix long and short roots, this establishes the invariance of $\cD^{\mathfrak{g}}_k$ under its action. 
   
   Note further that at $k=1$, \eqref{eq:denom_G} is independent of $\gs$.
    
    The power $n(k)$ of the Dedekind $\eta$ function occurring in \eqref{eq:ZkAnsatz} has been determined from topological string considerations \cite{Huang:2015sta} for the minimal singularities over the base surfaces $\IF_n$ equal to $\IF_0$, $\IF_1$ and $\IF_2$ to be $-12 k C_B \cdot K = 12k(n-2)$. Here, $C_B$ is the base curve of the Hirzebruch surface, $K$ is its canonical divisor. In \cite{DelZotto:2016pvm}, it was given for minimal (i.e. maximally Higgsed) models for $n>2$ as $n(k)= 4k h_G^\vee$, with $h_G^\vee$ the dual Coxeter number of the gauge group of the corresponding model, which happens to be given by $h_G^\vee = 3(n-2)$ for all occurring cases, as noted by \cite{Shimizu:2016lbw}. By matching to Gromov-Witten invariants, we find that expressing $n(k)$ in terms of the dual Coxeter number of the gauge group is misleading. In fact, it is the self-intersection number of the base curve which determines this quantity. The correct expression valid for all rank 1 Higgsing tree geometries is
	\be
	n(k) = 12 k |n-2| \,.
	\ee

    Turning now to the prefactor $c_k(\mathbf{Q})$ which enters in extracting $Z_k$ from $Z_{top}$ in \eqref{eq:Zk_from_Ztop}, it was given in \cite{Huang:2015sta} as $c_k(\mathbf{Q}) = q^{-\frac{k(n-2)}{2}}$ for the minimal models over $\IF_0, \IF_1, \IF_2$. For the minimal models over $\IF_n$, $n>2$, \cite{DelZotto:2016pvm} identified it as $c_k(\mathbf{Q}) = (\sqrt{q}/\prod_i Q_i^{a_i})^{k h_G^\vee/3}$\,. We find that this latter formula should be modified by adding a factor of $Q_0$ to the product in the denominator, and replacing $h_G^\vee$ by $3(n-2)$ for all geometries over a base $\IF_n$. This yields an expression valid for all bases $\IF_n$:
	\be
	c_k(\mathbf{Q}) = \left( \frac{\sqrt{q}}{\prod_i Q_0 Q_i^{a_i}}\right)^{k(n-2)} = q^{-\frac{k(n-2)}{2}} \,.
	\ee    
	
	The final ingredient is the numerator $\cN$ of $Z_k$. Its exact expression depends sensitively on the geometry considered. Upon determining the appropriate ring $J$ of holomorphic Jacobi forms in which it lies, an ansatz can be made in terms of the finite basis of $J$ at appropriate weight and index. The expansion coefficients must then be fixed by imposing appropriate boundary conditions, as we discuss in subsection \ref{ss:bcs}. Beyond the problem of providing sufficient boundary conditions, the sheer number of coefficients to be determined quickly becomes computationally untenable. By imposing the symmetries of the massless spectrum of the 5d theory on all of $Z_k$, the number of coefficients can be sufficiently reduced to render many more calculations feasible. In this work, we provide an a posteriori justification for this procedure by demonstrating that the constrained ansatz is consistent with Gromov-Witten invariants obtained via mirror symmetry.

	\subsection{The map between the K\"ahler cone and elliptic parameters} \label{ss:matching_elliptic_parameters}
	The exceptional fibral curves of the class of elliptic fibrations we are considering organize themselves in terms of representations of the corresponding Lie algebra $\mg$. As such, each curve can be identified with an element of the weight lattice $\LambdaW$ of $\mg$. It is therefore natural to identify the fiber components $m$ of the K\"ahler form with an element of the complexified dual lattice, the coroot lattice, such that the K\"ahler parameter $m_C$ associated to the curve $C$, obtained by integrating the complexified K\"ahler form against the curve class, is given by the pairing
	\be \label{eq:def_m_omega}
	m_\omega = (\omega, m)  \,,
	\ee
	with $\omega \in \LambdaW$ the weight identified with $C$.
	
	$-2$ rational curves in the fiber of the elliptic fibration organize themselves into the adjoint representation. Each such curve thus maps to a root $\alpha$ of $\mg$, which just as any other weight lies in $\LambdaW$. The corresponding K\"ahler parameters
	\be
	m_\alpha = (\alpha, m)
	\ee
	are identified with the gauge fugacities of the elliptic genus. They have already featured in the formula \eqref{eq:denom_SU2} above, while the exponentiated K\"ahler parameters
	\be \label{eq:definition_exponentiated_Kaehler_parameter}
	Q_i = e^{2\pi (\alpha_i,m)} \,,
	\ee
    with $\alpha_i$ a simple root, appear in equation \eqref{eq:generic_fiber} above. Note that the dependence of the elliptic genus on the gauge fugacities will generically contain fractional powers of $e^{2\pi i m_\alpha}$, as the weight lattice is generically finer than the root lattice.
	
	In theories with a Lagrangian description, $m$ can be identified with the VEV of the real scalar field $\phi$ in the 5d gauge multiplet, under identification of the complexified coroot lattice with the Cartan subalgebra of $\mg$. This gives rise to masses for hypermultiplets as follows. The scalar fields $(Q, \tilde{Q})$ of a hypermultiplet transforming in the irreducible representation $\rho$ of the gauge group couple to $\phi$ via 
	\be
	(\tilde{Q} , \rho(\phi) Q) \,.
	\ee 
	Recall that $Q$ and $\tilde{Q}$ transform in dual representations; in the above formula, $(\cdot,\cdot)$ indicates the pairing between the dual spaces. $\phi$ acquiring a VEV gives rise to the mass term
	\be
	(\tilde{Q} , \rho(m) Q) \,.
	\ee 
	Decomposing the vector $Q$ in terms of weight eigenspaces, $Q = \sum_\lambda Q_\lambda$ (assuming non-degenerate eigenspaces for notational simplicity), this yields
	\be \label{hypermultiplet_mass_term}
	(\tilde{Q} , \rho(m) Q) = \sum_{\lambda, \tilde{\lambda}} (\tilde{Q}_{\tilde{\lambda}}, (\lambda,m) Q_\lambda ) = \sum_{\lambda} (\lambda, m) (\tilde{Q}_{\tilde{\lambda}}, Q_\lambda) \,,
	\ee	
	where we have denoted by $\tilde{\lambda}$ the conjugate weight to $\lambda$. Note that if the irreducible representation $\rho$ has highest weight $\lambda_h$, all the weights $\lambda$ occurring in the mass term \eqref{hypermultiplet_mass_term} are of the form
	\be
	\lambda = \lambda_h - \sum_i n_i \alpha_i \,, \quad n_i \in \IN \,,
	\ee
	where the sum is over the simple roots of $\mg$. The hypermultiplet in representation $\rho$ hence introduces dependence on the parameter $(\lambda_h,m)$ in addition to the parameters $(\alpha_i,m)$.
	
	As explained in the previous subsection, the numerator of $Z_k$ as presented in \eqref{eq:ZkAnsatz} is a Weyl invariant Jacobi form with, aside from $\gs$, $m$ featuring as the elliptic parameter. Correctly identifying the dependence on $m$ requires some care. We will mostly take the generators of the ring $J(\mathfrak{g})$ of Weyl invariant Jacobi forms as derived in \cite{Bertola} as our starting point. These depend on $\rk \mg$ parameters $x_i$, which determine a point in a Euclidean lattice $\IE_n$. The action of $\weyl(\mathfrak{g})$, the Weyl group of $\mg$, on these parameters, as well as their behavior under shifts by elements of $\Lambda_r(\mathfrak{g})$, the root lattice of $\mg$, follows from the embedding of the root lattice into $\IE_n$ (note that $n$ can be larger than $\rk(\mg)$; as is the case e.g. for the $A$-series).
	
	From our identification of K\"ahler parameters with \eqref{eq:def_m_omega}, we conclude that it is shifts of $m$ via elements of the coroot lattice which should be symmetries of the theory. When studying the elliptic genus for the Lie algebra $\mg$, we hence need to consider Weyl invariant Jacobi forms of the Lie algebra $\tilde{\mg}$ whose root lattice equals the coroot lattice of $\mg$. For simply laced groups, we can identify roots with the corresponding coroots as elements of the orthogonal lattice, and this distinction is irrelevant.\footnote{Note that in the following, it will be convenient to refer to all the roots of simply laced root systems as long.} For $F_4$ and $G_2$, root and coroot lattice are isomorphic, the map between the two does not however preserve the inner product: short roots are mapped to long coroots and vice versa. E.g., in our conventions, the set of coroots for $G_2$ as embedded in $\IE_3$ is given by 
	\begin{equation*}
	    \pm (e_i-e_j), \quad i\neq j, \quad \quad \pm(2e_i-e_j-e_k), \quad i\neq j\neq k\neq i;
	\end{equation*}
	while the set of roots is given by
	\begin{equation*}
	    \pm (e_i-e_j), \quad i\neq j, \quad \quad \pm\frac{1}{3}(2e_i-e_j-e_k), \quad i\neq j\neq k\neq i.
	\end{equation*}
	Finally, the root lattice of $B_n$ is isomorphic to the coroot lattice of $C_n$, and vice versa. We must hence use the Weyl invariant forms assigned to the Lie algebra $C_n$ in the conventions of \cite{Bertola} to describe $Z_k$ on a background leading to gauge symmetry $B_n$, and vice versa.
	
	\subsection{Determining weight and index}
	
	$Z_k$ is a weight 0 Jacobi form whose index is determined by the anomaly polynomial of the elliptic genus or equivalently by the holomorphic anomaly of the topological string partition function \cite{Gu:2017ccq}. The anomaly polynomial for the elliptic genus of a string carrying charges $Q_i$ (not to be confused with the exponentiated K\"ahler parameters; in the type IIB picture, these charges encode the class of the base curves $C_i$ that the D3 brane giving rise to the string is wrapping) is given by \cite{Ohmori:2014kda,Shimizu:2016lbw} 
    \ba
    I_4 &=& \frac{\interMatrix^{ij} Q_i Q_j}{2} \left(c_2(L) - c_2(R) \right) + \\
    && Q_i \left( \frac{1}{4h^\vee} \interMatrix^{ia} \tr_{adj} F_a^2 - \frac{2- \interMatrix^{ii}}{4} \left(p_1(T) - 2 c_2(L) - 2 c_2(R)\right) + h_{G_i}^\vee c_2(I) \right)\,,  \nn
    \ea
    where $\interMatrix^{ij}=-C_{i}\cdot C_j$ is (minus) the intersection matrix of the curves in the base, $c_2(L)$ and $c_2(R)$ are the second Chern classes associated to the left and right parts of the Poincaré symmetry $SU(2)_L\times SU(2)_R$ of the normal bundle of the string in the 6d spacetime, $c_2(I)$ is the second Chern class for the $SU(2)$ R-symmetry bundle, $p_1(T)$ is the first Pontryagin class of the tangent bundle of the 6d spacetime, and $F_a$ is the field strength associated to the curve $C_a$ in the base. This latter index $a$ includes compact curves associated to gauge fields indexed by $i$ above and non-compact ones associated to global symmetries. 
    
    Specializing to rank 1, i.e. to the case of only one compact cycle in the base $B$, giving rise to one tensor multiplet, making the replacement $c_2(R)=c_2(I)=0,\,c_2(L)=-g_s^2$  for the unrefined string \cite{DelZotto:2016pvm, Gu:2017ccq,DelZotto:2017mee}, and introducing the norm
    \be
    ( \cdot, \cdot) = \frac{1}{2 h^\vee} \tr_{\!\!adj}   
    \ee
    on the Cartan subalgebra of the Lie algebra, which is normalized so that short coroots have norm squared 2, we obtain the index bilinear form for $Z_k$,
    \begin{equation}
    I_Z=i_{Z,top}g_{top}^2+i_{Z,gauge}(m,m)
    \end{equation}
    with
    \begin{align}
    &i_{Z,top}=-\frac{1}{2}(nk^2+(2-n)k) \,,\\
    &i_{Z,gauge}=-kn \,.
    \end{align}
    Here, $-n=\interMatrix^{11}=C_B\cdot C_B$ is the self-intersection number of the base curve, and we have replaced $F$ by $m$, which we will use from now on.
	
	To compute the index of the denominator in the presentation \eqref{eq:ZkAnsatz} for $Z_k$, we add the index bilinear form of each factor. For the universal part $\cD_{univ} $ in \eqref{eq:denom_uni}, it is given by
	\begin{align*}
		\sum_{m=1}^k m^2g_{top}^2 \,.
	\end{align*} 
	For the gauge group contribution $\cD_{k}^G$, the index bilinear form is
	\begin{equation*}
	\left(|\Delta_L|\sum_{\substack{ab\leq k\\ a,b>0}}(a-b)^2+|\Delta_S|\sum_{\substack{ab\leq k/c\\a,b>0}}(a-b)^2\right)g_{top}^2+\abSolLong 2h^\vee(m,m)_L+\abSolShort 2h^\vee(m,m)_S \,,
	\end{equation*}
	where 
	\begin{equation}
	    \abSolLong = \#\{ab\leq k \,|\, a,b>0\} \,, \quad \abSolShort = \#\{ab\leq k/c \,|\,a,b>0\}\,,
	\end{equation}
	and 
	\begin{equation}
	    (m,m)_{S/L}=\frac{1}{2h^\vee}\sum_{\alpha\in\Delta_{S/L}}m_\alpha \,.
	\end{equation}
	As the Weyl group does not mix short and long roots, the two inner products $(\cdot,\cdot)_{S/L}$ are Weyl invariant and therefore proportional to the inner product of the lattice,
	\begin{equation}
	(m,m)_S=\shortNorm_G(m,m), \quad (m,m)_L=(1-\shortNorm_G)(m,m).
	\end{equation}
    The value of $\shortNorm_G$ for all simple Lie algebras is given in table \ref{tab:j}.
    
    \begin{table}[ht]
    \centering
    \def\arraystretch{1.3}
    \begin{tabular}{c|ccccc}
    G& Simply laced & $ B_n$ & $ C_n$ & $ G_2$ & $ F_4$\\ \hline
    $\shortNorm_G$ & --- & $\frac{1}{2n-1}$ & $\frac{n-1}{n+1}$ & $\frac{1}{4}$ & $\frac{1}{3}$\\
    \end{tabular}
    \caption{$(m,m)_s=\shortNorm_G(m,m)$}
    \label{tab:j}
    \end{table}

    Combining these contributions, the index bilinear form of the denominator reads
    \begin{equation}
	I_{\cD}=i_{\cD,top} g_{top}^2+i_{\cD,gauge} (m,m) \,,
    \end{equation}
    where
    \begin{align}
    &i_{\cD,top}=\sum_{m=1}^k m^2+|\Delta_L|\sum_{\substack{ab\leq k\\ a,b>0}}(a-b)^2+|\Delta_S|\sum_{\substack{ab\leq k/c\\a,b>0}}(a-b)^2,\\
    &i_{\cD,gauge}=2h^\vee\left((1-\shortNorm_G)\abSolLong+\shortNorm_G\abSolShort\right).
    \end{align}

    By $I_{\mathcal N}=I_Z+I_\cD$, the index bilinear form of the numerator $\cN$ of \eqref{eq:ZkAnsatz} is then easily determined to be
    \begin{equation}
    I_{\mathcal N}=i_{\mathcal N,top}g_{top}^2+i_{\mathcal N,gauge}(m,m)\,,
    \end{equation}
    with
    \begin{align}
    &i_{\mathcal N,top}=-\frac{1}{2}(nk^2+(2-n)k)+\sum_{m=1}^k m^2+|\Delta_L|\sum_{\substack{ab\leq k\\ a,b>0}}(a-b)^2+|\Delta_S|\sum_{\substack{ab\leq k/c\\a,b>0}}(a-b)^2 \,, \label{eq:topStringIndex}\\
    &i_{\mathcal N ,gauge}=-kn+2h^\vee\left((1-\shortNorm_G)\abSolLong+\shortNorm_G\abSolShort\right)\,.\label{eq:gaugeIndexNumerator}
    \end{align}
    For the case $k=1$, $i_{\mathcal N,top}=0$ and we conclude that the numerator does not depend on $g_{top}$.
	
    To determine the weight $w_\cN$ of the numerator, note that the weights of $\cN$ and $\cD$ must be equal, as $Z_k$ has weight 0. The factor involving the Dedekind $\eta$ function in equation \eqref{eq:ZkAnsatz} contributes the weight $\frac{1}{2}n(k)=6k|n-2|$, the universal contribution $\cD_{univ}$ to the denominator has weight $-2k$, and the gauge contribution $D_k^G$ has weight $-|\Delta_L|\abSolLong-|\Delta_S|\abSolShort$. Adding these contributions, we obtain
    \begin{equation} \label{eq:weightNumerator}
        w_\cN=6k|n-2|-2k-|\Delta_L|\abSolLong-|\Delta_S|\abSolShort\,.
    \end{equation}

	\subsection{Imposing boundary conditions} \label{ss:bcs}
	As the numerator $\cN$ in the ansatz \eqref{eq:ZkAnsatz} for $Z_k$ is a holomorphic Jacobi form, it can be expanded in terms of a finite basis of forms of given weight \eqref{eq:weightNumerator} and index \eqref{eq:topStringIndex} and \eqref{eq:gaugeIndexNumerator}. Most of this work will be concerned with identifying the ring of Jacobi forms best adapted to a given gauge theory, i.e. maximally constrained by the symmetries of the problem. Once the ring is chosen and the expansion of $\cN$ in appropriate generators is performed, appropriate boundary conditions must be imposed on $Z_k$ to determine the expansion coefficients. 
	
	In \cite{Huang:2015sta,Gu:2017ccq,DelZotto:2017mee,Duan:2020cta}, these boundary conditions are imposed in the form of so-called vanishing conditions: the constraint that Gopakumar-Vafa invariants of a given curve class must vanish at sufficiently high genus. In \cite{Huang:2015sta}, it was argued that imposing generic vanishing conditions (i.e. requiring that these invariants vanish eventually) is sufficient to fix $Z_k$ for theories without gauge symmetry; this argument was extended to the refined context in \cite{Gu:2017ccq}. In \cite{DelZotto:2017mee}, it was argued that imposing generic vanishing conditions does not suffice to fix $Z_k$ for theories with gauge symmetry. Imposing sharp vanishing conditions, it was conjectured, does suffice. This was demonstrated in the case of the $A_2$ theory over the base $\IF_3$ up to base wrapping number 3 and the $D_4$ theory over $\IF_4$ for base wrapping 1.
	
	An alternative would be to compute Gopakumar-Vafa invariants for these geometries by imposing elliptic blow-up equations \cite{Gu:2018gmy,Gu:2019dan,Gu:2019pqj,Gu:2020fem}. For the purposes of this work, we rely on the technically less arduous path of mirror symmetry and impose genus 0 Gromov-Witten invariants as boundary conditions. This allows us to fix $Z_k$ at base wrapping $k=1$
	completely, and some coefficients in the expansion of $\cN$ for higher $k$ (only in the case of the E-string (of arbitrary rank) can $Z_k$ for all $k$ be determined solely by imposing genus 0 invariants \cite{Duan:2018sqe}).
	
	To determine the Gromov-Witten invariants for the various nodes of rank 1 Higgsing trees, we construct, where possible, the underlying geometry as hypersurfaces in toric varieties \cite{Candelas:1996su,Bershadsky:1996nh,Perevalov:1997vw,Kashani-Poor:2019jyo} to which we apply well-established mirror symmetry techniques \cite{Candelas:1990rm,Hosono:1993qy}. We organize the invariants in terms of a basis of curve classes adapted to the gauge theory interpretation by identifying the distinguished curves in the geometry as intersections of toric divisors with the hypersurface. We refer to \cite{Kashani-Poor:2019jyo} for a detailed exposition of these techniques.
	
	$C_n$ and $D_n$ singularities cannot be imposed torically on the elliptic fiber over base a Hirzebruch surface $\IF_n$ \cite{Bershadsky:1996nh, Kashani-Poor:2019jyo}. We hence cannot compute the elliptic genus for theories with these gauge groups directly using our techniques. However, the fact that all $D_n$ theories in rank 1 Higgsing trees arise via Higgsing of theories that we can solve provides us with an alternative path to obtaining their elliptic genus, as we explain in section \ref{sec:moving}. Reversing the arrow of dependencies, we can thus compute the genus 0 Gromov-Witten invariants for these geometries, invariants which are not available via traditional mirror symmetry techniques.
	
	\section{Specializing along Higgsing trees} \label{sec:moving}
	
	Descending via Higgsing from a theory with gauge group $\mathfrak{g}$ imposes constraints on the spectrum of charged matter of the resulting theory with gauge group $\mathfrak{g'}$. While the nature of these constraints generically strongly depends on the details of the Higgsing considered, some general statements can be made. E.g., when $\rk(\mathfrak{g}) = \rk(\mathfrak{g'})$, the Weyl symmetry of $\mathfrak{g}$ decomposes into $\weyl(\mathfrak{g'}) \ltimes \dynkinsym(\mathfrak{g'})$, with the Dynkin diagram symmetries continuing to act as an automorphism on the theory. The charged matter spectrum of the Higgsed theory must therefore be invariant under the action of $\dynkinsym(\mathfrak{g'})$ on the representations of $\mathfrak{g'}$. We see many examples of this phenomenon in the rank 1 Higgsing trees:
	\begin{itemize}
	    \item $\weyl(G_2) = \weyl(A_2) \ltimes \dynkinsym(A_2)$: the Dynkin diagram symmetry exchanges the fundamental representations $\mathbf{3}$ and $\mathbf{\bar{3}}$ of $A_2$. Symmetry under this exchange is however already required by CPT invariance, hence does not constrain the $A_2$ gauge theories further.
	    \item $\weyl(F_4) = \weyl(D_4) \ltimes \dynkinsym(D_4)$: the Dynkin diagram symmetry of $D_4$ famously permutes its vector and two spinor representations, and indeed, all $D_4$ theories occurring in rank 1 Higgsing trees (all descending from $F_4$ theories via Higgsing) have a charged matter spectrum invariant under such permutations.
	    \item $\weyl{(B_n)} = \weyl(D_n) \ltimes \dynkinsym(D_n)$: the $\IZ_2$ Dynkin diagram symmetry of $D_n$ ($n>4$) exchanges the two spinor representations. For $n$ odd, these are conjugate to each other, hence the $\IZ_2$ symmetry is imposed by CPT invariance. However, for $n$ even, the spinor representations are self-conjugate, allowing for the presence of half hypermultiplets in the spectrum of these theories. Here, $\dynkinsym(D_n)$ is an additional constraint on the spectrum. In perusing the rank 1 Higgsing trees, we indeed observe that this symmetry is not realized only in the case of the $D_6$ theories of the $\IF_2$ and $\IF_3$ Higgsing trees; these are the only $D_{2n}$ theories not descending from a $B_{2n}$ theory.
	\end{itemize}
	In section \ref{sec:enhanced}, we will see that these symmetries of the massless spectrum of the theory are inherited by the elliptic genus. What is more, at least at the level of the elliptic genus, symmetries of a Higgsing theory can be ``reverse inherited" by the unHiggsed theory. Thus e.g., via the $B_4$ and $F_4$ to $D_4$ branch of a Higgsing tree, $\weyl(F_4)$ has repercussions for the $B_4$ theory, even though it does not descend via Higgsing from a theory with gauge group $F_4$.
	
	In this section, we will study the relation between the elliptic genera of theories related by Higgsing $\mg \rightarrow \mg'$ in detail. We will define for each Higgsing a linear embedding
    \begin{equation} \label{eq:def_iota}
        \iota : \mathfrak{h}' \hookrightarrow \mathfrak{h}
    \end{equation}
    between the Cartan subalgebras. In practice, these maps have simple presentations in terms of the orthogonal coordinates $x_i$ on the Euclidean lattices in which $\LambdaCR(\mg')$ and $\LambdaCR(\mg)$ are embedded. The map $\iota$ induces a map $\iota^*$ between functions on the complexified Cartan algebra and in particular the Jacobi forms associated to the corresponding Lie algebras. Up to some possible change of coordinates, dictated by our explicit embeddings of $\LambdaCR(\mg)$ in $\Eucl_\mg$, the map $\iota^*$ corresponds to restricting the modular forms to the subspace $\iota(\mathfrak h')$. We give these restrictions for all simple Lie algebras, except $E_6,\,E_7$ and $E_8$.
    
    It is natural to ask whether the relation between $Z^{\mathfrak g}$ and $Z^{\mathfrak{g}'}$ is governed by $\iota$. In all the models we consider, we find this to be the case, at least at base wrapping 1:
	\begin{equation}
	    \iota^*(Z^{\mathfrak g}) = Z^{\mathfrak g'}.
	    \label{eq:specializing}
	\end{equation}
	In particular, when the Higgsed gauge algebra has fewer roots, the apparent extra divergences in equation \eqref{eq:denom_G} disappear. 
	
	We will discuss this specialization separately for the numerator and the denominator of the ansatz \eqref{eq:ZkAnsatz} for the elliptic genus.
	
	\subsection{Restriction maps between rings of Jacobi forms} \label{ss:restricting_Jacobi_forms}
	The specialization map $\iota^*$ takes Jacobi forms of the Lie algebra $\mathfrak g$ to Jacobi forms of the Lie algebra $\mathfrak g'$. We begin by studying how to specialize the generators of the ring of Jacobi modular forms along the Higgsing trees. Some technical details are relegated to appendix \ref{app:specializationformulas}.
		
	\subsubsection{$G_2$ to $A_2$} \label{ss:Jacobi_G2_to_A2}
	The (co)root lattices of the Lie algebras $A_2$ and $G_2$ are isomorphic. We can therefore choose $\iota$ to be the identity. The Weyl groups of the two algebras are related via 
	\begin{equation}
	    \weyl(G_2)=\weyl(A_2)\ltimes \text{DynkinSym}(A_2).
	    \label{eq:WeylG2}
	\end{equation}
	The ring $J(G_2)$ of $G_2$ Jacobi forms is thus equal to the subring of the ring $J(A_2)$ of $A_2$ Jacobi forms whose elements are Dynkin diagram symmetric. At the level of the standard basis of the orthogonal lattice $\Eucl_3$, the latter symmetry is realized by exchanging the lattice generators $e_1$ and $e_3$ and flipping the sign of all three generators. Note that $e_1 \leftrightarrow e_3$ is already an element of $\weyl(A_2)$. Of the three generators of $J(A_2)$, only $\phi_{-3,1}^{A_2}$ is not invariant under $x_i \rightarrow - x_i$; it changes by a sign. This observation fixes the relation between the two sets of generators, as summarized in figure \ref{fig:jacobi_G2_to_A2}. 
	
	The relation between the generators of $J(A_2)$ and $J(G_2)$ in particular implies that the numerator $\cN$ of the elliptic genus of all $A_2$ theories obtained via Higgsing from a $G_2$ theory must permit an expansion in an even power of the generator $\phi_{-3,1}^{A_2}$. This however does not impose an independent constraint, as the weight of the numerator in the expression \eqref{eq:ZkAnsatz} for the elliptic genus is even for all $k$, and $\phi_{-3,1}^{A_2}$ is the only generator of odd weight.
\begin{figure}
	    \centering
	    \begin{tikzcd}
	     \phi_{0,1}^{G_2}\arrow[d]& \phi_{-2,1}^{G_2}\arrow[d]&\phi_{-6,2}^{G_2}\arrow[d]\\
	     \phi_{0,1}^{A_2}&\phi_{-2,1}^{A_2} &(\phi_{-3,1}^{A_2})^2
	    \end{tikzcd}
	    \caption{The map between $J(G_2)$ and $J(A_2)$ generators. The vertical arrows denote equality.}
	    \label{fig:jacobi_G2_to_A2}
	\end{figure}

	\subsubsection{$B_3$ to $G_2$} 
	
	All rank 1 models with $G_2$ gauge group arise via Higgsing of theories with gauge group $B_3$. We recall that the relevant ring of Jacobi forms for the latter, in the conventions of \cite{Bertola}, is $J(C_3)$, as its elements are shift symmetric under coroots of $B_3$. 
	We identify the Cartan algebras of $G_2$ and $B_3$ with the subspace $x_1+x_2+x_3=0$ of $\mathbb R^3$ and $
	\mathbb R^3$ itself. The map $\iota$ is then the inclusion
	\begin{equation*}
	    \iota:\{x_1+x_2+x_3=0\}\hookrightarrow \mathbb R^3.
	\end{equation*}
	
	The root lattice of $C_3$, as embedded in the orthogonal lattice $\Eucl_3$, reduces to the root lattice of $G_2$ upon restriction to the subspace $x_1+x_2+x_3=0$; likewise, the Weyl group of $C_3$ (equal to the Weyl group of $B_3$) maps to the Weyl group of $G_2$. The restriction $\iota^*$ hence provides a map from $J(C_3)$ to $J(G_2)$. In fact, this map is surjective: it maps one of the generators of $J(\Bn_3)$ to 0, and the other three to the generators of $J(G_2)$, see figure \ref{fig:jacobiB3}.
	
	\begin{figure}
	    \centering
	    \begin{tikzcd}
	     \phi_{0,1}^{C_3}\arrow[d]& \phi_{-2,1}^{C_3}\arrow[d]&\phi_{-4,1}^{C_3}\arrow[d]&\phi_{-6,2}^{C_3}\arrow[d]\\
	     \phi_{0,1}^{G_2}&\phi_{-2,1}^{G_2} & 0 &\phi_{-6,2}^{G_2}
	    \end{tikzcd}
	    \caption{The specialization of $J(C_3)$ to $J(G_2)$ generators. The vertical lines correspond to setting $x_1+x_2+x_3=0$ and multiplying by a constant.}
	    \label{fig:jacobiB3}
	\end{figure}

    \subsubsection{$F_4$ to $D_4$}
    
    All rank 1 models with $D_4$ gauge group arise via Higgsing of theories with gauge group $F_4$. 

    The root lattices of $D_4$ and $F_4$ are isomorphic, but are embedded differently in $\IR^4$ in the conventions of \cite{Bourbaki}. We give a map $\iota$ between these two realizations in appendix \ref{sec:appF4}.
    The Weyl group of $F_4$ coincides with the semi-direct product of the Weyl group of $D_4$ with the Dynkin diagram symmetry of $D_4$,
    \begin{equation*}
		W(F_4)=O(F_4)=W(D_4)\ltimes \text{DynkinSym}(D_4)=W(D_4)\ltimes S_3 \,.
	\end{equation*}
    Therefore, $\iota^*$ embeds $J(F_4)$ as a subring into $J(D_4)$, its elements consisting of $\weyl(D_4)$ symmetric Jacobi forms which in addition exhibit $D_4$ Dynkin diagram symmetry. Imposing this symmetry on the generic elements of $J(D_4)$ of appropriate weight and index, we arrive at the generators of $J(F_4)$ given in figure \ref{fig:jacobiF4}.
    
    \begin{figure}
	    \centering
	    \begin{tikzcd}[scale cd=0.9, column sep=0.3, row sep=15]
	     \phi_{0,1}^{F_4}\arrow[d]& \phi_{-2,1}^{F_4}\arrow[d]&\phi^{F_4}_{-6,2}\arrow[d]&\phi_{-8,2}^{F_4}\arrow[d]&\phi_{-12,3}^{F_4}\arrow[d]\\
	     \phi_{0,1}^{D_4}-\frac{2}{3}E_4 \phi_{-4,1}^{D_4}  &\phi_{-2,1}^{D_4} & \phi_{-6,2}^{D_4}-\frac{1}{18}\phi_{-2,1}^{D_4}\phi_{-4,1}^{D_4} & \left(\phi_{-4,1}^{D_4}\right)^2+3\left(\omega_{-4,1}^{D_4}\right)^2 &\phi_{-4,1}^{D_4}\left(\omega_{-4,1}^{D_4}\right)^2-\frac{1}{9}\left(\phi_{-4,1}^{D_4}\right)^3
	    \end{tikzcd}
	    \caption{The specialization of $J(F_4)$ to $J(D_4)$ generators. The vertical lines correspond to composition via an isomorphism of the the two lattices which is given explicitly in appendix \ref{sec:appF4}.}
	    \label{fig:jacobiF4}
	\end{figure}

	\subsubsection{$A$-series}
	
	The Higgsing tree over $\IF_1$ and over $\IF_2$ both exhibit a branch of $A_n$ gauge theories for $n$ arbitrarily large, with a sequence of Higgsings $A_{n+1} \rightarrow A_n$ all the way down to a theory with gauge symmetry $A_1$.\footnote{The Higgsing of certain theories over the bases $\IF_1$ and $\IF_2$ is studied from the perspective of the 2D quiver theory living on the BPS strings in \cite{Kim:2015fxa}.}
	
\begin{figure}
	    \centering
	    	\begin{tikzcd}[scale cd=1, column sep=10, row sep=15]
	\vdots \arrow[d]&\vdots \arrow[d]&\vdots \arrow[d]&\dots \arrow[d]&\vdots \arrow[d]& \vdots \arrow[d]& \vdots \arrow[d]\\
	\phi_{0,1}^{A_n}\arrow[d]&\phi_{-2,1}^{A_n}\arrow[d]&\phi_{-3,1}^{A_n}\arrow[d]&\dots\arrow[d]&\phi_{-n,1}^{A_n}\arrow[d] &\phi_{-n-1,1}^{A_n}\arrow[d] & 0\\
	\phi_{0,1}^{A_{n-1}}\arrow[d]&\phi_{-2,1}^{A_{n-1}}\arrow[d]&\phi_{-3,1}^{A_{n-1}}\arrow[d]&\dots\arrow[d]&\phi_{-n,1}^{A_{n-1}}&0\\
	\vdots \arrow[d]&\vdots \arrow[d]&\vdots \arrow[d]&\dots \arrow[d]&\vdots \\
	\phi_{0,1}^{A_2}\arrow[d]&\phi_{-2,1}^{A_2}\arrow[d]&\phi_{-3,1}^{A_2}\arrow[d]& 0\\
	\phi_{0,1}&\phi_{-2,1}& 0
	\end{tikzcd}
	    \caption{Restriction of the $A$-series Jacobi forms. The vertical arrows correspond to setting the last coordinate to 0 and multiplying by a constant.}
	    \label{fig:jacobiA}
	\end{figure}
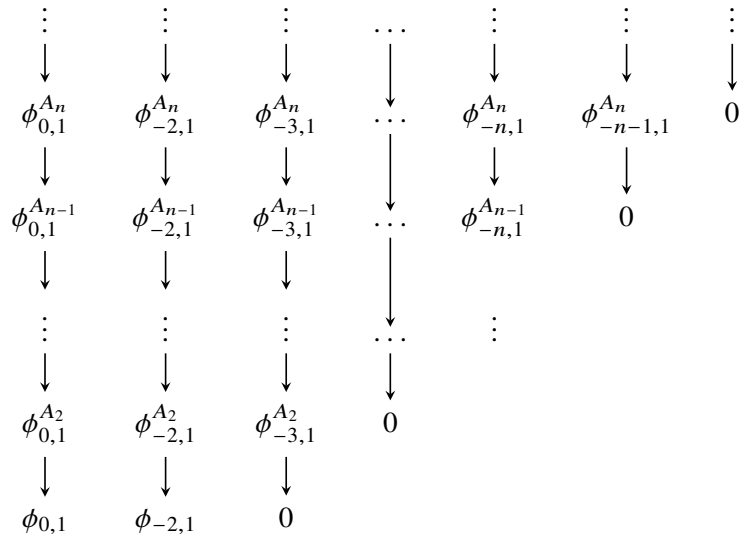

	The root lattice of $A_n$ is embedded in the orthogonal lattice $\Eucl_{n+1}$ via the constraint $\sum_{i=1}^{n+1} x_i = 0$. The Weyl group of $A_n$ is the group $S_{n+1}$ of permutations on the $n+1$ generators of $\Eucl_{n+1}$. The map $\iota$ for the $A_{n+1}\to A_n$ Higgsing is induced by the inclusion of the orthogonal lattices: $\Eucl_{n+1}\hookrightarrow \Eucl_{n+2}$. Restricting to the sublattice $x_{n+2}=0$ thus maps the root lattice and the Weyl group of $A_{n+1}$ to that of $A_{n}$. This restriction maps the lowest weight generator $\phi_{-(n+1),1}$ to 0, and otherwise preserves weight and index, leading to the relation between generators summarized in figure~\ref{fig:jacobiA}.
	
   \subsubsection{... $\rightarrow D_{n+1} \rightarrow B_n \rightarrow D_n \rightarrow$ ...}
   The Higgsing trees over $\IF_n$ for $n=0,1,2,3,4$ exhibit branches of $B_n$ and $D_n$ gauge theories for $n$ arbitrarily large. Along these branches, the pattern of Higgsing is $\ldots \rightarrow D_{n+1} \rightarrow B_n \rightarrow D_n \rightarrow \ldots$.\footnote{These branches over the bases $\IF_2$, $\IF_3$ and $\IF_4$ were studied from the perspective of brane-systems in \cite{Kim:2019dqn}.}
	
	Recall that in the conventions of \cite{Bertola}, $J(C_n)$ is the appropriate ring of Jacobi forms for the construction of the elliptic genera for $B_n$ gauge theories, as its elements are shift symmetric under $\LambdaCR(B_n) = \LambdaR(C_n)$.

    The orthogonal lattice for $C_n$ and $D_n$ is $\Eucl_n$. For the Higgising $B_n\to D_n$, the map $\iota$ is simply the identity. The root lattices of $C_n$ and $D_n$ coincide. Furthermore,
	\be
	W(C_n) = S_n \ltimes (\IZ_2)^n = W(D_n) \ltimes \text{DynkinSym}(D_n) \,;
	\ee
	in addition to permutations of the generators of the Euclidean lattice $\Eucl_n$, $W(C_n)$ includes arbitrary sign flips, whereas $W(D_n)$ includes only even numbers of sign flips. The generators of $J(D_n)$ and $J(C_n)$ can be chosen to reflect the close relation between these two groups: $n$ of the generators can be chosen to coincide (i.e. are in particular invariant under arbitrary sign flips). The final generator for $J(D_n)$ is odd under an odd number of sign flips. Its square provides the missing generator for $J(C_n)$.
	
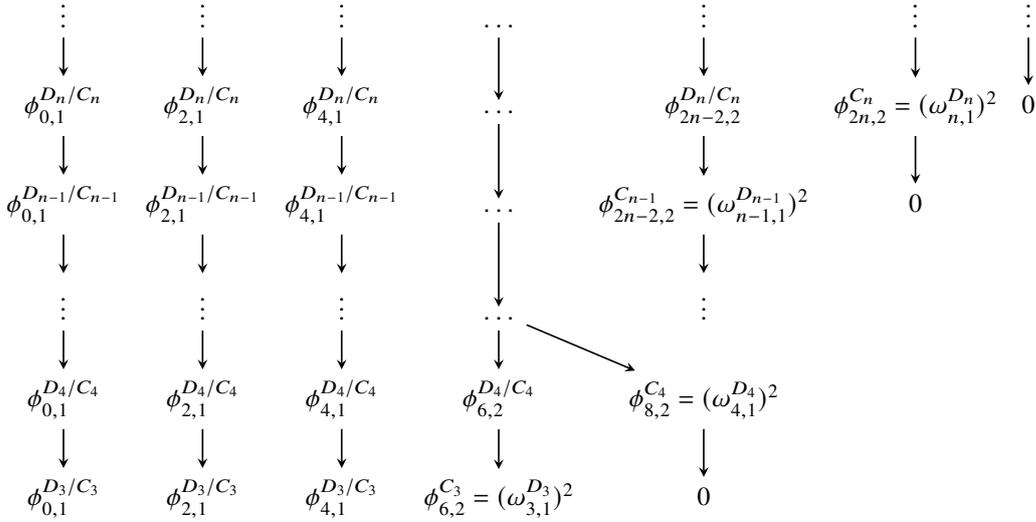
\begin{figure}
	    \centering
	    	\begin{tikzcd}[scale cd=0.9, column sep=0.3, row sep=15]
	\vdots \arrow[d]&\vdots \arrow[d]&\vdots \arrow[d]&\dots \arrow[d]&\vdots \arrow[d]& \vdots \arrow[d]& \vdots \arrow[d]\\
	\phi_{0,1}^{D_n/C_n}\arrow[d]&\phi_{2,1}^{D_n/C_n}\arrow[d]&\phi_{4,1}^{D_n/C_n}\arrow[d]&\dots\arrow[d]&\phi_{2n-2,2}^{D_n/C_n}\arrow[d] &\phi^{C_n}_{2n,2}=(\omega_{n,1}^{D_n})^2\arrow[d] & 0\\
	\phi_{0,1}^{D_{n-1}/C_{n-1}}\arrow[d]&\phi_{2,1}^{D_{n-1}/C_{n-1}}\arrow[d]&\phi_{4,1}^{D_{n-1}/C_{n-1}}\arrow[d]&\dots\arrow[d]&\phi_{2n-2,2}^{C_{n-1}}=(\omega_{n-1,1}^{D_{n-1}})^2\arrow[d] &0\\
	\vdots \arrow[d]&\vdots \arrow[d]&\vdots \arrow[d]&\dots \arrow[d]\arrow[rd]&\vdots \\
	\phi_{0,1}^{D_4/C_4}\arrow[d]&\phi_{2,1}^{D_4/C_4}\arrow[d]&\phi_{4,1}^{D_4/C_4}\arrow[d]&\phi_{6,2}^{D_4/C_4}\arrow[d]&\phi_{8,2}^{C_4}=(\omega_{4,1}^{D_4})^2\arrow[d] \\
	\phi_{0,1}^{D_3/C_3}&\phi_{2,1}^{D_3/C_3}&\phi_{4,1}^{D_3/C_3}&\phi_{6,2}^{C_3}=(\omega_{3,1}^{D_3})^2 & 0
	\end{tikzcd}
	    \caption{Restriction of the $D/C$ series Jacobi forms. The vertical arrows correspond to setting the last coordinate to 0 and multiplying by a constant (for $C_3$ we picked a different basis). This is the same table as in \cite{Adler_2020}.}
	    \label{fig:jacobiBD}
	\end{figure}

	For the $D_n\to B_{n-1}$ Higgsing, the map $\iota$ is induced by the inclusion $\Eucl_{n-1}\hookrightarrow \Eucl_{n}$. By restricting to $x_n=0$, the root lattice and Weyl group of $D_n$ are mapped to $\Lambda_r(C_{n-1})$ and $W(C_{n-1})$, respectively. Consequently, setting $x_n=0$ maps the generators of $J(D_n)$ to those of $J(C_{n-1})$. These relations are summarized in figure \ref{fig:jacobiBD}.
    
    For the case $n=4$, the restriction to $x_n=0$ does not yield the standard basis of $C_3$ Jacobi forms (as given for instance in \cite{Bertola}), which introduces some awkwardness in the reduction. The details are given in appendix \ref{app:specializationformulas}.
	
	\subsubsection{$C$-series}
	
	The numerator of the elliptic genus of models with $C_n$ gauge symmetry, which arise as nodes of the $\mathbb F_1$ Higgsing tree, take value in $J(B_n)$. As the necessary singularity enhancements of the elliptic fiber cannot be obtained torically \cite{Bershadsky:1996nh, Kashani-Poor:2019jyo}, we did not study these models in this work. However, as we will discuss in the next section, due to symmetry enhancement, the numerators of the elliptic genus of theories with gauge symmetry of $D$- and $B$-type over $\mathbb F_4$ are elements of these rings.
	
\begin{figure}
	    \centering
	    	\begin{tikzcd}
	\vdots \arrow[d]&\vdots \arrow[d]&\vdots \arrow[d]&\dots \arrow[d]&\vdots \arrow[d]& \vdots \arrow[d]& \vdots \arrow[d]\\
	\phi_{0,1}^{B_n}\arrow[d]&\phi_{-2,1}^{B_n}\arrow[d]&\phi_{-4,1}^{B_n}\arrow[d]&\dots\arrow[d]&\phi_{-(2n-2),1}^{B_n}\arrow[d] &\phi_{-2n,1}^{B_n}\arrow[d] & 0\\
	\phi_{0,1}^{B_{n-1}}\arrow[d]&\phi_{-2,1}^{B_{n-1}}\arrow[d]&\phi_{-4,1}^{B_{n-1}}\arrow[d]&\dots\arrow[d]&\phi_{-2(n-1),1}^{B_{n-1}}&0\\
	\vdots \arrow[d]&\vdots \arrow[d]&\vdots \arrow[d]&\dots \arrow[d]&\vdots \\
	\phi_{0,1}^{B_3}&\phi_{-2,1}^{B_3}&\phi_{-4,1}^{B_3}& 0
	\end{tikzcd}
	    \caption{Restriction of the $B$-series Jacobi forms. The vertical arrows correspond to setting the last coordinate to 0 and multiplying by a constant.}
	    \label{fig:jacobiB}
	\end{figure}
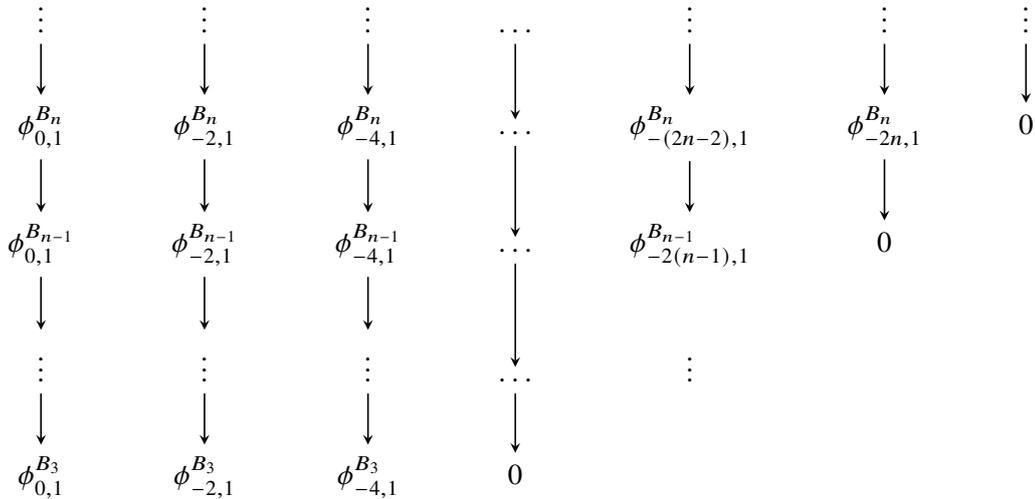

    The map $\iota$ for the $C_{n}\to C_{n-1}$ Higgsing is induced by the inclusion $\Eucl_{n-1}\hookrightarrow 
    \Eucl_n$. Once we set $x_n=0$, the Weyl group and root lattice of $B_n$ map to the Weyl group and root lattice of $B_{n-1}$, yielding the simple transformation law in figure \ref{fig:jacobiB}.
	
	\subsection{Specialization of the elliptic genus} \label{ss:specDen}
	
	The structure of the denominator as given in \eqref{eq:denom_G} is dictated by the roots of the Lie algebra. As explained in section \ref{ss:matching_elliptic_parameters}, we choose conventions for the roots in the orthogonal basis (i.e. the embedding of roots in a Euclidean lattice) such that the coroots of the Lie algebra $\mathfrak g$ coincide with the expressions as given by \cite{Bourbaki} for the roots of the dual Lie algebra. We recall that this is the natural normalization for us as the argument of the elliptic genus is an element of the (complexified) Cartan algebra $\mathfrak h_{\mathbb C}$ periodic under translation by elements of the coroot lattice (rather than the root lattice).
    
    In subsection \ref{ss:restricting_Jacobi_forms}, we defined for each Higgsing $\mathfrak{g} \rightarrow \mathfrak{g}'$ a restriction map $\iota^*$. Under this mapping, the image of a $\weyl(\mg)$ invariant function is $\weyl(\mg')$ invariant. Furthermore, the positive roots of $\mg'$ are mapped by $\iota$ onto a subset of positive roots of $\mg$. It follows that 
     \begin{equation*}
        \frac{\iota^*(\cD_k^{\mg})}{\cD_k^{\mg'}}
    \end{equation*}
    is an element of the ring $J(\mg') \otimes J(A_1)$ (the second factor having elliptic parameter $\gs$). Our prediction is that this element factors out of $\iota^*(\cN_k^\mg)$, thus establishing \eqref{eq:specializing}.\footnote{Note that the power of the Dedekind $\eta$ function in \eqref{eq:ZkAnsatz} only depends on the base of the elliptic fibration.} Below, we perform this reduction explicitly for various Higgsings.

    Once we have established \eqref{eq:specializing}, we can invoke the specialization mechanism to improve our ansatz for $Z_k$ as we move away from the root of a Higgsing tree. We distinguish between two cases.
    
    The first case is when the map induced by $\iota^*$ between $J(\mg)$ and $J(\mg')$ is one-to-one. In this case, we can impose the reduced denominator $(\iota^*)^{-1}(\cD_k^{\mg'})$ for the theory with $\mg$ gauge symmetry. In fact, the elliptic genera $Z_k^\mg$ and $Z_k^{\mg'}$ as functions on the Euclidean lattice (which coincides for both algebras) coincide. The two differ as functions of gauge fugacities or K\"ahler parameters, as the map between the Euclidean lattice and the K\"ahler cone of the underlying geometries (explained in subsection \ref{ss:matching_elliptic_parameters}) differ.\footnote{This is why \cite{Haghighat:2014vxa,DelZotto:2017mee} could extract Gopakumar-Vafa invariants of the $D_4$ geometry over base $\IF_4$ from the $B_4$ geometry one node up the Higgsing tree, resolving an issue raised in footnote 14 of \cite{Kashani-Poor:2019jyo}.} In section \ref{sec:enhanced}, we will discuss symmetry enhancements which allow to choose $\cN^{\mathfrak{g}}$ to lie in a smaller ring than $J(\mg)$. For such an ansatz to be sufficient, it is necessary to impose the reduced denominator; otherwise, factors not invariant under the enhanced symmetry are required in the numerator to cancel corresponding terms in the denominator. This effect becomes apparent in the base degree 2 example that we study in section \ref{sec:HigherBaseDegree}.

    The second case arises when Higgsing to a theory with smaller rank. In this case, the map $\iota^*:J(\mg)\to J(\mg')$ has a non-trivial kernel. We again have
    \begin{equation*}
        \iota^*(\cN^{\mathfrak{g}})=\cN^{\mathfrak{g}'}\frac{\iota^*(\cD_k^{\mathfrak{g}})}{\cD_k^{\mathfrak{g}'}} \,,
    \end{equation*}
    i.e. 
    \begin{equation*}
        \cN^{\mathfrak{g}}\in (\iota^*)^{-1}\left(\cN^{\mathfrak{g}'}\frac{\iota^*(\cD_k^{\mathfrak{g}})}{\cD_k^{\mathfrak{g}'}}\right) \,.
    \end{equation*}
    Now, any two elements in this preimage differ by an element of the kernel. The kernel of $\iota^*$ is a principal ideal $\psi J(\mg)$ generated by an element $\psi \in J(\mg)$ which can be read off from the figures in section \ref{ss:restricting_Jacobi_forms}. If we pick a particular element in the preimage $\hat \phi\in (\iota^*)^{-1}\left(\cN^{\mathfrak{g}'}\frac{\iota^*(\cD_k^{\mg})}{\cD_k^{\mg'}}\right)$ (for instance by going to section  \ref{ss:restricting_Jacobi_forms} and following the arrows in reverse), we have that 
    \begin{equation*}
        \cN^{\mathfrak{g}}=\hat \phi +\psi\phi.
    \end{equation*}
   If one has already computed $\cN^{\mathfrak{g}'}$, this reduces the calculation of $\cN^{\mathfrak{g}}$ to the determination of $\phi$, whose weight and index are (in absolute value) smaller than those of $\cN^{\mathfrak{g}}$, thus reducing the number of coefficients that must be determined. In practice however, this reduction is not substantial: the weight and index of $\psi$ are small against those of $\cN^\mg$ already at $k=1$; as $\psi$ is a fixed form while the weight and index of the numerator increase rapidly with base degree $k$, the reduction becomes even more marginal as $k$ increases.
    
    \subsubsection{$G_2$ to $A_2$}
    
    As the long roots of $A_2$ and $G_2$ coincide in our conventions (see the discussion at the end of subsection \ref{ss:matching_elliptic_parameters}),  
    \begin{equation*}
        \cD_{k,L}^{G_2}=\cD_{k,L}^{A_2} \,.
    \end{equation*}
    For the short roots, we have for $m \in \LambdaCR(G_2) \otimes \IC$ that by $\sum_{i=1}^3 x_i = 0$,
    \begin{equation*}
        \left(\frac{2}{3}e_i-\frac{1}{3}e_j-\frac{1}{3}e_k,m\right)=\frac{2}{3}x_i-\frac{1}{3}x_j-\frac{1}{3}x_k=x_i,\quad i\neq j\neq k \neq i.
    \end{equation*}
    Thus,
    \begin{equation*}
        \cD_{k,S}^{G_2}=\prod_{i=1}^3\cD^{A_1}_{k,e_i}.
    \end{equation*}
    
    As the map $\iota^*: J(G_2)\to J(A_2)$ is one-to-one, equation \eqref{eq:specializing} actually implies that we can refine our ansatz by taking the $A_2$ denominator in all of the theories with $G_2$ gauge symmetries, as all of them can be Higgsed to $A_2$. 
    
    We have computed $Z_1$ for all $A_2$ and $G_2$ nodes of rank 1 Higgsing trees (these occur over $\IF_n$, $n=0, \ldots, 3$). 
    At $k=1$, only the long roots contribute to the denominator. Hence,
	\be \label{eq:denominators_A2_G2_coincide}
	\cD_1^{A_2}=\cD_1^{G_2} \,.
	\ee
	We have verified that as functions of orthogonal coordinates $x_i$,
	\begin{equation*}
	    Z_1^{G_2}=Z_1^{A_2} \,.
	\end{equation*}

    \subsubsection{$B_3$ to $G_2$}
    The map $\iota^*$ associated to the Higgsing $B_3\to G_2$ is implemented by imposing the condition $x_1+x_2+x_3=0$ on the orthogonal coordinates of the Cartan algebra of $B_3$. The six long roots of $G_2$, given by $\pm(e_i- e_j)$, coincide with 6 of the 12 long roots of $B_3$, while the contribution of the remaining long roots maps under $\iota^*$ to that of the short roots of $G_2$. Thus,
    \begin{equation*}
        \cD_{k,L}^{B_3}|_{\sum x_i=0}=\cD_{k,L}^{G_2}\prod_{\alpha=e_i+e_j}\cD_{k,\alpha}^{A_1}=\cD_{k,L}^{G_2}\cD_{k,S}^{G_2}
    \end{equation*}
    Furthermore, $\iota^*$ maps the contribution of the short roots of $B_3$ to the contribution of the short roots of $G_2$. Hence,
    \begin{equation*}
        \cD_{k,S}^{B_3}|_{x_1+x_2+x_3=0}=\cD_{k,S}^{G_2} \,.
    \end{equation*}
    We thus expect that $\iota^*(\cN_k^{B_3})$ factorizes to cancel the contribution $\cD_{k,S}^{G_2}$.
    
    We have computed $Z_1$ for all $B_3$ nodes of rank 1 Higgsing trees (these occur over the base $\IF_n$ for $n= 0, \ldots, 3$).\footnote{The maximally Higgsed theories over the bases $\IF_0$ and $\IF_2$ are equivalent \cite{Morrison:1996na}. This however is no longer the case once one moves up from the root of the trees \cite{Kashani-Poor:2019jyo}.}

     At $k=1$, $\cD_{k,S}^{G_2}$ evaluates to
    \begin{equation}
      \prod_{\alpha=e_i+e_j}\cD_{1,\alpha}^{A_1}=(-1)^3\prod_i\phi_{-2,1}(x_i)=-\phi_{-6,2}^{G_2} \,,
    \end{equation}
    and we verify that $\iota^*(\cN_1^{B_3})$ is indeed divisible by $-\phi_{-6,2}^{G_2}$, such that
   \begin{equation*}
        \left.Z_1^{B_3}\right|_{\sum x_i=0}=\left.\frac{1}{\eta^{n(1)}}\frac{\mathcal N^{B_3}}{\mathcal D^{B_3}}\right|_{\sum x_i=0}=\frac{1}{\eta^{n(1)}}\frac{\mathcal N^{G_2}}{\mathcal D^{G_2}}=Z_1^{G_2}\,.
    \end{equation*}

	\subsubsection{$F_4$ and $B_4$ to $D_4$} \label{ss:F4 and B4 to D4: denom}
	The respective maps $\iota$ map the long roots of $D_4$ to those of $B_4$, $F_4$ respectively. As $\iota: \mathfrak{h}_{D_4} \rightarrow \mathfrak{h}_{B_4}$ is the identity, we suppress it (as well as the corresponding $\iota^*$) in this subsection to lighten notation. $\iota$ in the following will thus refer to the map \eqref{eq: F4 to D4}. With this understanding, 
	\begin{equation*}
	    \iota^*\cD_{k,L}^{F_4}=\cD_{k,L}^{B_4}=\cD_{k,L}^{D_4}=\cD_k^{D_4} \,.
	\end{equation*}
    For the specializations $\iota^*(Z_k^{F_4}) = Z_k^{D_4}$ and $Z_k^{B_4} = Z_k^{D_4}$ (the latter as an identity of functions of the orthogonal lattice coordinates) to be correct, we thus expect 
	\be \label{eq:F4/B4toD4Factorization}
	\iota^*(\cN_k^{F_4}) = \iota^*(\cD_{k/c,S}^{F_4}) \cN_k^{D_4} \,, \quad \cN_k^{B_4} = \cD_{k/c,S}^{B_4} \cN_k^{D_4} \,.
	\ee
    We have computed $Z_1$ for the $B_4$ and $F_4$ nodes of rank 1 Higgsing trees over the base $\IF_n$ for $n=0, \ldots, 4$ and checked the specialization equations by verifying
    \be \label{eq:z1F4=z1B4}
    \iota^*(Z_1^{F_4}) = Z_1^{B_4} \,.
    \ee
    As explained in section \ref{ss:bcs}, we could not directly compute the elliptic genera for $D_4$, as we cannot compute the required Gromov-Witten invariants using toric methods. But having demonstrated that specialization holds by verifying \eqref{eq:z1F4=z1B4}, we can use our results to compute $Z_1^{D_4}$ and extract the associated Gromov-Witten invariants.
    A non-trivial check on our results is that they specialize correctly (for $\IF_n$, $n=0,\ldots,3$) to the appropriate $B_3$ gauge theories, whose elliptic genera we compute independently.
    
    In section \ref{sec:HigherBaseDegree}, we test the factorization \eqref{eq:F4/B4toD4Factorization} at base degree 2 and genus 0.

	\subsubsection{$A$-series}
	The map $\iota^*$ for the Higgsing $A_n \rightarrow A_{n-1}$ is given by
	\be
	\iota^* = \left. \cdot \right|_{x_{n+1}=0} \,.
	\ee
	The roots of $A_n$ are given by the roots of $A_{n-1}$ together with the set $\pm(e_i-e_{n+1})$, $i=1,\dots,n$. Therefore, 
	\begin{equation*}
 \left.\cD_{k}^{A_n}\right|_{x_{n+1}=0}=\cD_k^{A_{n-1}}\prod_{i=1}^{n}D_{k,e_i}^{A_1} \,.
	\end{equation*}
	 For the specialization $\iota^*( Z_k^{A_n}) = Z_k^{A_{n-1}}$ to be correct, we thus expect 
	 \be  \label{eq:factorization A-series}
	\iota^*(\cN_k^{A_n}) = \iota^*(\prod_{i=1}^{n}D_{k,e_i}^{A_1})\cN_k^{A_{n-1}} \,.
	\ee
	At base degree 1, the prefactor of $\cN_k^{A_{n-1}}$ in the above equation evaluates to
	\begin{equation*}
    \left.\prod_{i=1}^{n}D_{1,e_i}^{A_1}\right|_{x_{n+1}=0}= \prod_{i=1}^{n}\left(-\phi_{-2,1}(x_i)\right)=(-1)^{n}(\phi_{-n,1}^{A_{n-1}})^2 \,.
	\end{equation*}
	We have checked the factorization \eqref{eq:factorization A-series} for the gaugings $A_3 \rightarrow A_2$ over the bases $F_1$ and $F_2$.

    \subsubsection{... $\rightarrow D_{n+1} \rightarrow B_n \rightarrow D_n \rightarrow$ ...}
    The map $\iota^*$ for the Higgsing $D_n \rightarrow B_{n-1}$ is given by
	\be
	\iota^* = \left. \cdot \right|_{x_{n}=0} \,.
	\ee
    The long roots of $D_n$ are given by the long roots of $B_{n-1}$ together with the set $\pm(e_i\pm e_n)$, $i=1,\dots,n-1$, which gives rise to the second factor on the RHS of the following equality,
    \begin{equation*}
        \left.\mathcal D_k^{D_n}\right|_{x_n=0}=\mathcal D_{k,L}^{B_{n-1}}\prod_{i=1}^{n-1}(D_{k,e_i}^{A_1})^2
    \end{equation*}
    For the specialization $\iota^*( Z_k^{D_n}) = Z_k^{B_{n-1}}$ to be correct, we thus expect 
	 \be  \label{eq:factorization D to B}
	\iota^*(\cN_k^{D_n}) = \iota^*\left(\prod_{i=1}^{n}D_{k,e_i}^{A_1}\right)^2\cN_k^{B_{n-1}} \,.
	\ee
    At base degree 1, we have
    \begin{equation*}
        \left.\mathcal D_1^{D_n}\right|_{x_n=0}=\mathcal D_1^{B_{n-1}}\left(\prod_{i=1}^{n-1} \phi_{-2,1}(x_i)\right)^2 = \mathcal D_1^{B_{n-1}}(\phi_{-2(n-1),2}^{C_{n-1}})^2.
    \end{equation*}
    For $n=4$, we checked the factorization \eqref{eq:factorization D to B} and thus the specialization
    \be
    \left. Z_1^{D_4}\right|_{x_4=0}=Z_1^{B_3}
    \ee
    for the corresponding nodes of the rank 1 Higgsing trees over bases $\mathbb F_0,\mathbb F_1,$ and $\mathbb F_2$, with $Z_1^{D_4}$ computed as explained in subsection \ref{ss:F4 and B4 to D4: denom}
	
	The map $\iota$ associated to the Higgsing $B_n \rightarrow D_n$ is the identity, the induced map $\iota^*:J(C_n)\hookrightarrow J(D_n)$ is injective. The long roots of both gauge algebras coincide, hence 
	\begin{equation*}
	    \mathcal D_{k,L}^{B_n}=\mathcal D_k^{D_n}.
	\end{equation*}
	For the specialization $\iota^*( Z_k^{B_n}) = Z_k^{D_{n}}$ to be correct, we thus expect 
	 \be  \label{eq:factorization B to D}
	\iota^*(\cN_k^{B_n}) = \iota^*(D_{k/c,S}^{B_n})\cN_k^{D_{n}} \,.
	\ee
	Furthermore, as $\iota^*$ is injective, we expect
	\be \label{eq:ZB = ZD}
	Z_k^{B_n} = Z_k^{D_n}
	\ee
	as functions of orthogonal coordinates. 
	
	Since we cannot compute $Z_k^{D_n}$ directly (see the discussion in subsection \ref{ss:bcs}), we cannot check \eqref{eq:factorization B to D} and \eqref{eq:ZB = ZD} directly. However, given the numerous checks that our specialization formulae have passed, we feel confident in invoking \ref{eq:ZB = ZD} to identify the elliptic genera of $D_n$ gauge theories.
	
   \subsubsection{$C$-series}
   The map $\iota^*$ for the Higgsing $C_n \rightarrow C_{n-1}$ is given by
	\be
	\iota^* = \left. \cdot \right|_{x_{n}=0} \,.
	\ee
    The denominators $\cD_k^{C_n}$ are mapped to zero under $\iota^*$, as the root $e_n$ leads to a factor $\phi(2x_n)$ which vanishes at $x_n=0$. However, one of the generators of $J(B_n)$ also lies in the kernel of $\iota^*$, and must factor out of the numerators $\cN_k^{C_n}$ to cancel these apparent poles.
   
   The long roots of $C_n$ are given by the long roots of $C_{n-1}$ together with the roots $\pm e_n$.\footnote{Note that $(m,e_n)=2x_n$ as the inner product here is $2 dx^2$, see the $B$-series column in table \ref{tab:rootSystems1}.} Then 
   \begin{equation*}
        \cD^{C_n}_{k,L} =\cD_{e_n,L}^{A_1}\cD^{C_{n-1}}_{k,L}=-\phi_{-2,0}(2x_n)\prod_{\substack{ab \le k, a,b >0 \\ ab\neq 0}} \phi_{-2,0}[(a-b)g_{top}]\cD^{C_{n-1}}_{k,L} + \cO(x_n^2)\,.
   \end{equation*}
    The short roots of $C_n$ are given by the short roots of $C_{n-1}$ together with $\pm e_i\pm e_n$. Therefore,
    \begin{equation*}
        \left. \cD_{k,S}^{C_n} \right|_{x_n=0}=\prod_{i=1}^{n-1}(D_{k,e_i}^{A_1})^2\cD_{k,S}^{C_{n-1}} \,.
    \end{equation*}
   At base degree 1, these relations reduce to
   \begin{equation*}
       \cD^{C_n}_{1}=-\phi_{-2,0}(2x_n)\cD^{C_{n-1}}_{1}
   \end{equation*}
   
   We did not study models with $C_n$ gauge symmetry, as the Gromov-Witten invariants are not accessible torically, see the discussion in subsection \ref{ss:bcs}. They could presumably be solved by imposing vanishing conditions on Gopakumar-Vafa invariants, but we did not pursue this approach here.
    
    \subsubsection{$E_6$, $E_7$, $E_8$}
    Rank 6 puts already the smallest of the Lie algebras $E_6,E_7,E_8$ out of computational reach. We do however want to briefly discuss the specialization of the denominator for these theories. Similar to the $C$-series, the denominator for a theory with $E_{n+1}$ gauge symmetry vanishes once we restrict to the Cartan algebra of $E_n$.
    
    In somewhat more detail, the contribution of the roots $\pm(e_7+e_8)$ of $E_8$ to  the denominator cause it to vanish once we impose the $E_7$ constraint $x_7+x_8=0$:
    \begin{equation*}
        \cD_1^{E_8}\propto \alpha^2(x_7+x_8)\xmapsto{x_7+x_8=0}\alpha^2(0)=0 \,.
    \end{equation*}
    We expect the $E_8$ numerator to be proportional to $\alpha^2(0)$ once we restrict to $x_7+x_8=0$. Checking this claim however lies beyond the scope of this work: the ring $J(E_8)$ being the only ring of $\weyl(\mg)$ invariant Jacobi forms for $\mg$ a simple Lie algebra which is not polynomially generated \cite{Wirthmuller:Jacobi,Wang:2018fil}, the strategy employed in this paper does not extend straightforwardly to this case.
    
    For the $E_7\to E_6$ Higgsing, the story is very similar. The roots $\pm(e_7-e_6)$ lead to a divergence once we impose the $E_6$ constraint $x_7-x_6=0$: 
    \begin{equation*}
        \cD_1^{E_8}\propto \alpha^2(x_7-x_6)\xmapsto{x_7-x_6=0}\alpha^2(0)=0.
    \end{equation*}
    Again, we expect this divergence to be cancelled by the numerator. Contrary to the $E_8$ case, theories with $E_6$ and $E_7$ gauge symmetry could be solved with our techniques, given more computational power, or more patience, and this claim could thus be checked explicitly.
    
    \subsection{Specializing Gromov-Witten invariants} \label{ss:GWinv}
    The specialization in equation $\eqref{eq:specializing}$ implies that the Gromov-Witten invariants of theories related by Higgsing $\mg \rightarrow \mg'$ are closely related. Naively,
    \begin{equation} \label{eq:GW_expansion}
    F^{\mg'}=\sum_{\kappa\in H_2(X^{\mg'})} \gs^{2g-2}r^{\mg'}_{g,\kappa}Q^\kappa=\log Z^{\mg'}=\iota^*\log Z^{\mg}=\iota^* F^{\mg}=\sum_{\kappa\in H_2(X^{\mg})}\gs^{2g-2}r^{\mg}_{g,\kappa}\iota^*Q^\kappa\,,
    \end{equation}
    such that
    \begin{equation}
    \label{eq:gwinv}
    r^{\mg'}_{g,\kappa'}=\sum_{\substack{\kappa\in H_2(X^\mg)\\ \iota^*(Q^\kappa)=Q^{\kappa'} }}r^{\mg}_{g,\kappa}\,.
    \end{equation}
    This reasoning must however be applied with care: the map $\iota$ was introduced in \eqref{eq:def_iota} to relate Weyl invariant expressions. As such, composition with any Weyl transformation yields an equally suitable map. However, Weyl transformations on the parameter $m$ introduced in section \ref{ss:matching_elliptic_parameters} map the set $\{Q_i=\exp(2\pi i (\alpha_i, m))\}$ of exponentiated fibral K\"ahler parameters to a new set $\{\tilde{Q}_i\}$ involving negative powers of the $Q_i$ (a consequence of the fact that any Weyl transformation maps a Weyl chamber to a distinct Weyl chamber). They hence do not commute with the Taylor expansion, indicated in \eqref{eq:GW_expansion}, required to extract Gromov-Witten invariants from $Z_g$.  When comparing Gromov-Witten invariants, the appropriate representative in the Weyl orbit of the map $\iota$ needs to be chosen in order to relate the Taylor expansions in terms of positive powers of the respective exponentiated K\"ahler parameters. Note that $\iota^*$ only acts on the exponentiated fibral K\"ahler parameters, not on those of the base curve $Q_b$ nor of the general fiber $q$. The action on the affine parameter $Q_0$ follows from $Q_0=q/\prod Q_i^{a_i}$. 

    How to determine the appropriate representative?
    \begin{enumerate}
    \item  \label{how_to_1} When a choice exists such that that $\iota^* Q^{\mathfrak g'}$ is a monomial in only positive powers of $Q^{\mathfrak g}$, this uniquely determines the map $\iota$ which commutes with the Taylor expansion.
    
    \item \label{how_to_2} Else, if in terms of orthogonal coordinates on the coroot lattice, $\iota^*$ identifies the denominator of $Z^\mathfrak{g}$, in the presentation \eqref{eq:ZkAnsatz}, with that of $Z^\mathfrak{g'}$, the appropriate representative is the one which allows the identification (up to a constant) of the prefactors $p_k^\mg$ and $p_k^{\mg'}$ in the presentation 
    \be
    \cD_k = p_k(Q) (1 + \sum_{\kappa>0} a_\kappa Q^\kappa)
    \ee
    of the denominator, as functions of orthogonal coordinates. 
    
    For $k=1$, the prefactor takes the simple form
    \begin{equation}
        p_1(Q) = \exp(-(2i\pi)m_{\rho_L}) \,,
    \end{equation}
    where $\rho_L$ is the sum of all positive long roots.
    
    \item Finally, only in the case of specializations $A_n \rightarrow A_{n-1}$, neither condition \ref{how_to_1} nor condition \ref{how_to_2} apply. It turns out that the composition of the Weyl transformation $x_n\leftrightarrow x_{n+1}$ with the map $\iota$ which sets $x_{n+1}=0$ gives the correct transformation in this case.
    
    \end{enumerate}

    \subsubsection{$G_2$ to $A_2$}
    For the specialization from $G_2$ to $A_2$, no choice of $\iota$ exists which maps the fibral K\"ahler parameters for the $G_2$ geometry to monomials of the K\"ahler parameters for the $A_2$ geometry. However, the denominators of the associated $Z_1$ coincide, see the discussion around equation \eqref{eq:denominators_A2_G2_coincide}. We are hence in case \ref{how_to_2}. With 
    \begin{equation}
        m_{\alpha_1}^{G_2} = -x_1 \,, \quad m_{\alpha_2}^{G_2} = x_1 - x_2
    \end{equation}
    and 
    \begin{equation}
        m_{\alpha_1}^{A_2} = x_1 - x_2 \,, \quad m_{\alpha_2}^{A_2} = x_1 + 2 x_2 \,,
    \end{equation}
    we obtain
    \begin{equation}
        m_{\rho_L}^{G_2}=-2x_1-2x_4\,, m_{\rho}^{A_2}=4x_1+2x_2\,.
    \end{equation}
    The map $\iota$ proposed in section \ref{ss:Jacobi_G2_to_A2} is the identity. To match the quantities $m_{\rho_L}^{G_2}$ and $m_{\rho}^{A_2}$, we thus perform the Weyl transformation $x_1\mapsto -x_2\,, x_2\mapsto -x_1$. This yields the map
    \begin{align}
        Q_1^{G_2}&\mapsto\left(\frac{Q_2^{A_2}}{Q_1^{A_2}}\right)^{1/3}\,,\nonumber\\
        Q_2^{G_2}&\mapsto Q_{1}^{A_2} \,,
    \end{align}
    so that 
    \begin{equation}
        Q_0^{G_2}=\frac{q}{(Q_1^{G_2})^3(Q_2^{G_2})^2} \mapsto \frac{q}{Q_1^{A_2}Q_2^{A_2}}=Q_0^{A_2}\,.
    \end{equation}
    
    We conclude that the $G_2$ Gromov-Witten invariant associated to the curve $a_BC_B+a_0C_0+a_1C_1+a_2C_2$ coincides with the $A_2$ Gromov-Witten invariant associated to the curve $a_BC_B+a_0C_0+(a_2-\frac{a_1}{3})C_1+\frac{a_2}{3} C_2$:
    \begin{equation}
        r^{G_2}_{a_B,a_0,a_1,a_2}=r^{A_2}_{a_B,a_0,a_2-\frac{a_1}{3},\frac{a_1}{3}}\,.
    \end{equation}
    
    \subsubsection{$B_3$ to $G_2$}
    For the specialization $B_3$ to $G_2$, we are in case \ref{how_to_1}. Performing the Weyl transformation
    \begin{align*}
        x_1&\mapsto x_3 \,,\\
        x_2& \mapsto -x_2 \,, \\
        x_3&\mapsto -x_1\,,
    \end{align*}
    we obtain the map
    \begin{align}
        Q_1^{B_3}&\mapsto Q_1^{G_2} \nonumber\\
        Q_2^{B_3}&\mapsto Q_2^{G_2} \nonumber\\
        Q_3^{B_3}&\mapsto Q_1^{G_2}\,,
    \end{align}
    and
    \begin{equation}
        Q_0^{B_3}=\frac{q}{Q_1^{B_3}(Q_2^{B3})^2(Q_3^{B_3})^2}\mapsto\frac{q}{(Q_1^{G_2})^3(Q_2^{G_2})^2}=Q_0^{G_2}\,,
    \end{equation}
    which involves only positive powers of the exponentiated K\"ahler parameters of the $G_2$ geometry. The relation among Gromov-Witten invariants which follows from this map is
    \begin{equation} \label{eq:gwB3G2}
        r^{G_2}_{a_B,a_0,a_1,a_2}=\sum_{a_3}r^{B_3}_{a_B,a_0,a_1-a_3,a_2,a_3} \,.
    \end{equation}
    Table \ref{tab:gwB3G2} in the appendix exemplifies this relation.
    
    \subsubsection{$F_4$ and $B_4$ to $D_4$}
    
    Both specializations $F_4 \rightarrow D_4$ and $B_4 \rightarrow D_4$ fall under case 2. For $D_4$ and $B_4$,
    \begin{equation*}
    m_{\rho}^{D_4}=6x_1+4x_2+2x_3=m_{\rho_L}^{B_4}\,,
    \end{equation*}
    while for $F_4$, the choice for $\iota$ corresponding to \eqref{eq: F4 to D4} yields
    \begin{equation}
        m_{\rho_L}^{F_4}\xmapsto{ \iota}6x_1+4x_2+2x_3\,.
    \end{equation}
    It follows that no additional Weyl transformations are needed. We obtain the maps
    \begin{equation}
        Q_1^{B_4} \mapsto Q_1^{D_4} \,,\,\, Q_2^{B_4} \mapsto Q_2^{D_4} \,,\,\, Q_3^{B_4} \mapsto Q_3^{D_4} \,,\,\, Q_4^{B_4} \mapsto \sqrt{\frac{Q_4^{D_4}}{Q_3^{D_4}}} \,, 
    \end{equation}
    and
    \begin{equation}
        Q_1^{F_4} \mapsto Q_2^{D_4} \,,\,\, Q_2^{F_4} \mapsto Q_1^{D_4} \,,\,\, Q_3^{F_4} \mapsto \sqrt{\frac{Q_3^{D_4}}{Q_1^{D_4}}}  \,,\,\, Q_4^{F_4} \mapsto \sqrt{\frac{Q_4^{D_4}}{Q_3^{D_4}}}  \,, 
    \end{equation}
    such that
    \begin{equation}
    \begin{array}{ccccc}
    \frac{q}{Q_1^{B_4}(Q_2^{B_4})^2(Q_3^{B_4})^2(Q_4^{B_4})^2}&=&\frac{q}{Q_1^{D_4}(Q_2^{D_4})^2Q_3^{D_4}Q_4^{D_4}}&=&\frac{q}{(Q_1^{F_4})^2(Q_2^{F_4})^3(Q_3^{F_4})^4(Q_4^{F_4})^2}\\
    Q_0^{B_4}&=&Q_0^{D_4}&=&Q_0^{F_4}
    \end{array}
    \end{equation}
    
    This implies the following relation between Gromov-Witten invariants:
    \begin{equation}
       r^{B_4}_{a_B,a_0,a_1,a_2,a_3,a_4}=r^{D_4}_{a_B,a_0,a_1,a_2,a_3-\frac{a_4}{2},\frac{a_4}{2}}=r^{F_4}_{a_B,a_0,a_2,a_1+a_3,2a_3,a_4} \,.
        \label{eq:GVB4D4F4}
    \end{equation}
    Table \ref{tab:B4F4gw} in the appendix exemplifies these relations.

    \subsubsection{$A$-series}
    
    For the $A_3\to A_2$ specialization, we find that composing $\iota$ with the Weyl transformation $x_3\leftrightarrow x_4$ yields the correct map between K\"ahler parameters:
    \begin{align}
        Q^{A_3}_1&\mapsto Q^{A_2}_1\,,\nonumber\\
        Q^{A_3}_2&\mapsto \frac{(Q^{A_2}_2)^{1/3}}{(Q^{A_2}_1)^{1/3}}\,,\nonumber\\
        Q^{A_3}_3&\mapsto (Q^{A_2}_1)^{1/3}(Q^{A_2}_2)^{2/3}
        \,,
    \end{align}
    so that,
    \begin{equation}
        Q_0^{A_3}=Q_0^{A_2}\,.
    \end{equation}
    
    This gives the relation among the invariants 
    \begin{equation}
        r^{A_2}_{a_B,a_0,a_1,a_2}=\sum_{a_3}r^{A_3}_{a_B,a_0,a_1+a_2-a_3,3a_2-2a_3,a_3}\,.
    \end{equation}
    
    We conjecture that this generalizes to the general case $A_{n+1} \rightarrow A_n$, yielding the map 
    \begin{align}
        Q_i^{A_n}& \mapsto Q_i^{A_{n-1}}, \quad i=1,\dots,n-2\,, \nonumber\\
        Q_{n-1}^{A_n}& \mapsto \frac{(Q_{n-1}^{A_{n-1}})^{1/n}}{\prod_{i=1}^{n-2}(Q_{i}^{A_{n-1}})^{i/n}}\,,\nonumber\\
        Q_n^{A_n}&\mapsto\prod_{i=1}^{n-1}(Q_{i}^{A_{n-1}})^{i/n}\,,
    \end{align}
    so that 
    \begin{equation}
        Q_0^{A_n}=Q_0^{A_{n-1}}\,.
    \end{equation}
    
    The relation among the invariants is then
    \begin{equation*}
        r^{A_{n-1}}_{a_B,a_0,a_1,\dots,a_{n-1}}=\sum_{a_n}r^{A_{n}}_{a_B,a_0,a_1+a_{n-1}-a_n,\dots,a_i+i(a_{n-1}-a_{n}),\dots,a_{n-2}+(n-2)(a_{n-1}-a_n),na_{n+1}-(n-1)a_n,a_n} \,. 
    \end{equation*}

    \subsubsection{... $\rightarrow D_{n+1} \rightarrow B_n \rightarrow D_n \rightarrow$ ...}
    For the $B_n \rightarrow D_n$ specialization, we are in case \ref{how_to_2}. As $m_{\rho_L}^{D_n}$ and $m_{\rho_L}^{B_n}$ coincide, no extra Weyl transformation is required. The map between K\"ahler parameters is
    \begin{align}
        Q_i^{B_n}&\mapsto Q_i^{D_n}\,, \quad\quad i=1,\dots,n-1\,, \nonumber \\
        Q_n^{B_n}&\mapsto \frac{(Q_n^{D_n})^{1/2}}{(Q_{n-1}^{D_{n}})^{1/2}}\,, \label{eq:gwBnDn}
    \end{align}
    so that 
    \begin{equation}
        Q_0^{B_n}=Q_0^{D_n}\,,
    \end{equation}
    implying the relation 
    \begin{equation}
        r^{B_n}_{a_B,a_0,a_1,\dots,a_n}=r^{D_n}_{a_B,a_0,a_1,\dots,a_{n-2},a_{n-1}-\frac{a_n}{2},\frac{a_n}{2}}
    \end{equation}
    among the Gromov-Witten invariants.
    
    Regarding the $D_n\to B_{n-1}$ Higgsing, the naive transformation reveals this to be a case 1 specialization, with map between K\"ahler parameters
    \begin{align}
        Q_i^{D_n}&\mapsto Q_i^{B_{n-1}}, \quad\quad i=1,\dots,n-1\,, \nonumber\\
        Q_n^{D_n}&\mapsto Q_{n-1}^{B_{n-1}} \,. \label{eq:gwDnBn-1}
    \end{align}
    
    The relation among the Gromov-Witten invariants that follows is
    \begin{equation}
         r^{B_{n-1}}_{a_B,a_0,a_1,\dots,a_{n-1}}=\sum_{a_n}r^{D_n}_{a_B,a_0,a_1,\dots,a_{n-2},a_{n-1}-a_n,a_n}=\sum_{a_n}r^{B_n}_{a_B,a_0,a_1,\dots,a_{n-2},a_{n-1},a_n} \,.
    \end{equation}
    We check this relation for the specialization $B_4 \rightarrow B_3$ in table \ref{tab:B4B3gw}.

	\section{Enhanced symmetries} 
	\label{sec:enhanced}
	The methods developed up to this point in this paper allow us to express the elliptic genus for any of the rank 1 models in terms of Weyl invariant Jacobi forms. However, depending on the matter representations present, some theories exhibit a higher symmetry than that implied merely by the gauge symmetry. Invoking this higher symmetry, the elliptic genus can be expanded in terms of a set of more restrictive Jacobi forms, in some cases drastically reducing the number of expansion coefficients which need to be determined.
	
	Symmetry enhancement due to the absence of certain matter representations is reminiscent of center symmetry in gauge theory: a given Lie algebra $\mg$ is compatible with a maximal center symmetry (the center of the simply connected Lie group among all Lie groups associated to $\mg$), but part or all of this symmetry is ruled out by certain matter representations. We will show that the symmetry enhancement we are seeing can indeed in some cases be interpreted in terms of center symmetry, though we also have examples in which the symmetry is enhanced beyond what full center symmetry would suggest.
	
	What is remarkable is that the symmetry enhancement that we see at the level of the massless matter spectrum, or equivalently, at the level of fibral curves, turns out to persist at the level of the entire elliptic genus. We present abundant evidence for this enhancement at base wrapping degree 1, and further evidence at base wrapping degree 2 in section \ref{sec:HigherBaseDegree}. To solve for the elliptic genus completely at higher base wrapping degree requires an alternative to genus 0 Gromov-Witten invariants as a source of boundary conditions. The elliptic genus for the $A_2$ theory over base $\IF_3$ was computed in \cite{DelZotto:2017mee} up to base degree 3 by invoking exact vanishing conditions on Gopakumar-Vafa invariants. Those results further corroborate the claim that symmetry enhancement persists beyond the massless spectrum. It would be important to establish this fact independently from explicit computations, both by arguing within the framework of gauge theory and, independently, geometrically.
	
	In terms of explicit computation, imposing the additional constraints dictated by enhanced symmetries puts otherwise burdensome models within computational reach. The extent of these simplifications is tabulated in table \ref{tab:reduction}.
	
	The symmetry enhancement in our computations rests on two pillars: the enhancement of the Weyl group, and the enhancement of the shift symmetry. We will discuss these two mechanisms in turn.
	
	\subsection{Enhancement of the shift symmetry by elements of the coweight lattice and center symmetry}
	Let $\omega$ lie in the subspace of $H_2(X,\IZ)$ spanned by the classes of exceptional fibral curves of $X$. The elliptic genus depends on the exponential of parameters
	\be
	(\omega, m) 
	\ee
	whenever the class $\omega$ is represented by a holomorphic curve. If a generating set of the entire weight lattice $\LambdaW$ is represented by such curves, the theory will possess the minimal amount of shift symmetry of $m$ compatible with the gauge symmetry, namely shift symmetry by the dual lattice to $\LambdaW$, the coroot lattice $\LambdaCR$. Conversely, the presence of enhanced gauge symmetry with Lie algebra $\mg$ implies that at least the classes $\alpha$ corresponding to roots of $\mg$ (in particular the simple roots $\alpha_i$) are represented by holomorphic curves. If only these classes are represented, the shift symmetry of $m$ will be enhanced from $\LambdaCR$ to the dual of the root lattice $\LambdaR$, the coweight lattice $\LambdaCW$ of $\mg$. The cases in between are when only some weights in $\LambdaW/\LambdaR$ are represented by holomorphic curves. Let $\lambda$ represent such a class. The shift symmetry of $m$ compatible with the presence of $\lambda$ will then be given by elements of the sublattice of $\LambdaCW$ for which
	\be
	\Lambda_{\mathrm{shift}}(\lambda) = \{ \omega^\vee \in \LambdaCW \,|\, (\lambda, \omega^\vee) \in \IZ \} \,. 
	\ee
	Now let us connect this discussion to center symmetry. Recall that the center of a Lie group equals the intersection of all possible choices of maximal tori, hence lies in the image of any choice $\mathfrak{h}$ of Cartan subalgebra of the Lie algebra under the exponential map. Recall also that for given Lie algebra $\mg$, the exponential map depends on the particular Lie group $G$ associated to $\mg$. For any $G$,
	\be
	Z = \exp(\LambdaCW) \,, \quad \LambdaCR \subset \ker(\exp) \,,
	\ee
    the identification $\mathfrak{h} \cong \LambdaCR \otimes \IC$ being understood. The kernel of the exponential map however depends on the choice of $G$. For $G$ simply connected, $\ker(\exp) = \LambdaCR$, hence
    \be
    Z \cong \LambdaCW/ \LambdaCR \,.
    \ee
	The question of the amount of center symmetry preserved by a representation $\boldsymbol{\lambda}$ of the group (we are using bold faced symbols to distinguish between representations of the group and of the algebra) hence amounts to deciding when an element of the center is represented trivially. The following lemma answers this question:
	\be
	\boldsymbol{\lambda}(\exp(\omega^\vee)) = 1 \quad \Leftrightarrow \quad (\lambda,\omega^\vee) \in \IZ \,.
	\ee
	We conclude that the lattice of shift symmetries of the elliptic genus due to the absence of certain matter representations is equal to $\exp^{-1}(Z)$, with the $Z$ the largest center compatible with the matter representations present.

	\subsection{Enhancement of the Weyl group by Dynkin diagram symmetries}
    At the level of the gauge theory, the Weyl symmetry acting on the Coulomb moduli of the theory is a remnant of the full gauge symmetry not fixed by the choice of which maximal torus the gauge symmetry is broken to. In contrast, symmetries of the Dynkin diagram of the gauge algebra, if respected by the matter representations present in the theory, give rise to an additional automorphism of the theory.

	\subsection{Further symmetry enhancements}
	We also encounter examples where the shift symmetry is enhanced even beyond the maximal amount of center symmetry compatible with the Lie algebra $\mg$, i.e. beyond $\LambdaCW(\mg)$, and the Weyl symmetry beyond the extension by the Dynkin diagram symmetry. All examples of such symmetry enhancements that we encounter arise when the specialization maps exhibited in the previous section are invertible (e.g. in the case $G_2 \rightarrow A_2$). In these cases, the theory with larger gauge group can "reverse inherit" the enhanced symmetry of the Higgsed theory. In such cases, the enhancement can be explained intrinsically (i.e. without reference to the Higgsed theory) by noting that certain weights associated to the matter representations present in the theory coincide with roots of $\mg$, and their contribution to the elliptic genus cancels. The cancellation is best described geometrically \cite{Kashani-Poor:2019jyo}: a root contributes the value -2 to the Gromov-Witten invariant of the associated curve class, and each half-hypermultiplet associated to a weight contributes +1. It is not at all evident that the enhanced symmetries of the lattice obtained due to this cancellation should be inherited by the full elliptic genus. In all of the examples we study, this however turns out to be the case. It would be interesting to study whether this symmetry enhancement has repercussions in the gauge theory beyond its effect on the elliptic genus.

	\subsection{Examples}

	\subsubsection{$A_2$ and $G_2$ over $\mathbb F_3$}
	The model with $A_2$ gauge symmetry over $\mathbb F_3$ has no matter. Hence, the shift symmetry is enhanced from $\LambdaCR(\mathfrak{a}_2)$ to $\LambdaCW(\mathfrak{a}_2)$, and the Weyl group symmetry is enhanced by the Dynkin diagram symmetry $\IZ_2$, yielding the Weyl group of $G_2$, see \eqref{eq:WeylG2}. Our task is thus to construct Jacobi forms invariant under the action of $W(G_2)$ and shifts by $\LambdaCW(\mathfrak{a}_2)$.
	
    Conveniently, the $A_2$ coweight lattice is isomorphic to the coroot lattice of $G_2$, upon multiplying the inner product of the former by a factor of 3.\footnote{We thank Haowu Wang for pointing this out to us, as well as further very useful comments both regarding the $A_2$ theory over base $\IF_3$ and the $D_4$ theory over base $\IF_4$ discussed further below.} With regard to the realizations of these two lattices in the conventions of Bourbaki \cite{Bourbaki}, an explicit isomorphism is given by 

    \begin{align*}
    \psi:&\left\{
    \begin{array}{rl}
         \frac{1}{3} (2 e_1^{A_2}-e_2^{A_2}-e_3^{A_2})&\mapsto e_1^{G_2}-e_2^{G_2} \\
         \frac{1}{3} (- e_1^{A_2}+2e_2^{A_2}-e_3^{A_2})&\mapsto e_2^{G_2}-e_3^{G_2}
    \end{array}
        \right.\\
         &\qquad \qquad \quad \quad  \big\Updownarrow \nonumber\\
    \psi:    x_1e_1^{A_2}+x_2e_2^{A_2}&+x_3e_3^{A_2} \mapsto (x_2-x_3)e_1^{G_2}+(x_1-x_2)e_2^{G_2}+(x_3-x_1)e_3^{G_2} \,.
    \end{align*}
	Note that this isomorphism pulls back the metric on $\LambdaCR(\mathfrak{g}_2)$ to 3 times the metric on $\LambdaCW(\mathfrak{a}_2)$, as announced: 
	\begin{equation*}
	    \psi^*\left(ds^2_{G_2}=\sum_i (dx_i)^2\right)=3\sum_i (dx_i)^2 \,.
	\end{equation*}
	The map $\psi$ thus induces via composition a map from $J(G_2)$ (whose elements for the purposes of the composition are considered as functions on $\LambdaCR(\mathfrak{g}_2)$) to $J(A_2)$ which triples the index. The image of this map is a subring of $J(A_2)$ generated by
    \begin{align} \label{eq:subring_of_W(G_2)}
    \phi_{0,3}^{A_2}&=\phi^{G_2}_{0,1}\circ \psi \,, \nn\\
    \phi_{-2,3}^{A_2}&=\phi^{G_2}_{-2,1}\circ \psi \,, \nn \\
    \phi_{-6,6}^{A_2}&=\phi^{G_2}_{-6,2}\circ \psi \,.
    \end{align}
    By the underlying assumption of this section that symmetries of the massless spectrum are inherited by the elliptic genus, the numerator $\cN$ in the ansatz \eqref{eq:ZkAnsatz} of the elliptic genus of the massless $A_2$ theory should be an element of this subring. The subring of Dynkin diagram and $\LambdaCW(\mathfrak{a}_2)$ shift symmetric forms was already identified in \cite{DelZotto:2017mee} as the appropriate ring in which $\cN$ should lie, and it was conjectured that this ring should be generated by the set of generators \eqref{eq:subring_of_W(G_2)}. We have now integrated the former observation into the larger context of enhanced symmetries of elliptic genera of rank 1 models, and provided a proof of the latter statement.

    In \cite{DelZotto:2017mee}, it was furthermore noticed that the function $\psi$ has a very simple expansion in terms of the coefficients $m_i=(m,\alpha_i)$, $m=m_1\omega_1^\vee+m_2\omega_2^\vee$:
    \begin{align*}
    m=&m_1\omega_1^\vee+m_2\omega_2^\vee=\frac{2m_1+m_2}{3}e_1^{A_2}+\frac{-m_1+m_2}{3}e_2^{A_2}+\frac{-m_1-2m_2}{3}e_2^{A_2}\\&\xmapsto{\psi} m_1 e_1^{G_2}+m_2 e_2^{G_2}+(-m_1-m_2)e_3^{G_2} \,.
    \end{align*}
	
	Now let us climb up one node in the Higgsing tree, and consider the theory with gauge group $G_2$ arising over the base $\IF_3$. As we discussed in detail in section \ref{sec:moving}, the elliptic genus specializes when moving from the branches towards the roots of the Higgsing tree. The specialization from a $G_2$ node to an $A_2$ node is particularly simple, see figure \ref{fig:jacobi_G2_to_A2}; in particular, it is invertible: the numerator $\cN$ of the elliptic genus of an $A_2$ theory obtained via Higgsing from a $G_2$ theory necessarily involves only even powers of the generator $\phi_{-3,1}^{A_2}$. Replacing the square of this generator by $\phi_{-6,2}^{G_2}$ and retaining the two other generators which coincide between the two gauge groups yields the numerator of the $G_2$ theory. This however poses a conundrum: the symmetry enhancement of the matterless $A_2$ theory just described must be reverse inherited by the $G_2$ theory with matter in the $\mathbf{7}$ representation. How does the enhancement of the shift symmetry to $\LambdaCW(\mathfrak{a_2})$ arise in this theory? The answer to this puzzle was already explained in general terms above: the weights of the $\mathbf{7}$ representation coincide with the short roots of $G_2$. The curves giving rise to the 2 half hypermultiplets in the $\mathbf{7}$ hence contribute to the same Gromov-Witten  invariants as those giving rise to the gauge fields associated to these roots, and in fact cancel the latter contribution. The remaining contributions are those due to the long roots. The lattice spanned by these is precisely the $A_2$ root lattice, with dual lattice $\LambdaCW(\mathfrak{a_2})$.
	
	Note that as the two fundamental representations of $A_2$ are complex conjugates, the spectrum of any $A_2$ gauge theory with $8$ supercharges will always be Dynkin diagram symmetric, hence cannot present an obstruction to the theory descending via Higgsing from a $G_2$ theory.

	\subsubsection{The $B_4$ and $F_4$ to $D_4$ branch}\label{ss:B_4_and_F_4_to_D_4_branch}
	
	All rank 1 Higgsing trees branch above a $D_4$ node, to a branch with adjacent node $B_4$, and another with adjacent node $F_4$. The matter content of the $D_4$ theory over the base $\IF_k$ is $ V^{\oplus (k-4)} \oplus S_+^{\oplus (k-4)} \oplus S_-^{\oplus (k-4)}$. In particular, it is invariant under the Dynkin diagram symmetry of $D_4$, consistent with the fact that the theory is obtained upon Higgsing an $F_4$ theory, given that the Weyl group of $D_4$ enhanced by the Dynkin diagram symmetry yields the Weyl group of $F_4$.
	
	By the hypothesis underlying this section, the numerator $\cN$ of the elliptic genus of the $D_4$ theory should therefore lie in the subring of $J(D_4)$ whose elements exhibit Dynkin diagram symmetry. As the root lattices of $D_4$ and $F_4$ are isomorphic, this subring is isomorphic to $J(F_4)$, as explained further in appendix \ref{sec:appF4}, hence spanned by $\{\phi_{-n,k}^{F_4} \circ i^{-1} \}$, the generators of $J(F_4)$ composed with the isomorphism $i^{-1} : \LambdaR(D_4) \rightarrow \LambdaR(F_4)$.
	
	The specialization of the elliptic genus for $B_n$ theories to that of $D_n$ theories is equally simple as that from $G_2$ to $A_2$ theories: all but one generator of the two rings $J(C_n)$ and $J(D_n)$ coincide.\footnote{Recall that $J(C_n)$ is the appropriate ring to expand the elliptic genus of $B_n$ theories in.} The remaining generator of the former, $\phi_{-2n,2}^{C_n}$, is the square of that of the latter, $\omega_{-n,1}^{D_n}$. This implies in particular that the expansion of the  numerator $\cN$ of the elliptic genus of the $D_4$ theory in $J(D_4)$ generators should exhibit only even powers of $\omega_{-4,1}^{D_4}$. However, as only even powers of $\omega_{-4,1}^{D_4}$ occur in the subring of $J(D_4)$ compatible with the $F_4$ Higgsing, this provides no further restriction on $\cN$.
	
	Inverting the specialization map from $B_4$ to $D_4$, we can conclude that $B_4$ must exhibit the same symmetry enhancement as the $D_4$ theory, i.e. from $W(B_4)$ to $W(F_4)$. To explain this enhancement intrinsically (i.e. without reference to the $D_4$ theory), we can invoke a more intricate realization of the mechanism at play in the transition from $G_2$ to $A_2$: the $B_4$ theories have matter content $V^{\oplus (4-k)+1}\oplus S^{\oplus (4-k)}$ over $\IF_k$. Two half hypermultiplets in the vector representation cancel the contribution of the short roots of $B_4$ to the Gromov-Witten invariants of the theory. The long roots of $B_4$ coincide with the roots of $D_4$, hence are invariant under $W(F_4)$. Furthermore, having cancelled one vector representation against the short roots,
	the remaining matter content, $(V \oplus S)^{\oplus (4-k)}$, is also invariant under $W(F_4)$, thus explaining the occurrence of this symmetry for these $B_4$ theories.

    \subsubsection{$D_4$, $B_4$ and $F_4$ over $\mathbb{F}_4$} \label{ss:enhanced symmetry D4 B4 F4}
    The symmetry enhancement of the $D_4$ theory over $\IF_4$ was already noted in \cite{DelZotto:2017mee}. The absence of matter means, once again, that the numerator $\cN$ of the elliptic genus is invariant under translations by the coweight lattice $\LambdaCW(D_4)$, and that the expansion is Dynkin-diagram symmetric. 

    The $D_4$ coweight lattice is isomorphic to the coroot lattice of $F_4$ upon multiplying the inner product of the former by a factor of 2. The $D_4$ Weyl group is enhanced by the Dynkin diagram symmetry of $D_4$ to the $F_4$ Weyl group. We conclude that the ring $J(F_4)$ can be mapped to the subring of $J(D_4)$ with precisely the enhancement of shift and Weyl symmetry that we require. Note that unlike the situation in section \ref{ss:B_4_and_F_4_to_D_4_branch}, we do not need to invoke the map $\iota^*$, as the two lattices $\LambdaCW(D_4)$ and $\LambdaCR(F_4)$ coincide as embedded in the Euclidean space $\IR^4$. In particular, this implies that due to the factor of 2 relating the inner products, elements of $J(F_4)$ interpreted as elements of $J(D_4)$ with enhanced shift symmetry have twice the index. Specifically, the subring of $J(D_4)$ in question is spanned by the generators
    \begin{align}
    \phi_{0,2}^{D_4}&=\phi_{0,1}^{F_4} \,, \label{eq:J(D_4)_with_enhanced_symmetry_generators}\\
    \phi_{-2,2}^{D_4}&=\phi_{-2,1}^{F_4}\,,\nn\\
    \phi_{-6,4}^{D_4}&=\phi_{-6,2}^{F_4}\,,\nn\\
    \phi_{-8,4}^{D_4}&=\phi_{-8,2}^{F_4}\,,\nn\\
    \phi_{-12,6}^{D_4}&=\phi_{-12,3}^{F_4}\,.\nn
    \end{align}

    We can move up the Higgsing tree to the $F_4$ node as in the previous subsection via the map $(\iota^*)^{-1}$. As only the generators \eqref{eq:J(D_4)_with_enhanced_symmetry_generators} arise, the $F_4$ theory will inverse inherit the enhanced symmetry of the $D_4$ theory. In particular, the numerator of the corresponding elliptic genus will permit an expansion in terms of the generators
    \begin{align*}
    \phi_{0,2}^{F_4}&= (\iota^*)^{-1}\phi_{0,2}^{D_4}   \,,\\
    \phi_{-2,2}^{F_4}&= (\iota^*)^{-1}\phi_{-2,2}^{D_4} \,,\\
    \phi_{-6,4}^{F_4}&= (\iota^*)^{-1}\phi_{-6,4}^{D_4} \,,\\
    \phi_{-8,4}^{F_4}&= (\iota^*)^{-1}\phi_{-8,4}^{D_4}\,,\\
    \phi_{-12,6}^{F_4}&=(\iota^*)^{-1}\phi_{-12,6}^{D_4} \,.
    \end{align*}
    To argue for this enhancement intrinsically, note that the contribution of the 2 half hypermultiplets in the $\mathbf{26}$ representation of $F_4$ cancel the contribution of the gauge fields associated to the short roots of $F_4$ to the Gromov-Witten invariants. Hence, following the logic of this section, we expect the numerator $\cN$ of the elliptic genus of this $F_4$ theory to have an expansion in the subring of $J(F_4)$ whose elements have a shift symmetry under the dual lattice to the sublattice of $\LambdaR(F_4)$ spanned by the long roots. This dual lattice is isomorphic to $\LambdaCW(D_4)$, as we can verify explicitly by considering the image of the sublattice of long roots of $F_4$ under the map $i$. Recall that we are identifying the embedding of the lattice $\LambdaR(F_4)$ in $\IZ^4$ as given in table \ref{tab:rootSystems2} (which reflects  the conventions of Bourbaki \cite{Bourbaki}) with the isomorphic lattice $\LambdaCR(F_4)$. We can obtain the dual lattice $\LambdaR(F_4)$ via the map $\alpha=\frac{2}{(\alpha^\vee,\alpha^\vee)}\alpha^\vee$, which in particular maps short coroots to long roots.
    The image of the long roots under the map $i$ is thus equal to the image of the short coroots, which we readily identify with $\LambdaR(D_4) = \LambdaCR(D_4)$. The dual lattice is hence $\LambdaCW(D_4)$, as we wished to argue.

    Choosing the other branch of the Higgsing tree, the next node up from the minimal $D_4$ theory is a theory with $B_4$ gauge symmetry and 2 half hypermultiplets in the vector representation. In addition to the symmetry enhancement discussed in section \ref{ss:B_4_and_F_4_to_D_4_branch}, this theory must also reverse inherit the enhanced shift symmetry under $\LambdaCW(D_4)$ which is particular to the massless $D_4$ theory. This enhancement follows the general pattern exhibited in this section: the contribution of the two half hypermultiplets to the Gromov-Witten invariants of the geometry cancels that of the gauge fields associated to the short roots of $B_4$. The remaining lattice of long roots of $B_4$ is the $D_4$ root lattice, with dual the $D_4$ co-weight lattice.

    \subsubsection{$D_n$ and $B_n$ over $\IF_4$}
    As already pointed out in \cite{Morrison:2020ool}, the only rank 1 models with matter present which does not break all of the center symmetry are the $D_n$ and $B_n$ models over $\IF_4$. These exhibit matter in the vector representation.
    
    We first consider the theories with $D_n$ gauge symmetry. The $\IZ_2$ Dynkin diagram symmetry (for $n>2$) exchanges the two spinor representations. As the only matter present in the $D_n$ series over $\IF_4$ is in the vector representation, this is a symmetry of these theories, enhancing the Weyl group symmetry to that of $B_n$:
    \be
    W(D_n) \ltimes \text{DynkinSym}(D_n) = W(B_n) \,.
    \ee
	Furthermore, the shift symmetry consists of all elements of the coweight lattice of $D_n$ which have integral pairing with the fundamental weight $\omega_1$ associated to the vector representation: in our conventions, this yields the lattice $\IZ^n$, which is the root lattice of $B_n$. The subring of $J(D_n)$ which exhibits the enhanced symmetry of the $D_n$ theory over $\IF_4$, is thus the ring $J(B_n)$. Note however that the inner product on the enhancement of the lattice $\LambdaCR(D_4)$ is the conventionally normalized Euclidean inner product $dx^2$. With regard to this norm, the indices of the generators of $J(B_n)$ (defined with regard to the inner product $2dx^2$) are doubled.
	
	Following the general strategy of this section, we expect the numerator of the elliptic genus for the $D_n$ theory over $\IF_4$ to lie in this subring. 
	
	We have computed the elliptic genus explicitly at base wrapping 1 for the model $D_5$. The ansatz in terms of the ring $J(B_5)$ with indices doubled indeed allows us to match all Gromov-Witten invariants as obtained via mirror symmetry. 
	
	For theories with $B_n$ gauge symmetry, the shift symmetry is enhanced from $\LambdaCR(B_n)$ (identified with $\LambdaR(C_n)$, as explained above) to the sublattice of $\LambdaCW(B_n)$ whose elements have integer pairing with $\omega_1$, the fundamental weight associated to the vector representation of $B_n$. As $\omega_1$ is also a root of $B_n$, this sublattice is indeed all of $\LambdaCW(B_n)$, i.e. $\IZ^n$ with inner product $dx^2$. We are hence seeking a subring of $J(C_n)$ with this shift symmetry. The ring $J(B_n)$ provides this subring, as $W(B_n) = W(C_n)$, and $\LambdaR(B_n) = \IZ^n$. The indices of the canonical generators are defined however with regard to the inner product $2dx^2$. As a subring of $J(C_n)$, they need to be doubled.

	In table \ref{tab:reduction} we summarize the reduction in the number of coefficients. 
	
	\begin{table}[ht]
    \centering
    \begin{tabular}{c|c|c|c}
         Base &Gauge algebra & Naive number of coefficients & Improved number of coefficients  \\\hline\hline
         
         \multirow{ 3}{*}{$\mathbb{F}_1$}
& $D_4$ & 902 &  \\
& $B_4$ & 495 & 197 \\
& $F_4$ & 197 &  \\\hline
         \multirow{ 4}{*}{$\mathbb{F}_2$} & $A_2/G_2$ & 4 & 1 \\\cline{2-4}
& $D_4$ & 295 &  \\
& $B_4$ & 161 & 64 \\
& $F_4$ & 64 &  \\\hline
        
         \multirow{ 3}{*}{$\mathbb{F}_3$} 
& $D_4$ & 310 &  \\
& $B_4$ & 171 & 69  \\
& $F_4$ & 69 &  \\\hline

 \multirow{ 6}{*}{$\mathbb{F}_4$} & $D_4$ & 287 &  \\
& $F_4$ & 69 & 2 \\
& $B_4$ & 163 &  \\\cline{2-4}
& $D_5/B_5$ & 1088 & 280 \\\cline{2-4}
& $B_6$ & 7086 & \multirow{ 2}{*}{1950}\\
& $D_6$ & 13248 &\\
    \end{tabular}
    \caption{By looking carefully at the symmetries of the low energy theory and assuming they hold for the full theory, we could reduce the number of coefficients for some models. The naive number is the number of coefficients one would have to fix using an arbitrary ansatz in $J(\tilde {\mathfrak{g}})$ and the improved number of coefficients is the number of coefficients one would have to fix if one refines the ansatz to the subring of $J(\tilde {\mathfrak{g}})$ invariant under the previously mentioned symmetries.}
    \label{tab:reduction}
\end{table}

	\subsection{A result at base wrapping 2}
	\label{sec:HigherBaseDegree}
	
	Several new features arise when we consider $Z_k$ at base wrapping $k>1$. Most importantly, the index of the numerator in the topological string coupling \eqref{eq:topStringIndex} is no longer 0, so the numerator depends on $g_{top}$. This dependence is through the Jacobi forms $\phi_{0,1}$ and $\phi_{-2,1}$. Expanding these in $x=\left(2\sin\left(\frac{g_{top}}{2}\right)\right)^2\sim \gs^2$ yields $\phi_{0,1}(g_{top})=2+o(x)$ and $\phi_{-2,0}=x+o(x^2)$.
	
	The universal part of the denominator \eqref{eq:denom_uni} at base degree $k$ scales as $x^k$. Therefore, a term of order $x^m$ in the numerator will contribute only to the Gopakumar-Vafa invariants of curves of genus $g_{m}$ or higher, with 
	\begin{equation*}
	    g_{m}-1=m-k.
	\end{equation*}
	Consequently, expanding the numerator as
	\begin{equation}
	    \cN_{i_{top},i_G,w}=\sum_{m=1}^{i_{top}}\cN^{(m)}_{i_G,w+2m}\phi_{0,1}(g_{top})^{i_{top}-m}\phi_{-2,1}(g_{top})^m,
	\end{equation}
	where $\cN^{(m)}$ is an element of $J(\mg)$ and we have indicated weights and indices by subscripts, the $\cN^{(m)}$ will contribute at genus $g_m$ and higher. For $m<k-1$, $\cN^{(m)}$ is completely fixed by requiring a cancellation of the contribution of order $x^{1+m-k}$ in the free energy $F=\sum_g F_g \gs^{2g-2}=\log Z_{top}$. $\cN^{(k-1)}$ can be fixed by imposing genus 0 Gromov-Witten invariants.
	
	As the dimension of the space of Jacobi forms of which $\cN$ is an element increases rapidly with $k$, results beyond $k>1$ are computationally expensive. We will here only discuss one example, the $F_4$ gauge theory over the base $\mathbb F_4$. For this model, invoking the enhanced shift symmetry discussed in subsection \ref{ss:enhanced symmetry D4 B4 F4} allows us to compute $\cN^{(1)}$ at $k=2$.
	
	From the expansions
	\begin{align*}
	   F&=\sum_g \gs^{2g-2} F_g=\log Z_{top}= \log Z_0 +Q_b \hat Z_1+Q_b^2\left(\hat Z_2-\frac{1}{2}\hat Z_1^2\right)+ \dots,\\
	    F_0&=F_0^{(0)}+F_0^{(1)}Q_b+F_0^{(2)}Q_b^2 + o(Q_b^3),
	\end{align*}
	where we have denoted $\hat Z_k=c_k(\mathbf{Q})Z_k$, we conclude that 
	\begin{equation}
	   \frac{F_0^{(2)}}{g_s^2}=\frac{F_0^{(2)}}{x}+o(x^0)=\hat Z_2-\frac{1}{2}\hat Z_1^2\,.
	   \label{eq:basedeg2}
	\end{equation}
	The universal contribution of the denominator for $Z_1$, $Z_2$ scale as $x$, $x^2$ respectively. The corresponding contributions must vanish in the linear combination appearing on the RHS of equation \eqref{eq:basedeg2}. This fixes $\cN^{(0)}$. 
	
	To fix $\cN^{(1)}$, we need to impose Gromov-Witten invariants obtained by mirror symmetry. The naive ansatz requires fixing 13189 coefficients. As explained in subsection \ref{ss:specDen}, we can impose the $D_4$ denominator in the ansatz \eqref{eq:ZkAnsatz}, i.e. we expect only divergences corresponding to the long roots of $F_4$. This reduces the number of coefficients to be fixed to 8620. Imposing the enhanced shift symmetry of this theory finally reduces the number of coefficients to be fixed to determine $\cN^{(1)}$ to 21. 
	
	We find that the ansatz with 21 coefficients is sufficient to match a large number of genus Gromov-Witten invariants at base degree $k=1$. We consider this strong evidence that our considerations in this and the previous section are also valid at higher base wrapping.

    \section*{Acknowledgements}
    We are grateful to Albrecht Klemm, Kimyeong Lee, Guglielmo Lockhart, Kaiwen Sun, Haowu Wang, Xin Wang, Timo Weigand for helpful discussions. ZD is supported by KIAS Individual Grant PG076901.
 
\chapter{Higgsing trees from the worldsheet}
\label{ch:paperWorldsheet}

\vskip 0.8 cm
		\centerline{David Jaramillo Duque and Amir-Kian Kashani-Poor }

		\vskip 0.2in
		
		\begin{center}{\footnotesize
				\begin{tabular}{c}
					{\em LPENS, CNRS, PSL Research University, Sorbonne Universit\'{e}s, UPMC, 75005 Paris, France}\\[0ex]
				\end{tabular}
		}\end{center}

		\setcounter{footnote}{0}
		\renewcommand{\thefootnote}{\arabic{footnote}}
		\vskip 60pt
		\begin{adjustwidth}{80pt}{80pt}
	       We study the elliptic genera of the non-critical strings of six dimensional superconformal field theories from the point of view of the strings' worldsheet theory. We formulate a general ansatz for these in terms of characters of the affine Lie algebra associated to the 6d gauge group at negative level, and provide ample evidence for the validity of this ansatz for 6d theories obtained via F-theory compactification on elliptically fibered Calabi-Yau manifolds over a Hirzebruch base. We obtain novel closed form results for many elliptic genera in terms of our ansatz, and show that our results specialize consistently when moving along Higgsing trees.
        
		\end{adjustwidth}
		
		{\let\thefootnote\relax
			\footnotetext{\tiny david.jaramillo-duque@phys.ens.fr,  amir-kian.kashani-poor@phys.ens.fr}}
\newpage

    \section{Introduction}
    The elliptic genus is a powerful tool to study two dimensional superconformal field theories for which an explicit description is not known. Due to its cohomological nature, it is robust under continuous deformations of the theory which do not change the asymptotics of field space. As such, it is invariant under renormalization group flow and can be computed if a Lagrangian UV description of the theory is available. Such descriptions exist e.g. for the (2,2) theories which arise upon considering Calabi-Yau compactifications of string theory \cite{Witten:1993yc} or for a small sample of the (0,2) theories which will be the object of this paper. Given such a description, it is possible to obtain exact results for the elliptic genus in terms of Jacobi $\theta$-functions and the Dedekind $\eta$-function \cite{Benini:2013a,Benini:2013xpa,Kim:2014dza,DelZotto:2018tcj,Haghighat:2014vxa,Kim:2018gjo,Kim:2016foj,Kim:2015fxa,Yun_2016,Gadde:2015tra,Haghighat:2013,Haghighat:2013tka,Hohenegger:2013ala,Putrov:2015jpa,Iqbal:2007ii,Agarwal:2018ejn}. While this solves the problem of computing the elliptic genus, the symmetries of the conformal fixed point of the theory will typically not be manifest. The scarceness of theories with available UV descriptions, and the obfuscation of the IR physics in the $\theta$-function form of the elliptic genus, motivate this work.
    
    We will study the elliptic genera of the non-critical strings of 6d superconformal field theories obtained via F-theory compactification on Calabi-Yau threefolds elliptically fibered over Hirzebruch surfaces $\IF_n$. We denote such theories as $G_n$, with $G$ specifying the 6d gauge symmetry.\footnote{With a very few exceptions at $n=1,2$, this specifies all theories in the class we are considering uniquely. For the exceptions, the flavor group $F$ must also be indicated.} The authors of \cite{DelZotto:2018tcj} propose that these genera take the universal form
    \begin{equation} \label{eq:affineERintro}
        \ER = \sum_{\lambda \in \dominantflavor} \hat{\chi}^F_\lambda(m_F,q) \, \xi^{n,G}_\lambda(m_G,q,v)
    \end{equation}
    with $\hat{\chi}_\lambda^F$ indicating the affine character of the integrable representation of the flavor group $F$ with highest affine dominant weight $\lambda$ at level $k_F$ -- we denote the set of such weights as $\dominantflavor$. Extending their work, we conjecture that $\xi^{n,G}_\lambda$ satisfies the ansatz 
    \begin{equation} \label{eq:affineAnsatzIntro}
    \xi_{\lambda}^{n,G}=\frac{1}{\eta(q)}\sum_{\omega \in \dominantgauge} \hat\chi_\omega(m_G,q) \sum_{k\in \mathbb Z} c^\lambda_{\omega,k} \,  q^{-\frac{k^2}{4\kappa}}v^k \,.
    \end{equation}
    with 
    \begin{equation} \label{eq:constraintIntro}
        c^\lambda_{\omega,k}\in \{0,\pm 1,\pm 2\} \,, \quad \#\{k \in \IZ\, | \, c_{\omega, k}^\lambda \neq 0 \} < \infty \,,
     \end{equation}
     i.e.,  the coefficients $c_{\omega, k}^\lambda$ are non-vanishing at given $\omega$ and $\lambda$ for only a finite number of integers $k$. A surprise, already pointed out in \cite{DelZotto:2016pvm,DelZotto:2018tcj}, is the occurrence of non-integrable representations of the affine Lie algebra associated to $G$ at negative level $-n$. The notation $\dominant(G)_{k}$ is meant to signify the set of affine weights at level $k$ with dominant finite part. Unlike $\affinedominant(F)_{k}$, this set is infinite. Such representations have only occurred sporadically in the physics literature, e.g. in \cite{Beem:2013sza, Dedushenko:2015opz,Dedushenko:2017osi}.\footnote{See \cite{Eager:2019zrc} for the occurrence of even more general non-integrable representations.} Their relevance for the elliptic genus of a handful of theories for which a UV description was known, namely $(C_r)_1$, $r=0,1,2$, $(D_4)_4$ and  $(G_2)_3$, was shown in \cite{DelZotto:2018tcj}. In this work, we demonstrate that the elliptic genera for a large number of theories for which no UV description is known can be expressed in terms of affine characters in the form \eqref{eq:affineERintro} with $\xi_\lambda^{n,G}$ given by our ansatz \eqref{eq:affineAnsatzIntro}, satisfying the constraints \eqref{eq:constraintIntro}. To be precise, we are able to compute elliptic genera for these theories up to some order in $q$ and $v$ by combining known constraints on the elliptic genus with the new constraints coming from our ansatz. The obtained results pass numerous consistency tests, leaving little doubt regarding their validity. 

    Our results can be used to provide boundary conditions for a universal ansatz for the elliptic genus in terms of Weyl invariant Jacobi forms \cite{Huang:2015sta, DelZotto:2016pvm, Gu:2017ccq, DelZotto:2017mee,Kim:2018gak} (referred to as the modular ansatz henceforth), leading to an exact expression for the genus. Mainly, though, we are interested in what they reveal about the nature of the worldsheet theories of the non-critical strings. The list of representations $\omega$ contributing to the elliptic genus, as well as the explicit form of the polynomials in $q$ and $v$ multiplying the characters $\hat \chi_\omega$ in equation \eqref{eq:affineAnsatzIntro} which we determine furnish non-trivial information regarding the structure of these theories. For the moment however, the nature of the non-integrable affine symmetry remains to be understood.

    Our original motivation for pursuing the calculations presented in this paper arose from our study in \cite{Duan:2020imo} of Weyl symmetry enhancement from $\cW_G$ to $\cW$ along  Higgsing trees as detected by the modular ansatz for the elliptic genus in terms of Jacobi forms. It is natural to ask whether the elliptic genus of a theory experiencing such an enhancement can beneficially be expanded in terms of affine characters $\hat \chi_\omega$ of the larger Lie algebra exhibiting $\cW$ as its Weyl symmetry. We will explain why this is always possible, and, sadly, why it does not appear to be beneficial, i.e. to lead to a simplification of the polynomial contribution in the ansatz \eqref{eq:affineAnsatzIntro} multiplying $\hat \chi_\omega(m_G,Q)$.
    
    This paper is organized as follows: section \ref{s:theories_considered} offers a quick review of the 6d spacetime theories under considerations and collects known and lesser known facts about the non-critical string in their spectrum and their worldsheet theory. We carefully introduce the $U(1)_v$ symmetry which will play an important role in our analysis. We introduce the constraints on the elliptic genera which will allow us to compute it in section \ref{s:four_constraints}. Some of the constraints that we wish to impose are on the elliptic genus with NS-NS boundary conditions. Section \ref{s:four_constraints} therefore also includes a subsection \ref{ss:intermezzo} in which we review and discuss the map introduced in \cite{DelZotto:2018tcj} between the R-R and the NS-NS elliptic genus. Section \ref{sec:algorithm} introduces the strategy to compute elliptic genera by imposing the constraints introduced in section \ref{s:four_constraints}. Much of the logic here follows \cite{DelZotto:2018tcj}, with the crucial addition of our constraints on the form of the function $\xi_\lambda^{n,G}$. In section \ref{s:closed_form_results}, we conjecture closed form results for the elliptic genera of multiple theories in terms of the ansatz \eqref{eq:affineERintro} and \eqref{eq:affineAnsatzIntro}. Section \ref{s:Higgs_tree_relations} discusses how, for multiple pairs of theories related by Higgsing, the elliptic genus of the parent theory specializes to the elliptic genus of the daughter theory. We also discuss the negative result regarding the expansion of the elliptic genus in terms of larger affine characters mentioned above. We conclude by summarizing the observations in this paper which would merit being understood purely from the point of view of the conformal field theory of the worldsheet theory of the non-critical string. Several appendices complete the paper. We review well-known facts regarding elliptic genera in appendix \ref{a:elliptic_genera}. Appendix \ref{a:Kazhdan-Lusztig} summarizes how we computed characters of negative level representations, an important technical ingredient in our calculations. Appendix \ref{a:polynomials} presents our results for the affine character expansion of the elliptic genus of a large number of theories. We end in appendix \ref{sec:HiggsingTrees} with a summary of the theories in the class we are considering, together with comments regarding the status of their elliptic genus. A file containing all negative level affine characters underlying the computations in this paper can be downloaded from \url{http://www.phys.ens.fr/~kashani/}.

    \section{6d theories and their non-critical strings} \label{s:theories_considered}
    \subsection{Spacetime theory} \label{ss:spacetime_theory}
    
    The theories we will consider in this paper are six dimensional rank 1 superconformal field theories with $(1,0)$ supersymmetry \cite{Gaiotto:2014lca,DelZotto:2014hpa,Bhardwaj:2015oru,DelZotto:2017pti,DelZotto:2018tcj,DelZotto:2016pvm,Losev:2003py,Lee:2018urn,Heckman:2018jxk}.\footnote{The rank indicates the number of tensor multiplets in the theory. $(1,0)$ is the minimal amount of supersymmetry possible in 6d: 8 real supercharges.}
    The existence of such theories as non-trivial infrared fixed points of 6d quantum field theories was one of the surprises that arose from the construction of 6d theories within the framework of string theory in the nineties \cite{Vafa:1996xn,Schwarz:1995dk}: the authors of \cite{Seiberg:1996vs} argued that singularities that occur in the moduli space of such theories must be due to IR dynamics, and concluded that the latter can therefore not be trivial.

    The theories we will consider can be constructed within F-theory by compactification on elliptically fibered Calabi-Yau manifolds $X$ over Hirzebruch bases $B=\IF_n$ \cite{Morrison:1996na,Morrison:1996pp,Morrison:2012np,Heckman:2013pva,Heckman:2015bfa}. These surfaces can be presented as $\IP^1$ fibrations over $\IP^1$, hence give rise to 6d supergravity theories with two tensor multiplets. Decompactifying the fiber $\IP^1$ (i.e. replacing $\IF_n$ by $\cO(-n) \rightarrow \IP^1$) decouples gravity and yields rank 1 theories with superconformal fixed points. 
    
    The base $\IF_n$ typically does not determine the elliptically fibered Calabi-Yau manifold above it uniquely, but determines a sometimes branched sequence of geometries called Higgsing trees. At the root of each tree lies the least singular fibration the base permits, leading to what is called the maximally Higgsed theory. Moving away from the root, one encounters increasingly singular fibrations, leading to theories with larger gauge symmetry and typically also larger charged matter content. Throughout this paper, we will refer to the gauge group as $G$, and a possible flavor symmetry of the charged matter as $F$ (or $F_i$, when $F = F_1 \times \ldots \times F_n$). Charged matter will transform in representations $(\omega_i, \lambda_i)$ of these symmetries, which can be determined e.g. by analyzing in detail the singularities occurring in the compactification geometry \cite{Bershadsky:1996nh, Morrison:2011mb} or extracted from its genus 0 Gromov-Witten invariants \cite{Kashani-Poor:2019jyo}. With a few exceptions, the gauge group $G$ only occurs once in the Higgsing tree above a given base $\IF_n$. We can hence identify these theories via the nomenclature $G_n$.\footnote{Specifying the flavor group is necessary only in the case of $n=1,2$, $G=D_6$ and $n=1$, $G=A_5$, see appendix \ref{a:higgsing_trees2}.}

    The unique tensor field $B$ of a rank 1 theory is sourced by instantons of the 6d gauge group via a coupling 
    \begin{equation} \label{eq:tensor_instanton_coupling}
        \int B \wedge \tr F \wedge F  \,.
    \end{equation}
    Quantization around an instanton solution in 6d leads to a 2d soliton in the spectrum of the theory: a non-critical string.
        
     \subsection{Worldsheet theory of the non-critical string} \label{ss:worldsheet_theory}
   
    The effective description of the non-critical string of 6d theories without charged matter is given by a non-linear sigma model on the worldsheet $\Sigma$ of the string (the trivial directions of the instanton background) with target space the one instanton moduli space $\cM_{G,1}$ (see e.g. \cite{Tong:2005un} for a review). The 6d coupling \eqref{eq:tensor_instanton_coupling} induces the coupling
    \begin{equation}
        \int_\Sigma B \,.
    \end{equation}
    on the worldsheet of the string.

    Factoring out the center of mass motion, the reduced instanton moduli space $\widetilde \cM_{G,1}$ is a hyperk\"ahler manifold of quaternionic dimension $h_G^\vee -1$. Both the gauge group $G$ and spacetime rotations $SO(4)$ act isometrically on $\tM_{G,1}$, the former by modifying the embedding of the instanton into the gauge group by a global factor, the latter by changing the orientation of the instanton in spacetime. Furthermore, the $G$ action is triholomorphic, i.e. commutes with the hyperk\"ahler structure. Writing the spacetime rotations $SO(4) \sim SU(2)^L \times SU(2)^R$, the first factor is triholomorphic, while the second rotates the three complex structures into each other.

    A non-linear sigma model with hyperk\"ahler target space permits $N=4$ supersymmetry \cite{Alvarez-Gaume:1981exv,Howe:1987qv,Howe:1988cj}, and indeed, a fermion in a one-instanton background exhibits $4h_G^\vee$ right-moving zero modes. The worldsheet theory of the non-critical string should therefore exhibit $(0,4)$ supersymmetry, implying that the string is a BPS object of the 6d theory which breaks half of its supersymmetry. The fact that $SU(2)^R$ rotates the complex structures of $\tM_{G,1}$ implies that it also acts as an R-symmetry on the corresponding supercharges. 

    To incorporate the effect of the presence of charged matter in the 6d theory, we need to introduce fields that transform both under the symmetry $G$ which acts on $\cM_{G,1}$ and the flavor symmetry $F$. A natural proposal \cite{DelZotto:2018tcj} for a minimal modification of the matterless case is to invoke fields in $(0,4)$ Fermi multiplets;\footnote{This is also how matter is introduced in theories for which a UV description is known.} this introduces left-moving fermions, which we take to transform in the appropriate $G$-equivariant vector bundle over $\cM_{G,1}$ determined by the representations $(\omega_i, \lambda_i)$ of the charged matter under the gauge and flavor symmetry of the 6d theory. 

    An elegant argument for obtaining the central charge of the theories along Higgsing trees for $n=3,4,5,6,8,12$ is presented in \cite{DelZotto:2018tcj}: for the matterless theories, the $4(h_G^\vee -1)$ bosons contribute $4(h_G^\vee -1)$ to both $c_L$ and $c_R$, while the right-moving fermions contribute $\frac{1}{2}4(h_G^\vee -1)$ to $c_R$. Hence, for theories without charged matter,
    \begin{equation}
        c_L = 4(h_G^\vee -1) \,, \quad c_R = 6 (h_G^\vee -1 ) \,.
    \end{equation}
    The right-moving central charge should retain this form also for theories with charged matter, as these differ only by a left-moving fermion bundle. To obtain the left-moving central charge of these theories, invoke the invariance of the gravitational anomaly $c_L-c_R$ under Higgsing to obtain
    \begin{equation}
        c_L = (c_L - c_R)(G_0) + c_R(G) \,.
    \end{equation}
    Here, $G_0$ denotes the gauge group of the theory without charged matter at the base of the Higgsing tree (this is where the constraint on $n$ enters). For such theories, $h_{G_0}^\vee = 3(n-2)$ \cite{Shimizu:2016lbw}. Hence,
    \begin{equation} \label{eq:central_charges}
        c_L = 6(h_G^\vee -n) +8 \,, \quad c_R = 6(h_G^\vee -1) \,.
    \end{equation}
    Reading off the number of left-moving bosons and fermions in the theory from this results allows us to infer the left-moving Casimir energy $E^0$ of these theories in the R- and the NS-sector. We follow the mnemonic presented e.g. in \cite{Polchinski:1998rq}: a contribution of $-\frac{1}{24}$ per boson, $\frac{1}{24}$ for a periodic fermion (R-sector), $-\frac{1}{48}$ for an anti-periodic fermion (NS-sector). This yields
    \begin{equation} \label{eq:casimir_energies}
        \casimirR = \frac{7}{6} - \frac{n}{2} \,, \quad \casimirNS = -\frac{c_L}{24} \,.
    \end{equation}
    The result \eqref{eq:central_charges} can alternatively be derived using anomaly arguments, as we review in section \ref{ss:anomaly}.
   
    \subsection{Symmetries of the string worldsheet} \label{ss:symmetries_of_string_worldsheet}

    As is familiar from the critical string, both the gauge symmetry $G$ and the flavor symmetry $F$ of the 6d spacetime theory manifest themselves as global symmetries of the worldsheet. As we argued in section \ref{ss:worldsheet_theory} above, the worldsheet theory also inherits half of the $(1,0)$ supersymmetry of the ambient theory. The latter can be decomposed with regard to the directions tangent and normal to the string worldsheet; a (1,0) spinor thus decomposes as
	\begin{equation} \label{eq:decomp_10}
	    (1,0) \rightarrow (\boldsymbol{2},\boldsymbol{1},+) \oplus (\boldsymbol{1},\boldsymbol{2},-) \,,
	\end{equation}
	with regard to $SO(4) \times SO(1,1)= \frac{SU(2)_L \times SU(2)_R}{\IZ_2} \times SO(1,1)$, where we have indicated the two spinor chiralities of $SO(1,1)$ with $\pm$. 

    Perhaps somewhat surprisingly for minimal supersymmetry, the $(1,0)$ supersymmetry algebra exhibits an R-symmetry, i.e. a linear action on its supercharges. One way to describe this action is by first noting that \cite{Kugo:1982bn}
	\begin{equation}
	    SL(2, \IH) \xrightarrow[]{2:1} SO(1,5) \,.
	\end{equation}
	The spin representation in 6d can thus be realized as a rank 2 module $S_6$ over the quaternion algebra $\IH$. The group $SL(1,\IH) \cong SU(2)$ acts on $S_6$ via multiplication on the right. This action gives rise to the R-symmetry which we shall denote as $SU(2)_I$. In terms of it, a (1,0) spinor transforms in the $\boldsymbol{2}$ representation:
	\begin{equation}
	    S_6 \ni \psi  = \begin{pmatrix} \mu \\ \nu \end{pmatrix} \,, \quad \mu, \nu \in \IH \,,
	\end{equation}
	with $\Lambda \in SL(2,\IH)$ and $M \in SL(1,\IH)$ acting as 
	\begin{equation}
	    \psi \mapsto \Lambda \cdot \psi \cdot M \,.
	\end{equation}
	To map to modules over $\IC$, we realize the quaternionic generators $i,j,k$ in terms of the matrices $i \sigma_1, i \sigma_2, i \sigma_3$. $\psi$ is then represented by a $2 \times 4$ matrix of complex numbers. Each of the four rows of this matrix transforms in the $\boldsymbol{2}$ of $SU(2)$. The diagonal action on all four rows commutes with the action of the Lorentz group.
	
	Returning to the decomposition \eqref{eq:decomp_10}, the representation under the R-symmetry $SU(2)_I$ goes along for the ride, yielding \cite{Shimizu:2016lbw}
	\begin{equation}
    	 (1,0,\boldsymbol{2}) \rightarrow (\boldsymbol{2},\boldsymbol{1},+,\boldsymbol{2}) \oplus (\boldsymbol{1},\boldsymbol{2},-,\boldsymbol{2}) \,.
	\end{equation}
	The string breaks half of the ambient supersymmetry. Taking the second summand to be the one that is conserved (this is a choice of instanton vs. anti-instanton), this gives rise to $(0,4)$ supersymmetry on the worldsheet, with the $SO(4)$ R-symmetry identified with $SU(2)_R \times SU(2)_I/\IZ_2$.\footnote{The $\IZ_2$ quotient reflects the fact that the negative of the diagonally embedded identity into $SU(2)_R \times SU(2)_I$ acts trivially on the spinor.} 
    
    In section \ref{ss:worldsheet_theory}, we identified a symmetry $SU(2)^L \times SU(2)^R$ of the reduced one-instanton moduli space $\widetilde \cM_{G,1}$. $SU(2)^L$ is naturally identified with $SU(2)_L$. $SU(2)^R$ however acts both as a right chiral spacetime rotation and an R-symmetry. It is hence naturally identified with a diagonal embedding into $SU(2)_R \times SU(2)_I$. This embedding will play a role in identifying the fugacities of the elliptic genus of the worldsheet theory with couplings of the topological string in section \ref{ss:fugacities} below.

    \subsection{What anomaly inflow teaches us about the worldsheet theory} \label{ss:anomaly}
    
    6D $\mathcal N=(1,0)$ theories are generically anomalous with an anomaly polynomial given by \cite{Shimizu:2016lbw,Sadov:1996zm,Sagnotti:1992qw}
    \begin{equation*}
        \frac{1}{2}\eta^{ij}I_i\wedge I_j\,,
    \end{equation*}
    where $\eta^{ij}$ is the charge pairing and
    \begin{equation*}
        \eta^{ij}I_j=\frac{1}{4}\left(\eta^{ia}\tr F_a^2-(2-\eta^{ii})p_1(T)\right)+h_{G_i}^\vee c_2(I)\,,
    \end{equation*}
    with $a$ indexing both the dynamical and the background fields, $c_2(I)$ the Chern class of the $SU(2)_I$ bundle and $p_1(T)$ the first Pontrjagin class of (the tangent bundle of) $M^{(6)}$ \cite{Shimizu:2016lbw}. This anomaly can be canceled via the addition of a Green-Schwarz term in the action,
    \begin{equation}
        S_{GS}=\int_{M^{(6)}}\eta^{ij} B_i\wedge I_j \,,
        \label{eq:GS_action}
    \end{equation}
    and a modification of the conservation law for $B_i$: the invariant field is not $dB_i$, but $H_i=dB_i+\alpha
    _i$ with $d\alpha_i=I_i$.
    
    This extra term needed to cancel the anomalies gives a contribution to the 2D theory of the non-critical string. To get its 2D contribution, one considers the 6D theory in the presence of the string. As the string sources the $B$-field, the Bianchi identity of $H$ gets modified to
    \begin{equation}
        dH_i=I_1+Q_i\prod_{j=2}^5\delta(x_j)dx^j\,,
        \label{eq:background}
    \end{equation}
     where $Q_i$ is the charge of the string, and we have put the string in the $x_0,\,x_1$ plane. In the background given by this $H$ field, the Green-Schwarz term \eqref{eq:GS_action} combines with the kinetic term of the $B_i$ fields to give a contribution to the 2D worldsheet action of the string. This contribution can be computed using anomaly inflow \cite{Sadov:1996zm,Shimizu:2016lbw} and, of course, is not gauge/diffeomorphism invariant. 
    
    The contribution of this Green-Schwarz term to the 2D anomaly polynomial is given by 
    \begin{equation}
        I_{4}=\frac{\eta^{ij}Q_iQ_j}{2}(c_2(L)-c_2(R))+\eta^{ij}Q_iI_j\,,
    \end{equation}
    where $c_2(L/R)$ are the Chern classes of the $SU(2)_{L/R}$ bundles introduced above. Of course in this last equation one needs to decompose the 6D characteristic classes appearing in $I_i$ in terms of their 2D counterparts:
    \begin{equation*}
        p^{6D}_1(T)=p_1^{(2D)}(T)+p_1^{(4D)}(N)=p_1^{(2D)}(T)-2 c_2(L)-2c_2(R)\,.
    \end{equation*}
    We then arrive at 
    \begin{equation}
        I_{4}=\frac{\eta^{ij}Q_iQ_j}{2}(c_2(L)-c_2(R))+Q_i\left(\frac{1}{4}\eta^{ia}\tr F_a^2-\frac{2-\eta^{ii}}{4}(p_1(T)-2c_2(L)-2c_2(R))+h^\vee_{G_i}c_2(I)\right)\,.
    \end{equation}
    
    Rank 1 theories exhibit a single tensor field $B$. In terms of the F-theory engineering geometry, the pairing $\eta^{11}=n$ is given by the negative self-intersection number of the only compact curve in the base of the elliptic fibration. Focusing on the elliptic genus of a single string, we set $Q=1$ to obtain
    \begin{align}
        I_4&=-\frac{2-n}{4}p_1(t)+\frac{n}{4}\tr F_G^2+\frac{1}{4}\eta^{1a}\tr F_a^2+(c_2(L)+c_2(R))+(h_G^\vee c_2(I)-n c_2(R))\,.
        \label{eq:AnomalyPol}
    \end{align}
    This equation encodes much non-trivial information regarding the 2d theory \cite{DelZotto:2018tcj}, as we now review. This is the anomaly polynomial of the non-linear sigma model on the moduli space of instantons $\mathcal M_{G,1}$. As the center of mass multiplet gives a universal contribution, it is convenient to factor it out and discuss the reduced theory on the reduced moduli space $\tM_{G,1}$. We use a superscript $^\cM$ when referring to quantities associated to the full theory.
\begin{itemize}
    \item \textbf{Central charges:}
    
    The difference $c_L^\cM-c^\cM_R$ of the central charges is fixed by the gravitational anomaly to be the coefficient of $-\frac{1}{24}p_1(T)$:
    \begin{equation} \label{eq:difference_c}
        c_L^\cM-c_R^\cM=6(2-n)\,.
    \end{equation}
    Moreover, the central charge on the right is linked to the level of the R-symmetry in the IR. This is identified with $SU(2)_I$. Its level is given by the coefficient of $c_2(I)$ in the anomaly polynomial: $k_R=h_G^\vee$. This fixes the right-moving central charge to $c_R=6k_R$ \cite{Putrov:2015jpa,Schwimmer:1987,Witten:1997yu}, allowing us to also extract the left-moving central charge from \eqref{eq:difference_c}, yielding
    \begin{equation}
        c_L^\cM=6( h_G^\vee-n+2)\,,\quad c_R^\cM=6h_G^\vee\,.
    \end{equation}
    Subtracting the $c=(4,6)$ contribution from the center of mass motion, we arrive at
    \begin{equation}
        c_L=6( h_G^\vee-n+2)-4\,,\quad c_R=6h_G^\vee-6\,.
    \end{equation}
    \item \textbf{Current levels:}
    
    From the WZW-models (for a review see \cite{francesco2012conformal}), we know that the gauge part of the anomaly polynomial is proportional to the level of the gauge current. We thus conclude that the gauge algebra current is at level $-n$ while the flavor algebra currents are at level $-\eta^{1a}\geq 0$.
    The authors of \cite{DelZotto:2018tcj} determine the flavor algebras and levels for most of the theories in the class we consider. These results are reproduced in appendix \ref{sec:HiggsingTrees}.
    \begin{equation*}
        \text{Gauge algebra level }=-n\,,\quad\text{flavor algebra level }=-\eta^{ia}\geq 0\,.
    \end{equation*}
    \item \textbf{$SU(2)^L$ and $SU(2)^R$ levels:}
    
    As we argue in section \ref{ss:symmetries_of_string_worldsheet}, the $SU(2)_L$ and $SU(2)^L$ symmetries coincide. We can hence identify the coefficient of $c_2(L)$ in the anomaly polynomial with the level of the $SU(2)^L$ current. As this coefficient is 1, we conclude that all dependence on the associated fugacity $x$ is captured by the center of mass contribution to the elliptic genus, i.e.
    \begin{center}
        the fugacity $x$ does not appear in $\ER$\,.\footnote{Note that this is no longer true for the elliptic genus of multiple strings.}
    \end{center}
    On the other hand, we identified the $SU(2)^R$ symmetry with the diagonal of $SU(2)_R\times SU(2)_I$. The level $\kappa^\cM$ of the corresponding current is therefore given by the sum of the coefficients of $c_2(I)$ and $c_2(R)$. This gives
    \begin{equation}
        \kappa^\cM=h^\vee_G-n+1\,,
    \end{equation}
    Removing the contribution of the center of mass multiplet yields
    \begin{equation}
        \kappa=h_G^\vee-n\,.
        \label{eq:kappa_def}
    \end{equation}

\end{itemize}

    \subsection{Fugacities of the elliptic genus and relation to topological string} \label{ss:fugacities}

    As reviewed in the appendix, the $(0,2)$ elliptic genus permits the inclusion of charges in the trace which commute with the two supercharges; the resulting index is sometimes referred to as a flavored elliptic genus. Natural choices are the exponentiated Cartan generators of the gauge and flavor symmetries $G$ and $F$, as well as of the chiral spacetime rotation $SU(2)^L$. We will call the associated fugacities as $m_G$, $m_F$, and $\epsilon_{-}$. For the exponentiated fugacities, we use $x=e^{2\pi i \epsilon_-}$ for the $SU(2)^L$ group and $Q_i=e^{2\pi i(\alpha_i,m)}$ or $X_i=e^{2\pi i (e_i,m)}$ for $G$ and $F$. Here, $\alpha_i$ are the simple roots and $e_i$ are the vectors providing the canonical basis of the associated Euclidean space. The conventions for the embedding of the coroot lattices of Lie groups in Euclidean lattices are the same as in \cite{Duan:2020imo}.
    
    As the worldsheet exhibits $(0,4)$ supersymmetry, an inclusion of generators for a subgroup of the R-symmetry with which two supercharges commute is also permissible. Over the reals, the best we can do is consider an embedding
    \begin{equation}
        U(1) \times U(1) \hookrightarrow SU(2)_R \times SU(2)_I \,.
    \end{equation}
    The insertion of the generator of one of the $U(1)$ factors into the trace determines which subset of 2d BPS states contributes to the elliptic genus. Let $J^3_R$ and $J^3_I$ indicating the infinitesimal generators of the Cartan of $SU(2)_R$, $SU(2)_I$ respectively. To connect to partition functions in the $\Omega$ background \cite{Nekrasov:2002qd} and the refined topological string partition function \cite{Iqbal:2007ii,CKK}, we would like to introduce a fugacity $v$ associated to the symmetry $SU(2)^R$ with Cartan $J^3_R + J^3_I$ by inserting
    \begin{equation}
        X_v = v^{J^3_R + J^3_I}
    \end{equation}
    into the elliptic genus. Without this insertion, only states whose right-moving factor is annihilated by all supercharges contribute. Upon insertion, only annihilation by supercharges which commute with $X_v$ suffices.

    \section{Three constraints on the elliptic genus and an intermezzo} \label{s:four_constraints}

    \subsection{The modular ansatz}  \label{ss:modular_ansatz}
    The (flavored) elliptic genus is essentially a meromorphic Jacobi form of vanishing weight and of index fixed by the 't Hooft anomalies governing the flavor symmetries \cite{Benini:2013a, Benini:2013xpa}. We say essentially, because in theories with a gravitational anomaly, i.e. for which left and right moving central charge do not coincide, the defining transformation properties for Jacobi forms under modular transformations are modified by a phase, see e.g. equation (2.16) in \cite{Benini:2013xpa}.
    
    For the theories under consideration, the R-R elliptic genus has been argued \cite{Huang:2015sta,  DelZotto:2016pvm, Gu:2017ccq, DelZotto:2017mee, DelZotto:2018tcj} to take the form 
    \begin{equation}
        \ER=\eta(q)^{24 \casimirR} \frac{N(q,v,m_G,m_F)}{D(q,v,m_G)}\,,
        \label{eq:modular_all_fugacities}
    \end{equation}
    with $N$ and $D$ polynomials in Weyl invariant holomorphic Jacobi forms \cite{Wirthmuller:Jacobi,Bertola},\footnote{We are considering weak Jacobi forms to be a special case of Weyl invariant holomorphic Jacobi forms, see e.g. the appendix of \cite{DelZotto:2017mee}.} and $\casimirR$ the left-moving Casimir energy in the Ramond sector.\footnote{\label{fn:shift}For theories over the Hirzebruch base $\IF_1$, the power of the $\eta$ function is actually the Casimir energy minus one. All statements in the rest of this paper remain true for the $n=1$ case if we interpret $\casimirR$ as the Casimir energy minus one in this case, and we will henceforth tacitly do so.} The denominator $D$ in \eqref{eq:modular_all_fugacities} for a given group $G$ is universal. The numerator $N$ is a holomorphic Jacobi form of weight equal the weight of $D$ minus $12 \casimirR$ (the contribution of the Dedekind $\eta$ function to the weight), and of index in the elliptic parameters $v$ and $G$ adjusted by the index of $D$. As the space of Weyl invariant holomorphic Jacobi forms of fixed weight and indices is a finite dimensional vector space,\footnote{This is true with the exception of the space of Weyl invariant Jacobi forms for the Weyl group of $E_8$ \cite{Wirthmuller:Jacobi}, see \cite{Sakai:2011xg,Sakai:2017ihc, Wang:2018fil,Sun:2021ije} for this case.} $N$ is fixed once a finite number of coefficients are determined. Unfortunately, the number of coefficients is typically very large: there are e.g. 236,509 terms for the $(F_4)_4$ theory. This number is in many cases considerably reduced by invoking multiple sources of symmetry enhancement \cite{Duan:2020imo} to express $N$ in terms of Jacobi forms of Weyl groups larger than that of $G$.

    The coefficients required to determine $N$ were fixed in \cite{Huang:2015sta, Gu:2017ccq, DelZotto:2017mee} by matching to boundary conditions provided by the topological string partition function. \cite{Kim:2018gak} determined them by matching to the elliptic genus expressed in terms of $\theta$-functions for theories for which a UV description is known. \cite{DelZotto:2018tcj} instead matched to universal features of the elliptic genus inspired by the IR theory (mainly for vanishing gauge and flavor fugacities). This latter approach is naturally integrated into our computation of the elliptic genus in the affine expansion \eqref{eq:affineAnsatzIntro}, and we shall review it below.

    Note that the presentation of $\ER$ in terms of Jacobi forms as in \eqref{eq:modular_all_fugacities} is exact, just like the $\theta$-function expressions arising from knowledge of a UV theory. To compare to the topological string partition function, this expression must be expanded for small $q$, and then for small exponentiated gauge and flavor fugacities. This leads (though non-trivially) to an expansion in only positive exponentiated K\"ahler classes, as behooves the topological string partition function \cite{Gu:2017ccq}. This expansion also preserves the $v \rightarrow \frac{1}{v}$ symmetry of \eqref{eq:modular_all_fugacities}. In contrast, the presentation in terms of affine characters that we are interested in requires the expansion in small $q$ to be followed up by an expansion in small $v$. The contribution to the elliptic genus that is thus sensitive to the expansion region is that stemming from the zero modes. Already in \cite{Witten:1993jg}, these were seen to require a careful treatment. It would be desirable to understand the relation between expansion region and physical interpretation of the elliptic genus better. 

    \subsection{The affine ansatz} \label{ss:affine_ansatz}
    The central contribution of this note is providing ample evidence for the conjecture that the R-R elliptic genus for the class of theories described in section \ref{s:theories_considered} can be parametrized as follows:
    \begin{equation} \label{eq:flavor_sum}
        \ER = \sum_{\lambda \in \dominantflavor} \hat{\chi}^F_\lambda(m_F,q) \, \xi^{n,G}_\lambda(m_G,q,v)
    \end{equation}
    with
    \begin{equation} \label{eq:affine_G_expansion}
    \xi_{\lambda}^{n,G}=\frac{1}{\eta(q)}\sum_{\omega \in \dominantgauge} \hat\chi_\omega(m_G,q) \sum_{k\in \mathbb Z} c^\lambda_{\omega,k} \,  q^{-\frac{k^2}{4\kappa}}v^k \,,
    \end{equation}
    with $c^\lambda_{\omega,k}\in \{0,\pm 1,\pm 2\}$ and non-vanishing at fixed $\omega$ and $\lambda$ for only finitely many integers $k$. This ansatz is heavily inspired by the work \cite{DelZotto:2018tcj}. Several comments are in order. 
    \begin{itemize}
        \item The sum in equation \eqref{eq:flavor_sum} is over all highest weight representations $L_\lambda$, $\lambda$ dominant, of the affine Lie algebra associated to $F$ at a fixed positive level $k_F$. As the level $k$ of a rank $r$ affine Lie algebra is related to the Dynkin labels $\lambda_i$ of its weights via
        \begin{equation} \label{eq:levelDynkin}
            k = \sum_{i=0}^r a_i^\vee \lambda_i \,,
        \end{equation}
        with the $a_i^\vee$ denoting the (non-negative) co-marks of the Lie algebra, this sum is necessarily finite. The notation $\hat \chi_{\lambda}^F$ indicates the affine character associated to the representation $\lambda$, normalized such that the leading power in $q$ is 
        \begin{equation}
            -\frac{c_F}{24}+h_\lambda^F \,,
        \end{equation}        
        with $c_F$ the central charge of the WZW model associated to the affine Lie algebra at this level, and $h_\lambda^F$ the conformal weight of a state with highest weight $\lambda$ in this theory. The explicit expressions are
        \begin{equation}
            c_F=\frac{k_F \dim(F)}{h^\vee_F+k_F}\,, \quad h_\lambda^F=\frac{\left<\lambda,\lambda+2\rho_F\right>}{2(h_F^\vee+k_F)}\,,
        \end{equation}
        where the inner product $\langle \cdot, \cdot \rangle$ is normalized such that long roots have length $2$.
        
        \item The sum over representations $\omega$ in equation \eqref{eq:affine_G_expansion} is over highest weight representations $L_\omega$ of the affine Lie algebra $G$ at negative level $-n$, constrained as follows: by the relation \eqref{eq:levelDynkin}, $\omega$ cannot be dominant. We require its finite projection to be dominant. This permits us to impose that the finite representations to which $L_\omega$ restrict at each grade be integrable, thus salvaging at least the symmetry under the finite part of the Weyl group.
        
        As $\lambda_0$ can become arbitrarily negative, the sum over $\omega$ is infinite. The affine characters $\hat \chi_\omega$ are normalized such that the leading power in $q$ is 
        \begin{equation}
            -\frac{c_G}{24}+h_\omega^G
        \end{equation}         
        with 
        \begin{equation} \label{eq:c_and_h_for_omega}
            c_G=\frac{\dim(G)(-n)}{\kappa}\,, \quad h_\omega^G=\frac{\left<\omega,\omega+2\rho_G\right>}{2\kappa}\,,
        \end{equation}
        and $\kappa$ given in equation \eqref{eq:kappa_def}. We discuss these characters and their computation further in appendix \ref{a:Kazhdan-Lusztig}.

        \item We will argue below that for fixed $\omega$, the sum over $k$ is finite. $\xi_{\lambda}^{n,G}$ thus has the structure
        \begin{equation} \label{eq:xi_in_terms_of_p}
        \xi_{\lambda}^{n,G}=\frac{1}{\eta(q)}\sum_{\omega \in \dominantgauge} \hat\chi_\omega(m_G,q) \,p_\omega^\lambda(q^{h_v^1},v) \,,
        \end{equation}
        where we have defined
        \begin{equation} 
            h_v^k = -\frac{k^2}{4\kappa} 
        \end{equation}
        and $p_\omega^\lambda(y,v)$ denotes a polynomial of the form
        \begin{equation} \label{eq:poly_multiplying_omega}
            p_\omega^\lambda(y,v) = \sum_k c^\lambda_{\omega,k} \,v^k y^{k^2} \,.
        \end{equation}
        \end{itemize}
    The computational task required to obtain $E_R$ is to determine which representations $\omega$ contribute, and for each such $\omega$, to compute the polynomials $p_\omega^\lambda(y,v)$.

    \subsection{Intermezzo: from $\ER$ to $\ENS$} \label{ss:intermezzo}
    Both for deriving a lower bound on the negative powers of $v$ that occur in $\ER$, and to obtain some constraints on the expansion coefficients, it will prove useful to be able to relate $\ER$ to the NS-NS elliptic genus. Surprisingly (recall that we have fermions in the left-moving, non-supersymmetric sector), this relation is (conjecturally) delightfully simple: the following relation is conjectured in \cite{DelZotto:2016pvm,DelZotto:2018tcj}:\footnote{The authors of \cite{DelZotto:2018tcj,DelZotto:2016pvm} conjecture this map to yield the elliptic genus in the NS-R sector. This is not consistent with level matching, due to the ensuing occurrence of both integral and half-integral powers of $q$ relative to the Casimir energy. This structure is consistent with the NS-NS elliptic genus, due to the shift \eqref{eq:anticommutation_NS} by the right-moving R-current  $\bar J_0$ in the power of  $\bar q$ in that case. We thank Michele Del Zotto for a discussion regarding this point.}
    \begin{equation}
        \ENS (m_G,m_F,v,q)=q^{-\kappa/4}v^\kappa \ER(m_G,m_F,q^{1/2}/v,q), \quad \kappa=h_G^\vee - n. \label{eq:NSR-R relation}
    \end{equation}
    Note that this is a priori not a manifestation of spectral flow of $N=2$ theories, as the left-hand side of the theories we are considering do not exhibit supersymmetry. However, in a theory of bosons and fermions with a $U(1)$ charge with generator $J_0$, one may try to relate traces over R and NS sectors by shifting $L_0$ with a multiple of the generator $J_0$ to account for the half-integral moding of fermions in the NS sector compared to the R sector, and shifting both $L_0$ and $J_0$ to account for the different weight and charge of the vacuum in the NS vs. the R sector. For the $N=2$ theory of free fermions and bosons, this procedure indeed reproduces spectral flow. Surprisingly, the same strategy reproduces \eqref{eq:NSR-R relation}. The shifts that lead to \eqref{eq:NSR-R relation} are
    \begin{equation} \label{eq:shifts_of_weight_and_charge}
        h_{\mathrm{NS}}=h_\mathrm{R}+\frac{1}{2}l_\mathrm{R}+\frac{1}{2}k\,, \quad l_{\mathrm{NS}}=l_\mathrm{R}+2 k \,,
    \end{equation}
    with $k= -\kappa/2$. Indeed, composing this shift with the $v \to 1/v$ symmetry of the elliptic genus in the form \eqref{eq:modular_all_fugacities}, we have 
    \begin{equation} \label{eq:RtoNS_monomial}
        q^{h_\mathrm{R}}v^{l_\mathrm{R}}\mapsto q^{h_\mathrm{R} + \frac{l_\mathrm{R}}{2} + \frac{k}{2}} v^{-l_\mathrm{R} -2k} =q^{k/2}v^{-2k}\left(\left.q^{h_\mathrm{R}} v^{l_\mathrm{R}}\right|_{v\mapsto q^{1/2}/{v}}\right) \,,
    \end{equation}
    which induces the transformation \eqref{eq:NSR-R relation}.

    It will be convenient to introduce the following notation: we will have $\mathcal F_\kappa[f]$ denote the function $f$ upon acting with the transformation \eqref{eq:RtoNS_monomial}, i.e. 
    \begin{equation} \label{eq:def_F}
        \mathcal F_\kappa [f](m_G,m_F,v,q):=q^{-\kappa/4}v^\kappa f(m_G,m_F,q^{1/2}/v,q) \,,
    \end{equation}
    such that 
    \begin{equation}
        \ENS = \mathcal F_\kappa[\ER] \,.
    \end{equation}
    
    We will see below that $\ER$ has an expansion in positive powers of $q/v^2$ and $v$. Disregarding the prefactor, the transformation \eqref{eq:def_F} on a monomial of this expansion is
    \begin{equation}
        \left(\frac{q}{v^2}\right)^j v^l \mapsto v^{2j} \left(\frac{q}{v^2}\right)^{l/2} \,.
    \end{equation}
    The expansion region, small $q^2/v$, small $v$, is hence preserved by $\mathcal F_{\kappa}$. To obtain a transformation with this property, we had to compose the shifts \eqref{eq:shifts_of_weight_and_charge} with the symmetry $v \rightarrow 1/v$ of the unexpanded elliptic genus.
    
The fact that $\ENS=  \mathcal F_\kappa(\ER)$ implies that the same ansatz we had for $\ER$ holds for $\ENS$:\footnote{Note that $q^{-\kappa/4}v^\kappa \left.\left(v^\ell q^{-\frac{\ell^2}{4\kappa}}\right)\right|_{v\to q^{1/2}/v}= q^{-\frac{(\kappa-\ell)^2}{4\kappa}}v^{\kappa-\ell}$}
\begin{equation}
    \ENS = \sum_{\lambda \in \dominantflavor} \hat{\chi}^F_\lambda \, \xi^{\mathrm{NS}}_\lambda=
      \frac{1}{\eta} \sum_{\lambda \in \dominantflavor} \hat{\chi}^F_\lambda \, \sum_{\omega \in \dominantgauge} c^{\mathrm{NS},\lambda}_{\omega,z}\hat \chi_\omega(m_G,q) \sum_{\ell\in \mathbb Z} q^{-\frac{\ell^2}{4\kappa}}v^\ell.
      \label{eq:affineAnsatzNS}
\end{equation}

    For the cases without massless matter, the NS-NS and the R-R elliptic genus coincide (up to an irrelevant sign choice). Hence, the $\xi$ functions must be $\mathcal{F}_\kappa$ invariant. For the general case\footnote{$(E_7)_7$, the only theory with massless matter but no flavor group, is an exception to this rule, see subsection \ref{subsec:E7_7}. }, we will find that $\mathcal F_\kappa$ permutes the different $\xi$ functions, i.e.
\begin{equation}
    \mathcal F_\kappa (\xi^G_\lambda)=\pm\xi^G_{\lambda'} \quad \text{ for some } \lambda' \,.
    \label{eq:Fpermutation}
\end{equation}

    \subsection{The low lying spectrum in the NS sector} \label{ss:low_lying_spectrum}
    \cite{DelZotto:2016pvm} argue, building on previous observations in \cite{Benvenuti:2010pq} and \cite{Keller:2011ek,Keller:2012da,Hanany:2012dm,Rodriguez-Gomez:2013dpa}, that the leading contribution at $q^{-c_L/24}$ to the NS elliptic genus should essentially coincide with the Hilbert series of the corresponding one-instanton moduli spaces. \cite{DelZotto:2018tcj} observe that the contribution (in the presence of charged matter in the spacetime theory) at the next level, $q^{-c_L/24+1/2}$, also has a universal form in terms of (finite) characters of the groups $G$ and $F$.\footnote{This is observed for the examples $(C_r)_1$ for $r=1,2,3,4$, $(G_2)_3$, $(D_r)_4$ for $r=4,5$, and $(B_4)_4$.} We conjecture that this form is universally valid for the theories that we can consider and impose the ansatz
    \begin{equation} \label{eq:ansatz_low_lying}
        \ENS=q^{-\frac{c_L}{24}} v^{h_G^\vee-1}\left(\sum_{k}v^{2k}\chi^G_{k\theta}(m_G)-q^{\frac{1}{2}}\sum_{k}v^{2k+1}\sum_{i}\chi^G_{k\theta+\omega_i}(m_G)\chi^{F}_{\lambda_i}(m_F)+O(q)\right)\,,
    \end{equation}
    as a constraint on the elliptic genera. Here, $\theta$ is the highest root of $G$, and the pairs ($\omega_i$, $\lambda_i$) of representations of $G$ and $F$ were discussed in section \ref{ss:spacetime_theory}. Let us comment on these two contributions: considering the NS-NS rather than the R-R elliptic genus disentangles the fermionic and bosonic contributions at the ground state energy $\casimirNS = -c_L/24$ calculated in \eqref{eq:casimir_energies}. The leading contribution thus arises from symmetric tensor products of bosonic zero modes. These carry the adjoint representation of $G$. The $k$-th such product gives rise to the contribution $\chi_{k\theta}^G(m_G)$. As
    \begin{equation}
        \left(\theta^{\otimes k} \right)_{\mathrm{Sym}} = k \theta + \ldots \,,
    \end{equation}
    the form \eqref{eq:ansatz_low_lying} encodes that all additional contributions in the decomposition of the tensor product do not contribute independently \cite{Benvenuti:2010pq}.\footnote{We thank Noppadol Mekareeya for explaining to us how all representations beside $k\theta$ are set to zero by $D$-term constraints in the case $G=A_n$.} We can further read off the $U(1)_v$ charge of the NS vacuum to be $h_G^\vee -1$, and the $U(1)_v$ charge of the bosonic fields of the non-linear sigma model to be 2. The contributions at energy $-c_L/24 + 1/2$ arise when $1/2$ modes of the left moving fermion fields act on the bosonic ground states. These carry the representations $(\omega_i, \lambda_i)$ under the groups $(G,F)$, and we can read off their $U(1)_v$ charge to be 1. We again see that only the leading term $k\theta \oplus \omega_i$ in the decomposition of $\theta^{\otimes k} \otimes \omega_i$ in irreducible representations contributes. 

\subsection{Imposing $|c^\lambda_{\omega,k}|\leq 2$}

We have already discussed the three main constraints on the elliptic genus: (1) the modular ansatz \eqref{eq:modular_all_fugacities}, (2) compatibility with the NS elliptic genus, and (3) the affine ansatz \eqref{eq:affineAnsatzIntro}. The modular ansatz has finitely many unknowns; hence, by providing a finite amount of initial data or vanishing conditions, it is possible, in principle, to solve for the elliptic genus \cite{Huang:2015sta,Gu:2017ccq,Duan:2020imo,DelZotto:2016pvm,DelZotto:2018tcj}. In this work, we bypass the computation of initial data (except for the data coming from \eqref{eq:ansatz_low_lying}), and conjecture that imposing the other constraints is enough to determine the entire elliptic genus. However, even for small rank groups, the dimension of the Jacobi ring to which the numerator $N$ belongs is very large, making the problem of solving for all coefficients computationally intractable. Therefore, we opt to specialize some of the elliptic parameters to 0, thus greatly reducing the dimension of the space of Jacobi forms. This process, of course, removes information, so we are not able to determine the complete set of coefficients $c^{\lambda}_{\omega,k}$ uniquely. We find experimentally, however, that imposing the constraint
\begin{equation}
|c^{\lambda}_{\omega,k}|\leq 2\
\label{eq:constraintC}
\end{equation}
allows us to do so.\footnote{Up to small subtleties. See section \ref{sec:algorithm}}

We believe that the constraint \eqref{eq:constraintC} holds generally, for the following reasons:
\begin{itemize}
\item It holds for the cases we can compute exactly: For a handful of theories, including $(C_r)_1$, $(B_r/D_r)_4$, and $(G_2)_3$, we know the full elliptic genus with all fugacities turned on \cite{Kim:2014dza,DelZotto:2018tcj,Haghighat:2014vxa,Kim:2018gjo,Kim:2016foj,Kim:2015fxa,Kim:2018gak}. For these theories, we did not impose \eqref{eq:constraintC}, but we observed that it was satisfied.
\item The constraint is compatible with Higgsing: As we will explain in section \ref{s:Higgs_tree_relations}, there is a simple rule to obtain the elliptic genus of a Higgsed theory from the elliptic genus of its parent. We consider the fact that Higgsing preserves this constraint as strong evidence for its validity.
\item Simplicity: The affine characters reflect the structure of a multiplet plus all of its descendants. The affine ansatz already captures the structure of the gauge and flavor groups, so the only remaining source for structure is the relatively simple supersymmetry algebra (only 4 generators). We therefore expect $c^\lambda_{\omega,k}$ to be relatively small numbers. For several cases, we slightly increased the bound (up to 8) and found no additional solutions. In some cases in which we did find solutions violating this bound, the bound was violated for some coefficients by many orders of magnitude.
\item Existence of a solution: Using the algorithm in section \ref{sec:algorithm}, we found a solution that satisfies the bound \ref{eq:constraintC} in every theory we studied. The equations satisfied by the $c$'s are linear equations with large coefficients (approximately $10^8$ in the larger cases). We consider the fact that solutions exist that satisfy the bound to be strong evidence that the constraint should hold in general.
\end{itemize}

    \section{Putting the constraints to work}    
    \label{sec:algorithm}
    \subsection{The strategy} \label{ss:strategy}
    In the expressions \eqref{eq:flavor_sum} and \eqref{eq:affine_G_expansion}, $E_R$ is obtained as a sum over irreducible highest weight representations $L_\lambda$ of the symmetry $F$ at positive level $k_F$, a sum of highest weight representations $L_\omega$ of the symmetry $G$ at negative level $-n$, and a sum over powers $k$ of the fugacity $v$. The sum over $\lambda$ is finite, as only finitely many irreducible integrable highest weight representations exist at given level. The sum over $\omega$ is infinite. We will now argue that at fixed $\lambda$ and $\omega$, the sum over $k$ is finite. In each summand, the powers of $q$ are manifestly integrally spaced, with the leading power $\fpq_0$ of $q$ in the summand indexed by $\lambda$, $\omega$, and $k$ given by
    \begin{equation}
        \lpq(\lambda,\omega,k):=-\frac{c_F}{24}+h_\lambda^F-\frac{c_G}{24}+h_\omega^G+h^v_{k}-\frac{1}{24} \,.
        \label{eq:qExponentEquation}
    \end{equation}
    Note that $h_k^v$ is a negative definite quadratic form of $k$. Thus, at given $\lambda$ and $\omega$, $\lpq$ can be bounded by the Casimir energy $\casimirR$ (computed in \eqref{eq:casimir_energies}) only for a finite set of $k$. This is why at given $\omega$, only a finite number of terms can occur in the sum over $k$ in the expression \eqref{eq:affine_G_expansion} for $\xi_\lambda^{n,G}$, i.e. $c_{\omega,k}^\lambda = 0$ for all but finitely many $k$. 
    
    The modular ansatz \eqref{eq:modular_all_fugacities} implies that all occurring powers of $q$ (i.e. not merely restricted to a given summand) be integrally spaced relative to the lower bound on this power provided by the Casimir energy $E_0^R$,\footnote{Recall our conventions regarding the notation $\casimirR$ for the case $n=1$ as stated in footnote \ref{fn:shift}.}
    \begin{equation}
        \lpq(\lambda, \omega, k) - E_0^R\,\in\, \mathbb N  \quad \forall \lambda, \omega, k \,.
    \label{eq:constraint1}
    \end{equation}
     This constraint also follows from level matching at the level of the worldsheet theory. It constrains the representations $\omega$ that can contribute to the expansion \eqref{eq:affine_G_expansion} at fixed order in $q$ and $v$. As $h_\omega^G$, given in equation \eqref{eq:c_and_h_for_omega} above, is a positive definite quadratic form of the Dynkin labels of $\omega$, the number of such representations compatible with the constraint \eqref{eq:constraint1} is finite. 

    Next, we will argue that the power of $v$ that occurs in the expansion of $\ER$ at a given order in $q$ is also bounded below. This argument relies on considering the elliptic genus in the NS-NS sector, and invoking the lower bound on the power of $q$ there. Recall that, as discussed in section \ref{ss:intermezzo}, the monomial $q^\fpq v^k$ contribution to $\ER$ is mapped to the contribution $q^{\fpq+k/2-\kappa/4}v^{\kappa-k}$ to $\ENS$. The NS-NS elliptic genus permits both integral and half-integral energy levels relative to the Casimir energy $\casimirNS=-\frac{c_L}{24}=-\frac{\kappa}{4}-\frac{1}{3}$,
    \begin{equation} \label{eq:bound_on_k}
        \fpq + \frac{k}{2} - \frac{\kappa}{4}-(\casimirNS)=(\fpq-\casimirR)+\frac{k}{2}-\frac{n-3}{2}\, \in\, \frac{1}{2}\mathbb N_0  \,,
    \end{equation}
    thus implying the sought after lower bound on the power of $v$. Introducing the non-negative integers $j$ and $\ell$ via  
    \begin{equation}
       \ell = k- (n-3) + 2(\fpq- \casimirR) = k- (n-3) + 2j
    \end{equation}
    allows us to write $\ER$ as
    \begin{equation}
       \ER=q^{\casimirR}v^{n-3}\sum_{j,\ell \ge 0} b_{j,\ell}(m_G,m_F)\left(\frac{q}{v^2}\right)^jv^\ell \,.
     \label{eq:bexpansion}
    \end{equation}

    The relation \eqref{eq:bound_on_k} also implies a lower bound on the power $k$ of $v$ in the polynomials \eqref{eq:poly_multiplying_omega}:
    \begin{equation}
    k\geq (n-3) -2(\lpq(\lambda, \omega,k) - \casimirR)
        \label{eq:constraint2}
    \end{equation}
    recall that the $k$ dependence in $\lpq(\lambda, \omega, k)$ is via the summand $-\frac{k^2}{4\kappa}$.

    The discussion above clarifies how the expansion \eqref{eq:affine_G_expansion} is to be understood: at any fixed order in $q$ and in $v$, all occurring sums, in particular the sum of non-integrable representations $\omega$, are finite.

    Now that we understand the nature of the expansion, we can go about solving for the coefficients $c_{\omega,k}^\lambda$ in the ansatz \eqref{eq:affine_G_expansion} by equating the expansion \eqref{eq:flavor_sum} to the modular ansatz \eqref{eq:modular_all_fugacities}. To render the computation feasible, we set, following \cite{DelZotto:2018tcj}, the fugacities $m_G = m_F =0$. This greatly reduces the number of unknown coefficients in the numerator $N$ of the modular ansatz, as the Jacobi forms depending on these fugacities specialize to integers. The constraint on the coefficients $c_{\omega,k}^\lambda$ thus takes the form 
    \begin{equation}
        \eta^{24 \casimirR}\frac{N(q,v,m_G=0,m_F=0)}{D(q,v,m_G=0)}=\left.\sum_{\lambda \in \dominantflavor} \hat{\chi}^F_\lambda \, \xi^G_\lambda\right|_{m_F=m_G=0} \,,
    \label{eq:tosolve}
    \end{equation}
    where the LHS is expanded first in $q$ and then in $v$. This gives rise to a linear homogeneous equation on both the coefficients of the basis of Jacobi forms at appropriate weight and level contributing to $N$ as well as the expansion coefficients $c_{\omega,k}^\lambda$ occurring in $\xi_\lambda^G$. To introduce inhomogeneities, we impose the knowledge of the low lying spectrum in the NS sector as described in section \ref{ss:low_lying_spectrum}.

    In terms of the expansion given in equation \eqref{eq:bexpansion}, the low lying spectrum in equation \eqref{eq:ansatz_low_lying} fixes\footnote{For $n=1$, the conditions are slightly different due to the presence of the tachyon and the shift in $\casimirR$ we introduced in footnote \ref{fn:shift}. They read $b_{0,0}=1\,,b_{0,\ell}=0\,,b_{k,2}=\chi_{k\theta}^G\,,b_{k+1,3}=\sum_i\chi^G_{\omega_i+k\theta}\chi_{\lambda_i}^F$}
    \begin{equation}
        b_{k+n-2,0}=\chi^G_{k\theta}
    \quad \text{and}\quad b_{k+n-1,1}=\sum_{i}\chi^G_{\omega_i+k\theta}\chi^F_{\lambda_i}
        \label{eq:contraint3} \,;
    \end{equation}
    equivalently,
    \begin{align*}
        c^{\lambda}_{\omega,1-n-2k}&=\delta_{\lambda,0}\delta_{\omega,k\theta} &\text{ if } k-\left((n-3)-2(\lpq(\lambda,\omega,k)-\ER^0)\right)=0,\\
        c^{\lambda}_{\omega,-n-2k}&=-\sum_i\delta_{\lambda_i,\lambda}\delta_{\omega,\omega_i+ k\theta} &\text{ if } k-\left((n-3)-2(\lpq(\lambda,\omega,k)-\ER^0)\right)=1/2.
    \end{align*}
    Note that in all of the examples that we consider, we can solve for the coefficients of $N$ first, just by taking advantage of coefficients of monomials in $q$ and $v$ that we know to vanish on the RHS of equation \eqref{eq:tosolve}. In most cases, the remaining constraints have a unique solution (once Dynkin symmetry is addressed, see immediately below) upon imposing $|c^\lambda_{\omega,k}|\leq 2$.

    Finally, we wish to discuss a complication due to possible Dynkin symmetry when setting the gauge and flavor fugacities in equation \eqref{eq:tosolve} to zero : characters associated to weights related by Dynkin symmetry are equal in the $m\to 0$ limit. Explicitly, if we have a Dynkin symmetry $s\in \text{Dyn}(G)$ for any character $\hat \chi_{w}^G$, 
    \begin{equation*}
       \hat \chi^G_{s\omega}(m_G)=\hat \chi^G_{\omega}(s m_G) \Rightarrow  \hat \chi^G_{s\omega}(0)=\hat \chi^G_{\omega}(0).
    \end{equation*}
    Therefore, equation \eqref{eq:tosolve} can only be solved for the sum $c^\lambda_{\omega,k}+c^\lambda_{s\omega,k}$ rather than for the individual coefficients. In many cases, we expect the elliptic genus to be Dynkin symmetric, as this symmetry is inherited via Higgsing from a theory further up the Higgsing tree \cite{Duan:2020imo}.\footnote{The cases for which we do not expect Dynkin symmetry are $(D_6)_{3}\,,\,\,(D_5)_{3}$, the two theories $(D_6)_1$, and $(D_6)_{2}$ with flavor group $C_6\times(\mathrm{Ising})\times (\mathrm{Ising})$.} In such cases, we can set $c^\lambda_{\omega,k}=c^\lambda_{s \omega,k}$ to resolve the ambiguity. Dynkin symmetry with regard to the flavor group can also occur, as  for $s\in \text{Dyn}(F)$, we cannot differentiate between $c^\lambda_{\omega,k}$ and $c^{s\lambda}_{\omega,k}$ upon setting $m_F = 0$. In these cases, we can only solve for $\xi_\lambda+\xi_{s\lambda}$. In the case of $U(1)$ flavor groups, the $m_{U(1)}\to -m_{U(1)}$ leads to the same style of ambiguity as Dynkin symmetry. 

    We exemplify this latter ambiguity at the hand of the example $(E_6)_5$ in appendix \ref{aa:polynomialsE65}.
    
    \subsection{An example: $\Fff$}
    
    As an example, consider the theory $(F_4)_4$. The flavor group is $A_1$ at level $3$ \cite{DelZotto:2018tcj}. At this level, there exist 4 dominant highest weight representations. Their highest weights are
    \begin{equation}
        \lambda_i = (3-i) \,\Lambda_0^{A_1} + i\, \Lambda_1^{A_1} \,, \quad i = 0, \ldots, 3 \,,
    \end{equation}
    where we have denoted the fundamental weights of $\hat A_1$ by $\Lambda_0^{A_1}$ and $\Lambda_1^{A_1}$.

    We will consider the expansion of the $\xi$ functions in equation \eqref{eq:affine_G_expansion} to order 
    \begin{equation}
        O\left(q^{\casimirR}\left(\frac{q}{v^2}\right)^{M+1},v^{m+(4-3)+1}\right) \,, \quad M=3 \,,\,\,m=8 \,.
    \end{equation}
    From our discussion in section \ref{ss:strategy}, we know that at a given power in $q$, the powers of $v$ are bounded below by \eqref{eq:constraint2}. Furthermore, at each power of $q$ and $v$, the new representations that contribute to the $\xi$ functions, i.e. that did not already contribute at lower power of $q$, are given by solving \eqref{eq:constraint1}. We organize the calculation in terms of orders of $q$:

    \hspace{1cm}
    
    {\bf Leading order $q^{\casimirR}$}
    
    Specialized to the leading order $q^{\casimirR}$, \eqref{eq:constraint2} which bounds the power of $v$ from below reduces to $k\geq (4-3)=1$. Hence, the powers $k$ of $v$ that can arise at order $O\left(q^{\casimirR}\left(\frac{q}{v^2}\right)^1,v^{m+2}\right)$ are $-1\leq k \leq m=9$. To find the affine characters of $F_4$ that contribute at this order, we thus find the admissible weights $\omega \in \dominant(F_4)_{-4}$ which solve \eqref{eq:constraint1}, 
    \begin{equation}
        \lpq(\lambda_i,\omega,k) - \casimirR = 0 \,,
        \label{eq:0orderF_4_4}
    \end{equation}
    for these values of $k$ and each of the four flavor weights $\lambda_i$. The solutions are given in table \ref{tab:0orderF_4_4}.
    
    \hspace{0.5cm}

    {\bf Order $q^{\casimirR+1} $}
    
    At the next order in $q$, $q^{\ER^0+1}$, equation \eqref{eq:constraint2} reads $k\geq (4-3)-2=-1$, so for every value of $k$ with $-1\leq k \leq (m+1)-2 = 7$\footnote{$(m+1)-2$, as $(q/v^2) v^{m+1}=qv^{m+1-2}$.} we invoke \eqref{eq:constraint1},
    \begin{equation}
        \lpq(\lambda_i,\omega,k) - \casimirR = 1 \,,
        \label{eq:1orderF_4_4}
    \end{equation}
    to identify the new  representations $\omega$ that contribute at this order: we find that there are no solutions (those identified in table \ref{tab:0orderF_4_4} of course contribute at all powers of $q$ above the ground state).  
    
\begin{table}[]
        \centering
        \begin{tabular}{c|ccccccccc}
 \diagbox[]{$\lambda$}{k} & 1 & 2 & 3 & 4 & 5 & 6 & 7 & 8 & 9 \\\hline
 $\lambda_0$ & - & - & - & - & - & - & \text{(0000)} & - & - \\
  $\lambda_1$ & - & - & - & - & - & - & - & - & - \\
  $\lambda_2$& - & - & - & - & - & - & - & - & \text{(0001)} \\
  $\lambda_3$& - & - & - & - & - & - & - & \text{(0000)} & - \\    
        \end{tabular}
        \caption{Solutions of equation \eqref{eq:0orderF_4_4}}
        \label{tab:0orderF_4_4}
    \end{table}

In general at order $q^{E_R^0+l}$ the equations read:
\begin{equation}
    e^{4,F_4}_{\lambda_i}(\omega,k)-E_R^0=l, \quad  (4-3)-2l\leq k \leq m+1-2l \,.
    \label{eq:lOrderF_4_4}
\end{equation}
The solutions for $\omega$ for these equations for $l=2,3$ are given in tables \ref{tab:2orderF_4_4} and \ref{tab:3orderF_4_4}.

\begin{table}[]
        \centering
        \begin{tabular}{c|ccccccccc}
 \diagbox[]{$\lambda$}{k} & -3 & -2 & -1 & 0 & 1 & 2 & 3 & 4 & 5 \\\hline
 $\lambda_0$ & \text{(0000)} & - & - & - & - & - & \text{(0000)} & - & - \\
  $\lambda_1$  & - & - & - & - & - & - & - & - & - \\
  $\lambda_2$ & - & - & - & - & - & - & - & - & - \\
 $\lambda_3$ & - & - & - & - & - & - & - & - & - \\
        \end{tabular}
        \caption{Solutions of equation \eqref{eq:lOrderF_4_4} for $l=2$} 
        \label{tab:2orderF_4_4}
    \end{table}        

\begin{table}[]
        \centering
        \begin{tabular}{c|ccccccccc}
 \diagbox[]{$\lambda$}{k} & -5 & -4 & -3 & -2 & -1 & 0 & 1 & 2 & 3 \\\hline
 $\lambda_0$ & \text{(1000)} & - & - & - & - & - & - & - & - \\
  $\lambda_1$ & - & \text{(0001)} & - & - & - & - & - & - & - \\
  $\lambda_2$ & - & - & - & - & - & - & - & - & - \\
  $\lambda_3$ & - & - & - & \text{(0000)} & - & - & - & \text{(0000)} & - \\
        \end{tabular}
        \caption{Solutions of equation \eqref{eq:lOrderF_4_4} for $l=3$} 
        \label{tab:3orderF_4_4}
    \end{table}

    At this order, the ansatz for the $\xi$ functions therefore reads
\begin{align}
\label{eq:F_4_4xi}
\eta\xi^{4,F_4}_{(0)}&=\hat{\chi }_{\text{(0000)}} \left(v^7 y^{49} c^{\text{(0)}}_{\text{(0000)},7}+\frac{y^9 c^{\text{(0)}}_{\text{(0000)},-3}}{v^3}+v^3 y^9
   c^{\text{(0)}}_{\text{(0000)},3}\right)+\frac{y^{25} \hat{\chi }_{\text{(1000)}} c^{\text{(0)}}_{\text{(1000)},-5}}{v^5}+\dots,\\
\eta\xi^{4,F_4}_{(1)}&=   \frac{y^{16} \hat{\chi }_{\text{(0001)}} c^{\text{(1)}}_{\text{(0001)},-4}}{v^4}+\dots,\\
\eta\xi^{4,F_4}_{(2)}&=   v^9 y^{81} \hat{\chi }_{\text{(0001)}} c^{\text{(2)}}_{\text{(0001)},9}+\dots,\\
\eta\xi^{4,F_4}_{(3)}&=  \hat{\chi }_{\text{(0000)}} \left(v^8 y^{64} c^{\text{(3)}}_{\text{(0000)},8}+\frac{y^4 c^{\text{(3)}}_{\text{(0000)},-2}}{v^2}+v^2 y^4 c^{\text{(3)}}_{\text{(0000)},2}\right)+\dots\,,
\label{eq:F_4_4xilast}
\end{align}
where $y=q^{-\frac{1}{4\kappa}}=q^{-1/20}$ as in equation \eqref{eq:poly_multiplying_omega}. To lighten the notation, we have indexed the affine characters and the coefficients $c_{\omega,k}^\lambda$ only with the finite Dynkin labels; the zeroth Dynkin label is then determined by the level.

From the expressions \eqref{eq:F_4_4xi}-\eqref{eq:F_4_4xilast} it is difficult to read off at a glance the power of $q$ at which each term contributes to the elliptic genus, as both the affine characters associated to the flavor and to the gauge group exhibit a non-trivial leading $q$-power. Consider e.g. $\xi_{(1)}^{4,F_4}$:
\begin{align}
    x_0(\lambda=(1),\omega=(0001),k=-4) &= -\frac{c_{A_1}}{24}+h_{\lambda}^{A_1}-\frac{c_{F_4}}{24}+h_{\omega}^{F_4}+h^v_{-4}-\frac{1}{24}\\
&= -\frac{9/5}{24}+\frac{3}{20}-\frac{-208/5}{24}+\frac{6}{5}-\frac{16}{20}-\frac{1}{24}\\
&=\frac{13}{6}\\
&=3+\casimirR
\end{align}
where we have used $\casimirR=-\frac{5}{6}$. The first two terms come from the flavor character, the following two come from the gauge character, the $h^v_{-4}$ comes from $y^{16}$, and the $-\frac{1}{24}$ is the contribution from the Dedekind $\eta$-function.

    Before imposing equality with the modular ansatz as in \eqref{eq:tosolve}, we can fix some of the unknown constants from the knowledge of the low energy spectrum \eqref{eq:ansatz_low_lying} in the NS sector.

For the ${(F_4)}_4$ example \eqref{eq:F_4_4xi} these conditions fix three coefficients:
\begin{equation*}
    c^{\text{(0)}}_{\text{(0000)},-3}=1,\quad c^{\text{(0)}}_{\text{(1000)},-5}=1,\quad c^{\text{(1)}}_{\text{(0001)},-4}=-1.
\end{equation*}

For the ${(F_4)}_4$, the modular numerator $N$ has 61 undetermined coefficients, i.e. the Jacobi ring at the desired weight and index has dimension 61. Using the fact that small/negative powers of $v$ are constrained by \eqref{eq:bexpansion} we can solve for 42 coefficients and using \eqref{eq:contraint3} we solve for an extra 12 coefficients. The remaining 7 coefficients can be fixed by comparing the modular and the affine ansatz. Once the modular ansatz is fixed, we find a series of equations for the coefficients $c^\lambda_{\omega,k}$. Imposing $c^\lambda_{\omega,k}\in \mathbb Z$ with $|c^\lambda_{\omega,k}|\leq 2$ gives a unique solution.

\section{Closed form results for $\ER$} \label{s:closed_form_results}
For a subset of the theories we consider, we have conjectural closed form results for the elliptic genus $\ER$  in the form \eqref{eq:flavor_sum}: we enumerate all of the representations $\omega$ that contribute, and the associated polynomials $p_\omega^\lambda$ as defined in equation \eqref{eq:poly_multiplying_omega}. We present these results in this section. We have computed the representations $\omega$ that contribute to $\xi^{n,G}_\lambda$ to a certain fixed order in $q$ and the associated polynomials $p_\omega^\lambda$ for a host of other examples $G_n$, but we are not confident that to the order achieved, our results reflect the complete structure of $\ER$. These results are presented in appendix \ref{a:polynomials}.

Note that explicit results for the elliptic genera of the theories $(C_r)_1$, $(B_r/D_r)_4$, $(G_2)_3$ among others are known in terms of $\theta$ functions or modular forms are known \cite{Kim:2014dza,DelZotto:2018tcj,Haghighat:2014vxa,Kim:2018gjo,Kim:2016foj,Kim:2015fxa,Kim:2018gak}. Expressing these results in terms of our affine ansatz \eqref{eq:affineAnsatzIntro} (for $(C_r)_1$ below, and $(C_3)_1$, $(C_4)_1$, $(B_4)_4$, and $(D_5)_4$ in the appendix) hence provides evidence for the validity of this ansatz.\footnote{In practice, we only check the result with completely arbitrary fugacities to low orders due to the computational complexity of the problem. To arrive at the results we give in appendix \ref{a:polynomials}, we expand the $\theta$-function expression at several (computer generated) random points for the gauge and flavor fugacities.}.

\subsection{Theories without charged matter}
\label{sub:matter-less}
Theories without matter necessarily exhibit trivial flavor symmetry $F$: the sum over $\lambda$ is absent for these theories, such that 
\begin{equation}
    \ER=\xi^{n,G} \,.
\end{equation}
The absence of the left-moving fermionic bundle in the non-linear sigma model description of the worldsheet theory significantly simplifies the structure of the elliptic genus.  

Indeed, for the matterless models $(D_4)_4,\,(F_4)_5,\,(E_6)_6,\,(E_7)_8$, we find 
\begin{equation}
    \xi^{n,G}= \frac{1}{\eta(q)}\sum_n\hat \chi_{n\theta}\sum_{m=0}^2a_m\left(v^{-b_{n,m}+\kappa/2}q^{-\frac{(b_{n,m}-\kappa/2)^2}{4\kappa}}-(-1)^s v^{b_{n,m}+\kappa/2}q^{-\frac{(b_{n,m}+\kappa/2)^2}{4\kappa}}
    \right) \,,
    \label{eq:massless_result}
\end{equation}
where 
\begin{equation*}
    b_{n,m}=2n+\kappa m-(\kappa-2),\quad \kappa=h^\vee_G-n\,,
\end{equation*}
and $a,s$ are given in table \ref{tab:tab_as}.
The $(D_4)_4$ result was already pointed out in \cite{DelZotto:2018tcj}.
\begin{table}[]
    \centering
    \begin{tabular}{c|cccc}
         & $a_0$&$a_1$&$a_2$&$s$ \\\hline
        $(D_4)_4$ & 1&2&1&0\\
        $(F_4)_5$ & -1&0&1&1\\
        $(E_6)_6 $& 1&2&1&0\\
        $(E_7)_8$&1&2&1&0
    \end{tabular}
    \caption{Coefficients and relative sign in expansion of the elliptic genus of non-Higgsable models in affine characters}
    \label{tab:tab_as}
\end{table}
We note that the operator $\mathcal F_\kappa$ simply permutes the two terms in parentheses and the sign $(-1)^s$ determines whether $\ER$ is periodic or anti-periodic under this transformation. 

Note that for $(D_4)_4$, the UV theory was derived in \cite{Haghighat:2014vxa} and $\ER$ as a function of the gauge and flavor fugacities determined in terms of $\theta$-functions. Obtaining the result \eqref{eq:massless_result} for this theory hence does not require the algorithm presented in section \ref{sec:algorithm}. 

\subsection{$(E_7)_7$ and $(B_4)_4$}
\label{subsec:E7_7} 
$\mathbf{(E_7)_7}$

$(E_7)_7$ is the only theory with matter and without flavor group. We find that the elliptic genus can be written as 

\begin{align}
\footnotesize
\begin{split}
        \xi^{7,E_7}=\frac{1}{\eta(q)}\Bigg(\sum_n \hat \chi_{n\theta}\sum_{m=0}^2a_m v^{-b^+_{n-1,m}+\kappa/2}q^{-\frac{(b^+_{n-1,m}-\kappa/2)^2}{4\kappa}}+\sum_{n}\hat \chi_{n\theta+(0000001)}\sum_{m=0}^2a_m v^{ b^+_{n,m}+\kappa/2}q^{-\frac{(b^+_{n,m}+\kappa/2)^2}{4\kappa}}\\
        -\sum_n \hat \chi_{n\theta+(0000010)}\sum_{m=0}^2a_m v^{-b^-_{n,m}+\kappa/2}q^{-\frac{(b^-_{n,m}-\kappa/2)^2}{4\kappa}}-\sum_{n}\hat \chi_{n\theta+(0100000)}\sum_{m=0}^2a_m v^{ b^-_{n+1,m}+\kappa/2}q^{-\frac{(b^-_{n+1,m}+\kappa/2)^2}{4\kappa}}\\
        -\hat \chi_{(0000100)}\sum_{m=0}^2a_m v^{b^-_{0,m}+\kappa/2}q^{-\frac{(b^-_{0,m}+\kappa/2)^2}{4\kappa}}+\hat \chi_{(0001000)}\sum_{m=0}^2a_m v^{b^+_{-1,m}+\kappa/2}q^{-\frac{(b^+_{-1,m}+\kappa/2)^2}{4\kappa}}\Bigg)\,,
\end{split}
\end{align}

where 
\begin{align}
    b^\pm_{m,n}=(-1)^m\left(2n+\kappa m-(\kappa-2)\pm \frac{1}{2}\right),\, a_0=1,\, a_1=-2,\, a_2=1\,.
\end{align}

Due to the presence of matter, we expect $\mathcal{F}_\kappa$ to not act trivially on $\xi$; however, as there is no flavor group, $\mathcal F_\kappa$ cannot act as a permutation of the flavor weights. We observe that $\mathcal F_\kappa$ permutes the contributions of the different characters. In terms of the polynomials in equation \eqref{eq:poly_multiplying_omega},
\begin{equation}
    \mathcal{F}_\kappa (p_\omega)(y,v)=p_{\sigma(\omega)}(y,v)\,,
\end{equation}
where $\sigma^2=1$ and 
\begin{align*}
    \sigma(n\theta)&=(n+1)\theta+(0000001),\quad n\geq 0\,,\\
    \sigma(0)&=(0001000)\,,\\
    \sigma(n\theta+(0100000))&=(n+1)\theta+(0000010)\,, \quad n\geq 0\\
    \sigma(0000010)&=(0000100) \,.
\end{align*}

$\mathbf{(B_4)_4}$

For $(B_4)_4$, the flavor group is $A_1$ at level $1$. We observe that the $\xi$ functions can be written as

\begin{align}
\begin{split}
        \xi^{4,B_4}_{(0)}&=\frac{1}{\eta(q)}\Bigg(\sum_{n>0} \hat \chi_{n\theta}\sum_{m=0}^3a_m v^{b^-_{n,m}+\kappa/2}q^{-\frac{(b^-_{n,m}+\kappa/2)^2}{4\kappa}}+\sum_{n}\hat \chi_{n\theta+(1000)}\sum_{m=0}^3a_m v^{ -b^+_{n,m}+\kappa/2}q^{-\frac{(b^+_{n,m}-\kappa/2)^2}{4\kappa}}\\&
        +\hat \chi_{(0000)}\sum_{m=2,3}a_m v^{b^-_{0,m}+\kappa/2}q^{-\frac{(b^-_{0,m}+\kappa/2)^2}{4\kappa}}+\hat \chi_{(2000)}\sum_{m=0,1}a_m v^{b^-_{0,m}+\kappa/2}q^{-\frac{(b^-_{0,m}+\kappa/2)^2}{4\kappa}}\Bigg)\,,
\end{split}\\
\xi^{4,B_4}_{(1)}&=-\mathcal{F}_\kappa(\xi^{4,B_4}_{0})
\end{align}

with

\begin{equation*}
    b_{n,m}^\pm=(-1)^m\left(2n+\kappa m+(\kappa-2) -6 \pm \frac{1}{2}\right), \, a_0=1,\,a_1=-1,\,a_2=-1,\,a_3=1\,.
\end{equation*}

\subsection{The $C_r$ branch of the E-string Higgsing tree}
\label{sub:E-string results}
The elliptic genera of the theories of the $E$-string Higgsing tree ($n=1$) with gauge group $C_r$ where given in \cite{DelZotto:2018tcj} for $r=1,2$ . We conjecture a general result for arbitrary $r>1$:
\begin{align} \label{eq:Cr}
    \eta\, \xi_0^{C_r}&=\hat\chi_0+\hat\chi_{\Lambda_2}+\sum_{k\geq 1}  \hat\chi_{2k\Lambda_1} q^{-\frac{(2k)^2}{4\kappa}}(v^{2k}+v^{-2k})\,,\\
    \eta \,\xi_v^{C_r}&= - \sum_{k\geq 1}  \hat\chi_{(2k-1)\Lambda_1} q^{-\frac{(2k-1)^2}{4\kappa}} (v^{2k-1}+v^{-(2k-1)})\,,\nonumber\\
    \xi_s^{C_r}&=\mathcal{F}_\kappa\xi_1^{C_r}\,, \nonumber\\
    \xi_c^{C_r}&=\mathcal{F}_\kappa\xi_v^{C_r}\,. \nonumber
\end{align}
Recall that the operators $\mathcal{F}_\kappa$ were introduced in equation \eqref{eq:def_F}. Also, the flavor symmetry of the $(C_r)_1$ theory is $D_{8+2r}$ at level 1. Only the representations with fundamental weight associated to the two extremities of the affine Dynkin diagram contribute; we retain the same name for these as in the case of $D_4$. The result at $r=1$ takes a slightly different form:
\begin{align} \label{eq:C1}
    \eta\, \xi_0^{C_1}&=\sum_{k\geq 1}  \hat\chi_{2k\Lambda_1} \sum_{m=-k}^k q^{-m^2}v^{2m}\,,\\
    \eta \,\xi_v^{C_1}&= - \sum_{k\geq 1}  \hat\chi_{(2k-1)\Lambda_1} \sum_{m=-k}^k q^{-\frac{(2m+1)^2}{4}}v^{-1-2m}, \nonumber\\
    \xi_s^{C_1}&=\mathcal{F}_1\xi_1^{C_1}\,,\nonumber\\
    \xi_c^{C_1}&=\mathcal{F}_1\xi_v^{C_1}\,, \nonumber
\end{align}
where we have explicitly substituted $\kappa=1$.

Note that just as for the $(D_4)_4$ theory, a UV description of the $(C_r)_1$ theories is known; the elliptic genera can hence be computed exactly in terms of $\theta$-functions \cite{Kim:2014dza}. The results \eqref{eq:Cr} and \eqref{eq:C1} were consequently obtained without recourse to the algorithm presented in section \ref{sec:algorithm}. The fact that the coefficients $c^\lambda_{\omega,k}$ all equal $\pm 1$ thus provides additional evidence for the conjectured form \eqref{eq:affineAnsatzIntro} of the elliptic genus which lies at the heart of this work. 

We remark that for these theories, we can combine the knowledge of $\ER$ in terms of $\theta$-functions with the affine ansatz \eqref{eq:affine_G_expansion} to derive explicit formulae for (some) level $-1$ characters of $C_r$.

Let us consider $r>1$, as only a single affine character contributes at each power of $v$ in this case (except for $v^0$ in $\xi_0$). The functions $\xi_0,\,\xi_v$ in terms of $\theta$-functions are given by \cite{Kim:2014dza,DelZotto:2018tcj}
\begin{equation}
     \xi_{0/v}^{C_r}=\frac{1}{2}\left(\prod_{\substack{1\leq i\leq r\\s=\pm}}^r\frac{\eta}{\theta_3(v (X_i^{C_1})^s)}\pm\prod_{\substack{1\leq i\leq r\\s=\pm}}^r\frac{\eta}{\theta_4(v (X_i^{C_1})^s)}\right)\,,
     \label{eq:xi_n=1_theta}
\end{equation}
where the $+$ sign corresponds to $0$ and the $-$ to $v$. We then have that
\begin{align}
\hat\chi_{\ell\Lambda_1}&=q^\frac{\ell^2}{4r}\left[\frac{1}{2}\left((-1)^\ell\prod_{\substack{1\leq i\leq r\\s=\pm}}^r\frac{\eta}{\theta_3(v (X_i^{C_1})^s)}+\prod_{\substack{1\leq i\leq r\\s=\pm}}^r\frac{\eta}{\theta_4(v (X_i^{C_1})^s)}\right)\right]_{v^{\ell}}\,,
\end{align}
where $[\cdot]_{v^\ell}$ signifies the order $v^\ell$ term of $[\cdot]$.\footnote{We could of course equally well choose the term $v^{-\ell}$.} This expression can be further simplified using the explicit forms or the modular transformation properties of the $\theta$- functions to give
\begin{equation}
    \hat\chi_{\ell\Lambda_1}=q^\frac{\ell^2}{4r}\left[\prod_{\substack{1\leq i\leq r\\s=\pm}}^r\frac{\eta}{\theta_4(v (X_i^{C_1})^s)}\right]_{v^{\ell}}\,.
\end{equation}
To our knowledge, this expression for level $-1$ characters of $C_r$ has not appeared previously in the literature. It would be interesting to compare it to a recent result by Kac and Wakimoto \cite{Kac2018} on these characters.

\section{Relations along the Higgsing tree} \label{s:Higgs_tree_relations}
When explicit results for the elliptic genera of a pair of theories related by Higgsing are known in terms of $\theta$-functions, it has been found in the literature \cite{Okuda_2012,DelZotto:2018tcj,Gu:2020fem} that the elliptic genus of the Higgsed theory can be obtained by a specialization of the parameters of the parent theory. Notably, some flavor fugacities of the parent theory must be replaced by the fugacity $v$ of the $SU(2)_v$ symmetry. In section \ref{ss:specializing}, we determine this specialization map for a number of theories, notably for multiple theories for which no expressions for the elliptic genus in terms of $\theta$-functions is known. The fact that also in these cases, the specialization map is of the expected form provides additional evidence for the various assumptions that enter in our derivation of the affine presentation of $\ER$.

In section \ref{sub:constrainguptree}, we revisit an idea of \cite{Duan:2020imo}, which demonstrated that the computation of elliptic genera can sometimes be simplified by imposing a larger Weyl symmetry, one which occurs ``further up the tree'' from the theory $G_n$ in question, in choosing the ring of Jacobi forms than the one suggested by the group $G$. We will consequentially consider the problem of expanding the elliptic genera with characters of a Lie algebra that correspond to a possible unHiggsing up the tree. Note that such an expansion is necessarily possible as long as additional $q$ dependence outside that occurring in the group and flavor characters is permitted: it imposes a larger (finite) Weyl symmetry on the $m_G$ dependence of $\ER$, which is permissible following the results of \cite{Duan:2020imo}. However, we will see that the constrained form of the ansatz \eqref{eq:affineAnsatzIntro} no longer holds for this larger Lie algebra. The possibility of unHiggsing can however be used to constrain the representations $\omega$ occurring in the ansatz \eqref{eq:affineAnsatzIntro} further. We will illustrate these points using the example of the Higgsing $(F_4)_4\to (D_4)_4$.

\subsection{Higgsing via specialization of fugacities} \label{ss:specializing}

Given a theory with gauge and flavor groups $G,\,F$ and a Higgsing down the Higgsing tree to a theory with gauge and flavor groups $G',\,F'$, we find that there are two maps \cite{Okuda_2012,Gu:2020fem,DelZotto:2018tcj}
\begin{equation} \label{eq:specialization_maps}
    \iota_G:\mathfrak h'\to\mathfrak h,\quad \text{and }\quad \iota_F:\mathfrak f'\oplus \mathfrak h_v\to\mathfrak{f}\,,
\end{equation}
with $\mathfrak h',\,\mathfrak h,\,\mathfrak f',\,\mathfrak f,$ and $\mathfrak h_v$ denoting the Cartans of the Lie algebras associated to $G',\,G,\,F',\,F, $ and $SU(2)_v$ respectively, such that
\begin{equation}
   \ER^{(G',F')}(m_G',m_F',v,q)=\iota^*\left(\ER^{(G,F)}\right)(m_G',m_F',v,q):=\ER^{(G,F)}(\iota_G (m_G'),\iota_F(m_F'),v,q)\,.
\end{equation}
The map $\iota_G$ was described for most gauge groups in \cite{Duan:2020imo}, where the same question was studied from the point of view of the modular ansatz for $\ER$. The map $\iota_F$ is given by the same transformation for the flavor part $\mathfrak f$, and an inclusion of $\mathfrak{h}_v$ in an orthogonal direction. 

We will describe these transformations explicitly in several examples in the following. We give the functions $\iota^*$ acting in either the exponentiated fugacities of the Euclidean lattice $X_i=e^{2\pi i (e_i,m_G)}$ or in terms of the exponentiated fugacities corresponding to the Lie algebra roots $Q_i=e^{2\pi i(\alpha_i,m_G)}$. We follow the same normalization conventions outlined in appendix C of \cite{Duan:2020imo}.


{\bf $\mathbf C_r$ tower}

The only theories with $C_r$ gauge symmetry appear on a branch of the E-string Higgsing tree. The specialization maps \eqref{eq:specialization_maps} can be read off \cite{DelZotto:2018tcj} from the explicit expression of the elliptic genera in terms of $\theta$-functions \cite{Kim:2014dza}: the Cartan algebra of $C_r$ is mapped to the hyperplane $x_{r+1}=0$ of the Euclidean space in which the Cartan algebra of $C_{r+1}$ is embedded. For the flavor group, one identifies the Cartan algebra of $D_{8+2r}$ with the co-dimension 2 space $x_{8+2r+1}=x_{8+2r+2}=0$ of the corresponding higher-dimensional Euclidean space. The fugacities corresponding to these two directions are then replaced by $v$. Explicitly,
\begin{equation}
   \iota_G^*: X_i^{C_{r+1}}\mapsto\left\{\begin{array}{cc}
        X_i^{C_r} & i=1,\dots, r  \\
        1 & i=r+1
    \end{array}\right.\!,\, \iota_F^*: X_i^{D_{8+2(r+1)}}\mapsto \left\{\begin{array}{cc}
        X_i^{D_{8+2r}} & i=1,\dots, 8+2r  \\
        v & i=8+2r+1, 8+2r+2
    \end{array}\right..
\end{equation}

{\bf $\mathbf{B/D\sim SO(N)}$ towers}

The Higgsing trees at $n=1,2,3,4$ exhibit a (finite or infinite) branch of alternating $B$ and $D$ theories.

We study the $n=4$ case for which the exact elliptic genus is known in terms of $\theta$-functions \cite{Kim:2014dza,DelZotto:2018tcj}. From this expressions one can get the fugacity transformation for the different Higgsings. For the $B_r\to D_r$ Higgsing we identify the two Cartan sub-algebras through their embedding in Euclidean space and the $C_{r-3}\to C_{r-4}$ flavor fugacities as in the previous example, except that we set the last fugacity to $v$. Explicitly,
\begin{equation}
    \iota_G^*:X_i^{B_{r}}\mapsto X_i^{D_r},\quad \iota_F^*:X_i^{C_{r-3}}\mapsto \left\{\begin{array}{cc}
        X_i^{C_{r-4}} & i=1,\dots, r-4  \\
        v & i=r-3
    \end{array}\right.\,.
\end{equation}

For the Higgsing $D_{r+1}\to B_{r}$, the flavor fugacities transform in the same fashion, while the $(r+1)^{st}$ $D_{r+1}$ fugacity must be set to 1:
\begin{equation}
    \iota_G^*: X_i^{D_{r+1}}\mapsto\left\{\begin{array}{cc}
        X_i^{B_r} & i=1,\dots, r  \\
        1 & i=r+1,
        \end{array}\right.
        \quad 
        \iota_F^*:X_i^{C_{r-2}}\mapsto \left\{\begin{array}{cc}
        X_i^{C_{r-3}} & i=1,\dots, r-3  \\
        v & i=r-2
    \end{array}\right.\,.
\end{equation}

{$\mathbf{F_4\to D_4}$}

\label{sub:F4toD4Qtransformation}
The Higgsing trees with $n=1,2,3,4$ each exhibit a Higgsing from a theory with gauge group $F_4$ to one with gauge group $D_4$. Only for the $(D_4)_4$ theory is the elliptic genus known exactly in terms of $\theta$-functions.

We study the $n=4$ case. The transformation of the gauge fugacities is the same as in \cite{Duan:2020imo}:
\begin{equation}
        Q_1^{F_4} \mapsto Q_2^{D_4} \,,\,\, Q_2^{F_4} \mapsto Q_1^{D_4} \,,\,\, Q_3^{F_4} \mapsto \sqrt{\frac{Q_3^{D_4}}{Q_1^{D_4}}}  \,,\,\, Q_4^{F_4} \mapsto \sqrt{\frac{Q_4^{D_4}}{Q_3^{D_4}}}  \,, 
        \label{eq:F4toD4Qtransformation}
    \end{equation}
while for the flavor fugacity we have
\begin{equation*}
    X^{C_1}\mapsto v.
\end{equation*}
We will revisit this example in section \ref{sub:constrainguptree}.

{$\mathbf{E_6 \to F_4}$}

For the $n=5$ tree, we consider the elliptic genera of $(F_4)_5$ and $(E_6)_5$. To find the appropriate replacement of the gauge parameters, we use the transformation \eqref{eq:F4toD4Qtransformation} to identify the $D_4$ and $F_4$ root lattices. Then, the $D_4$ root lattice can be embedded in the root lattice of $E_6$ simply by identifying the $D_4$ Euclidean space with the space $x_5=0,x_6=0$ in the Euclidean space of $E_6$. This gives the $\iota_G^*$ transformation 
\begin{equation}
    \begin{array}{ccc}
 Q_1^{E_6}\to \frac{1}{Q_3^{F_4} Q_4^{F_4}}\,,&Q_2^{E_6}\to \frac{1}{Q_2^{F_4}}\,,&Q_3^{E_6}\to \frac{1}{Q_1^{F_4}},
     \\Q_4^{E_6}\to
   \frac{1}{Q_2^{F_4} \left(Q_3^{F_4}\right)^2}\,,&Q_5^{E_6}\to \frac{1}{Q_4^{F_4}}\,,&Q_6^{E_6}\to \left(Q_1^{F_4}\right)^2
   \left(Q_2^{F_4}\right)^3 \left(Q_3^{F_4}\right)^4 \left(Q_4^{F_4}\right)^2\,.
    \end{array}
    \label{eq:E6toF5}
\end{equation}
The flavor fugacity transformation is simply $\iota_F^*:Q^{U_1}\mapsto v$. 

Recall that we can fix the functions $\xi$ for $(E_6)_5$ only up to a Dynkin symmetry, as explained at the end of section \ref{ss:strategy}. We unfortunately cannot resolve this ambiguity using the Higgsing, as the finite representations of $E_6$ related by the Dynkin symmetry have the same image under the transformation \eqref{eq:E6toF5},
\begin{equation}
   \iota_G^* \chi^{E6}_{(a_1,a_2,a_3,a_4,a_5,a_6)}=\iota_G^* \chi^{E6}_{(a_5,a_4,a_3,a_2,a_1,a_6)}
\end{equation}

\subsection{Enhanced Weyl symmetry $\nRightarrow$ enhanced affine symmetry}
\label{sub:constrainguptree}
In \cite{Duan:2020imo}, a variety of circumstances were found under which the Weyl symmetry $\cW_G$ of a $G_n$ theory is enhanced.\footnote{In many cases, we also observe an enhancement of the coroot translational invariance of the elliptic genus to finer lattices. This is explained by the absence of certain representations in the charged spectrum or delicate cancellation between matter and gauge contributions \cite{Kashani-Poor:2019jyo}.} The possible enhancements include $\cW_{D_4}\to \cW_{F_4}$, $\cW_{B_4}\to \cW_{F_4} $, $\cW_{A_2}\to \cW_{G_2}$, and $\cW_{D_n}\to \cW_{B_n}$. This observation was put to good use in choosing to expand the elliptic genus in terms of Jacobi forms invariant under the enhanced Weyl symmetry. It is thus natural to ask whether in the context of our affine ansatz \eqref{eq:affineAnsatzIntro}, enhanced Weyl symmetry implies a simplified ansatz in terms of affine characters of the Lie algebra associated to it. This sadly does not appear to be the case. We will use the example $(F_4)_4 \rightarrow (D_4)_4$ which already appeared above to illustrate this point.

Let us thus look at this example in somewhat more detail. The  transition $\iota$ in this case is implemented by the identification of the $D_4$ and $F_4$ coweight lattices and the transformation of the unique flavor fugacity to $v$ as explained in subsection \ref{sub:F4toD4Qtransformation}. At leading order in $q$, the functions $\xi$ and the corresponding flavor characters (specialized to the value $v$) are given by
\begin{equation}
    \begin{array}{ccc}
     \lambda& \hat\chi_\lambda(v) &\eta\xi  \\\hline
    (0) &1&-v^7 \chi^{F_4}_{(0000)}+v^{11}\sum v^{2j}\chi^{F_4}_{(j100)}\\
     (1) &v^{-1}+v& -v^{10}\sum_j v^{2j}\chi^{F_4}_{(j010)}\\
     (2) &v^{-2}+1+v^2& v^{9}\sum_j v^{2j}\chi^{F_4}_{(j001)}\\
     (3) &v^{-3}+v^{-1}+v+v^3&-v^{8}\sum_j v^{2j}\chi^{F_4}_{(j000)}
\end{array}
\end{equation}
To recover the elliptic genus of $(D_4)_4$, we must multiply the two columns, add the resulting rows, and express the $F_4$ characters in terms of $D_4$ characters. This yields
\begin{equation} \label{eq:leadingD4inF4}
\begin{aligned}
        &-v^5\chi^{F_4}_{(0000)}-v^7(\chi^{F_4}_{(1000)}-\chi^{F_4}_{(0001)}+2\chi^{F_4}_{(0000)})\\
        &-v^9(\chi^{F_4}_{(2000)}-\chi^{F_4}_{(1001)}+\chi^{F_4}_{(1000)}+\chi^{F_4}_{(0010)}-\chi^{F_4}_{(0001)}+\chi^{F_4}_{(0000)})\\
        &-\sum_{j\geq 3} v^{5+2j}(\chi^{F_4}_{(j000)}-\chi^{F_4}_{(j-1,001)}+\chi^{F_4}_{(j-1,000)}+\chi^{F_4}_{(j-2,010)}-\chi^{F_4}_{(j-2,001)}\\&+\chi^{F_4}_{(j-2000)}+\chi^{F_4}_{(j
        -3,010)}-\chi^{F_4}_{(j-3,001)}+\chi^{F_4}_{(j-3,000)})\\
        &=\sum_jv^{5+2j}\chi^{D_4}_{0j00}\,,
\end{aligned}\end{equation}
where in the last line we have implemented \eqref{eq:F4toD4Qtransformation}. We draw two conclusions from this calculation: firstly, that it is possible to write the $(D_4)_4$ elliptic genus in terms of affine characters of $F_4$: \eqref{eq:leadingD4inF4} demonstrates this for the finite characters at leading order in $q$. Replacing all characters by their affine counterparts preserves the equality at leading order in $q$. We can correct the expression at the next order in $q$ by subtracting affine characters of $F_4$ that are induced but do not occur in the $(D_4)_4$ elliptic genus at this level, and by adding in affine characters of $D_4$ that do not arise by affinizing the $F_4$ characters at lower level. For this procedure to work, it is of course crucial that the gauge fugacities of the $(D_4)_4$ elliptic genus exhibit $\cW_{F_4}$ symmetry. This was observed in \cite{Duan:2020imo}, and is manifest in the affine expansion, as the sum over $\omega$ is over Dynkin symmetric representations $(0j00)$ only (at each order in $q$, the finite representations contributing to $\hat \chi^{D_4}_{(0j00)}$ occur in Dynkin symmetric combinations). Note that we have permitted ourselves to introduce $q$ dependence beyond that carried by the characters. The second conclusion we draw from the result \eqref{eq:leadingD4inF4} is that expressing the leading contribution to the $(D_4)_4$ elliptic genus in terms of $F_4$ characters has rendered the result more cumbersome. In particular, the structure \eqref{eq:affineAnsatzIntro} is not preserved.

The source of this complication is of course the additional $v$ dependence contributed by $\hat \chi_\lambda(v)$. Consider e.g. the $F_4$ character $\chi^{F_4}_{(j000)}$. It appears at leading order in $q$ in the $(F_4)_4$ elliptic genus because 
\begin{equation*}
    x_0((3),(j000),8+2j)=\casimirR\,.
\end{equation*}
However, in the $(D_4)_4$ elliptic genus, there is a contribution from the same character at orders $8+2j+\delta$ where $\delta={-3,-1,1,3}$ (the powers of $v$ appearing in $\chi^{C_1}_{(3)}$). Therefore, any prescription for expressing $\ER^{D_4}$ in terms of $F_4$ characters would need to predict a term at order $k$ in $v$ for every $F_4$ dominant weight $\omega $ that satisfies 
\begin{equation*}
    x_0((3),\omega,k-\delta)=\casimirR \quad \text{ for any choice of } \delta={-3,-1,1,3}\,,
\end{equation*}
or that satisfies the analogous equation for any of the other $\lambda$'s. At higher order in $q$ the problem gets amplified because there are $q$-dependent contributions coming from $\hat \chi_\lambda(v)$. 

    To end on a positive note, we point out that expressing the characters of the Higgsed gauge group in terms of the characters of the parent gauge group does teach us something about the parent theory: which characters must occur. Thus, the relation \eqref{eq:leadingD4inF4} tells us that characters of the form $(j000),\,(j100),\,(j010),\,(j001)$ must appear in the $(F_4)_4$ elliptic genus, to make the specialization to the $(D_4)_4$ elliptic genus possible.

    \section{Conclusions} \label{s:conclusions}
    We have put forth in this paper a conjecture on the form of the elliptic genus of the non-critical strings of 6d field theories obtained from F-theory compactifications on elliptically fibered Calabi-Yau manifolds over Hirzebruch bases, given in equations \eqref{eq:affineERintro}, \eqref{eq:affineAnsatzIntro} and \eqref{eq:constraintIntro}, and provided ample evidence for its validity. Our results match expressions for the elliptic genus when these are known, and fit into specialization sequences in Higgsing trees containing known results. 

    It would be interesting to extend our ansatz beyond the class of theories discussed here, e.g. to the class F-theory compactifications on elliptic fibrations without section \cite{Cota:2019cjx} or theories obtained via twisted compactification \cite{Duan:2021ges}. Both constructions lead to elliptic genera involving Jacobi forms of congruence subgroups of $SL(2,\IZ)$. Also, studying the elliptic genera of multiple strings is a natural next step.

    Already for the non-critical strings whose elliptic genera we explored in this paper, an important challenge remains to be met: to explain our ansatz and the results of our computation from the worldsheet theory of the non-critical string. What is the origin of the non-integrable affine symmetry of $G$? More ambitiously still, can the closed form results presented in section \ref{s:closed_form_results} and the polynomials $p_\omega^\lambda$ introduced in equation \eqref{eq:poly_multiplying_omega} and computed for many examples in appendix \ref{a:polynomials} be computed from the vantage point of the worldsheet theory? We hope to return to these question elsewhere.

    \section*{Acknowledgements}

    We would like to thank Michele Del Zotto, Gugliemo Lockhart, Lionel Mason, Noppadol Mekareeya, Thorsten Schimannek and Timo Weigand for useful conversations. Special thanks to Balt van Rees for sharing the Mathematica code from the project \cite{Beem:2013sza} for computing negative level characters of $D_4$ with us. This helped speed up the development of our own SageMath code for arbitrary simple Lie algebras substantially.

    A.K.\ acknowledges support under ANR grant ANR-21-CE31-0021.

\chapter*{Conclusions}
\addchaptertocentry{Conclusions} 
\label{cha:conclusion}

The primary objective of this thesis is to compute the topological partition function on families of elliptically fibered Calabi-Yau spaces and to explore the associated 6D, 5D, and 2D theories. The initial three chapters cover standard material in topological strings and F-theory, while the final two chapters discuss the findings presented in \cite{Duan:2020imo} and \cite{Duque:2022tub}.

In Chapter \ref{ch:paperJacobi}, we leverage the modular properties of the topological string partition function to compute the spectrum of several theories. We introduce a set of rules for obtaining the partition function of a daughter theory from its parent, corresponding naturally to the embedding of the weight lattice of the daughter theory into the weight lattice of its mother. These rules grant access to the spectrum of theories whose partition function cannot be obtained through traditional mirror symmetry computation and unveil new symmetries. These symmetries are frequently linked to the center of the corresponding Lie group. It is therefore plausible that the observed symmetry enhancement is associated with a higher form symmetry, potentially explaining the extension of symmetries from massless modes to the entire tower. 

Chapter \ref{ch:paperWorldsheet} introduces a new ansatz \eqref{eq:affineAnsatzIntro} for part of the topological string partition function. Motivated by the study of the worldsheet theory of the non-critical string, this ansatz utilizes the WZW characters of the flavor, gauge, and R-symmetry current algebras of the 6D theories. The robust nature of this ansatz leaves minimal room for a handful of unknown constants, $c_{\omega,\lambda}^k$, restricted to values of $0,\pm 1, \pm 2$. Employing this ansatz, we computed the elliptic genus of several theories and obtained closed-form results for a select few. These findings not only validate the proposed Higgsing rules, with slight adjustments for the refinement parameter, but also pave the way for further exploration. Extending these results to other F-theory models, such as genus-1 fibrations without sections, or to the elliptic genus of multiple strings, is a natural step forward. Another natural question to pursue is to find a pure worldsheet explanation of the ansatz and the results in Appendix \ref{a:polynomials}, aiming to explain the appearance of negative level characters.


\appendix 



\chapter{Toric geometry} 
\label{ap:ToricGeometry}
The toric construction provides a powerful framework for describing many Calabi-Yau threefolds in terms of fans and polytopes. This description allows us to understand these varieties to a great extent by considering the polytopes associated with them. Furthermore, from the perspective of mirror symmetry, the polytope description is highly enlightening, as the association of a polytope with its polar counterpart is equivalent to the association of a threefold with its mirror,
\begin{equation*}
    \Delta\leftrightarrow\Delta^\circ \quad \iff \quad X_\Delta \leftrightarrow X_{\Delta^\circ}=\tilde{X}_\Delta.
\end{equation*}

The basic idea of the toric construction, as described for instance in \cite{cox_toric_2011,cox_mirror_1999,fulton_introduction_1993,oda_convex_2012}, is to mimic the description of projective spaces as
\begin{equation*}
    \P^n=\frac{\C^{n+1}-\{0\}}{\C^*}\,.
\end{equation*}
A fan $\Sigma$ provides all the information needed to construct the variety. The fan encodes the group action (corresponding to the action $\mathbb{C}^*\acts \mathbb{C}^{n+1}$ in projective space), the points that need to be removed (corresponding to $\{0\}$), and the dimension of the variety (corresponding to $n$).

\section{The lattices \texorpdfstring{$N,M$}{N,M}, the cones, and the fan}
In the toric construction, we start with a lattice $N \simeq \mathbb{Z}^n$ and its dual lattice $M = \text{Hom}(N,\mathbb{Z})$. A subset $\sigma \subset N_\mathbb{R} = N \otimes \mathbb{R} \simeq \mathbb{R}^n$ is a \emph{strongly convex rational polyhedral cone} if there exist vectors $u_1, \dots, u_s \in N$ such that
\begin{equation*}
    \sigma=\R_{\geq 0}u_1+\dots\R_{\geq 0}u_s =\left\{\sum_{i=1}^s\lambda_i u_i: \lambda_i\in \R,\, \lambda_i\geq 0\right\}
\end{equation*}
and $\sigma\cap (-\sigma)=\{0\}$. For our purposes, we will be considering only strongly convex rational polyhedral cones, so whenever we mention cones, it implies these specific types of cones.\footnote{Strongly convex means $\sigma\cap (-\sigma)=\{0\}$ and rational means $u_i$ are vectors in $N$, rather than in $N_\R$.}

A fan $\Sigma$ is then defined as a collection of cones that satisfy two conditions:
\begin{itemize}
    \item If $\sigma\in \Sigma$ and $\tau$ is a \emph{face} of $\sigma$ then $\tau\in \Sigma$.
    \item If $\sigma,\tau\in\Sigma$, then $\tau\cap\sigma$ is a face of both $\tau$ and $\sigma$.
\end{itemize}
To understand the concept of faces more explicitly, we can define a face as follows. Using the natural pairing between $N$ and $M$, we can construct the dual cone $\check{\sigma}$ of a cone $\sigma$ as
\begin{equation}
\begin{split}
        \check\sigma&=\{m\in M_\R=M\otimes\R: \left<m,u\right>\geq 0,\, \forall u\in \sigma\}\\
    &=\{m\in M_\R=M\otimes\R: \left<m,u_i\right>\geq 0,\,i=1,\dots s\}.
\end{split}
    \label{eq:dual}
\end{equation}
We say that $\tau \subset \sigma$ is a face of $\sigma$ if there exists $m \in M \cap \check{\sigma}$ such that
\begin{equation*}
    \tau=\{u\in \sigma:\left<m,u\right>=0\}.
\end{equation*}

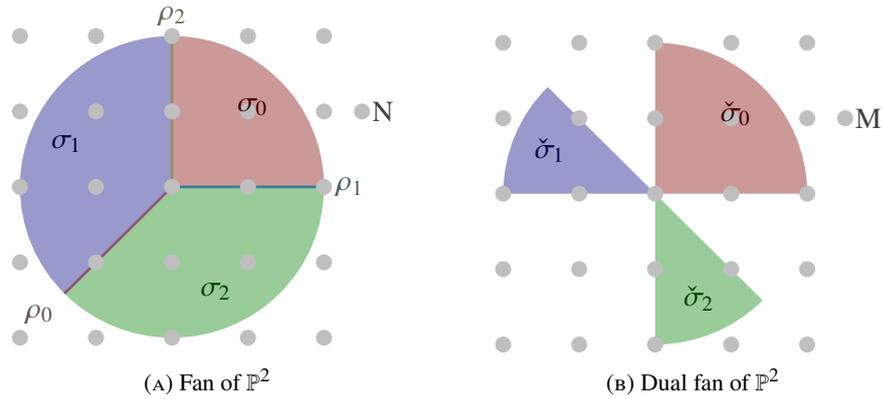
\begin{figure}
    \centering
    \begin{subfigure}[b]{0.5\textwidth}
        \centering
        \begin{tikzpicture}
\fill[mdtRed!40] (0,0)--(2,0) arc (0:90:2)--cycle;
         \fill[mdtBlue!40] (0,0)--(90:2) arc (90:225:2)--cycle;
         \fill[mdtGreen!40] (0,0)--(225:2) arc (225:360:2)--cycle;
        
        \draw[line width=1pt, mdtYellow] (0,0)--(90:2)node[anchor=south,mdtYellow!50!black]{$\rho_2$};
        \draw[line width=1pt, mdtCyan] (0,0)--(0:2)node[anchor=west,mdtCyan!50!black]{$\rho_1$};
        \draw[line width=1pt, mdtMagenta] (0,0)--(225:2)node[anchor=north east,mdtMagenta!50!black]{$\rho_0$};
        
        \foreach \i in {-2,-1,0,1,2}
        {\foreach \j in {-2,-1,0,1,2}
        {\fill[gray!50] (\i,\j) circle (3pt);
        }
        }
        \fill[gray!50] (2.5,1) circle (3pt) node[anchor=west,gray!50!black]{N};
        
        \node[mdtRed!50!black] at (45:1.5) {$\sigma_0$};
        \node[mdtBlue!50!black] at (157.5:1.5) {$\sigma_1$};
        \node[mdtGreen!50!black] at (292.5:1.5) {$\sigma_2$};
        \end{tikzpicture}
        \caption{Fan of $\P^2$}
         \label{fig:fanP2}
    \end{subfigure}%
    \begin{subfigure}[b]{0.4\textwidth}
        \centering
        \begin{tikzpicture}
         \fill[mdtRed!40] (0,0)--(2,0) arc (0:90:2)--cycle;
         \fill[mdtBlue!40] (0,0)--(135:2) arc (135:180:2)--cycle;
         \fill[mdtGreen!40] (0,0)--(270:2) arc (270:315:2)--cycle;
         
         \foreach \i in {-2,-1,0,1,2}
        {\foreach \j in {-2,-1,0,1,2}
        {\fill[gray!50] (\i,\j) circle (3pt);
        }
        }
        \fill[gray!50] (2.5,1) circle (3pt) node[anchor=west,gray!50!black]{M};
        
        \node[mdtRed!50!black] at (45:1.5) {$\check\sigma_0$};
        \node[mdtBlue!50!black] at (157.5:1.5) {$\check\sigma_1$};
        \node[mdtGreen!50!black] at (292.5:1.5) {$\check\sigma_2$};
        \end{tikzpicture}
        \caption{Dual fan of $\P^2$}
        \label{fig:dualfanP2}
    \end{subfigure}%
    \caption{Fan an dual fan of the projective space $\P^2$}
\end{figure}

To illustrate these concepts, let's consider the fan of the projective space $\mathbb{P}^2$, shown in Figure \ref{fig:fanP2}. In this fan, there are three 2-dimensional cones, three 1-dimensional cones, and one 0-dimensional cone. We can also compute the dual fan, which consists of the dual big cones and their faces. For example, the dual cone of $\sigma_1 = \mathbb{R}_{\geq 0} (0,1)^T + \mathbb{R}_{\geq 0} (-1,-1)^T$ is the subset of $M_\mathbb{R}$ defined by the conditions
\begin{equation}
    m_2 \geq 0 \quad \text{and}\quad -m_1 - m_2 \geq 0
\end{equation}
The dual fan is shown in Figure \ref{fig:dualfanP2}. This serves as an example to help illustrate the general ideas, and we will use it to explain the general construction in a concrete context.

\section{Homogeneous coordinates}

Toric varieties, as constructed from fans, can be described in several equivalent ways. The homogeneous coordinates description is one in which the group action takes a particularly simple form and divisors can be described in a relatively simple fashion. In this approach, we associate a coordinate $x_\rho$ to each element $\rho \in \Sigma(1)$, where $\Sigma(1)$ is the set of 1-dimensional cones in the fan. The space we want to take the quotient of is $\mathbb{C}^{\Sigma(1)}$, which consists of the coordinates $x_\rho$.

To construct the group action, we use the fact that for each $\rho \in \Sigma(1)$, there exists a unique vector $v_\rho \in N$ such that $\rho \cap N = v_\rho \mathbb{Z}$. Using these vectors, we can define the Chow group $A_{n-1}$ through the short exact sequence
\begin{equation}
    0\to M\to \Z^{\Sigma(1)}\to A_{n-1}\to 0,
    \label{eq:chowseq}
\end{equation}
where the second map sends $m$ to $(\left<m,v_\rho\right>)_\rho$.

Taking the $\text{Hom}(\cdot,\mathbb{C}^*)$ of this sequence, we obtain
\begin{equation}
    1\to G=\Hom(A_{n-1},\C^*)\to (\C^*)^{\Sigma(1)}\to T_N=N\otimes \C^*\to 1\,.
    \label{eq:actseq}
\end{equation}
In particular, we have a natural action on $G\acts\C^{\Sigma(1)}$ via
\begin{equation*}
    g(a_\rho)_\rho=(g(v_\rho)a_\rho)_\rho\,.
\end{equation*}
The \emph{toric action} of $T_N$ is compatible with the quotient by $G$.

To complete the description of the toric variety, we define the points we want to remove. A subset $\mathcal{S} \subset \Sigma(1)$ is called a primitive collection if it satisfies two conditions: (1) $\mathcal{S}$ is not the set of 1-dimensional cones of any cone in $\Sigma$, and (2) every proper subset of $\mathcal{S}$ is contained in some cone in $\Sigma$. We then define
\begin{equation*}
    Z(\Sigma)=\bigcup_{\Sc \text{ primitive}}V(\Sc),\quad V(\Sc)=\{(x_\rho)_\rho\in \C^{\Sigma(1)}: x_\rho=0,\forall \rho \in \Sc\}\,.
\end{equation*}
Finally, the toric variety $\mathbb{P}(\Sigma)$ is defined as\footnote{This definition only holds when $\Sigma(1)$ spans $N_\R$ and the fan is simplicial, that is to say, the generators of each cone are linearly independent over $\R$. In general we will blow up singularities and work with smooth resolutions, hence the development in the text is sufficient.}
\begin{equation}
    \P(\Sigma)=\frac{\C^{\Sigma(1)}-Z(\Sigma)}{G}\,.
\end{equation}

Let's use the fan in Figure \ref{fig:fanP2} to construct the associated toric variety and show that it is indeed $\mathbb{P}^2$. There are three 1-dimensional cones, so we have three coordinates $x_0, x_1, x_2$. The Chow group is found by noticing that $M\to \Z^{\Sigma(1)}\simeq \Z^3$ is given by 
\begin{equation*}
    m=\left(\begin{array}{c}
         m_1\\m_2  
    \end{array}\right)\mapsto\left(\begin{array}{c}
         -m_1-m_2\\m_1\\m_2 
    \end{array}\right)\,.
\end{equation*}
The image of this map consists of the vectors $(a_0, a_1, a_2)^T\in\mathbb{Z}^3$ with $a_0 + a_1 + a_2 = 0$. Thus, $A_{2}$ is one-dimensional and isomorphic to $\mathbb{Z}$. The map $\mathbb{Z}^{\Sigma(1)}\to A_{2}$ can be explicitly described as $(a_0, a_1, a_2)^T \mapsto a_0 + a_1 + a_2$ which shows $G=\Hom(\Z,\C^*)\simeq\C^*$, where $g(v_0) = g(1,0,0) = g^{1+0+0}$, and similarly for the other two 1-dimensional cones.  In general, the action of $G$ on the coordinates is given by
\begin{equation*}
    g\cdot\left(\begin{array}{c}
         x_0  \\x_1\\x_2
    \end{array}\right)=\left(\begin{array}{c}
         gx_0  \\gx_1\\gx_2
    \end{array}\right).
\end{equation*}

In this case, the only primitive collection is $\Sc = \{\rho_1, \rho_2, \rho_2\}$, and $V(\Sc) = \{(0,0,0)\}$. Therefore, the set $Z(\Sigma)$ is given by ${(0,0,0)}$ and the associated variety is then given by:
\begin{equation*}
    \P(\Sigma)=\frac{\C^3-\{0\}}{\C^*}=\P^2
\end{equation*}
This confirms that $\mathbb{P}(\Sigma)$ is indeed the projective plane.

Regarding the toric action, the torus $T_N$ is isomorphic to $(\mathbb{C}^*)^2$ and the short exact sequence \ref{eq:actseq} can be used to describe the action explicitly. The action of $T_N$ on itself is given by $(t_1,t_2) \cdot (v_1,v_2) = (t_1v_1, t_2v_2)$. To obtain the action on the coordinates $[x_0,x_1,x_2]$, we can pull it back from $T_N=(\C^*)^3/\C^*=(\C^*)^2$. This gives $(t_1,t_2) \cdot [x_0,x_1,x_2] = [t_1^{-1}t_2^{-1}x_0, t_1x_1, t_2x_2]$. The exponents in this expression are fixed by the coordinates of $v_\rho$. \footnote{The asymmetry in the coordinates comes from a choice of basis. For instance, for the basis $u_1=t_1^{-1}t_2^{-1},u_2=t_2$ of $T_N$, we get $(u_1,u_2)[x_0,x_1,x_2]=[u_1x_0,u_1^{-1}u_2^{-1}x_1,u_2x_2]$ where the exponents of the $u$'s multiplying $x_1$ correspond to $(1,0)^T=-1(-1,-1)^T-1(0,1)^T$.}.


\section{Divisors, other sub-varieties, and compactifications}
In addition to reconstructing the toric variety from the fan, we can extract more information from the fan by associating $n-k$ dimensional subvarieties to each $k$-dimensional cone $\sigma \in \Sigma$. The subvariety corresponding to $\sigma$ is given by
\begin{equation*}
    \sigma\in \Sigma\leftrightarrow \{x_\rho=0:\rho\subset \sigma\}\subset \P(\Sigma).
\end{equation*}
Of particular interest are the divisors, which are the subvarieties associated with 1-dimensional cones. Divisors play a special role as they are in correspondence with line bundles on the manifold.

In our example of $\mathbb{P}^2$, we can compute the corresponding subvarieties as follows:
\begin{equation*}
    \begin{array}{ccc}
         0\to \P^2,&
         \rho_i\to \{x_i=0\}\simeq \P^1,&
         \sigma_0\to [1,0,0], \dots.
    \end{array}
\end{equation*}
We can also observe the role of higher-dimensional cones. The variety $P(\Sigma)$ depends on higher dimensional cones only through he set $Z(\Sigma)$, which determines the points we remove. If we remove the cone $\sigma_0$ from the fan, as shown in Figure \ref{fig:fanP2mod}, the collection ${\rho_0, \rho_1, \rho_2}$ is no longer primitive, but $\{\rho_1, \rho_2\}$ is. The associated variety is then given by
\begin{equation*}
    V(\rho_1,\rho_2)=\{(\lambda,0,0)\in \C^3\}\,,
\end{equation*}
and the associated toric variety is
\begin{equation*}
    \P(\Sigma)=\P^2-\{[1,0,0]\}\,.
\end{equation*}
We thus see that removing the 2-dimensional cone $\sigma_0$ corresponds to removing the associated variety. Moreover, we see that the resulting variety is no longer compact. In general, the toric variety $\mathbb{P}(\Sigma)$ is compact if and only if $\Sigma$ covers the entire $N_\mathbb{R}$.

\begin{figure}
    \centering
    \begin{tikzpicture}
         \fill[mdtBlue!40] (0,0)--(90:2) arc (90:225:2)--cycle;
         \fill[mdtGreen!40] (0,0)--(225:2) arc (225:360:2)--cycle;
        
        \draw[line width=1pt, mdtYellow] (0,0)--(90:2)node[anchor=south,mdtYellow!50!black]{$\rho_2$};
        \draw[line width=1pt, mdtCyan] (0,0)--(0:2)node[anchor=west,mdtCyan!50!black]{$\rho_1$};
        \draw[line width=1pt, mdtMagenta] (0,0)--(225:2)node[anchor=north east,mdtMagenta!50!black]{$\rho_0$};
        
        \node[mdtBlue!50!black] at (157.5:1.5) {$\sigma_1$};
        \node[mdtGreen!50!black] at (292.5:1.5) {$\sigma_2$};
        \end{tikzpicture}
    \caption{Fan of $\P^2-\{[1,0,0]\}$}
    \label{fig:fanP2mod}
\end{figure}

In similar fashion, we can add, for instance, 1-dimensional cones to the fan. This would would result in a variety that looks like $\mathbb{P}^2$ but with an additional line $\mathbb{P}^1$ corresponding to the new 1-dimensional cone. This is the toric description of a blow-up.

\section{Polytopes}
To construct the mirror Calabi-Yau threefold associated with toric varieties, we can use the polytope approach. Given a rational polytope $\Delta$, there are two ways to obtain a fan: the normal fan and the face fan.

In the case of the normal fan, for each vertex $m$ of $\Delta$, we consider the cone $\check\sigma_m$ defined by
\begin{equation*}
    \check\sigma_m=\{\lambda(p-m):\lambda\in \R_{\geq 0},\,p\in \Delta\}.
\end{equation*}
These cones can be visualized as the angles of the polytope. By taking the dual of each cone $\check\sigma_m$, denoted as $\sigma_m$, and collecting all such cones and their faces for every vertex of $\Delta$, we obtain the normal fan denoted by $\Sigma(\Delta)$. In this case, we write $\mathbb{P}_\Delta = \mathbb{P}(\Sigma(\Delta))$, which represents the variety given by the normal fan of $\Delta$.

There is an equivalent way of defining the normal fan. Consider a vertex of $\Delta$ and consider all the edges that meet at that vertex. Each edge defines a normal hyperplane and the cone $-\sigma$ bounded by all such hyperplanes gives rise to a cone $\sigma=-(-\sigma)$ of the normal fan. This description is useful in lower dimension but gets cumbersome when the number of dimensions increases.

The face fan, on the other hand, is more intuitive. Given a polytope $\Delta$ with $0$ in its interior, we construct the fan whose cones are given by
\begin{equation*}
    \sigma_F=\{\lambda p: p\in F\}, \quad F \text{ a face of } \Delta.
\end{equation*}
These two types of fans and the algebraic varieties they describe are closely related through mirror symmetry. For any polytope $\Delta$ with $0$ in its interior, we define the dual polytope as
\begin{equation*}
    \Delta^\circ=\{v\in \R^n:\left<v,u\right>\geq -1,\,\forall u\in \Delta\}.
\end{equation*}
We say that a rational polytope $\Delta$ is reflexive if $\Delta^\circ$ is also a rational polytope. In the context of mirror symmetry, we specifically consider reflexive rational polytopes. The normal fan $\Sigma(\Delta)$ is exactly equal to the face fan of $\Delta^\circ$, and the polytope $\Delta$ gives rise to a family of Calabi-Yau threefolds that are mirror to the family given by $\Delta^\circ$.

To find the appropriate polytope for the example of $\mathbb{P}^2$, it is easier to identify a polytope whose face fan matches $\Sigma$. We can then compute the dual polytope. Refer to Figure \ref{fig:polP2}, which shows the polytope that corresponds to $\mathbb{P}^2$. Note that in the figure, the lines joining 0 to the vertices are added, which provides information about the blown-up singularities in more general cases.

\begin{figure}
\centering
    \begin{tikzpicture}
\fill[mdtRed!40] (0,0)--(2,0) arc (0:90:2)--cycle;
         \fill[mdtBlue!40] (0,0)--(90:2) arc (90:225:2)--cycle;
         \fill[mdtGreen!40] (0,0)--(225:2) arc (225:360:2)--cycle;
        
        \draw[line width=3pt,mdtRed] (1,0)--(0,1)--(-1,-1)--cycle;
        
        \foreach \i in {-2,-1,0,1,2}
        {\foreach \j in {-2,-1,0,1,2}
        {\fill[gray!50] (\i,\j) circle (3pt);
        }
        }
        \fill[gray!50] (2.5,1) circle (3pt) node[anchor=west,gray!50!black]{N};

        \begin{scope}[xshift=5cm]
         \draw[line width=3pt,mdtRed] (1,0)--(0,1)--(-1,-1)--cycle;
         \draw[line width=1pt,mdtRed] (0,0)--(0,1);
         \draw[line width=1pt,mdtRed] (0,0)--(-1,-1);
          \draw[line width=1pt,mdtRed] (0,0)--(1,0);
        \end{scope}
        \begin{scope}[xshift=8cm]
         \draw[line width=3pt,mdtBlue] (-1,-1)--(2,-1)--(-1,2)--cycle;
         \draw[line width=1pt,mdtBlue] (0,0)--(-1,2);
         \draw[line width=1pt,mdtBlue] (0,0)--(2,-1);
         \draw[line width=1pt,mdtBlue] (0,0)--(-1,-1);
          \end{scope}
          \begin{scope}[xshift=10cm]
          
         \draw[line width=3pt,mdtBlue] (0,2)--(0.5,2)node[anchor=west]
         {$\Delta$};
         \draw[line width=3pt,mdtRed] (0,1.5)--(0.5,1.5)node[anchor=west]
         {$\Delta^\circ$};
          \end{scope}
         \end{tikzpicture}
                \caption{Polytope for $\P^2$ and its dual.}
         \label{fig:polP2}
         \end{figure}

\section{Hypersurfaces}
In our exploration of toric varieties, our ultimate goal is to construct Calabi-Yau threefolds. It turns out that many Calabi-Yau spaces can be realized as hypersurfaces in toric varieties. To begin our discussion of hypersurfaces, we note that the polytope $\Delta$ associated with a toric variety can also be related to a divisor. In particular, it corresponds to the divisor associated with the anticanonical bundle, which is a divisor with vanishing curvature and thus gives rise to a Calabi-Yau manifold.

For any formal linear combination of divisors $D = \sum a_\rho D_\rho$\footnote{This is the space $\Z^{\Sigma(1)}$ in \ref{eq:chowseq}.}, where $D_\rho$ is the divisor corresponding to the 1-dimensional cone $\rho$, we can associate a polytope $\Delta_D$ defined as
\begin{equation}
    \Delta_D=\{m\in M_\R:\left<m,v_\rho\right>\geq -a_\rho\,\forall\rho\in\Sigma(1)\}\,.
\end{equation}
It is possible to show that the classes of holomorphic sections of the line bundle $[D]$ associated with $D$ are in one-to-one correspondence with the points in $\Delta_D \cap M$ \cite{cox_toric_2011}. This correspondence is given by the map
\begin{equation*}
    m\mapsto x^{m+D}:=\prod_{\rho\in \Sigma(1)}x_\rho^{\left<m,v_\rho\right>+a_\rho}.
\end{equation*}


For a divisor $D$, the adjunction formula reads
\begin{equation*}
    K_D=(K_M\otimes[D])|_D,
\end{equation*}
where $|_D$ stands for restrict the bundle to $D$ \cite{griffiths_principles_2014}.
For $\mathbb{P}_\Delta$, the canonical bundle is given by $K_{\mathbb{P}_\Delta} = \left[-\sum_\rho D_\rho\right]$. By choosing $D = \sum_\rho D_\rho$, we find that $K_D$ is the trivial line bundle, and therefore, the Chern class of $D$ vanishes and the associated variety is a Calabi-Yau manifold. The polytope associated $D$ is precisely the polytope $\Delta$ we described in the previous section. The correspondence between $\Delta \cap M$ and holomorphic sections is given by

\begin{equation}
    m\mapsto x^{m+D}:=\prod_{\rho\in \Sigma(1)}x_\rho^{\left<m,v_\rho\right>+1}.
    \label{eq:mon-div}
\end{equation}

With this in mind, we can associate the polytope $\Delta$ with the variety $X_\Delta$, which is obtained as the zero locus of the polynomial
\begin{equation*}
    f_\Delta :=\sum_{m\in \Delta\cap M}a_mx^{m+D}=\sum_{m\in \Delta\cap M}a_m\prod_{\rho\in \Sigma(1)}x_\rho^{\left<m,v_\rho\right>+1}.
    \label{eq:ToricFunction}
\end{equation*}
To be more precise, we would need to specify each coefficient $a_m$ to obtain a particular hypersurface. However, we are interested in considering the entire family of hypersurfaces for all possible values of $a_m$.

A simple mnemonic to determine if the zero locus of a function $f_\Delta$ associated with an arbitrary polynomial $\Delta$ corresponds to a Calabi-Yau manifold is the following: it is a Calabi-Yau if and only if the only point in $\Delta \cap M$ that lies in the interior of $\Delta$ is the origin \cite{Greene_string_1997}.

Returning to the example of $\mathbb{P}^2$, we have the polytope $\Delta$ shown in Figure \ref{fig:polP2}. The corresponding monomials are shown in Figure \ref{fig:monomials}. We observe that our polytope provides enough information to reconstruct the normal fan and its associated toric variety, as well as the family of hypersurfaces that are the zero loci of degree-three polynomials. Moreover, we notice that there is only a single point in $\Delta \cap M$ lying in the interior of $\Delta$, which is in agreement with the aforementioned characterization. Thus, we have constructed the simplest example of a Calabi-Yau manifold: The solution to a cubic equation in $\mathbb P^2$.

\begin{figure}
    \centering
    \begin{tikzpicture}[scale=1.5]
         \draw[line width=3pt,mdtBlue!20] (-1,-1)--(2,-1)--(-1,2)--cycle;
         \draw[line width=1pt,mdtBlue!20] (0,0)--(-1,2);
         \draw[line width=1pt,mdtBlue!20] (0,0)--(2,-1);
         \draw[line width=1pt,mdtBlue!20] (0,0)--(-1,-1);
         \fill[mdtBlue] (-1,-1) circle (1pt) node[anchor=north,black]{$x_0^3$};
         \fill[mdtBlue] (-1,0) circle (1pt) node[anchor=north,black]{$x_0^2x_2$};
         \fill[mdtBlue] (0,-1) circle (1pt) node[anchor=north,black]{$x_0^2x_1$};
         \fill[mdtBlue] (1,-1) circle (1pt) node[anchor=north,black]{$x_0x_1^2$};
         \fill[mdtBlue] (-1,1) circle (1pt) node[anchor=north,black]{$x_0x_2^2$};
         \fill[mdtBlue] (0,0) circle (1pt) node[anchor=north,black]{$x_0x_1x_2$};
         \fill[mdtBlue] (2,-1) circle (1pt) node[anchor=north,black]{$x_1^3$};
        \fill[mdtBlue] (-1,2) circle (1pt) node[anchor=north,black]{$x_2^{3}$};
        \fill[mdtBlue] (0,1) circle (1pt) node[anchor=north,black]{$x_1x_2^2$};
        \fill[mdtBlue] (1,0) circle (1pt) node[anchor=north,black]{$x_1^2x_2$};
    \end{tikzpicture}
    \caption{Cubic surface in $\P^2$.}
    \label{fig:monomials}
\end{figure}

\section{The K\"ahler and Mori cones}
Having developed the tools of toric geometry, it is useful to revisit the concept of the Kähler cone. It can be proven that
\begin{equation*}
    A_{n-1}\otimes \R=H^2=H^{1,1}\,,
\end{equation*}
where the last equality follows from the Hodge diamond (refer to Figure \ref{fig:HodgeDiamond}). This means that the Kähler cone can be viewed as a cone in $A_{n-1}\otimes \mathbb{R}$. Its dual, in the sense of equation \ref{eq:dual}, is the Mori cone denoted by $L$. By taking the dual of the sequence in equation \ref{eq:chowseq}, we obtain the sequence:
\begin{equation*}
    0\to A_1=A_{n-1}^*=H_2\to (\Z^{\Sigma(1)})^*\to N\to 0 \,
\end{equation*}
which implies that the Mori cone is a cone in $H_2\otimes \mathbb{R}$. We can describe $H_2$ explicitly. The exactness of the sequence implies that:
\begin{equation*}
    A_1=\left\{(l_\rho)\in (Z^{\Sigma(1)})^*: \sum_\rho l_\rho v_\rho =0\right\}.
\end{equation*}
Furthermore, the Mori and the Kähler cones have the same dimension because they are dual to each other. This observation is crucial because a basis $(l^{(k)})_k$ of the Mori cone, denoted by
\begin{equation}
    L=\sum_k\R_{\geq 0}l^{(k)},
    \label{eq:basismori}
\end{equation}
provides a coordinate system $(z_k)_k$ in the complex structure moduli of the mirror of $X_\Delta$ as explained in section \ref{sec:MirrorSymToricHyper}.

	\chapter{The numerator of the $F_4$, $B_4$, and $D_4$ theory over $\IF_4$}
	\label{app:explicit}
    To give a flavor for the form of our results, we give one explicit example in this appendix. Further explicit results are available upon request.

    The numerator $\cN_1$ for the $B_4$ and $D_4$ theories over the base $\IF_4$ is given by
    \begin{equation*}
        \cN=\frac{1}{4} \phi _{-12,3}^{F_4} \phi _{-2,1}^{F_4}-\frac{1}{2} \phi _{-8,2}^{F_4} \phi _{-6,2}^{F_4} \,.
    \end{equation*}
    The numerator for the $F_4$ theory can be obtained from this result by application of the map $(\iota^*)^{-1}$.
    
    To exemplify the power of imposing enhanced symmetries, we also give the result in term of standard $D_4$ forms:
    
    \hspace{1cm}
    
    \resizebox{\linewidth}{!}{\setstretch{1.5}
  $
   \begin{array}{l}
 \frac{E_4 \phi_{-6,2}^{D_4} \phi_{-4,1}^{D_4} \phi_{0,1}^{D_4} \left(\phi_{-2,1}^{D_4}\right){}^4}{1492992}-\frac{E_6 \phi_{-6,2}^{D_4} \phi_{-4,1}^{D_4} \left(\phi_{-2,1}^{D_4}\right){}^5}{6718464}-\frac{\phi_{-6,2}^{D_4} \phi_{-4,1}^{D_4} \left(\phi_{0,1}^{D_4}\right){}^3 \left(\phi_{-2,1}^{D_4}\right){}^2}{497664}+\frac{\phi_{-6,2}^{D_4} \left(\phi_{-4,1}^{D_4}\right){}^2 \left(\phi_{0,1}^{D_4}\right){}^4}{82944} \\
 -\frac{5 E_4^2 \phi_{-6,2}^{D_4} \left(\phi_{-4,1}^{D_4}\right){}^2 \left(\phi_{-2,1}^{D_4}\right){}^4}{8957952}-\frac{E_4 \phi_{-6,2}^{D_4} \left(\phi_{-4,1}^{D_4}\right){}^2 \left(\phi_{0,1}^{D_4}\right){}^2 \left(\phi_{-2,1}^{D_4}\right){}^2}{995328}-\frac{7 E_4 \phi_{-6,2}^{D_4} \left(\phi_{-4,1}^{D_4}\right){}^3 \left(\phi_{0,1}^{D_4}\right){}^3}{248832}+\frac{7 E_6 \phi_{-6,2}^{D_4} \left(\phi_{-4,1}^{D_4}\right){}^2 \phi_{0,1}^{D_4} \left(\phi_{-2,1}^{D_4}\right){}^3}{4478976} \\+
 \frac{23 E_4^2 \phi_{-6,2}^{D_4} \left(\phi_{0,1}^{D_4}\right){}^2 \left(\phi_{-4,1}^{D_4}\right){}^4}{995328}+\frac{E_4^2 \phi_{-6,2}^{D_4} \left(\phi_{-2,1}^{D_4}\right){}^2 \phi_{0,1}^{D_4} \left(\phi_{-4,1}^{D_4}\right){}^3}{186624}-\frac{E_4 E_6 \phi_{-6,2}^{D_4} \left(\phi_{-2,1}^{D_4}\right){}^3 \left(\phi_{-4,1}^{D_4}\right){}^3}{839808}-\frac{E_6 \phi_{-6,2}^{D_4} \phi_{-2,1}^{D_4} \left(\phi_{0,1}^{D_4}\right){}^2 \left(\phi_{-4,1}^{D_4}\right){}^3}{248832} \\
 -\frac{35 E_4^3 \phi_{-6,2}^{D_4} \phi_{0,1}^{D_4} \left(\phi_{-4,1}^{D_4}\right){}^5}{4478976}-\frac{35 E_4^3 \phi_{-6,2}^{D_4} \left(\phi_{-2,1}^{D_4}\right){}^2 \left(\phi_{-4,1}^{D_4}\right){}^4}{13436928}+\frac{E_4 E_6 \phi_{-6,2}^{D_4} \phi_{-2,1}^{D_4} \phi_{0,1}^{D_4} \left(\phi_{-4,1}^{D_4}\right){}^4}{165888}-\frac{E_6^2 \phi_{-6,2}^{D_4} \left(\phi_{-2,1}^{D_4}\right){}^2 \left(\phi_{-4,1}^{D_4}\right){}^4}{26873856} \\
 -\frac{5 E_4^2 E_6 \phi_{-6,2}^{D_4} \phi_{-2,1}^{D_4} \left(\phi_{-4,1}^{D_4}\right){}^5}{2239488}+\frac{25 E_4^4 \phi_{-6,2}^{D_4} \left(\phi_{-4,1}^{D_4}\right){}^6}{26873856}-\frac{5 E_4 E_6^2 \phi_{-6,2}^{D_4} \left(\phi_{-4,1}^{D_4}\right){}^6}{26873856}+\frac{E_6^2 \phi_{-6,2}^{D_4} \phi_{0,1}^{D_4} \left(\phi_{-4,1}^{D_4}\right){}^5}{4478976} \\
 -\frac{E_4 \left(\phi_{-6,2}^{D_4}\right){}^2 \phi_{0,1}^{D_4} \left(\phi_{-2,1}^{D_4}\right){}^3}{165888}+\frac{E_6 \left(\phi_{-6,2}^{D_4}\right){}^2 \left(\phi_{-2,1}^{D_4}\right){}^4}{746496}-\frac{E_6 \left(\phi_{-6,2}^{D_4}\right){}^2 \phi_{-4,1}^{D_4} \phi_{0,1}^{D_4} \left(\phi_{-2,1}^{D_4}\right){}^2}{41472}+\frac{\left(\phi_{-6,2}^{D_4}\right){}^2 \left(\phi_{0,1}^{D_4}\right){}^3 \phi_{-2,1}^{D_4}}{55296} \\+
 \frac{E_4^2 \left(\phi_{-6,2}^{D_4}\right){}^2 \phi_{-4,1}^{D_4} \left(\phi_{-2,1}^{D_4}\right){}^3}{124416}-\frac{5 E_4^2 \left(\phi_{-6,2}^{D_4}\right){}^2 \left(\phi_{-4,1}^{D_4}\right){}^2 \phi_{0,1}^{D_4} \phi_{-2,1}^{D_4}}{82944}+\frac{11 E_4 E_6 \left(\phi_{-6,2}^{D_4}\right){}^2 \left(\phi_{-4,1}^{D_4}\right){}^2 \left(\phi_{-2,1}^{D_4}\right){}^2}{497664}+\frac{E_6 \left(\phi_{-6,2}^{D_4}\right){}^2 \left(\phi_{-4,1}^{D_4}\right){}^2 \left(\phi_{0,1}^{D_4}\right){}^2}{18432} \\+
 \frac{E_4^2 E_6 \left(\phi_{-6,2}^{D_4}\right){}^2 \left(\phi_{-4,1}^{D_4}\right){}^4}{41472}+\frac{7 E_4^3 \left(\phi_{-6,2}^{D_4}\right){}^2 \phi_{-2,1}^{D_4} \left(\phi_{-4,1}^{D_4}\right){}^3}{186624}-\frac{E_4 E_6 \left(\phi_{-6,2}^{D_4}\right){}^2 \phi_{0,1}^{D_4} \left(\phi_{-4,1}^{D_4}\right){}^3}{13824}-\frac{E_6^2 \left(\phi_{-6,2}^{D_4}\right){}^2 \phi_{-2,1}^{D_4} \left(\phi_{-4,1}^{D_4}\right){}^3}{373248} \\
 -\frac{E_4^2 \left(\phi_{-2,1}^{D_4}\right){}^2 \left(\phi_{-6,2}^{D_4}\right){}^3}{41472}+\frac{E_4^2 \phi_{-4,1}^{D_4} \phi_{0,1}^{D_4} \left(\phi_{-6,2}^{D_4}\right){}^3}{3456}-\frac{E_4 \left(\phi_{0,1}^{D_4}\right){}^2 \left(\phi_{-6,2}^{D_4}\right){}^3}{4608}+\frac{E_6 \phi_{-2,1}^{D_4} \phi_{0,1}^{D_4} \left(\phi_{-6,2}^{D_4}\right){}^3}{6912} \\
 -\frac{\left(\phi_{-2,1}^{D_4}\right){}^3 \left(\phi_{0,1}^{D_4}\right){}^3 \left(\omega_{-4,1}^{D_4}\right){}^2}{5971968}-\frac{E_4^3 \left(\phi_{-4,1}^{D_4}\right){}^2 \left(\phi_{-6,2}^{D_4}\right){}^3}{6912}-\frac{E_4 E_6 \phi_{-4,1}^{D_4} \phi_{-2,1}^{D_4} \left(\phi_{-6,2}^{D_4}\right){}^3}{10368}+\frac{E_6^2 \left(\phi_{-4,1}^{D_4}\right){}^2 \left(\phi_{-6,2}^{D_4}\right){}^3}{20736} \\+
 \frac{E_4 \phi_{0,1}^{D_4} \left(\phi_{-2,1}^{D_4}\right){}^5 \left(\omega_{-4,1}^{D_4}\right){}^2}{17915904}+\frac{E_4 \phi_{-4,1}^{D_4} \left(\phi_{0,1}^{D_4}\right){}^2 \left(\phi_{-2,1}^{D_4}\right){}^3 \left(\omega_{-4,1}^{D_4}\right){}^2}{373248}-\frac{E_6 \left(\phi_{-2,1}^{D_4}\right){}^6 \left(\omega_{-4,1}^{D_4}\right){}^2}{80621568}-\frac{\phi_{-4,1}^{D_4} \left(\phi_{0,1}^{D_4}\right){}^4 \phi_{-2,1}^{D_4} \left(\omega_{-4,1}^{D_4}\right){}^2}{124416} \\
 -\frac{E_4^2 \phi_{-4,1}^{D_4} \left(\phi_{-2,1}^{D_4}\right){}^5 \left(\omega_{-4,1}^{D_4}\right){}^2}{13436928}+\frac{37 E_4 \left(\phi_{-4,1}^{D_4}\right){}^2 \left(\phi_{0,1}^{D_4}\right){}^3 \phi_{-2,1}^{D_4} \left(\omega_{-4,1}^{D_4}\right){}^2}{1492992}-\frac{5 E_6 \phi_{-4,1}^{D_4} \phi_{0,1}^{D_4} \left(\phi_{-2,1}^{D_4}\right){}^4 \left(\omega_{-4,1}^{D_4}\right){}^2}{13436928}+\frac{7 E_6 \left(\phi_{-4,1}^{D_4}\right){}^2 \left(\phi_{0,1}^{D_4}\right){}^2 \left(\phi_{-2,1}^{D_4}\right){}^2 \left(\omega_{-4,1}^{D_4}\right){}^2}{5971968} \\
 -\frac{47 E_4^2 \left(\phi_{-4,1}^{D_4}\right){}^2 \phi_{0,1}^{D_4} \left(\phi_{-2,1}^{D_4}\right){}^3 \left(\omega_{-4,1}^{D_4}\right){}^2}{8957952}-\frac{41 E_4^2 \left(\phi_{-4,1}^{D_4}\right){}^3 \left(\phi_{0,1}^{D_4}\right){}^2 \phi_{-2,1}^{D_4} \left(\omega_{-4,1}^{D_4}\right){}^2}{1492992}+\frac{101 E_4 E_6 \left(\phi_{-4,1}^{D_4}\right){}^2 \left(\phi_{-2,1}^{D_4}\right){}^4 \left(\omega_{-4,1}^{D_4}\right){}^2}{161243136}-\frac{E_4 E_6 \left(\phi_{-4,1}^{D_4}\right){}^3 \phi_{0,1}^{D_4} \left(\phi_{-2,1}^{D_4}\right){}^2 \left(\omega_{-4,1}^{D_4}\right){}^2}{497664} \\+
 \frac{739 E_4^3 \phi_{-2,1}^{D_4} \phi_{0,1}^{D_4} \left(\phi_{-4,1}^{D_4}\right){}^4 \left(\omega_{-4,1}^{D_4}\right){}^2}{53747712}+\frac{53 E_4^3 \left(\phi_{-2,1}^{D_4}\right){}^3 \left(\phi_{-4,1}^{D_4}\right){}^3 \left(\omega_{-4,1}^{D_4}\right){}^2}{20155392}-\frac{E_4 E_6 \left(\phi_{0,1}^{D_4}\right){}^2 \left(\phi_{-4,1}^{D_4}\right){}^4 \left(\omega_{-4,1}^{D_4}\right){}^2}{497664}-\frac{E_6^2 \left(\phi_{-2,1}^{D_4}\right){}^3 \left(\phi_{-4,1}^{D_4}\right){}^3 \left(\omega_{-4,1}^{D_4}\right){}^2}{5038848} \\+
 \frac{E_4^2 E_6 \phi_{0,1}^{D_4} \left(\phi_{-4,1}^{D_4}\right){}^5 \left(\omega_{-4,1}^{D_4}\right){}^2}{373248}+\frac{E_4^2 E_6 \left(\phi_{-2,1}^{D_4}\right){}^2 \left(\phi_{-4,1}^{D_4}\right){}^4 \left(\omega_{-4,1}^{D_4}\right){}^2}{1679616}-\frac{115 E_4^4 \phi_{-2,1}^{D_4} \left(\phi_{-4,1}^{D_4}\right){}^5 \left(\omega_{-4,1}^{D_4}\right){}^2}{40310784}+\frac{37 E_6^2 \phi_{-2,1}^{D_4} \phi_{0,1}^{D_4} \left(\phi_{-4,1}^{D_4}\right){}^4 \left(\omega_{-4,1}^{D_4}\right){}^2}{53747712} \\
 -\frac{5 E_4^3 E_6 \left(\phi_{-4,1}^{D_4}\right){}^6 \left(\omega_{-4,1}^{D_4}\right){}^2}{5971968}-\frac{13 E_4 E_6^2 \phi_{-2,1}^{D_4} \left(\phi_{-4,1}^{D_4}\right){}^5 \left(\omega_{-4,1}^{D_4}\right){}^2}{40310784}-\frac{E_6^3 \left(\phi_{-4,1}^{D_4}\right){}^6 \left(\omega_{-4,1}^{D_4}\right){}^2}{17915904}+\frac{\phi_{-6,2}^{D_4} \left(\phi_{0,1}^{D_4}\right){}^4 \left(\omega_{-4,1}^{D_4}\right){}^2}{27648} \\+
 \frac{E_4^2 \phi_{-6,2}^{D_4} \left(\phi_{-2,1}^{D_4}\right){}^4 \left(\omega_{-4,1}^{D_4}\right){}^2}{2985984}-\frac{E_4 \phi_{-6,2}^{D_4} \left(\phi_{0,1}^{D_4}\right){}^2 \left(\phi_{-2,1}^{D_4}\right){}^2 \left(\omega_{-4,1}^{D_4}\right){}^2}{110592}-\frac{11 E_4 \phi_{-6,2}^{D_4} \phi_{-4,1}^{D_4} \left(\phi_{0,1}^{D_4}\right){}^3 \left(\omega_{-4,1}^{D_4}\right){}^2}{82944}+\frac{E_6 \phi_{-6,2}^{D_4} \phi_{0,1}^{D_4} \left(\phi_{-2,1}^{D_4}\right){}^3 \left(\omega_{-4,1}^{D_4}\right){}^2}{1492992} \\+
 \frac{E_4^2 \phi_{-6,2}^{D_4} \phi_{-4,1}^{D_4} \phi_{0,1}^{D_4} \left(\phi_{-2,1}^{D_4}\right){}^2 \left(\omega_{-4,1}^{D_4}\right){}^2}{20736}+\frac{E_4^2 \phi_{-6,2}^{D_4} \left(\phi_{-4,1}^{D_4}\right){}^2 \left(\phi_{0,1}^{D_4}\right){}^2 \left(\omega_{-4,1}^{D_4}\right){}^2}{6144}-\frac{E_4 E_6 \phi_{-6,2}^{D_4} \phi_{-4,1}^{D_4} \left(\phi_{-2,1}^{D_4}\right){}^3 \left(\omega_{-4,1}^{D_4}\right){}^2}{139968}-\frac{E_6 \phi_{-6,2}^{D_4} \phi_{-4,1}^{D_4} \left(\phi_{0,1}^{D_4}\right){}^2 \phi_{-2,1}^{D_4} \left(\omega_{-4,1}^{D_4}\right){}^2}{27648} \\
 -\frac{E_4^3 \phi_{-6,2}^{D_4} \left(\phi_{-4,1}^{D_4}\right){}^2 \left(\phi_{-2,1}^{D_4}\right){}^2 \left(\omega_{-4,1}^{D_4}\right){}^2}{27648}-\frac{61 E_4^3 \phi_{-6,2}^{D_4} \left(\phi_{-4,1}^{D_4}\right){}^3 \phi_{0,1}^{D_4} \left(\omega_{-4,1}^{D_4}\right){}^2}{746496}+\frac{13 E_4 E_6 \phi_{-6,2}^{D_4} \left(\phi_{-4,1}^{D_4}\right){}^2 \phi_{-2,1}^{D_4} \phi_{0,1}^{D_4} \left(\omega_{-4,1}^{D_4}\right){}^2}{248832}+\frac{11 E_6^2 \phi_{-6,2}^{D_4} \left(\phi_{-4,1}^{D_4}\right){}^2 \left(\phi_{-2,1}^{D_4}\right){}^2 \left(\omega_{-4,1}^{D_4}\right){}^2}{1492992} \\
 -\frac{7 E_4^2 E_6 \phi_{-6,2}^{D_4} \phi_{-2,1}^{D_4} \left(\phi_{-4,1}^{D_4}\right){}^3 \left(\omega_{-4,1}^{D_4}\right){}^2}{373248}+\frac{149 E_4^4 \phi_{-6,2}^{D_4} \left(\phi_{-4,1}^{D_4}\right){}^4 \left(\omega_{-4,1}^{D_4}\right){}^2}{8957952}-\frac{13 E_4 E_6^2 \phi_{-6,2}^{D_4} \left(\phi_{-4,1}^{D_4}\right){}^4 \left(\omega_{-4,1}^{D_4}\right){}^2}{8957952}-\frac{E_6^2 \phi_{-6,2}^{D_4} \phi_{0,1}^{D_4} \left(\phi_{-4,1}^{D_4}\right){}^3 \left(\omega_{-4,1}^{D_4}\right){}^2}{746496} \\
 -\frac{E_4^2 \left(\phi_{-6,2}^{D_4}\right){}^2 \phi_{-2,1}^{D_4} \phi_{0,1}^{D_4} \left(\omega_{-4,1}^{D_4}\right){}^2}{9216}+\frac{E_4 E_6 \left(\phi_{-6,2}^{D_4}\right){}^2 \left(\phi_{-2,1}^{D_4}\right){}^2 \left(\omega_{-4,1}^{D_4}\right){}^2}{55296}-\frac{E_4 E_6 \left(\phi_{-6,2}^{D_4}\right){}^2 \phi_{-4,1}^{D_4} \phi_{0,1}^{D_4} \left(\omega_{-4,1}^{D_4}\right){}^2}{4608}+\frac{E_6 \left(\phi_{-6,2}^{D_4}\right){}^2 \left(\phi_{0,1}^{D_4}\right){}^2 \left(\omega_{-4,1}^{D_4}\right){}^2}{6144} \\+
 \frac{E_4^2 E_6 \left(\phi_{-6,2}^{D_4}\right){}^2 \left(\phi_{-4,1}^{D_4}\right){}^2 \left(\omega_{-4,1}^{D_4}\right){}^2}{13824}-\frac{E_4^3 \left(\phi_{-6,2}^{D_4}\right){}^3 \left(\omega_{-4,1}^{D_4}\right){}^2}{6912}+\frac{E_4^3 \left(\phi_{-6,2}^{D_4}\right){}^2 \phi_{-4,1}^{D_4} \phi_{-2,1}^{D_4} \left(\omega_{-4,1}^{D_4}\right){}^2}{6912}-\frac{E_6^2 \left(\phi_{-6,2}^{D_4}\right){}^2 \phi_{-4,1}^{D_4} \phi_{-2,1}^{D_4} \left(\omega_{-4,1}^{D_4}\right){}^2}{13824} \\
 -\frac{E_4^2 \left(\phi_{-2,1}^{D_4}\right){}^3 \phi_{0,1}^{D_4} \left(\omega_{-4,1}^{D_4}\right){}^4}{2985984}+\frac{E_4 \phi_{-2,1}^{D_4} \left(\phi_{0,1}^{D_4}\right){}^3 \left(\omega_{-4,1}^{D_4}\right){}^4}{497664}+\frac{E_6^2 \left(\phi_{-6,2}^{D_4}\right){}^3 \left(\omega_{-4,1}^{D_4}\right){}^2}{6912}-\frac{E_6 \left(\phi_{-2,1}^{D_4}\right){}^2 \left(\phi_{0,1}^{D_4}\right){}^2 \left(\omega_{-4,1}^{D_4}\right){}^4}{1990656} \\+
 \frac{E_4^3 \phi_{-4,1}^{D_4} \left(\phi_{-2,1}^{D_4}\right){}^3 \left(\omega_{-4,1}^{D_4}\right){}^4}{2239488}-\frac{E_4^2 \phi_{-4,1}^{D_4} \phi_{-2,1}^{D_4} \left(\phi_{0,1}^{D_4}\right){}^2 \left(\omega_{-4,1}^{D_4}\right){}^4}{497664}+\frac{5 E_4 E_6 \left(\phi_{-2,1}^{D_4}\right){}^4 \left(\omega_{-4,1}^{D_4}\right){}^4}{53747712}-\frac{E_4 E_6 \phi_{-4,1}^{D_4} \left(\phi_{-2,1}^{D_4}\right){}^2 \phi_{0,1}^{D_4} \left(\omega_{-4,1}^{D_4}\right){}^4}{1492992} \\+
 \frac{5 E_4^2 E_6 \left(\phi_{-4,1}^{D_4}\right){}^2 \left(\phi_{-2,1}^{D_4}\right){}^2 \left(\omega_{-4,1}^{D_4}\right){}^4}{4478976}-\frac{7 E_4^3 \left(\phi_{-4,1}^{D_4}\right){}^2 \phi_{-2,1}^{D_4} \phi_{0,1}^{D_4} \left(\omega_{-4,1}^{D_4}\right){}^4}{2985984}+\frac{E_4 E_6 \left(\phi_{-4,1}^{D_4}\right){}^2 \left(\phi_{0,1}^{D_4}\right){}^2 \left(\omega_{-4,1}^{D_4}\right){}^4}{248832}-\frac{E_6^2 \left(\phi_{-4,1}^{D_4}\right){}^2 \phi_{-2,1}^{D_4} \phi_{0,1}^{D_4} \left(\omega_{-4,1}^{D_4}\right){}^4}{2985984} \\+
 \frac{11 E_4^3 E_6 \left(\phi_{-4,1}^{D_4}\right){}^4 \left(\omega_{-4,1}^{D_4}\right){}^4}{5971968}-\frac{E_4^2 E_6 \left(\phi_{-4,1}^{D_4}\right){}^3 \phi_{0,1}^{D_4} \left(\omega_{-4,1}^{D_4}\right){}^4}{186624}+\frac{13 E_4^4 \left(\phi_{-4,1}^{D_4}\right){}^3 \phi_{-2,1}^{D_4} \left(\omega_{-4,1}^{D_4}\right){}^4}{6718464}+\frac{E_4 E_6^2 \left(\phi_{-4,1}^{D_4}\right){}^3 \phi_{-2,1}^{D_4} \left(\omega_{-4,1}^{D_4}\right){}^4}{6718464} \\
 -\frac{E_4^3 \phi_{-6,2}^{D_4} \left(\phi_{-2,1}^{D_4}\right){}^2 \left(\omega_{-4,1}^{D_4}\right){}^4}{497664}-\frac{E_4^2 \phi_{-6,2}^{D_4} \left(\phi_{0,1}^{D_4}\right){}^2 \left(\omega_{-4,1}^{D_4}\right){}^4}{110592}+\frac{E_4 E_6 \phi_{-6,2}^{D_4} \phi_{-2,1}^{D_4} \phi_{0,1}^{D_4} \left(\omega_{-4,1}^{D_4}\right){}^4}{165888}-\frac{E_6^3 \left(\phi_{-4,1}^{D_4}\right){}^4 \left(\omega_{-4,1}^{D_4}\right){}^4}{17915904} \\
 -\frac{E_4^2 E_6 \phi_{-6,2}^{D_4} \phi_{-4,1}^{D_4} \phi_{-2,1}^{D_4} \left(\omega_{-4,1}^{D_4}\right){}^4}{248832}+\frac{E_4^3 \phi_{-6,2}^{D_4} \phi_{-4,1}^{D_4} \phi_{0,1}^{D_4} \left(\omega_{-4,1}^{D_4}\right){}^4}{55296}+\frac{E_6^2 \phi_{-6,2}^{D_4} \left(\phi_{-2,1}^{D_4}\right){}^2 \left(\omega_{-4,1}^{D_4}\right){}^4}{995328}-\frac{E_6^2 \phi_{-6,2}^{D_4} \phi_{-4,1}^{D_4} \phi_{0,1}^{D_4} \left(\omega_{-4,1}^{D_4}\right){}^4}{165888} \\+
 \frac{E_4^3 \phi_{-2,1}^{D_4} \phi_{0,1}^{D_4} \left(\omega_{-4,1}^{D_4}\right){}^6}{1990656}-\frac{11 E_4^4 \phi_{-6,2}^{D_4} \left(\phi_{-4,1}^{D_4}\right){}^2 \left(\omega_{-4,1}^{D_4}\right){}^4}{995328}+\frac{7 E_4 E_6^2 \phi_{-6,2}^{D_4} \left(\phi_{-4,1}^{D_4}\right){}^2 \left(\omega_{-4,1}^{D_4}\right){}^4}{995328}-\frac{E_4 E_6 \left(\phi_{0,1}^{D_4}\right){}^2 \left(\omega_{-4,1}^{D_4}\right){}^6}{497664} \\
 -\frac{E_4^2 E_6 \left(\phi_{-2,1}^{D_4}\right){}^2 \left(\omega_{-4,1}^{D_4}\right){}^6}{4478976}+\frac{E_4^2 E_6 \phi_{-4,1}^{D_4} \phi_{0,1}^{D_4} \left(\omega_{-4,1}^{D_4}\right){}^6}{373248}-\frac{E_4^4 \phi_{-4,1}^{D_4} \phi_{-2,1}^{D_4} \left(\omega_{-4,1}^{D_4}\right){}^6}{1492992}+\frac{5 E_6^2 \phi_{-2,1}^{D_4} \phi_{0,1}^{D_4} \left(\omega_{-4,1}^{D_4}\right){}^6}{5971968} \\
 -\frac{7 E_4^3 E_6 \left(\phi_{-4,1}^{D_4}\right){}^2 \left(\omega_{-4,1}^{D_4}\right){}^6}{5971968}+\frac{E_4^4 \phi_{-6,2}^{D_4} \left(\omega_{-4,1}^{D_4}\right){}^6}{331776}-\frac{E_4 E_6^2 \phi_{-4,1}^{D_4} \phi_{-2,1}^{D_4} \left(\omega_{-4,1}^{D_4}\right){}^6}{4478976}+\frac{5 E_6^3 \left(\phi_{-4,1}^{D_4}\right){}^2 \left(\omega_{-4,1}^{D_4}\right){}^6}{17915904} \\+
 \frac{E_4^3 E_6 \left(\omega_{-4,1}^{D_4}\right){}^8}{5971968}-\frac{E_4 E_6^2 \phi_{-6,2}^{D_4} \left(\omega_{-4,1}^{D_4}\right){}^6}{331776}-\frac{E_6^3 \left(\omega_{-4,1}^{D_4}\right){}^8}{5971968} \\
\end{array}$}

\chapter{Tables of Gromov-Witten invariants and their specializations}
\label{app:GWinv}

In tables \ref{tab:B4F4gw},\ref{tab:B4B3gw},\ref{tab:gwB3G2} we provide some data to check the relations in section \ref{ss:GWinv}.

\begin{table}[]
    \centering
    \footnotesize
    \setlength{\tabcolsep}{2pt}
    \begin{tabular}{c|c|ccccccccccc||ccccccccccc}
         $(a_0,a_3^{B_4},a_4^{B_4})$&\multicolumn{12}{c||}{$B_4$ and $D_4$ invariants}&  \multicolumn{11}{c}{$F_4$ invariants}\\\hline\hline
         \multirow{11}{*}{(0,0,0)}& \diagbox{$a_1$}{$a_2$}
          & 0 & $\frac{1}{2}$ & 1 & $\frac{3}{2}$ & 2 & $\frac{5}{2}$ & 3 & $\frac{7}{2}$ & 4 & $\frac{9}{2}$ & 5 & 0 & $\frac{1}{2}$ & 1 & $\frac{3}{2}$ & 2 & $\frac{5}{2}$ & 3 &
   $\frac{7}{2}$ & 4 & $\frac{9}{2}$ & 5 \\\cline{2-24}&
 0 & $-$2 & \zero & $-$2 & \zero & $-$4 & \zero & $-$6 & \zero & $-$8 & \zero & $-$10 & $-$2 & \zero & \zero & \zero & \zero
   & \zero & \zero & \zero & \zero & \zero & \zero \\&
 $\frac{1}{2}$ & \zero & \zero & \zero & \zero & \zero & \zero & \zero & \zero & \zero & \zero &
   \zero & \zero & \zero & \zero & \zero & \zero & \zero & \zero & \zero & \zero & \zero &
   \zero \\&
 1 & \zero & \zero & $-$2 & \zero & $-$6 & \zero & $-$10 & \zero & $-$14 & \zero & $-$18 & $-$2 & \zero & $-$2 & \zero &
   \zero & \zero & \zero & \zero & \zero & \zero & \zero \\&
 $\frac{3}{2}$ & \zero & \zero & \zero & \zero & \zero & \zero & \zero & \zero & \zero & \zero &
   \zero & \zero & \zero & \zero & \zero & \zero & \zero & \zero & \zero & \zero & \zero &
   \zero \\&
 2 & \zero & \zero & \zero & \zero & $-$6 & \zero & $-$12 & \zero & $-$18 & \zero & $-$24 & $-$4 & \zero & $-$6 & \zero &
   $-$6 & \zero & $-$4 & \zero & $-$6 & \zero & $-$8 \\&
 $\frac{5}{2}$ & \zero & \zero & \zero & \zero & \zero & \zero & \zero & \zero & \zero & \zero &
   \zero & \zero & \zero & \zero & \zero & \zero & \zero & \zero & \zero & \zero & \zero &
   \zero \\&
 3 & \zero & \zero & \zero & \zero & $-$4 & \zero & $-$12 & \zero & $-$20 & \zero & $-$28 & $-$6 & \zero & $-$10 & \zero &
   $-$12 & \zero & $-$12 & \zero & $-$10 & \zero & $-$14 \\&
 $\frac{7}{2}$ & \zero & \zero & \zero & \zero & \zero & \zero & \zero & \zero & \zero & \zero &
   \zero & \zero & \zero & \zero & \zero & \zero & \zero & \zero & \zero & \zero & \zero &
   \zero \\&
 4 & \zero & \zero & \zero & \zero & $-$6 & \zero & $-$10 & \zero & $-$20 & \zero & $-$30 & $-$8 & \zero & $-$14 & \zero &
   $-$18 & \zero & $-$20 & \zero & $-$20 & \zero & $-$18 \\&
 $\frac{9}{2} $& \zero & \zero & \zero & \zero & \zero & \zero & \zero & \zero & \zero & \zero &
   \zero & \zero & \zero & \zero & \zero & \zero & \zero & \zero & \zero & \zero & \zero &
   \zero \\&
 5 & \zero & \zero & \zero & \zero & $-$8 & \zero & $-$14 & \zero & $-$18 & \zero & $-$30 & $-$10 & \zero & $-$18 & \zero &
   $-$24 & \zero & $-$28 & \zero & $-$30 & \zero & $-$30 \\\hline\hline
   
    \multirow{11}{*}{$(1,\frac{1}{2},0)$}& \diagbox{$a_1$}{$a_2$}&  0 & $\frac{1}{2}$ & 1 & $\frac{3}{2}$ & 2 & $\frac{5}{2}$ & 3 & $\frac{7}{2}$ & 4 & $\frac{9}{2}$ & 5 & 0 & $\frac{1}{2}$ & 1 & $\frac{3}{2}$ & 2 & $\frac{5}{2}$ & 3 & $\frac{7}{2}$
   & 4 & $\frac{9}{2}$ & 5 \\\cline{2-24}&
 0 & \zero & \zero & \zero & \zero & \zero & \zero & \zero & \zero & \zero & \zero & \zero &
   \zero & \zero & \zero & \zero & \zero & \zero & \zero & \zero & \zero & \zero & \zero \\&
 $\frac{1}{2}$ & \zero & \zero & 12 & \zero & 16 & \zero & 32 & \zero & 48 & \zero & 64 & \zero & \zero & \zero &
   \zero & \zero & \zero & \zero & \zero & \zero & \zero & \zero \\&
 1 & \zero & \zero & \zero & \zero & \zero & \zero & \zero & \zero & \zero & \zero & \zero &
   \zero & \zero & 12 & \zero & \zero & \zero & \zero & \zero & \zero & \zero & \zero \\&
 $\frac{3}{2}$ & \zero & \zero & \zero & \zero & 16 & \zero & 48 & \zero & 80 & \zero & 112 & \zero & \zero &
   \zero & \zero & \zero & \zero & \zero & \zero & \zero & \zero & \zero \\&
 2 & \zero & \zero & \zero & \zero & \zero & \zero & \zero & \zero & \zero & \zero & \zero &
   \zero & \zero & 16 & \zero & 16 & \zero & \zero & \zero & \zero & \zero & \zero \\&
 $\frac{5}{2}$ & \zero & \zero & \zero & \zero & \zero & \zero & 48 & \zero & 96 & \zero & 144 & \zero &
   \zero & \zero & \zero & \zero & \zero & \zero & \zero & \zero & \zero & \zero \\&
 3 & \zero & \zero & \zero & \zero & \zero & \zero & \zero & \zero & \zero & \zero & \zero &
   \zero & \zero & 32 & \zero & 48 & \zero & 48 & \zero & 32 & \zero & 48 \\&
 $\frac{7}{2}$ & \zero & \zero & \zero & \zero & \zero & \zero & 32 & \zero & 96 & \zero & 160 & \zero &
   \zero & \zero & \zero & \zero & \zero & \zero & \zero & \zero & \zero & \zero \\&
 4 & \zero & \zero & \zero & \zero & \zero & \zero & \zero & \zero & \zero & \zero & \zero &
   \zero & \zero & 48 & \zero & 80 & \zero & 96 & \zero & 96 & \zero & 80 \\&
 $\frac{9}{2}$ & \zero & \zero & \zero & \zero & \zero & \zero & 48 & \zero & 80 & \zero & 160 & \zero &
   \zero & \zero & \zero & \zero & \zero & \zero & \zero & \zero & \zero & \zero \\&
 5 & \zero & \zero & \zero & \zero & \zero & \zero & \zero & \zero & \zero & \zero & \zero &
   \zero & \zero & 64 & \zero & 112 & \zero & 144 & \zero & 160 & \zero & 160 \\\hline\hline
   
   \multirow{11}{*}{$(0,\frac{1}{2},1)$}& \diagbox{$a_1$}{$a_2$} & 0 & $\frac{1}{2}$ & 1 & $\frac{3}{2}$ & 2 & $\frac{5}{2}$ & 3 & $\frac{7}{2}$ & 4 & $\frac{9}{2}$ & 5 & 0 & $\frac{1}{2}$ & 1 & $\frac{3}{2}$ & 2 & $\frac{5}{2}$ & 3 & $\frac{7}{2}$
   & 4 & $\frac{9}{2}$ & 5 \\\cline{2-24}&
 0 & \zero & \zero & \zero & \zero & \zero & \zero & \zero & \zero & \zero & \zero & \zero &
   \zero & \zero & \zero & \zero & \zero & \zero & \zero & \zero & \zero & \zero & \zero \\&
 $\frac{1}{2}$ & \zero & \zero & 4 & \zero & 12 & \zero & 20 & \zero & 28 & \zero & 36 & \zero & \zero & \zero &
   \zero & \zero & \zero & \zero & \zero & \zero & \zero & \zero \\&
 1 & \zero & \zero & \zero & \zero & \zero & \zero & \zero & \zero & \zero & \zero & \zero &
   \zero & \zero & 4 & \zero & \zero & \zero & \zero & \zero & \zero & \zero & \zero \\&
 $\frac{3}{2}$ & \zero & \zero & \zero & \zero & 16 & \zero & 32 & \zero & 48 & \zero & 64 & \zero & \zero &
   \zero & \zero & \zero & \zero & \zero & \zero & \zero & \zero & \zero \\&
 2 & \zero & \zero & \zero & \zero & \zero & \zero & \zero & \zero & \zero & \zero & \zero &
   \zero & \zero & 12 & \zero & 16 & \zero & 12 & \zero & 20 & \zero & 28 \\&
 $\frac{5}{2}$ & \zero & \zero & \zero & \zero & 12 & \zero & 36 & \zero & 60 & \zero & 84 & \zero & \zero &
   \zero & \zero & \zero & \zero & \zero & \zero & \zero & \zero & \zero \\&
 3 & \zero & \zero & \zero & \zero & \zero & \zero & \zero & \zero & \zero & \zero & \zero &
   \zero & \zero & 20 & \zero & 32 & \zero & 36 & \zero & 32 & \zero & 48 \\&
 $\frac{7}{2}$ & \zero & \zero & \zero & \zero & 20 & \zero & 32 & \zero & 64 & \zero & 96 & \zero & \zero &
   \zero & \zero & \zero & \zero & \zero & \zero & \zero & \zero & \zero \\&
 4 & \zero & \zero & \zero & \zero & \zero & \zero & \zero & \zero & \zero & \zero & \zero &
   \zero & \zero & 28 & \zero & 48 & \zero & 60 & \zero & 64 & \zero & 60 \\&
 $\frac{9}{2}$ & \zero & \zero & \zero & \zero & 28 & \zero & 48 & \zero & 60 & \zero & 100 & \zero & \zero &
   \zero & \zero & \zero & \zero & \zero & \zero & \zero & \zero & \zero \\&
 5 & \zero & \zero & \zero & \zero & \zero & \zero & \zero & \zero & \zero & \zero & \zero &
   \zero & \zero & 36 & \zero & 64 & \zero & 84 & \zero & 96 & \zero & 100 \\
    \end{tabular}
    \caption{Some genus 0 Gromov-Witten invariants of the $B_4,\,D_4$ and $F_4$ theories over $\mathbb F_2$. On the left, we give the values of $a_0,a_3^{B_4},a_4^{B_4}$, which determine the values of $a_3,a_4$ for $F_4$ and $D_4$ via  $2a_3^{B_4},a_4^{B_4}$ and $a_3^{B_4}-\frac{a_4^{B_4}}{2},\frac{a_4^{B_4}}{2}$ respectively. Inside each block, $a_1$ (resp. $a_2$) grows from $0$ to $5$ in steps of $1/2$ from top to bottom (resp. left to right); this growth in steps of $1/2$ is artificial for $F_4$, as all the possible weights have integer indices, but it makes it simpler to visually compare the invariants. In accordance with \eqref{eq:GVB4D4F4}, this presentation allows us to record the values of the Gromov-Witten invariants for the $B_4$ and $D_4$ geometries in a single table. Regarding the $F_4$ geometry, we note that if we reflect the $B_4$ invariants through the diagonal and shift them horizontally by the value of $a_3$, we recover the $F_4$ invariants, again in agreement with \eqref{eq:GVB4D4F4}.}
    \label{tab:B4F4gw}
\end{table}

\begin{table}[]
\setlength{\tabcolsep}{5pt}
    \centering
\begin{tabular}{c|c|ccccccccccc||c}
    $(a_0,a_1,a_3)$&\multicolumn{12}{c||}{$B_4$ invariants}&$B_3$ invariants\\\hline\hline
        \multirow{12}{*}{$(0,0,0)$}& \diagbox{$a_2$}{$a_4$} & 0 & $\frac{1}{2}$
        & 1 & $\frac{3}{2}$ & 2 & $\frac{5}{2}$ & 3 & $\frac{7}{2}$ & 4 & $\frac{9}{2}$ & 5 &\\\cline{2-14}&
 0 & $-$2 & \zero & \zero & \zero & \zero & \zero & \zero & \zero & \zero & \zero & \zero &$-$2\\&
 $\frac{1}{2}$ & \zero & \zero & \zero & \zero & \zero & \zero & \zero & \zero & \zero & \zero &
   \zero &\zero\\&
 1 & $-$2 & \zero & \zero & \zero & \zero & \zero & \zero & \zero & \zero & \zero & \zero &$-$2\\&
 $\frac{3}{2}$ & \zero & \zero & \zero & \zero & \zero & \zero & \zero & \zero & \zero & \zero &
   \zero &\zero\\&
 2 & $-$4 & \zero & \zero & \zero & \zero & \zero & \zero & \zero & \zero & \zero & \zero &$-$4\\&
 $\frac{5}{2}$ & \zero & \zero & \zero & \zero & \zero & \zero & \zero & \zero & \zero & \zero &
   \zero &\zero\\&
 3 & $-$6 & \zero & \zero & \zero & \zero & \zero & \zero & \zero & \zero & \zero & \zero &$-$6\\&
 $\frac{7}{2}$ & \zero & \zero & \zero & \zero & \zero & \zero & \zero & \zero & \zero & \zero &
   \zero &\zero\\&
 4 & $-$8 & \zero & \zero & \zero & \zero & \zero & \zero & \zero & \zero & \zero & \zero&$-$8 \\&
 $\frac{9}{2}$ & \zero & \zero & \zero & \zero & \zero & \zero & \zero & \zero & \zero & \zero &
   \zero &\zero\\&
 5 & $-$10 & \zero & \zero & \zero & \zero & \zero & \zero & \zero & \zero & \zero & \zero &$-$10\\\hline\hline
 
 \multirow{12}{*}{$(0,1,1)$}&\diagbox{$a_2$}{$a_4$} & 0 & $\frac{1}{2}$ & 1 & $\frac{3}{2}$ & 2 & $\frac{5}{2}$ & 3 & $\frac{7}{2}$ & 4 & $\frac{9}{2}$ & 5 \\\cline{2-14}&
 0 & \zero & \zero & \zero & \zero & \zero & \zero & \zero & \zero & \zero & \zero & \zero&\zero \\&
 $\frac{1}{2}$ & \zero & \zero & \zero & \zero & \zero & \zero & \zero & \zero & \zero & \zero &
   \zero &\zero \\&
 1 & $-$2 & \zero & 4 & \zero & $-$2 & \zero & \zero & \zero & \zero & \zero & \zero&\zero \\&
 $\frac{3}{2}$ & \zero & \zero & \zero & \zero & \zero & \zero & \zero & \zero & \zero & \zero &
   \zero &\zero\\&
 2 & $-$20 & \zero & $-$16 & \zero & $-$20 & \zero & \zero & \zero & \zero & \zero & \zero&$-56$\\&
 $\frac{5}{2}$ & \zero & \zero & \zero & \zero & \zero & \zero & \zero & \zero & \zero & \zero &
   \zero&\zero \\&
 3 & $-$40 & \zero & $-$32 & \zero & $-$40 & \zero & \zero & \zero & \zero & \zero & \zero &$-112$\\&
 $\frac{7}{2}$ & \zero & \zero & \zero & \zero & \zero & \zero & \zero & \zero & \zero & \zero &
   \zero&\zero \\&
 4 & $-$60 & \zero & $-$48 & \zero & $-$60 & \zero & \zero & \zero & \zero & \zero & \zero &$-168$\\&
 $\frac{9}{2}$ & \zero & \zero & \zero & \zero & \zero & \zero & \zero & \zero & \zero & \zero &
   \zero &\zero\\&
 5 & $-$80 & \zero & $-$64 & \zero & $-$80 & \zero & \zero & \zero & \zero & \zero & \zero&$-224$ \\\hline\hline
 
   \multirow{12}{*}{$(1,\frac{1}{2},\frac{3}{2})$}&\diagbox{$a_2$}{$a_4$}& 0 & $\frac{1}{2}$ & 1 & $\frac{3}{2}$ & 2 & $\frac{5}{2}$ & 3 & $\frac{7}{2}$ & 4 & $\frac{9}{2}$ & 5 \\\cline{2-14}&
 0 & \zero & \zero & \zero & \zero & \zero & \zero & \zero & \zero & \zero & \zero & \zero &\zero\\&
 $\frac{1}{2}$ & \zero & \zero & \zero & \zero & \zero & \zero & \zero & \zero & \zero & \zero &
   \zero &\zero\\&
 1 & \zero & \zero & 12 & \zero & 12 & \zero & \zero & \zero & \zero & \zero & \zero &24\\&
 $\frac{3}{2}$ & \zero & \zero & \zero & \zero & \zero & \zero & \zero & \zero & \zero & \zero &
   \zero &\zero\\&
 2 & 16 & \zero & \zero & \zero & \zero & \zero & 16 & \zero & \zero & \zero & \zero&32 \\&
 $\frac{5}{2}$ & \zero & \zero & \zero & \zero & \zero & \zero & \zero & \zero & \zero & \zero &
   \zero &\zero\\&
 3 & 48 & \zero & $-$36 & \zero & $-$36 & \zero & 48 & \zero & \zero & \zero & \zero&24\\&
 $\frac{7}{2}$ & \zero & \zero & \zero & \zero & \zero & \zero & \zero & \zero & \zero & \zero &
   \zero &\zero \\&
 4 & 80 & \zero & $-$60 & \zero & $-$60 & \zero & 80 & \zero & \zero & \zero & \zero&40 \\&
 $\frac{9}{2}$ & \zero & \zero & \zero & \zero & \zero & \zero & \zero & \zero & \zero & \zero &
   \zero&\zero \\&
 5 & 112 & \zero & $-$84 & \zero & $-$84 & \zero & 112 & \zero & \zero & \zero & \zero &56\\
    \end{tabular}
    \caption{Genus 0 Gromov-Witten invariants for $B_4$ and $B_3$ theories over $\mathbb F_2$. We see that adding the rows of $B_4$, we get the $B_3$ invariants in agreement with equations \eqref{eq:gwBnDn} and  \eqref{eq:gwDnBn-1}.}
    \label{tab:B4B3gw}
\end{table}

\begin{table}[]
    \footnotesize
        \setlength{\tabcolsep}{2pt}
    \centering
    \begin{tabular}{c|c|ccccccccccc||c}
    $(a_0,a_2)$&\multicolumn{12}{c||}{$B_3$ invariants}&$G_2$ invariants\\\hline\hline

     \multirow{12}{*}{$(0,0)$}&\diagbox{$a_1$}{$a_3$}& 0 & $\frac{1}{2}$ & 1 & $\frac{3}{2}$ & 2 & $\frac{5}{2}$ & 3 & $\frac{7}{2}$ & 4 & $\frac{9}{2}$ & 5 &  \\\hline&
 0 & $-$2 & \zero & \zero & \zero & \zero & \zero & \zero & \zero & \zero & \zero & \zero & $-$2 \\&
 $\frac{1}{2}$ & \zero & \zero & \zero & \zero & \zero & \zero & \zero & \zero & \zero & \zero & \zero
   & \zero \\&
 1 & \zero & \zero & \zero & \zero & \zero & \zero & \zero & \zero & \zero & \zero & \zero &
   \zero \\&
 $\frac{3}{2}$ & \zero & \zero & \zero & \zero & \zero & \zero & \zero & \zero & \zero & \zero & \zero
   & \zero \\&
 2 & \zero & \zero & \zero & \zero & \zero & \zero & \zero & \zero & \zero & \zero & \zero &
   \zero \\&
 $\frac{5}{2}$ & \zero & \zero & \zero & \zero & \zero & \zero & \zero & \zero & \zero & \zero & \zero
   & \zero \\&
 3 & \zero & \zero & \zero & \zero & \zero & \zero & \zero & \zero & \zero & \zero & \zero &
   \zero \\&
 $\frac{7}{2}$ & \zero & \zero & \zero & \zero & \zero & \zero & \zero & \zero & \zero & \zero & \zero
   & \zero \\&
 4 & \zero & \zero & \zero & \zero & \zero & \zero & \zero & \zero & \zero & \zero & \zero &
   \zero \\&
 $\frac{9}{2}$ & \zero & \zero & \zero & \zero & \zero & \zero & \zero & \zero & \zero & \zero & \zero
   & \zero \\&
 5 & \zero & \zero & \zero & \zero & \zero & \zero & \zero & \zero & \zero & \zero & \zero &
   \zero \\\hline\hline

  \multirow{12}{*}{$(1,2)$}&\diagbox{$a_1$}{$a_3$}& 0 & $\frac{1}{2}$ & 1 & $\frac{3}{2}$ & 2 & $\frac{5}{2}$ & 3 & $\frac{7}{2}$ & 4 & $\frac{9}{2}$ & 5 \\\hline &
 0 & $-$6 & \zero & \zero & \zero & $-$8 & \zero & \zero & \zero & $-$6 & \zero & \zero &$-$6\\&
 $\frac{1}{2}$ & \zero & 32 & \zero & 32 & \zero & 32 & \zero & 32 & \zero & \zero & \zero &\zero\\&
 1 & $-$8 & \zero & $-$56 & \zero & $-$160 & \zero & $-$56 & \zero & $-$8 & \zero & \zero &24\\&
 $\frac{3}{2}$ & \zero & 32 & \zero & 32 & \zero & 32 & \zero & 32 & \zero & \zero & \zero &\zero\\&
 2 & $-$6 & \zero & \zero & \zero & $-$8 & \zero & \zero & \zero & $-$6 & \zero & \zero&$-$6 \\&
 $\frac{5}{2}$ & \zero & \zero & \zero & \zero & \zero & \zero & \zero & \zero & \zero & \zero &
   \zero & \zero\\&
 3 & \zero & \zero & \zero & \zero & \zero & \zero & \zero & \zero & \zero & \zero & \zero&$-$96 \\&
 $\frac{7}{2}$ & \zero & \zero & \zero & \zero & \zero & \zero & \zero & \zero & \zero & \zero &
   \zero&\zero \\&
 $4$ & \zero & \zero & \zero & \zero & \zero & \zero & \zero & \zero & \zero & \zero & \zero &$-$6\\&
 $\frac{9}{2}$& \zero & \zero & \zero & \zero & \zero & \zero & \zero & \zero & \zero & \zero &
   \zero &\zero\\&
 5 & \zero & \zero & \zero & \zero & \zero & \zero & \zero & \zero & \zero & \zero & \zero &24\\\hline\hline
 
   \multirow{12}{*}{$(1,3)$}&\diagbox{$a_1$}{$a_3$}& 0 & $\frac{1}{2}$ & 1 & $\frac{3}{2}$ & 2 & $\frac{5}{2}$ & 3 & $\frac{7}{2}$ & 4 & $\frac{9}{2}$ & 5 &  \\\hline&
 0 & $-$10 & \zero & \zero & \zero & $-$6 & \zero & \zero & \zero & $-$6 & \zero & \tikz[remember picture]\node[inner sep=0pt](f){\zero}; & $-$10 \\&
 $\frac{1}{2}$ & \zero & 64 & \zero & 24 & \zero & 32 & \zero & 32 & \zero & 24 & \zero & \zero \\&
 1 & $-$16 & \zero & $-$168 & \zero & $-$166 & \zero & $-$112 & \zero & $-$166 & \zero & $-$168 & 48 \\&
 $\frac{3}{2}$ & \zero & 96 & \zero & 256 & \zero & 392 & \zero & 392 & \zero & 256 & \zero & \zero \\&
 2 & $-$18 & \zero & $-$224 & \zero & $-$308 & \zero & $-$224 & \zero & $-$308 & \zero & $-$224 & $-$72 \\&
 $\frac{5}{2}$ & \zero & 96 & \zero & 248 & \zero & 288 & \zero & 288 & \zero & 248 & \zero & \zero \\&
 3 & $-$16 & \zero & $-$168 & \zero & $-$432 & \zero & $-$336 & \zero & $-$432 & \zero & $-$168 & $-$22 \\&
 $\frac{7}{2}$ & \zero & 64 & \zero & 448 & \zero & 512 & \zero & 512 & \zero & 448 & \zero & \zero \\&
 4 & $-$10 & \zero & $-$280 & \zero & $-$710 & \zero & $-$560 & \zero & $-$710 & \zero & $-$280 & 132 \\&
 $\frac{9}{2}$ & \zero & 96 & \zero & 672 & \zero & 768 & \zero & 768 & \zero & 672 & \zero & \zero \\&
 5 & $-$\tikz[remember picture]\node[inner sep=0pt](i){14}; & \zero & $-$392 & \zero & $-$994 & \zero & $-$784 & \zero & $-$994 & \zero & $-$392 &\tikz[remember picture]\node[inner sep=0pt](c){132}; \\
\end{tabular}

\begin{tikzpicture}[remember picture,overlay]
\draw[line width=4mm,mdtRed,opacity=0.2] (i.west)--(f.east);
\draw[fill=mdtRed,opacity=0.2] (c) circle (0.3);
\end{tikzpicture}

    \caption{Genus 0 Gromov-Witten invariants of $B_3$ and $G_2$ over $\mathbb F_2$. Adding the entries in the diagonals given by $a_1+a_3$ constant gives the $G_2$ invariant with $a_1^{G_2}=a_1^{B_3}+a_3^{B_3}$, in accordance with equation \eqref{eq:gwB3G2}. For example, the sum of the $B_3$ invariants in the blue box gives the $G_2$ invariant in the blue circle.}
    \label{tab:gwB3G2}
\end{table}

\chapter{Root systems} \label{app:root_systems}
Root systems, and consequently finite simple Lie algebras, enjoy a classification into four infinite families, the $A$, $B$, $C$, and $D$ series, and five exceptional cases $G_2$, $F_4$, $E_6$, $E_7$, $E_8$. In this section, we will summarize the properties of these root systems, following the conventions of \cite{Bourbaki}. Note in particular that in these conventions, short roots are normalized to have norm squared 2; roots of simply laced Lie algebras are considered as short.

It is natural to embed the root systems in a Euclidean space $\mathbb{R}^n$. We will we denote the standard basis of this space by $e_i,\, i=1,\dots n$ and the coordinates in this basis by $x_i$. We refer to this coordinate system as the orthogonal basis. In most cases, $n$ equals the rank of the gauge group, i.e. the dimension of the root system. When it does not, the root system lies in a subspace of $\mathbb{R}^n$ and the last $n-r$ coordinates of this space can be expressed in terms of the first $r$ of them.

In tables \ref{tab:rootSystems1}, \ref{tab:rootSystems2}, and \ref{tab:rootSystems3} we summarize the properties of all root systems.

As we explain in section \ref{ss:matching_elliptic_parameters}, the elliptic genus of a gauge theory with gauge algebra $\mathfrak{g}$ is a meromorphic Jacobi forms whose elliptic parameters (aside from $\gs$) take values in the complexified coroot lattice $\LambdaCR(\mathfrak{g}) \otimes \IC$. The Weyl invariant Jacobi forms defined in the literature \cite{Bertola, Adler_2020,adler2020structure} have elliptic parameters taking values in  $\LambdaR(\mathfrak{g})\otimes \IC$ (in the conventions of \cite{Bourbaki}). Rather than redefine these forms, we use conventions in this paper in which the root systems as defined by \cite{Bourbaki} are interpreted as the coroots of the dual algebra $\tilde{\mathfrak{g}}$. In these conventions, short coroots have norm squared 2. The roots that enter in the definition of the denominator of the elliptic genus \eqref{eq:ZkAnsatz} are obtained from these coroots, hence are normalized such that all long roots have norm squared 2; with regard to these normalization conventions, roots of simply laced algebras are considered long. The roots of the Lie algebra $G_2$ in our conventions are worked out as an example in section \ref{ss:matching_elliptic_parameters}.

\begin{table}
    \centering
    \small
    
\newcolumntype{A}{ >{\centering\arraybackslash} m{1.75cm} }

\newcolumntype{B}{ >{\centering\arraybackslash} m{2.5cm} }
    
\renewcommand{\arraystretch}{2}
    \begin{tabular}{A||B|B|B|B}
        $\mg$& $A_n$& $B_n$& $C_n$ &$D_n$ \\\hline
        $\tilde{\mg}$& $A_n$& $C_n$&$B_n$& $D_n$\\\hline
        Embedding space &$V=\{\sum_{i=1}^{n+1}x_i=0\}\subset \mathbb{R}^{n+1}$& $\mathbb{R}^n$& $\mathbb{R}^n$& $\mathbb{R}^n$ \\\hline
        Roots& $e_i-e_j, i\neq j$& $\pm e_i $ \newline $\pm e_i\pm e_j,\, i\neq j$&$\pm 2e_i $ \newline $\pm e_i\pm e_j,\, i\neq j$&$\pm e_i\pm e_j,\, i\neq j$\\\hline
        Simple roots& $\alpha_i=e_i-e_{i+1},\, i=1,\dots,n$&$\alpha_i=e_i-e_{i+1}$\newline$i=1,\dots,n-1$ \newline $\alpha_n=e_n$&$\alpha_i=e_i-e_{i+1}$\newline$i=1,\dots,n-1$ \newline $\alpha_n=2 e_n$&$\alpha_i=e_i-e_{i+1}$\newline$i=1,\dots,n-1$ \newline $\alpha_n=e_{n-1}+e_n$
        \\\hline
        $\Lambda_r$&$\Z^{n+1}\cap V$&$\Z^n$&$x\in \Z^n$ such that $\sum x_i\in 2\Z$&$x\in \Z^n$ such that $\sum x_i\in 2\Z$\\\hline
        Fund. Weights & $\omega_i=e_1+\cdots+e_i-\frac{i}{n+1}\sum e_i$& $\omega_i=e_1+\cdots+e_i,\,i\neq n$\newline$\omega_n=\frac{1}{2}\sum e_i$&$\omega_i=e_1+\cdots+e_i$&$\omega_i=e_1+\cdots+e_i, \, i\neq n-1,n$\newline
        $\omega_{n-1}=\frac{1}{2}(e_1+\dots+e_{n-1}-e_n)$\newline
        $\omega_{n}=\frac{1}{2}(e_1+\dots+e_{n-1}+e_n)$\\\hline
        $\Lambda_w$&$\left<\Lambda_r,e_1-\frac{1}{n+1}\sum e_i\right>$&$\Z^{n}+\left(\frac{1}{2}\sum e_i\right)\IZ$& $\Z^n$&$\Z^{n}+\left(\frac{1}{2}\sum e_i\right)\IZ$\\\hline
        $ds^2$&$dx^2$&$2dx^2$& $dx^2$&$dx^2$\\\hline
        $W$& $S_n$ permuting the $x_i$&$S_{n}\ltimes \Z_2^n$ permuting $x_i$
        and multiplying each coordinate by $\pm 1$&$S_{n}\ltimes \Z_2^n$ permuting $x_i$
        and multiplying each coordinate by $\pm 1$&$S_{n}\ltimes \Z_2^{n-1}$ permuting $x_i$
        and multiplying an even number of coordinates by $-1$\\\hline

    \end{tabular}
    \caption{Description and properties of the A, B, C, and D type root systems}
    \label{tab:rootSystems1}

\end{table}
\begin{table}
    \centering
    \small
    
\newcolumntype{A}{ >{\centering\arraybackslash} m{1.75cm} }

\newcolumntype{B}{ >{\centering\arraybackslash} m{5cm} }
    
\renewcommand{\arraystretch}{2}
    \begin{tabular}{A||B|B}
        $\mg$& $G_2$& $F_4$ \\\hline
        $\tilde{\mg}$& $G_2$& $F_4$\\\hline
        Embedding space &$V=\{x_1+x_2+x_3=0\}\subset \mathbb{R}^{3}$& $\mathbb{R}^4$ \\\hline
        Roots& $e_i-e_j, i\neq j$ \newline
        $ \pm(2e_i-e_j-e_k),\,i\neq j\neq k\neq i$& $\pm e_i$\newline
        $\pm e_i \pm e_j$\newline
        $\frac{1}{2}(\pm e_1\pm e_2\pm e_3\pm e_4)$
    \\\hline
        Simple roots& $\alpha_1=e_1-e_{2}$ \newline
        $\alpha_2=-2e_1+e_2+e_3$& $\alpha_1=e_2-e_3$\newline
        $\alpha_2=e_3-e_4$\newline
        $\alpha_3=e_4$\newline
        $\alpha_4=\frac{1}{2}(e_1-e_2-e_3-e_4)$
        \\\hline
        $\Lambda_r$&$\Z^{3}\cap V$&$\Z^4+\left(\frac{1}{2}\sum e_i\right)\Z$\\\hline
        Fund. Weights & $\omega_1=-e_2+e_3$\newline
        $\omega_2=-e_1-e_2+2e_3$& $\omega_1=e_1+e_2$\newline
        $\omega_2=2e_1+e_2+e_3$\newline$\omega_3=\frac{1}{2}(3e_1+e_2+e_3+e_4)$\newline$\omega_4=e_4$\\\hline
        $\Lambda_w$&$\Z^{3}\cap V$&$\Z^4+\left(\frac{1}{2}\sum e_i\right)\Z$\\\hline
        $ds^2$&$dx^2$&$2dx^2$\\\hline
        $W$& $D_4=S_2\ltimes \Z_2$ permuting $x_1,x_2$ and taking $x\to -x$ &$S_3\ltimes(S_4\ltimes \Z_2^3)$ where $S_4$ permutes the coordinates $\Z_2^3$ changes the sign of an even number of coordinates and $S_3$ is generated by the reflection along $\frac{1}{2}\sum e_i$\\\hline

    \end{tabular}
    \caption{Description and properties of the $G_2$ and $F_4$ root systems}
    \label{tab:rootSystems2}

\end{table}

\begin{table}
    \centering
    \small
    
\newcolumntype{A}{ >{\centering\arraybackslash} m{2.0cm} }

\newcolumntype{B}{ >{\centering\arraybackslash} m{3.2cm} }
    
\renewcommand{\arraystretch}{2}
    \begin{tabular}{A||B|B|B}
    \tiny
        $\mg$& $E_6$& $E_7$& $E_8$\\\hline
        $\tilde{\mg}$& $E_6$& $E_7$&$E_8$\\\hline
        Embedding space & $V = \{x_6 = x_7 = -x_8\} \subset \mathbb{R}^8$ & $V = \{x_7 = -x_8\} \subset \mathbb{R}^8$ & $\mathbb{R}^8$ \\\hline
        Roots& $\{\pm e_i \pm e_j | i < j \leq 5\} \cup \{\frac{1}{2} \sum_{i=1}^8 (-1)^{n(i)} e_i|\newline \sum_{i=1}^8 n(i) \in 2\mathbb{Z}\}$   & $\{\pm e_i \pm e_j | i < j \leq 6\} \cup \{\pm(e_7-e_8)\}\cup \{\frac{1}{2} \sum_{i=1}^8 (-1)^{n(i)} e_i|\newline \sum_{i=1}^8 n(i) \in 2\mathbb{Z}\} $ &$\{\pm e_i \pm e_j | i < j\} \cup \{\frac{1}{2} \sum_{i=1}^8 (-1)^{n(i)} e_i|\newline \sum_{i=1}^8 n(i) \in 2\mathbb{Z}\}$ \\\hline
        Simple roots& $\alpha_1 = \frac{1}{2}(e_8 - e_7 -e_6-e_5-e_4-e_3-e_2+e_1)$\newline $\alpha_2 = e_2 + e_1$ \newline $\alpha_3 = e_2 - e_1 $\newline $\alpha_4 = e_3 - e_2$\newline $\alpha_5 = e_4 - e_3$\newline $\alpha_6 = e_5 -e_4$\newline  &$\alpha_1 = \frac{1}{2}(e_8 - e_7 -e_6-e_5-e_4-e_3-e_2+e_1)$\newline $\alpha_2 = e_2 + e_1$ \newline $\alpha_3 = e_2 - e_1 $\newline $\alpha_4 = e_3 - e_2$\newline $\alpha_5 = e_4 - e_3$\newline $\alpha_6 = e_5 -e_4$\newline $\alpha_7 = e_6-e_5$\newline &$\alpha_1 = \frac{1}{2}(e_8 - e_7 -e_6-e_5-e_4-e_3-e_2+e_1)$\newline $\alpha_2 = e_2 + e_1$ \newline $\alpha_3 = e_2 - e_1 $\newline $\alpha_4 = e_3 - e_2$\newline $\alpha_5 = e_4 - e_3$\newline $\alpha_6 = e_5 -e_4$\newline $\alpha_7 = e_6-e_5$\newline $\alpha_8 = e_7-e_6$\newline 
        \\\hline
        $\Lambda_r$&$\Lambda_r(E_8)\cap V$& $\Lambda_r(E_8)\cap V$ & $x\in \mathbb R^8$ such that $2x_i\in \Z, \,x_i-x_j\in \Z, \,\sum x_i\in 2\Z$  \\\hline
        Fund. Weights & $\omega_1= \frac{2}{3}(-e_6-e_7+e_8)$\newline $\omega_2 = \frac{1}{2}(e_1 + e_2 + e_3 + e_4 +e_5 -e_6 -e_7 +e_8)$\newline $\omega_3 =  \frac{1}{2}(-e_1+e_2+e_3+e_4+e_5) + \frac{5}{6}(-e_6-e_7 +e_8)$\newline $\omega_4 = e_3+e_4+e_5 -e_6 -e_7+e_8$\newline $\omega_5 =  e_4 + e_5 + \frac{2}{3}(-e_6 - e_7 +e_8)$\newline $\omega_6 = e_5 + \frac{1}{3}(-e_6 -e_7 +e_8)$ &
        $\omega_1= -e_7 +e_8$\newline $\omega_2 = \frac{1}{2}(e_1 + e_2 + e_3 + e_4 +e_5 +e_6 - 2e_7 +2e_8)$\newline $\omega_3 = \frac{1}{2}(-e_1+e_2+e_3+e_4+e_5 + e_6-3e_7+3e_8)$\newline $\omega_4 = e_3+e_4+e_5 +e_6 +2(-e_7 + e_8)$\newline $\omega_5 =  e_4 + e_5 + e_6 + \frac{3}{2}(-e_7 + e_8)$\newline $\omega_6 = e_5 +e_6 -e_7 + e_8$\newline $\omega_7 = e_6 + \frac{1}{2}(-e_7 + e_8)$  & $\omega_1= 2e_8$\newline $\omega_2 = \frac{1}{2}(e_1 + e_2 + e_3 + e_4 +e_5 +e_6 +e_7 +5e_8)$\newline $\omega_3 = \frac{1}{2}(-e_1+e_2+e_3+e_4+e_5 + e_6+e_7+7e_8)$\newline $\omega_4 = e_3+e_4+e_5 +e_6 + e_7 + 5e_8$\newline $\omega_5 = e_4 + e_5 + e_6 + e_7 + 4e_8$\newline $\omega_6 = e_5 +e_6 +e_7 + 3e_8$\newline $\omega_7 = e_6 + e_7 + 2e_8$\newline $\omega_8 = e_7+e_8$  \\\hline
        $\Lambda_w$ &$<\Lambda_r,\frac{1}{3}(e_1 + e_2 +e_3$ \newline $+ e_4 -2e_5-2e_6)> $ & $<\Lambda_r,\frac{1}{4}(e_1 + e_2 +e_3$ \newline $+ e_4 +e_5+e_6-3e_7-3e_8)>$ & $\Z^8$ \\\hline
        $ds^2$&$dx^2$& $dx^2$ & $dx^2$\\\hline

    \end{tabular}
    \caption{Description and properties of the E type root systems}
    \label{tab:rootSystems3}

\end{table}

\chapter{The rings of Jacobi forms} \label{app:Jacobi_forms}

To simplify notation, we set 
\begin{equation*}
    e(z)=\exp(2\pi i z).
\end{equation*}

We use the following conventions for the Jacobi $\theta$ functions,

\begin{equation}
    \Theta\left[
    \begin{array}{c}
         a\\b
    \end{array}\right](\tau,z)=\sum_{n\in \Z}e^{2\pi i b n}q^{\frac{1}{2}(n+a)^2}Z^{n+a},\quad \text{where } q=e(\tau), \, Z=e(z) \,,
\end{equation}

\begin{equation}
    \theta_1=i\Theta\left[
    \begin{array}{c}
         1/2\\1/2
    \end{array}\right],\quad\theta_2=\Theta\left[
    \begin{array}{c}
         1/2\\0
    \end{array}\right],\quad\theta_3=\Theta\left[
    \begin{array}{c}
         0\\0
    \end{array}\right],\quad\theta_4=\Theta\left[
    \begin{array}{c}
         0\\1/2
    \end{array}\right],
\end{equation}

for the Dedekind $\eta$ function,

\begin{equation}
    \eta(\tau)=q^{1/24}\prod_n (1-q^n), \quad  \text{where }  q=e(\tau),
\end{equation}

and for the Eisenstein series,
\begin{equation}
    E_n(\tau)=1+\frac{2}{\zeta(1-k)}\sum_{n\in \mathbb N^*}\frac{n^{k-1}q^n}{1-q^n}, \quad \text{where } \mathbb N^*=\{1,2,\dots\} \text{ and } q=e(\tau).
    \label{eq:EisensteinSeries}
\end{equation}

A weak Jacobi modular form of weight $w$ and index $n$ of the Lie algebra $\mg$ is a holomorphic function $\phi_{w,n}:\mathbb H\times \mathfrak{h}^*_{\mathbb{C}}\to \mathbb C$, where $\mathbb H$ is the upper-half plane, and $\mathfrak{h}^*_{\mathbb C}$ is the complexified root system, satisfying the following conditions:
\begin{itemize}
    \item Modularity:
    \begin{equation*}
        \phi_{w,n}\left(\frac{a\tau+b}{c\tau+d},\frac{z}{c\tau+d}\right)=(c\tau+d)^w e\left[\frac{nc}{2(c\tau+d)}(z,z)\right]\phi_{w,n}(\tau,z), \quad \forall \left(\begin{array}{cc}
             a& b \\
            c & d
        \end{array}\right)\in SL(2,\Z)
    \end{equation*}
    \item Quasi-periodicity:
    \begin{equation*}
        \phi_{w,n}(\tau,z+\lambda \tau +\mu)= e\left[-n\left(\frac{(\lambda,\lambda)}{2}+(\lambda,z)\right)\right]\phi_{w,n}(\tau,z), \quad \forall \lambda,\mu \in \Lambda_r
    \end{equation*}
    \item Weyl Symmetry:
    \begin{equation*}
    \phi_{w,n}(\tau,wz)=\phi_{w,n}(\tau,z),\quad \forall w \in W 
    \end{equation*}
    \item Fourier expansion: $\phi_{w,n}$ can be expanded as
    \begin{equation*}
        \phi_{w,n}=\sum_{l\in \mathbb{N},\gamma \in \Lambda_{cw}=(\Lambda_r)^*}c(l,\gamma)q^l\zeta^\gamma, \quad \text{where } \zeta^\gamma=e[(z,\gamma)]
    \end{equation*}
\end{itemize}
Note that no negative powers of $q$ occur in the Fourier expansion. Other classes of Jacobi forms can be defined upon dropping this requirement or imposing conditions on the coefficients $c(l,\gamma)$ \cite{EZ}. For every finite simple Lie algebra, except $E_8$, the space of Jacobi modular forms is a freely generated algebra over $\mathbb{C}[E_4,E_6]$ (the space of holomorphic modular forms) \cite{Wirthmuller:Jacobi,wang2020weyl}. Below, we will give explicit expressions for the generators of the rings of Jacobi forms we consider in the text. The forms are compactly written in terms of 
\begin{equation}
    \alpha:=\phi_{-1,1/2}=i\frac{\theta_1}{\eta^3} \,.
\end{equation}

\section{$A_n$ Jacobi forms}
We can construct the Jacobi forms of the Lie algebras $A_n$ by repeated use of the differential operator \cite{Bertola}
\begin{equation}
    \cZ=\frac{1}{2\pi i}\left(\sum_{i=1}^{n+1}\frac{\partial}{\partial x_i}+\frac{\pi^2}{3}E_2\sum_{i=1}^{n+1}x_i\right)
    \label{eq:Zoperator}
\end{equation}
on the lowest weight form. Let
\begin{equation}
    \Phi^{A_n}=\prod_{i=1}^{n+1}\alpha(x_i).
\end{equation}
then 
\begin{equation}
    \phi^{A_n}_{-k,1}=\left.\left(\cZ^{n+1-k}\Phi^{A_n}\right)\right|_{\sum x_i=0}, \quad k=0,2,3,4,\dots,n+1
\end{equation}
where the $\sum x_i=0$ condition is imposed after acting by the operator $\cZ$.

\section{$B_n$ Jacobi forms}
The Jacobi forms were given in \cite{Bertola}. We follow the conventions in \cite{Kim:2018gak}:

In terms of the Weierstrass function 
\begin{equation}
    \wp(z)=\frac{\theta_3(0)\theta_2(0)^2}{4}\frac{\theta_4(z)}{\theta_1(z)}-\frac{1}{12}\left(\theta_3(0)^4+\theta_2(0)^4\right),
\end{equation}
we have 
\begin{equation}
    \sum_{i=0}^n\wp^{(2i-2)}(v)\phi_{-2i,1}(x)=-\frac{1}{2^{n-2}(n-1)!}
    \frac{
    \left|\begin{array}{cccc}
         1& \wp(v)&\cdots&\wp^{(2n-2)}(v) \\
         1& \wp(x_1)&\cdots&\wp^{(2n-2)}(x_1) \\
         \vdots&\vdots &&\vdots\\
         1& \wp(x_n)&\cdots&\wp^{(2n-2)}(x_n) \\
    \end{array}\right|
    }{
    \left|\begin{array}{cccc}
         1& \wp(x_1)&\cdots&\wp^{(2n-4)}(x_1) \\
         \vdots&\vdots &&\vdots\\
         1& \wp(x_n)&\cdots&\wp^{(2n-4)}(x_n) \\
    \end{array}\right|
    }\prod_{i=1}^n\alpha^2(x_i),
    \label{eq:Bforms}
\end{equation}
where $\wp^{(-2)}$ is to be understood as $1$.
\section{$D_n$ Jacobi Forms}
	
	The Jacobi forms for $D_n$ ($n\leq 8$) were built in \cite{Adler_2020}.
	 
	 In $D_n$ there is a form of weight $-n$ and index 1 given by
	 \begin{equation}
	 	\omega_{-n,1}^{D_n}
	 	=\prod_{i=1}^{n}\alpha(x_i).
	 \end{equation}
	 The remaining forms of index 1 can be obtained from the lowest weight form by the repeated use of Hecke operator. On forms of weight $k$ and index $m$, the Hecke operator is given by
	 \begin{equation}
 H_{k}=q\partial_q -\frac{1}{2m}\left(X_i\partial_{X_i}+X_i^2\partial_{X_i}^2\right)-\frac{2k-8}{24}E_2,\quad X_i=e(x_i).
 \end{equation}
 The forms of index 2 follow from the inclusion $D_n(2)\leq nA_1$.
	 
	 Explicitly, all the $D_8$ forms are given by 
	 \begin{align}
	   	\omega_{-8,1}^{D_8}&=\prod_{i=1}^{8}\alpha(x_i),\\
	   	\phi_{-4,1}^{D_8}&=\frac{1}{\eta^{24}}\left(E_4\sum_{j=1}^4\prod_{i=1}^8 \theta_j(x_i)-\sum_{j=2}^4\theta_j(0)^8\prod_{i=1}^8\theta_j(x_i)\right)-E_4\omega_{8,1}^{D_8},\\
	   	 \phi_{-2,1}^{D_8}&=3 H_4 \phi_{4,1}^{D_8},\\
 \phi_{0,1}^{D_8}&=\frac{1}{32}\left(2 H_2 \phi_{-2,1}^{D_8} -E_4 \phi_{-4,1}^{D_8}\right),\\
	   	 \phi_{-2k,2}^{D_8}&=\frac{1}{k!(n-k)!}\sum_{\sigma\in S_n}\phi_{-2,1}^{A_1}(\tau,z_{\sigma(1)})\cdots \phi_{-2,1}^{A_1}(\tau,z_{\sigma(k)})\phi_{0,1}^{A_1}(\tau,z_{\sigma(k+1)})\cdots \phi_{0,1}^{A_1}(\tau,z_{\sigma(1)})\\
	   	&\quad k=3,4,5,6,7.
	 \end{align}
	
	Besides $\omega^{D_n}_{n,1}$, the $D_n$ forms with $n\leq 8$ can be obtained by setting $x_i=0, \text{ for } i>n$. 
For $D_4$, to agree with previous conventions \cite{Bertola,DelZotto:2017mee}, we use the forms $\omega_{-4,1}^{D_4}$ and
\begin{align}
\phi_{-4,1}^{D_4}&=\left.-\frac{1}{16}\phi_{-4,1}^{D_8}\right|_{x_{i>4}=0}\,,\\
\phi_{-2,1}^{D_4}&=\left.-\frac{1}{8}\phi_{-2,1}^{D_8}\right|_{x_{i>4}=0}\,,\\
\phi_{0,1}^{D_4}&=\left.2\phi_{0,1}^{D_8}\right|_{x_{i>4}=0}\,,\\
\phi_{-6,2}^{D_4}&=\left.\frac{1}{32}\phi_{-6,2}^{D_8}\right|_{x_{i>4}=0}\,.
\end{align}

\section{$C_n$ Jacobi forms }

For $n\geq 4$, we have 
\begin{equation}
    \text{Weyl}(C_n)=\text{Weyl}(D_n)\ltimes \mathbb {Z}_2.
\end{equation}

From the basis of $D_n$ Jacobi forms, the only one that is not invariant under the extra involution is $\omega_{n,1}^{D_n}$ which switches sign. We thus get a basis of $C_n$ Jacobi forms by squaring this last form
\begin{align*}
    \phi_{-k,1}^{C_n}&=\phi^{D_n}_{-k,1},\quad k=0,2,4\\
    \phi_{-k,2}^{C_n}&=\phi^{D_n}_{-k,2},\quad k=6,8,\dots,2n-2\\
    \phi_{-2n,2}^{C_n}&=(\omega_{-n,1}^{D_n})^2.
\end{align*}

 For $C_3$, we can get a basis of the forms by going down from $D_4$ as sketched in figure \ref{fig:jacobiBD} but, we decided to use the convention from \cite{Bertola} that uses the fact that the $C_3$ and $A_3$ root lattices are isomorphic and the Weyl group of $C_3$ compared to the one of $A_3$ just has an extra involution \cite{Bertola}.

Consider the map 
\begin{align}
		j:\left\{\begin{array}{ccc}
		C_3&\to& A_3\\
		\left(
		\begin{array}{c}
		x_1\\x_2\\x_3
		\end{array}\right)&\mapsto&\left(
		\begin{array}{c}
		\frac{x_1-x_2-x_3}{2}\\\frac{-x_1+x_2-x_3}{2}\\\frac{-x_1-x_2+x_3}{2}\\\frac{x_1+x_2+x_3}{2}
		\end{array}\right)
		\end{array}\right.
		\label{eq: C3 to A3}
	\end{align}
we then set 
\begin{align}
    \phi^{C_3}_{-k,1}&=\phi^{A_3}_{-k,1}\circ j,\quad k=0,2,4\\
    \phi^{C_3}_{-6,2}&=(\phi^{A_3}_{-3,1})^2\circ j
\end{align}

	\section{$F_4$ Jacobi forms}
	\label{sec:appF4}
	The generators of the ring $J(F_4)$ can be built from the results in \cite{Bertola,Wirthmuller:Jacobi} and were already given explicitly in \cite{adler2020structure}.
	
	The $F_4$ and $D_4$ lattices are isomorphic\footnote{If one takes the Euclidean norm, the exact statement is that $F_4(2)$, the $F_4$ lattice with norm scaled by 2 ( $(\cdot,\cdot)_{F_4(2)}=2(\cdot,\cdot)$), is isomorphic to $D_4$. In the present paper, we fix the norm by imposing the condition that short coroots have norm squared 2.}. An explicit isomorphism is given by\footnote{We choose this particular isomorphism as it maps the vector $\rho_L=\sum_{\alpha \in \Delta_L^+}\alpha$ of $F_4$ to the corresponding vector of $D_4$. This is only relevant for the considerations of section \ref{ss:GWinv}.}
	\begin{align}
		i:\left\{\begin{array}{ccc}
		F_4&\to& D_4\\
		\left(
		\begin{array}{c}
		x_1\\x_2\\x_3\\x_4
		\end{array}\right)&\mapsto&\left(
		\begin{array}{c}
		x_3-x_4\\-x_3-x_4\\x_1+x_2\\x_1-x_2
		\end{array}\right)
		\end{array}\right..
		\label{eq: F4 to D4}
	\end{align}
	
	The Weyl groups of the two lattices are different, however. The $F_4$ Weyl group is the full orthogonal group of the lattice which is given by the semi-direct product of the $D_4$ Weyl group and the $D_4$ Dynkin diagram symmetry, $S_3$:
	\begin{equation*}
		W(F_4)=O(F_4)=W(D_4)\ltimes \text{DynkinSym}(D_4)=W(D_4)\ltimes S_3.
	\end{equation*}
The generators of the ring $J(F_4)$ can then be obtained from the $D_4$ forms by imposing the extra $S_3$ symmetry. 

In terms of Weyl invariant polynomials, see appendix \ref{app:Weyl_invariant_polynomials}, this symmetry is straightforward to impose. As $S_3$ permutes the three external legs of the $D_4$ Dynkin diagram, this symmetry simply permutes the Weyl invariant polynomials $p_1,p_3,p_4$. Hence, $J(F_4)\subset J(D_4)$ is the sub-ring of Jacobi forms invariant under permutations of $p_1,p_3,p_4$. 

By looking directly at the expansion in terms of Weyl invariant polynomials, the following are the generators of $J(F_4)$:
\begin{align}
\label{eq:F4forms}
\phi_{0,1}^{F_4}&=\left(\phi_{0,1}^{D_4}-\frac{2}{3}E_4\phi_{-4,1}^{D_4}\right)\circ i\\
\phi_{-2,1}^{F_4}&=\phi_{-2,1}^{D_4}\circ i\\
\phi_{-6,2}^{F_4}&=\left(\phi_{-6,2}^{D_4}-\frac{1}{18}\phi_{-2,1}^{D_4}\phi_{-4,1}^{D_4}\right)\circ i\\
\phi_{-8,2}^{F_4}&=\left(\left(\phi_{-4,1}^{D_4}\right)^2+3\left(\omega_{-4,1}^{D_4}\right)^2\right)\circ i\\
\phi_{-12,3}^{F_4}&=\left(\phi_{-4,1}^{D_4}\left(\omega_{-4,1}^{D_4}\right)^2-\frac{1}{9}\left(\phi_{-4,1}^{D_4}\right)^3\right)\circ i
\end{align}

\section{$G_2$ Jacobi forms}
The $G_2$ and $A_2$ root lattices are the same and the Weyl groups differ only by the involution $x\mapsto -x$ which is in $W(G_2)$ but not in $W(A_2)$. The only generator of $J(A_2)$ not invariant under this transformation is $\phi_{-3,1}$ which gets maps to minus itself. Therefore we can get a set of generators of $J(G_2)$ by squaring this form.

\begin{align}
    \phi_{0,1}^{G_2}&=\phi_{0,1}^{A_2}\\
    \phi_{-2,1}^{G_2}&=\phi_{-2,1}^{A_2}\\
    \phi_{-6,2}^{G_2}&=(\phi_{-3,2}^{A_2})^2.
\end{align}

\chapter{Specialization formulas}
\label{app:specializationformulas}
\section{$A$ series}
The derivative of a modular form is only quasi-modular form. The term proportional to $E_2$ in \ref{eq:Zoperator} is precisely what is needed to cancel the anomalous modular transformation of the other term. This second term is important because if it weren't there, the action of $\cZ$ in a Jacobi modular form wouldn't give another Jacobi modular form. However, at this point, we only want to find the proportionality constant between the different forms in figure \ref{fig:jacobiA}, therefore it is enough to focus on the first term. 

Consider then, the form $\phi_{-k,1}^{A_n}$ with $k=0,2,\dots,n$ restricted to the subspace $x_{n+1}=0$
\begin{align*}
    \phi_{-k,1}^{A_n}|_{x_{n+1}=0}&=\left.\frac{1}{(2\pi i)^{n+1-k}}\left(\sum_{i=1}^{n+1}\frac{\partial}{\partial x_i}\right)^{n+1-k}\prod_{i=1}^{n+1}\alpha(x_i)\right|_{x_{n+1}=0}+\dots\\
    &=(n+1-k)\frac{1}{(2\pi i)^{n+1-k}}\alpha'(0)\left(\sum_{i=1}^{n}\frac{\partial}{\partial x_i}\right)^{(n-1)+1-k}\prod_{i=1}^{n+1}\alpha(x_i)+\dots\\
    &=-(n+1-k)\phi_{-k,1}^{A_{n-1}},
\end{align*}
where $\dots$ means a term proportional to $E_2$ and we used $\alpha(0)=0$ and $\alpha'(0)=-2\pi i$.

We repeat the formula for reference
\begin{equation}
    \phi_{-k,1}^{A_n}|_{x_{n+1}=0}=\left\{\begin{array}{cc}
     0    &k=n+1  \\
    -(n+1-k)\phi_{-k,1}^{A_{n-1}}     & k=0,2,3,\dots,n
    \end{array}\right.
\end{equation}

\section{$B$ series}
The Weierstrass function is divergent at the origin. Then one has to be careful when going to $x_n=0$. The series expansion of the Weierstrass function is given by
\begin{equation}
    \wp(x_n)=-x_n^{-2}+o(x_n^2),
\end{equation}
then
\begin{equation*}
    x_n^{2k}\wp^{(2k-2)}(x_n)|_{x_n=0}=-(2k-1)!.
\end{equation*}
Using this, and $\left.\frac{\alpha^2(x_n)}{x_n^2}\right|_{x_n=0}=1$ in equation \ref{eq:Bforms} we find that
\begin{equation*}
    \phi^{B_n}_{-2i,1}|_{x_n=0}=
    \left\{\begin{array}{cc}
     0    &i=n  \\
    (2n-1)\phi^{B_{n-1}}_{-2i,1}   & i=0,\dots,n-1
    \end{array}\right.
\end{equation*}

Because the denominator for theories with $C_n$ gauge algebra goes to 0 after Higgsing, we give the reduction of the lowest weight form of $B_n$ to the smallest order in $x_n$

\begin{equation*}
    \phi_{-2n,1}^{B_n}=\frac{1}{2(n-1)}\alpha^2(x_n)\phi_{-2(n-1)}^{B_{n-1}}=\frac{x_n^2}{2(n-1)}\phi_{-2(n-1)}^{B_{n-1}}+o(x_n^3)
\end{equation*}

\section{$C/D$ series}
For $n\geq 4$, by definition of the forms there is no extra factor and the results in figure \ref{fig:jacobiBD} are exact. For $n=3$, the last line in the figure, our bases for $D_4$ and $C_3$ forms are different. A calculation shows:
\begin{align*}
    \phi^{D_4}_{0,1}|_{x_4=0}&=\frac{4}{3} E_4 \phi_ {-4,1}^{C_3}+2 \phi _{0,1}^{C_3}\\
    \phi^{D_4}_{-2,1}|_{x_4=0}&=-6\phi^{C_3}_{0,1}|_{x_4=0}\\
    \phi^{D_4}_{-4,1}|_{x_4=0}&=2\phi^{C_3}_{-4,1}|_{x_4=0}\\
    \phi^{D_4}_{-6,2}|_{x_4=0}&=-\phi^{C_3}_{-6,2}|_{x_4=0}\\
    \omega^{D_4}_{-4,1}|_{x_4=0}&=0
\end{align*}

\section{$F_4, C_4$ and $D_4$} 
The root lattices for $F_4,C_4$ and $D_4$ are isomorphic; the Weyl groups on the other hand decrease in order from $F_4$ via $C_4$ to $D_4$,
\begin{equation*}
    O(D_4)=\text{Weyl}(F_4)\supset\text{Weyl}(B_4)\supset\text{Weyl}(D_4)\,.
\end{equation*}
The formulas going from $F_4$ to $D_4$ and $B_4$ were already given, but we repeat them explicitly:

\begin{align*}
\phi_{0,1}^{F_4}&=\left(\phi_{0,1}^{C_4}-\frac{2}{3}E_4\phi_{-4,1}^{C_4}\right)\circ i=\left(\phi_{0,1}^{D_4}-\frac{2}{3}E_4\phi_{-4,1}^{D_4}\right)\circ i\\
\phi_{-2,1}^{F_4}&=\phi_{-2,1}^{B_4}\circ i=\phi_{-2,1}^{D_4}\circ i\\
\phi_{-6,2}^{F_4}&=\left(\phi_{-6,2}^{C_4}-\frac{1}{18}\phi_{-2,1}^{C_4}\phi_{-4,1}^{C_4}\right)\circ i=\left(\phi_{-6,2}^{D_4}-\frac{1}{18}\phi_{-2,1}^{D_4}\phi_{-4,1}^{D_4}\right)\circ i\\
\phi_{-8,2}^{F_4}&=\left(\left(\phi_{-4,1}^{C_4}\right)^2+3
\phi_{-8,2}^{C_4}\right)\circ i=\left(\left(\phi_{-4,1}^{D_4}\right)^2+3\left(\omega_{-4,1}^{D_4}\right)^2\right)\circ i\\
\phi_{-12,3}^{F_4}&=\left(\phi_{-4,1}^{D_4}\phi_{-8,2}^{C_4}-\frac{1}{9}\left(\phi_{-4,1}^{C_4}\right)^3\right)\circ i=\left(\phi_{-4,1}^{D_4}\left(\omega_{-4,1}^{D_4}\right)^2-\frac{1}{9}\left(\phi_{-4,1}^{D_4}\right)^3\right)\circ i.
\end{align*}

\section{$C_3$ to $G_2$}
We just have some coefficients in the relations in figure \ref{fig:jacobiB3}. An explicit calculation gives 

\begin{align*}
    \phi_{0,1}^{C_3}|_{\sum x=0}&=-4\phi_{0,1}^{G_2}\\
    \phi_{-2,1}^{C_3}|_{\sum x=0}&=-2\phi_{-2,1}^{G_2}\\
    \phi_{-4,1}^{C_3}|_{\sum x=0}&=0\\
    \phi_{-6,2}^{C_3}|_{\sum x=0}&=\phi_{-6,2}^{G_2}\\
\end{align*}

    \chapter{Weyl invariant polynomials} \label{app:Weyl_invariant_polynomials}
    
    The most straightforward path towards determining the coefficients in the expansion of the numerator $\cN$ of $Z_k$ in the ansatz \eqref{eq:ZkAnsatz} is to expand $Z_k$ in the exponentiated K\"ahler parameters and compare coefficients to the corresponding expansion of the topological string partition function. We have found it computationally advantageous to first express all quantities at a given order in $q$ and $\gs$ in terms of Weyl invariant polynomials.
    
    A Weyl invariant polynomial for the algebra $\mg$ is a function $p:\mathfrak h^*\to \mathbb C$ such that
    \begin{equation*}
        p(x+\alpha^\vee)=p(x), \quad \forall \alpha^\vee \in \LambdaCR(\mg)
    \end{equation*}
    and
    \begin{equation*}
        p(w x)=p(x), \quad \forall w \in \weyl( \mg).
    \end{equation*}
    It is shown in \cite{lorenz2006multiplicative,Bourbaki} that a set of generators for the ring of $\weyl(\mg)$ invariant polynomials is given by
    \begin{equation*}
        p_i(x)=\sum_{\omega \in \weyl(\mg)\{\omega_i\}}\exp[2\pi i (\omega,x)], \quad \omega_i \text{ a fundamental weight}.
    \end{equation*}
   
    The Weyl invariant polynomials are also useful in imposing Dynkin diagram symmetry. This symmetry permutes simple roots of the gauge algebra or, equivalently, fundamental weights. It hence acts by permutation on the $p_i$, rendering the construction of invariant polynomials straightforward. For instance, Weyl and Dynkin diagram symmetric polynomials for the Lie algebra $D_4$, consist of all polynomials in the corresponding $p_i$ which are invariant under permutations of $p_1,\,p_3$ and $p_4$ (the index 2 being assigned to the central node).

    \chapter{Rank 1 Higgsing trees} \label{app:Higgsing_trees}
     
    The manifolds underlying rank 1 6d theories are elliptic fibrations
     \begin{equation*}
        \xymatrix{ E \ar[r]&X\ar[d]^\pi\\ 
        &B\\}
    \end{equation*}
    We assume that the fibration contains a global section, so that $X$ can be described by a Weierstrass model. Defining $X$ as the zero set of an anti-canonical section of the weighted projective space
    \begin{equation} \label{eq:ambient_space}
        \mathbb{P}^{2,3,1} (2 K_B \oplus 3 K_B \oplus \cO)\,,
    \end{equation}
    where $\mathcal{O}$ and $K_B$ denote the trivial and the canonical line bundle respectively of the base $B$ respectively, guarantees that it is both Calabi-Yau and an elliptic fibration. With $x$, $y$, $z$ denoting sections of the vector bundle being projectivized, a generic such section can be written as 
    \begin{equation} \label{eq:weierstrass}
        y^2 = x^3 + f x z^4 + g z^6\,,
    \end{equation}
    with $f$ and $g$ sections of $\mathcal{O}(-4K_B)$ and $\mathcal{O}(-6K_B)$ respectively. The zero set of the section \eqref{eq:weierstrass} is generically singular. The singular locus lies along the zero set of the discriminant of the fibration, given by $\Delta = 4 f^3 + 27 g^3 \in \mathcal{O}(-12K_B)$. Along the locus $\Delta = 0$, the elliptic fiber degenerates. In F-theory parlance, this signals the presence of D7-branes. The singularities can be resolved by successive blow-ups. For this process to preserve the Calabi-Yau condition, the vanishing order of $(f, g, \Delta)$ along any divisor in the base must be strictly smaller than $(4,6,12)$. As a result, the possible singularities along a divisor in $B$ must be of Kodaira type. The resolution of the singularity can be worked out explicitly using Tate's algorithm \cite{Bershadsky:1996nh,Katz:2011qp}.
  
    For rank 1 models, the base can be chosen as the total space of the line bundle $\cO(-n) \rightarrow \IP^1$ (which coincides with the normal bundle of the base curve of the Hirzebruch surface $\mathbb{F}_n$). The bases that lead to good F-theory models have $0 \leq n \leq 8$ or $n=12$ \cite{Morrison:2012js}. The zero set of the generic section of the anti-canonical bundle of the projective bundle \eqref{eq:ambient_space} is singular for $n>2$. The generic singularity leads to the maximally Higgsed or non-Higgsable model, in the terminology of \cite{Morrison:2012js}. Upon specializing to a subset of sections (this is the process of specialization of complex structure referred to in the body of this paper), the singularity can be enhanced, leading to gauge theories with higher rank gauge symmetry and charged matter.
    
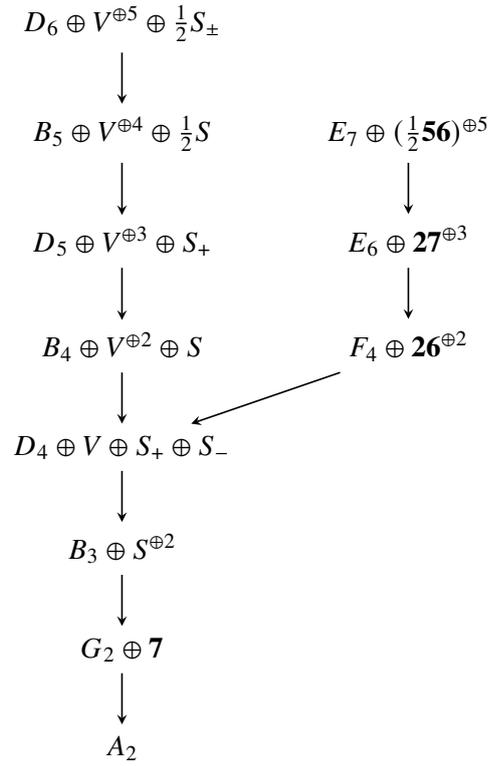
\begin{figure}[ht]
        \centering
        \begin{tikzcd}
        D_6\oplus V^{\oplus 5} \oplus \frac{1}{2} S_\pm\ar[d]&\\
        B_5\oplus V^{\oplus 4} \oplus \frac{1}{2} S\ar[d]& E_7  \oplus (\tfrac{1}{2} {\bf 56})^{\oplus 5}\ar[d]\\
        D_5 \oplus V^{\oplus 3} \oplus S_+\ar[d]& E_6  \oplus {\bf 27}^{\oplus 3} \ar[d]\\
        B_4 \oplus V^{\oplus 2} \oplus S\ar[d] & F_4 \oplus {\bf 26}^{\oplus 2} \ar[dl]\\
        D_4 \oplus V \oplus S_+ \oplus S_-\arrow[d]&\\
        B_3 \oplus S^{\oplus 2} \arrow[d]&\\
        G_2 \oplus {\bf 7} \arrow[d]&\\
         A_2&
        \end{tikzcd}
        \caption{Finite length Higgsing tree over the base $\cO(-3) \rightarrow \IP^1$, with the maximally Higgsed gauge group $A_2$ at its root.}\label{fig:higgsA3}
    \end{figure}

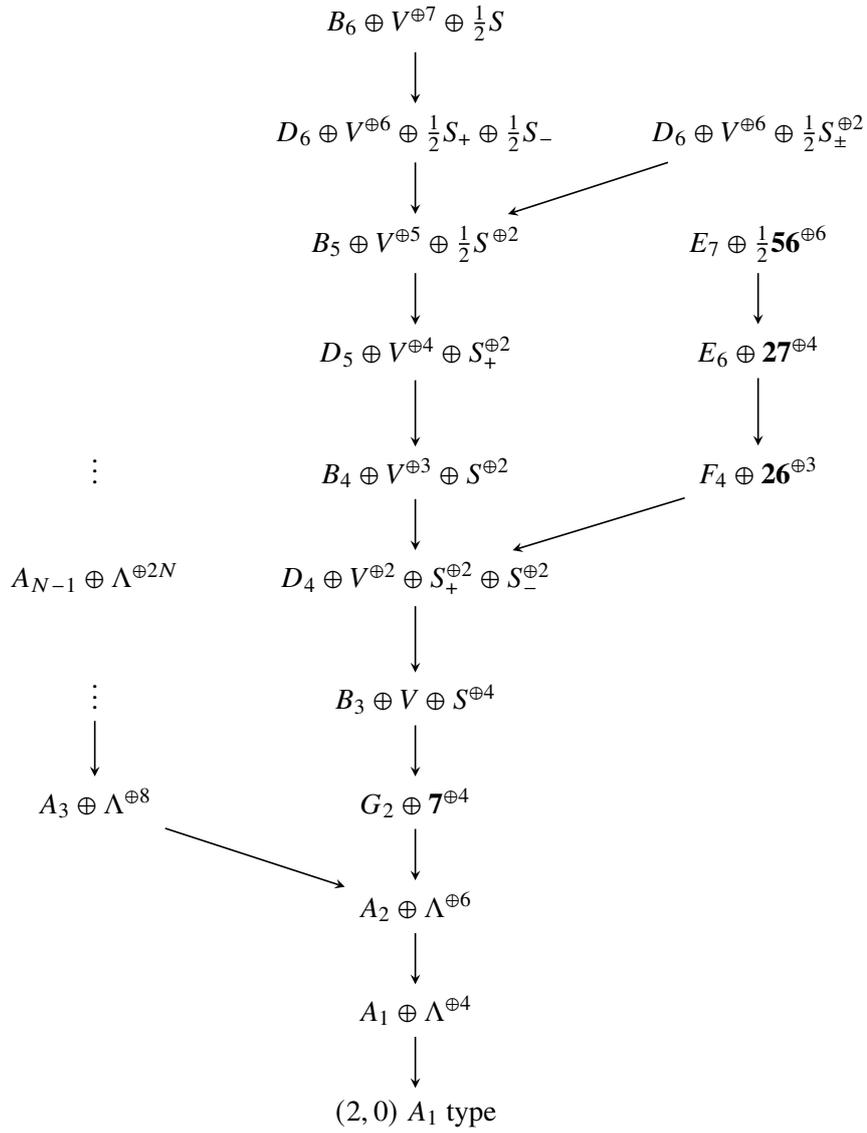
\begin{figure}[ht]
        \centering
        \begin{tikzcd}
        &B_6 \oplus V^{\oplus 7} \oplus \tfrac12 S \ar[d]&\\
        &D_6 \oplus V^{\oplus 6} \oplus \tfrac{1}{2} S_+ \oplus \tfrac12 S_-\ar[d]&D_6 \oplus V^{\oplus 6} \oplus \tfrac{1}{2} S_\pm^{\, \oplus 2}\ar[dl]\\
        &B_5 \oplus V^{\oplus 5} \oplus \tfrac{1}{2} S^{\,\oplus 2}\ar[d]&E_7 \oplus \tfrac{1}{2}{\bf 56}^{\oplus 6}\ar[d]\\
        &D_5 \oplus V^{\oplus 4} \oplus S_+^{\oplus 2} \ar[d]&E_6 \oplus {\bf 27}^{\oplus 4}\ar[d]\\
        \vdots&B_4 \oplus V^{\oplus 3} \oplus S^{\oplus 2}\ar[d] &F_4\oplus {\bf 26}^{\oplus 3}\ar[dl]\\
        A_{N-1} \oplus \Lambda^{\oplus 2N}&D_4 \oplus V^{\oplus 2} \oplus S_+^{\oplus 2}\oplus S_-^{\oplus 2}\ar[d]&\\
        \vdots\ar[d]&B_3 \oplus V \oplus S^{\oplus 4}\ar[d]&\\
        A_3 \oplus \Lambda^{\oplus 8}\ar[dr]&G_2 \oplus {\bf 7}^{\oplus 4}\ar[d]&\\
       &A_2 \oplus \Lambda^{\oplus 6}\ar[d]&\\
       &A_1 \oplus \Lambda^{\oplus 4}\ar[d]&\\
         &(2,0)\ A_1\text{\ type}
        \end{tikzcd}
        \caption{Infinite length Higgsing tree over the base $\cO(-2) \rightarrow \IP^1$, with the M-string theory at its root.}\label{fig:higgsM}
    \end{figure}

    We give a few examples of the resulting Higgsing trees in figure \ref{fig:higgsA3}, \ref{fig:higgsD4} and \ref{fig:higgsM}, following \cite{Bershadsky:1996nh, DelZotto:2018tcj}. The base of the trees corresponds to the maximally Higgsed models, and each successive node corresponds to a further specialization of the complex structure and resolution of the ensuing singularity. The nodes are labelled by the gauge group and matter content of the corresponding F-theory compactification. We denote, following \cite{DelZotto:2018tcj}, the fundamental representation of $A_{N-1}$ by $\Lambda$ stands for; $V$ and $S$ ($S_{\pm}$) denote the vector and spinor (Weyl spinors) representation of $B_N$ ($D_N$) respectively. For exceptional Lie algebras, we label the irreducible representation by its dimension. 

    \begin{figure}[ht]
        \centering
        \begin{tikzcd}
        \vdots\\
         D_{N+1}\oplus V^{\oplus (2N-6)}\ar[d]&\\
        B_N  \oplus V^{\oplus (2N-7)}&\\
        \vdots\ar[d]& E_7 \oplus \tfrac{1}{2}{\bf 56}^{\oplus 4}\ar[d]\\
        D_5\oplus V^{\oplus 2}\ar[d]&E_6 \oplus {\bf 27}^{\oplus 2}\ar[d]\\
        B_4\oplus V \ar[d]&F_4\oplus {\bf 26}\ar[dl]\\
         D_4&
        \end{tikzcd}
        \caption{Infinite length Higgsing tree over the base $\cO(-4) \rightarrow \IP^1$, with the maximally Higgsed gauge group $D_4$ at its root.}\label{fig:higgsD4}
    \end{figure}

    \chapter{Elliptic genera} \label{a:elliptic_genera}
    
    An $N=1$ right-moving supersymmetry $Q_R$ in two dimensions together with a right-moving fermion charge operator $F_R$ permits defining a trace \cite{Witten:1982df}
    \begin{equation}
        \tr_{\cH} (-1)^{F_R} \bar{q}^{H_R}
    \end{equation}
    which receives contributions only from states at $H_R=0$. When the kernel of $H_R$ is infinite dimensional, additional operators must be included in the trace to render it well-defined, 
    \begin{equation} \label{eq:flavored_trace}
        \tr_\cH (-1)^{F_R} X q^{H_L} \bar{q}^{H_R} \,.
    \end{equation}
    To maintain the property that the trace receives contributions only at $H_R =0$, two states $\psi$ and $Q_R \psi $ must have the same $X$ eigenvalue, i.e. only operators $X$ that commute with $Q_R$ are permissible insertions.
    
    An additional right-moving supersymmetry $\bar{Q}_R$ yielding an $\cN=2$ algebra
    \begin{equation} \label{eq:N2}
        Q_R^2 = \bar{Q}_R^2 = 0 \,, \quad \{Q_R , \bar{Q}_R\} = 2H_R
    \end{equation}
    permits interpreting the elements of the kernel of $H_R$ as representatives of the cohomology of $\bar{Q}_R$. Central to this argument is that $\bar{Q}_R$ be the adjoint operator to $Q_R$, such that\footnote{In two dimensions, the Weyl and Majorana conditions can simultaneously be imposed on spinors. Consequently, supercharges can be chosen to be chiral and hermitian \cite{Polchinski:1998rq}. Multiple such supercharges satisfy the algebra
    \begin{eqnarray} \label{eq:Majorana_Weyl_supercharges}
        \{Q_L^A , Q_L^B \} &=& \delta^{AB} (P^0 - P^1) = \delta^{AB} H_L \,,\\
        \{Q_R^A , Q_R^B \} &=& \delta^{AB} (P^0 + P^1) = \delta^{AB} H_R \,, \nn \\
        \{Q_L^A , Q_R^B \} &=& Z^{AB}\,. \nn
    \end{eqnarray}
    For the purposes of defining the elliptic genus, it is more convenient to work with a pair of charges that are adjoint to each other, to mimic the behavior of the $\bar{\partial}$ and $\bar{\partial}^\dagger$ operators in Hodge theory. These can be defined as
    \begin{equation}
        Q_R = Q_R^1 + i Q_R^2 \,, \quad Q_R^\dagger = Q_R^1 - i Q_R^2
    \end{equation}
    with anti-commutation relations
    \begin{equation} \label{eq:adjoint_supercharges}
        Q_R^2 = (Q_R^\dagger)^2 = 0 \,, \quad \{Q_R, Q_R^\dagger\} = 2 H_R \,.
    \end{equation}
    Note that the $SO(2)$ symmetry of the right moving sector in the presentation \eqref{eq:Majorana_Weyl_supercharges} (restricted to a pair of charges) is realized as a $U(1)$ symmetry of the algebra \eqref{eq:adjoint_supercharges}.}
    \begin{equation}
        \langle \{Q_R , \bar{Q}_R \} \psi , \psi \rangle = || Q_R \psi ||^2 + || \bar{Q}_R \psi ||^2 \,.
    \end{equation}
    If $\psi \in \ker H_R$, it is therefore both $Q_R$ and $\bar{Q}_R$ closed. On the other hand, $\bar{Q}_R \psi \in \ker H_R \Rightarrow \bar{Q}_R \psi = 0$ by a similar argument. Acting by $\bar{Q}_R$ hence either annihilates the state or maps out of the kernel of $H_R$. In both cases, the image does not contribute to the trace \eqref{eq:flavored_trace}; it is therefore only sensitive to the $\bar{Q}_R$ cohomology. In the context of $\cN=2$ theories, the trace \eqref{eq:flavored_trace} is referred to as the elliptic genus \cite{Witten:1986bf, Witten:1993jg}.
    
    This cohomological underpinning gives the elliptic genus its stability under sufficiently mild variations of the Lagrangian.

    In the case of $\cN=2$ superconformal symmetry, the anti-commutation \eqref{eq:N2} holds in the Ramond sector with $H_R = \bar L_0 - \frac{c_R}{24}$, while in the NS sector, it becomes
    \begin{equation}
        \{Q_R, \bar Q_R \} = \bar L_0 + \frac{1}{2} \bar J_0 \,, \label{eq:anticommutation_NS}
    \end{equation}
    with $\bar J_0$ the $U(1)$ charge generator of the $\cN=2$ superconformal algebra. All the consideration which were formulated above thus go through for $H_R$ replaced by the RHS of the relation \eqref{eq:anticommutation_NS}.

    Evaluation of the elliptic genus in Landau-Ginzburg models, by deforming away the potential, boils down to multiplying the universal contributions from each multiplet, followed, in the case of gauge theories, by an integration over holonomies \cite{Benini:2013a, Benini:2013xpa}.

\chapter[Computing affine characters at negative level via Kazhdan-Lusztig\\ polynomials]{Computing affine characters at negative level via Kazhdan-Lusztig polynomials} 
\label{a:Kazhdan-Lusztig}

The character of a representation $R$ of an affine Lie algebra $\affineg$ is given by 
\begin{equation*}
    \cha R=\sum_{\mu\in \LambdaW} m_R(\mu) e^{2\pi i\mu}
\end{equation*}
where the sum is over the weight lattice $\LambdaW$ and $m_R(\mu)$ denotes the multiplicity of the weight $\mu$ in the representation.

The characters $\hat \chi_\lambda$ for the highest weight representation $L_\lambda$ that appear in the main text differ from $\cha L_\lambda$ by a shift in the powers of $q$ given by the ground state energy level of the Wess-Zumino-Witten model associated to $G$:
\begin{equation}
    \hat \chi_\lambda =q^{-\frac{c}{24}+h_\lambda}\cha L_\lambda\,,
\end{equation}
with 
\begin{equation*}
    c=\frac{\dim(G)k}{h_G^\vee+k}\,, \quad h_\omega^G=\frac{\left<\lambda,\lambda+2\rho\right>}{h_G^\vee+k}\,,
\end{equation*}
where $k$ is the level of $\lambda$, $h^\vee_G$ the dual Coxeter number of $G$, $\rho$ the Weyl vector, and the inner product $\langle \cdot , \cdot \rangle$ is normalized to $2$ for long roots.

For positive level representations of affine Lie algebras, the Weyl-Kac formula
\begin{equation}
    \cha L_\lambda = \sum_{\omega\in W}\text{sign}(w)\cha M_{\omega \cdot \lambda}
    \label{eq:weyl-kac}
\end{equation}
gives the character of an irreducible module $L_\lambda$ of highest weight $\lambda$ in terms of Verma module characters $M_\mu$, whose characters can easily be computed as
\begin{equation}
    \cha M_\mu=\frac{e^\mu}{\prod_{\alpha\in \Delta_+}(1-e^\alpha)^{r_\alpha}}, \quad r_\alpha =\left\{\begin{array}{cc}
         \text{Rank}(\mathfrak{g})& \text{ if } \alpha\in \mathbb N\delta  \\
         1& \text{otherwise}
    \end{array}\right.\,.
\end{equation}
The representations of the flavor group which appear in the ansatz \eqref{eq:affineERintro} are at positive level. There exist efficient algorithms to compute such characters: Freudenthal's formula e.g. computes $m_\lambda(\mu)$, the multiplicity $m_\lambda(\mu)$ of the weight $\mu$ in the highest weight representation $L(\lambda)$ as
    \begin{equation} \label{eq:Freudenthal}
        (||\lambda+\rho||^2-||\mu+\rho||^2)m_\lambda(\mu)=2\sum_{\alpha \in \Delta^+}\sum_{j\geq 1}(\lambda+j\alpha,\alpha)m_\lambda(\lambda+j\alpha)\,,
    \end{equation}
    with $\Delta^+$ the set of positive roots.

The theory underlying characters of negative level representations is more intricate. In particular, the representation theory is no longer invariant under the action of the affine Weyl group $\affineWeyl$ (though the symmetry under the action of the finite Weyl group $\finiteWeyl$ is preserved if all finite Dynkin labels of the highest weight are non-negative). The Kazhdan-Lusztig conjecture states that for such representations, the Weyl-Kac like relation between the sought after character and Verma module characters is 
\begin{equation}
    \cha L_{\lambda}=\sum_{w\leq w'}m_{w,w'}\,\cha M_{w'\cdot \Lambda}\,.
    \label{eq:klchar}
\end{equation}
The ingredients in this formula are the following: $\Lambda$ is the unique weight in the Weyl orbit $\affineWeyl \cdot \lambda$ under the dotted action
\begin{equation}
    \affineWeyl \cdot \lambda= \{ w(\lambda+\rho)-\rho \, | \, w \in \affineWeyl \} 
\end{equation}
such that $\Lambda+\rho$ is a dominant weight. The affine Weyl group element $w \in \affineWeyl$ satisfies $\lambda=w\cdot \Lambda$. This condition does not fix $w$ uniquely, but the RHS of equation \eqref{eq:klchar} depends on $w$ only modulo $\hat{\cW}_0$, the stabilizer of $\Lambda$ in $\affineWeyl$. The inequality $\le$ defining the sum is with regard to Bruhat ordering (see e.g. \cite{HumphreysCoxeter}). Finally, the coefficients $m_{w,w'}$ which replace the sign in the Weyl-Kac formula \eqref{eq:weyl-kac} are specializations of linear combinations of Kazhdan-Lusztig polynomials. Specifically,
\begin{equation*}
    m_{w,w'}=\left\{\begin{array}{cc}
        \tilde Q_{w,w'}(1) & \text{if } h^\vee_G-n>0 \\
        \tilde P_{w,w'}(1) & \text{if } h^\vee_G-n<0
    \end{array}\right.\,.
\end{equation*}
We will define only the polynomials $\tilde Q$ relevant for our computations at $h_G^\vee-n>0$ (see e.g. \cite{DeVos:1995an} for the definition of the polynomials $\tilde P$). They are given by

\begin{equation*}
    \tilde Q_{w,w'}=\sum_{z \in [w']}(-1)^{l(\bar w)+l(z)} Q_{\bar w,z} \,.
\end{equation*}
Here, $\bar w$ is a representative of maximal length of the class of $w$ in $\affineWeyl/\hat{\cW}_0$, the sum is over all $z$ in the class $[w']\in \affineWeyl/\hat{\cW}_0$ of $w'$, and $Q$ are the inverse Kazhdan-Lusztig polynomials. They can be computed recursively as follows:

\begin{enumerate}
    \item Initial data:
    \begin{align}
        Q_{w,w}&=1\,, \forall w \in \affineWeyl\,,\\ Q_{w_1,w_2}&=0 \,, \text{ if } w_1\nleq w_2\,.
    \end{align}
    \item Recursion step: For any simple reflection $s$
    \begin{align}
       Q_{w_1,w_2 s}&=Q_{w_1,w_2}\,,\quad \text{ if } w_1 s<w_1\,,  \\
       Q_{w_1,w_2 s}&=Q_{w_1 s,w_2}-q Q_{w_1,w_2}+q \sum_{\substack{w_1\leq w <w_2\\ws>w}}\slashed Q_{w_1,w}Q_{w,w_2}\,,\quad\text{ if } w_1 s>w_1\,,
    \end{align}
    where $q$ is the polynomial variable $Q$ depends on,
    and $\slashed Q_{w_1,w}$ is the highest order monomial in $Q_{w_1,w}$.
\end{enumerate}
Once all of this is in place, we can calculate the multiplicities $m_\lambda(\mu)$ of the weight $\mu$ in the highest weight representation $L_\lambda$ at negative level by comparing the coefficients of $e^\mu$ in equation \eqref{eq:klchar}. This yields
\begin{equation}
        m_\lambda(\mu)=\sum_{w\leq w'}m^M_{w'\cdot \Lambda}(\mu)\tilde Q_{w,w'}\,,
        \label{eq:klMult}
    \end{equation}
    where $m^M_{w'\cdot \Lambda}(\mu)$ denotes the multiplicity of $\mu$ in the Verma module of $w'\cdot \Lambda$. Note that the sum is finite because $m_{w'\cdot \Lambda}^M(\mu)$ is non-zero only if $\mu\leq w'\cdot \Lambda$.

    Based on the above, we implemented an algorithm to compute $\cha(L_\lambda)$ for $\lambda \in \dominant(\mathfrak{g})_{-n}$ in SageMath. The computation proceeds in two steps: first, we determine all weights up to a given grade which can appear in $L_\lambda$:
    \begin{enumerate}
        \item We consider the list of all weights of the form
    \begin{equation} \label{eq:weighttree}
        \lambda-k\alpha_i\,, \quad 0 <k \leq \lambda_i\,, \quad i=1,\dots,r
    \end{equation}
    with $\lambda_i$ the Dynkin labels of $\lambda$. For every weight $\omega$ in this list, we generate a new list following \eqref{eq:weighttree} with $\lambda$ replaced by $\omega$. We repeat this process until we find no further new weights. This gives all the weights at grade 0.
        \item We seed the above process at each new grade by subtracting the $0^{th}$ root $\alpha_0$ from all weights at the previous grade.
    \end{enumerate}
    To compute multiplicities, we decompose the weights at each grade into weight orbits of the finite Weyl group $\cW$, as multiplicities are invariant under the action of $\cW$. We choose the highest weight in each orbit. If it is not a null weight, we compute its multiplicity using Freudenthal's formula \eqref{eq:Freudenthal}. This is substantially faster than invoking \eqref{eq:klMult}, to which we resort in the case of null weights.

\chapter{The polynomials $p^\lambda_\omega$} \label{a:polynomials}

In this section, we list the polynomials $p^\lambda_\omega$ defined in equation \eqref{eq:poly_multiplying_omega} for numerous theories. The order to which we compute is encoded in the integers $M,m$ introduced in section \ref{sec:algorithm}. The polynomials are given in terms of $v$ and $y=q^{-\frac{1}{4\kappa}}$.

\section{$(C_r)_1$}
For $(C_r)_1$ the elliptic the functions $\xi$ can be written in terms of theta functions, as in equation \eqref{eq:xi_n=1_theta}. For the trivial and the vector representations, the $\xi$ functions only have $\theta_{3}=1+O(\sqrt{q})$ and $\theta_4=1+O(\sqrt{q})$ factors in the denominator therefore it is not necessary to make an expansion in $v$. We only need to expand in $q$ and we obtain results exact in $v$. We thus just truncate our expressions in $q$ and not in $v$. A conjecture for arbitrary $r$ is given in section \ref{sub:E-string results}. 

We use $0,v$ for the trivial and vector representation of $D_{8+2r}$ to ease notation. We only include the $0,v$ results because for these theories we know that applications $\mathcal{F}_\kappa$ give us the two spin contributions as explained in section \ref{sub:E-string results}.

For $r=3$ we computed the result to order $5+E_0$ in $q$. $\kappa=2$, so $y=q^{-\frac{1}{8}}$.
\begin{center}
     
    \scalebox{1}{$
\begin{array}{c|c}
 \xi _0 & \xi _v \\\hline
\begin{array}{cc}
 \hat{\chi }_{\text{(000)}} & 1 \\
 \hat{\chi }_{\text{(010)}} & 1 \\
 \hat{\chi }_{\text{(200)}} & v^2 y^4+\frac{y^4}{v^2} \\
 \hat{\chi }_{\text{(400)}} & v^4 y^{16}+\frac{y^{16}}{v^4} \\
 \hat{\chi }_{\text{(600)}} & v^6 y^{36}+\frac{y^{36}}{v^6} \\
 \hat{\chi }_{\text{(800)}} & v^8 y^{64}+\frac{y^{64}}{v^8} \\
 \hat{\chi }_{\text{(10,00)}} & v^{10} y^{100}+\frac{y^{100}}{v^{10}} \\
 \hat{\chi }_{\text{(12,00)}} & v^{12} y^{144}+\frac{y^{144}}{v^{12}} \\
\end{array}
& 
\begin{array}{cc}
 \hat{\chi }_{\text{(100)}} & -v y-\frac{y}{v} \\
 \hat{\chi }_{\text{(300)}} & -v^3 y^9-\frac{y^9}{v^3} \\
 \hat{\chi }_{\text{(500)}} & -v^5 y^{25}-\frac{y^{25}}{v^5} \\
 \hat{\chi }_{\text{(700)}} & -v^7 y^{49}-\frac{y^{49}}{v^7} \\
 \hat{\chi }_{\text{(900)}} & -v^9 y^{81}-\frac{y^{81}}{v^9} \\
 \hat{\chi }_{\text{(11,00)}} & -v^{11} y^{121}-\frac{y^{121}}{v^{11}} \\
\end{array}

\end{array}
$
}

\end{center}

For $r=4$ we computed the result to order $5+E_0$ in q. $\kappa=4$, so $y=q^{-\frac{1}{16}}$.
\begin{center}
     
\scalebox{1}{$
\begin{array}{c|c}
 \xi _0 & \xi _v \\\hline
\begin{array}{cc}
 \hat{\chi }_{\text{(0000)}} & 1 \\
 \hat{\chi }_{\text{(0100)}} & 1 \\
 \hat{\chi }_{\text{(2000)}} & v^2 y^4+\frac{y^4}{v^2} \\
 \hat{\chi }_{\text{(4000)}} & v^4 y^{16}+\frac{y^{16}}{v^4} \\
 \hat{\chi }_{\text{(6000)}} & v^6 y^{36}+\frac{y^{36}}{v^6} \\
 \hat{\chi }_{\text{(8000)}} & v^8 y^{64}+\frac{y^{64}}{v^8} \\
 \hat{\chi }_{\text{(10,000)}} & v^{10} y^{100}+\frac{y^{100}}{v^{10}} \\
 \hat{\chi }_{\text{(12,000)}} & v^{12} y^{144}+\frac{y^{144}}{v^{12}} \\
\end{array}
&
\begin{array}{cc}
 \hat{\chi }_{\text{(1000)}} & -v y-\frac{y}{v} \\
 \hat{\chi }_{\text{(3000)}} & -v^3 y^9-\frac{y^9}{v^3} \\
 \hat{\chi }_{\text{(5000)}} & -v^5 y^{25}-\frac{y^{25}}{v^5} \\
 \hat{\chi }_{\text{(7000)}} & -v^7 y^{49}-\frac{y^{49}}{v^7} \\
 \hat{\chi }_{\text{(9000)}} & -v^9 y^{81}-\frac{y^{81}}{v^9} \\
 \hat{\chi }_{\text{(11,000)}} & -v^{11} y^{121}-\frac{y^{121}}{v^{11}} \\
\end{array}
\end{array}
$
}

\end{center}

\section{$(B_3)_3$}
${(B_3)}_3$ was checked to order $M=10,m=13$. 
$\kappa=2$, so $y=q^{-\frac{1}{8}}$.

\begin{center}
     
\scalebox{1}{$
\begin{array}{c|c|c}
 \xi _{\text{(00)}} & \xi _{\text{(01)}} & \xi _{\text{(10)}} \\\hline
 
\begin{array}{cc}
 \hat{\chi }_{\text{(000)}} & \frac{y^4}{v^2}-v^2 y^4 \\
 \hat{\chi }_{\text{(002)}} & v^6 y^{36}-v^2 y^4 \\
 \hat{\chi }_{\text{(010)}} & \frac{y^{16}}{v^4}-1 \\
 \hat{\chi }_{\text{(012)}} & v^8 y^{64} \\
 \hat{\chi }_{\text{(020)}} & \frac{y^{36}}{v^6}-\frac{y^4}{v^2} \\
 \hat{\chi }_{\text{(022)}} & v^{10} y^{100} \\
 \hat{\chi }_{\text{(030)}} & \frac{y^{64}}{v^8} \\
 \hat{\chi }_{\text{(032)}} & v^{12} y^{144} \\
 \hat{\chi }_{\text{(040)}} & \frac{y^{100}}{v^{10}} \\
 \hat{\chi }_{\text{(050)}} & \frac{y^{144}}{v^{12}} \\
 \hat{\chi }_{\text{(060)}} & \frac{y^{196}}{v^{14}} \\
 \hat{\chi }_{\text{(070)}} & \frac{y^{256}}{v^{16}} \\
 \hat{\chi }_{\text{(080)}} & \frac{y^{324}}{v^{18}} \\
 \hat{\chi }_{\text{(090)}} & \frac{y^{400}}{v^{20}} \\
 \hat{\chi }_{\text{(100)}} & v^4 y^{16}-1 \\
\end{array}
 & 
\begin{array}{cc}
 \hat{\chi }_{\text{(000)}} & v^4 y^{16}-1 \\
 \hat{\chi }_{\text{(002)}} & \frac{y^{16}}{v^4}-1 \\
 \hat{\chi }_{\text{(010)}} & v^6 y^{36}-v^2 y^4 \\
 \hat{\chi }_{\text{(012)}} & \frac{y^{36}}{v^6} \\
 \hat{\chi }_{\text{(020)}} & v^8 y^{64} \\
 \hat{\chi }_{\text{(022)}} & \frac{y^{64}}{v^8} \\
 \hat{\chi }_{\text{(030)}} & v^{10} y^{100} \\
 \hat{\chi }_{\text{(032)}} & \frac{y^{100}}{v^{10}} \\
 \hat{\chi }_{\text{(040)}} & v^{12} y^{144} \\
 \hat{\chi }_{\text{(042)}} & \frac{y^{144}}{v^{12}} \\
 \hat{\chi }_{\text{(052)}} & \frac{y^{196}}{v^{14}} \\
 \hat{\chi }_{\text{(062)}} & \frac{y^{256}}{v^{16}} \\
 \hat{\chi }_{\text{(072)}} & \frac{y^{324}}{v^{18}} \\
 \hat{\chi }_{\text{(100)}} & \frac{y^4}{v^2}-v^2 y^4 \\
\end{array}
 & 
\begin{array}{cc}
 \hat{\chi }_{\text{(001)}} & -v^5 y^{25}-\frac{y^9}{v^3}+2 v y \\
 \hat{\chi }_{\text{(011)}} & -v^7 y^{49}-\frac{y^{25}}{v^5}+v^3 y^9+\frac{y}{v} \\
 \hat{\chi }_{\text{(021)}} & -v^9 y^{81}-\frac{y^{49}}{v^7} \\
 \hat{\chi }_{\text{(031)}} & -v^{11} y^{121}-\frac{y^{81}}{v^9} \\
 \hat{\chi }_{\text{(041)}} & -v^{13} y^{169}-\frac{y^{121}}{v^{11}} \\
 \hat{\chi }_{\text{(051)}} & -\frac{y^{169}}{v^{13}} \\
 \hat{\chi }_{\text{(061)}} & -\frac{y^{225}}{v^{15}} \\
 \hat{\chi }_{\text{(071)}} & -\frac{y^{289}}{v^{17}} \\
 \hat{\chi }_{\text{(081)}} & -\frac{y^{361}}{v^{19}} \\
\end{array}
 \\
\end{array}
$
}

\end{center}
We note that $\mathcal F_\kappa(\xi_{(00)})=\xi_{(01)}$ and $\mathcal{F}_\kappa(\xi_{10})=\xi_{(10)}$.

\section{$(F_4)_3$}
${(F_4)}_3$ was checked to order $M=5,m=13$. 
$\kappa=6$, so $y=q^{-\frac{1}{24}}$.

\begin{center}
     
\scalebox{0.7}{$
\begin{array}{c|c|c|c}
 \xi _{\text{(00)}} & \xi _{\text{(01)}} & \xi _{\text{(02)}} & \xi _{\text{(03)}} \\\hline
 
\begin{array}{cc}
 \hat{\chi }_{\text{(0000)}} & \frac{y^4}{v^2} \\
 \hat{\chi }_{\text{(0001)}} & v^2 y^4 \\
 \hat{\chi }_{\text{(0100)}} & \frac{y^4}{v^2}-v^{10} y^{100} \\
 \hat{\chi }_{\text{(1000)}} & \frac{y^{16}}{v^4}-v^8 y^{64} \\
 \hat{\chi }_{\text{(2000)}} & \frac{y^{36}}{v^6}-v^6 y^{36} \\
 \hat{\chi }_{\text{(3000)}} & \frac{y^{64}}{v^8} \\
 \hat{\chi }_{\text{(4000)}} & \frac{y^{100}}{v^{10}} \\
\end{array}
 & 
\begin{array}{cc}
 \hat{\chi }_{\text{(0000)}} & v^6 y^{36} \\
 \hat{\chi }_{\text{(0001)}} & v^6 y^{36} \\
 \hat{\chi }_{\text{(0002)}} & \frac{y^{16}}{v^4}-v^8 y^{64} \\
 \hat{\chi }_{\text{(0020)}} & v^{12} y^{144} \\
 \hat{\chi }_{\text{(1002)}} & \frac{y^{36}}{v^6} \\
 \hat{\chi }_{\text{(1100)}} & \frac{y^{36}}{v^6} \\
 \hat{\chi }_{\text{(2002)}} & \frac{y^{64}}{v^8} \\
\end{array}
 & 
\begin{array}{cc}
 \hat{\chi }_{\text{(0000)}} & 1 \\
 \hat{\chi }_{\text{(0001)}} & 1 \\
 \hat{\chi }_{\text{(0002)}} & v^{10} y^{100}-\frac{y^4}{v^2} \\
 \hat{\chi }_{\text{(0020)}} & \frac{y^{36}}{v^6} \\
 \hat{\chi }_{\text{(1002)}} & v^{12} y^{144} \\
\end{array}
 & 
\begin{array}{cc}
 \hat{\chi }_{\text{(0000)}} & v^8 y^{64} \\
 \hat{\chi }_{\text{(0001)}} & v^4 y^{16} \\
 \hat{\chi }_{\text{(0100)}} & -\frac{y^{16}}{v^4} \\
 \hat{\chi }_{\text{(1000)}} & v^{10} y^{100}-\frac{y^4}{v^2} \\
 \hat{\chi }_{\text{(2000)}} & v^{12} y^{144} \\
\end{array}
 \\\hline
 \xi _{\text{(10)}} & \xi _{\text{(11)}} & \xi _{\text{(12)}} & \xi _{\text{(20)}} \\\hline
 
\begin{array}{cc}
 \hat{\chi }_{\text{(0000)}} & -v^3 y^9 \\
 \hat{\chi }_{\text{(0001)}} & -\frac{y^9}{v^3} \\
 \hat{\chi }_{\text{(0010)}} & v^9 y^{81}-\frac{y^9}{v^3} \\
 \hat{\chi }_{\text{(0012)}} & -v^{11} y^{121} \\
 \hat{\chi }_{\text{(0110)}} & -v^{13} y^{169} \\
 \hat{\chi }_{\text{(1001)}} & v^7 y^{49}-\frac{y^{25}}{v^5} \\
 \hat{\chi }_{\text{(2001)}} & -\frac{y^{49}}{v^7}-\frac{y^{25}}{v^5} \\
 \hat{\chi }_{\text{(3001)}} & -\frac{y^{81}}{v^9} \\
\end{array}
 & 
\begin{array}{cc}
 \hat{\chi }_{\text{(0011)}} & -v^{11} y^{121}+v^7 y^{49}-\frac{y^{25}}{v^5}+\frac{y}{v} \\
 \hat{\chi }_{\text{(1011)}} & -v^{13} y^{169}-\frac{y^{49}}{v^7} \\
\end{array}
 & 
\begin{array}{cc}
 \hat{\chi }_{\text{(0000)}} & -v^3 y^9 \\
 \hat{\chi }_{\text{(0001)}} & -v^9 y^{81} \\
 \hat{\chi }_{\text{(0010)}} & \frac{y^9}{v^3}-v^9 y^{81} \\
 \hat{\chi }_{\text{(1001)}} & \frac{y}{v}-v^{11} y^{121} \\
 \hat{\chi }_{\text{(2001)}} & -v^{13} y^{169} \\
\end{array}
 & 
\begin{array}{cc}
 \hat{\chi }_{\text{(0000)}} & -v^4 y^{16} \\
 \hat{\chi }_{\text{(0001)}} & -v^8 y^{64} \\
 \hat{\chi }_{\text{(0003)}} & v^{10} y^{100}-\frac{y^4}{v^2} \\
 \hat{\chi }_{\text{(0010)}} & \frac{y^{16}}{v^4}-v^8 y^{64} \\
 \hat{\chi }_{\text{(0101)}} & v^{12} y^{144} \\
 \hat{\chi }_{\text{(1002)}} & \frac{y^{16}}{v^4} \\
 \hat{\chi }_{\text{(1010)}} & \frac{y^{36}}{v^6} \\
 \hat{\chi }_{\text{(2010)}} & \frac{y^{64}}{v^8} \\
\end{array}
 \\\hline
 &\xi _{\text{(21)}} & \xi _{\text{(30)}}&\\\hline
 &
\begin{array}{cc}
 \hat{\chi }_{\text{(0000)}} & -v^2 y^4 \\
 \hat{\chi }_{\text{(0001)}} & -\frac{y^4}{v^2} \\
 \hat{\chi }_{\text{(0003)}} & \frac{y^{16}}{v^4} \\
 \hat{\chi }_{\text{(0010)}} & v^{10} y^{100}-\frac{y^4}{v^2} \\
 \hat{\chi }_{\text{(0101)}} & \frac{y^{36}}{v^6} \\
 \hat{\chi }_{\text{(1010)}} & v^{12} y^{144} \\
\end{array}
 & 
\begin{array}{cc}
 \hat{\chi }_{\text{(0000)}} & v^7 y^{49}+\frac{y}{v} \\
 \hat{\chi }_{\text{(0001)}} & v^5 y^{25}+v y \\
 \hat{\chi }_{\text{(0100)}} & -v^{11} y^{121}+v^7 y^{49}-\frac{y^{25}}{v^5}+\frac{y}{v} \\
 \hat{\chi }_{\text{(1002)}} & -v^{11} y^{121}-\frac{y^{25}}{v^5} \\
 \hat{\chi }_{\text{(1100)}} & -v^{13} y^{169}-\frac{y^{49}}{v^7} \\
\end{array}&
 
\end{array}
$
}

\end{center}

\section{$(D_4)_4$}
$(D_4)_4$ was checked to order $M=5,m=8$. A conjecture for the general form is given in \ref{sub:matter-less}. $\kappa=2$, so $y=q^{-\frac{1}{8}}$.
\begin{center}
  
\scalebox{1}{$
\begin{array}{c}
 \xi _0 \\\hline
\begin{array}{cc}
 \hat{\chi }_{\text{(0000)}} & v^5 y^{25}+2 v^3 y^9-\frac{y^9}{v^3}-\frac{2 y}{v} \\
 \hat{\chi }_{\text{(0100)}} & v^7 y^{49}+2 v^5 y^{25}-\frac{y^{25}}{v^5}+v^3 y^9-\frac{2 y^9}{v^3}-\frac{y}{v} \\
 \hat{\chi }_{\text{(0200)}} & v^9 y^{81}+2 v^7 y^{49}-\frac{y^{49}}{v^7}+v^5 y^{25}-\frac{2 y^{25}}{v^5} \\
 \hat{\chi }_{\text{(0300)}} & 2 v^9 y^{81}-\frac{y^{81}}{v^9} \\
 \hat{\chi }_{\text{(0400)}} & -\frac{y^{121}}{v^{11}} \\
 \hat{\chi }_{\text{(0500)}} & -\frac{y^{169}}{v^{13}} \\
 \hat{\chi }_{\text{(0600)}} & -\frac{y^{225}}{v^{15}} \\
\end{array}
\end{array}$
}

\end{center}

We note $\mathcal F_\kappa(\xi_0)=-\xi_0$.

\section{${(B_4)}_4$}
${(B_4)}_4$ was checked to order $M=10,m=20$. A conjecture for the general form is given in section \ref{subsec:E7_7}.
$\kappa=3$, so $y=q^{-\frac{1}{12}}$.

\begin{center}
    
\scalebox{1}{$
\begin{array}{c|c}
 \xi _{\text{(0)}} & \xi _{\text{(1)}} \\\hline
 
\begin{array}{cc}
 \hat{\chi }_{\text{(0000)}} & \frac{y^9}{v^3}-v^3 y^9 \\
 \hat{\chi }_{\text{(0100)}} & -v^5 y^{25}+\frac{y^{25}}{v^5}-v y+\frac{y}{v} \\
 \hat{\chi }_{\text{(0200)}} & -v^7 y^{49}+\frac{y^{49}}{v^7}+v y-\frac{y}{v} \\
 \hat{\chi }_{\text{(0300)}} & \frac{y^{81}}{v^9}-v^9 y^{81} \\
 \hat{\chi }_{\text{(0400)}} & \frac{y^{121}}{v^{11}} \\
 \hat{\chi }_{\text{(0500)}} & \frac{y^{169}}{v^{13}} \\
 \hat{\chi }_{\text{(0600)}} & \frac{y^{225}}{v^{15}} \\
 \hat{\chi }_{\text{(0700)}} & \frac{y^{289}}{v^{17}} \\
 \hat{\chi }_{\text{(0800)}} & \frac{y^{361}}{v^{19}} \\
 \hat{\chi }_{\text{(1000)}} & v^7 y^{49}+v^5 y^{25}-v y-\frac{y}{v} \\
 \hat{\chi }_{\text{(1100)}} & v^9 y^{81}-\frac{y^9}{v^3} \\
 \hat{\chi }_{\text{(1200)}} & v^{11} y^{121}-v^5 y^{25}-\frac{y^{25}}{v^5}+v y \\
 \hat{\chi }_{\text{(1300)}} & v^{13} y^{169}-\frac{y^{49}}{v^7} \\
 \hat{\chi }_{\text{(1400)}} & v^{15} y^{225} \\
 \hat{\chi }_{\text{(1500)}} & v^{17} y^{289} \\
 \hat{\chi }_{\text{(1600)}} & v^{19} y^{361} \\
 \hat{\chi }_{\text{(1700)}} & v^{21} y^{441} \\
 \hat{\chi }_{\text{(2000)}} & \frac{y^9}{v^3}-v^3 y^9 \\
\end{array}
 & 
\begin{array}{cc}
 \hat{\chi }_{\text{(0000)}} & 1-v^6 y^{36} \\
 \hat{\chi }_{\text{(0100)}} & -v^8 y^{64}-v^4 y^{16}+v^2 y^4+\frac{y^4}{v^2} \\
 \hat{\chi }_{\text{(0200)}} & -v^{10} y^{100}+v^4 y^{16}+\frac{y^{16}}{v^4}-v^2 y^4 \\
 \hat{\chi }_{\text{(0300)}} & \frac{y^{36}}{v^6}-v^{12} y^{144} \\
 \hat{\chi }_{\text{(0400)}} & -v^{14} y^{196} \\
 \hat{\chi }_{\text{(0500)}} & -v^{16} y^{256} \\
 \hat{\chi }_{\text{(0600)}} & -v^{18} y^{324} \\
 \hat{\chi }_{\text{(0700)}} & -v^{20} y^{400} \\
 \hat{\chi }_{\text{(1000)}} & v^4 y^{16}-\frac{y^{16}}{v^4}+v^2 y^4-\frac{y^4}{v^2} \\
 \hat{\chi }_{\text{(1100)}} & v^6 y^{36}-\frac{y^{36}}{v^6} \\
 \hat{\chi }_{\text{(1200)}} & v^8 y^{64}-\frac{y^{64}}{v^8}+\frac{y^4}{v^2} \\
 \hat{\chi }_{\text{(1300)}} & -\frac{y^{100}}{v^{10}} \\
 \hat{\chi }_{\text{(1400)}} & -\frac{y^{144}}{v^{12}} \\
 \hat{\chi }_{\text{(1500)}} & -\frac{y^{196}}{v^{14}} \\
 \hat{\chi }_{\text{(1600)}} & -\frac{y^{256}}{v^{16}} \\
 \hat{\chi }_{\text{(1700)}} & -\frac{y^{324}}{v^{18}} \\
 \hat{\chi }_{\text{(2000)}} & 1-v^6 y^{36} \\
\end{array}
 \\
\end{array}$
}

\end{center}

We note that $\mathcal F_\kappa(\xi_{(0)})=-\xi_{(1)}$.

\section{${(D_5)}_4$}
${(D_5)}_4$ was checked to order $M=6,m=13$. 
$\kappa=4$, so $y=q^{-\frac{1}{16}}$.

\begin{center}
     
\scalebox{0.8}{$
\begin{array}{c|c|c}
 \xi _{\text{(00)}} & \xi _{\text{(01)}} & \xi _{\text{(10)}} \\\hline
 
\begin{array}{cc}
 \hat{\chi }_{\text{(00000)}} & v^5 y^{25}-2 v^3 y^9+\frac{y^9}{v^3} \\
 \hat{\chi }_{\text{(01000)}} & -2 v^5 y^{25}+\frac{y^{25}}{v^5}+v^3 y^9 \\
 \hat{\chi }_{\text{(02000)}} & \frac{y^{49}}{v^7}-v^9 y^{81} \\
 \hat{\chi }_{\text{(03000)}} & \frac{y^{81}}{v^9} \\
 \hat{\chi }_{\text{(04000)}} & \frac{y^{121}}{v^{11}} \\
 \hat{\chi }_{\text{(20000)}} & -v^9 y^{81}-v^7 y^{49}+v y+\frac{y}{v} \\
 \hat{\chi }_{\text{(21000)}} & \frac{y^{25}}{v^5}-v^{11} y^{121} \\
 \hat{\chi }_{\text{(22000)}} & -v^{13} y^{169} \\
\end{array}
 & 
\begin{array}{cc}
 \hat{\chi }_{\text{(00000)}} & -v^7 y^{49}+2 v y-\frac{y}{v} \\
 \hat{\chi }_{\text{(01000)}} & -v^9 y^{81}-v y+\frac{2 y}{v} \\
 \hat{\chi }_{\text{(02000)}} & \frac{y^{25}}{v^5}-v^{11} y^{121} \\
 \hat{\chi }_{\text{(03000)}} & -v^{13} y^{169} \\
 \hat{\chi }_{\text{(20000)}} & -v^5 y^{25}+\frac{y^{25}}{v^5}-v^3 y^9+\frac{y^9}{v^3} \\
 \hat{\chi }_{\text{(21000)}} & \frac{y^{49}}{v^7} \\
 \hat{\chi }_{\text{(22000)}} & \frac{y^{81}}{v^9} \\
\end{array}
 & 
\begin{array}{cc}
 \hat{\chi }_{\text{(10000)}} & v^8 y^{64}+v^4 y^{16}-\frac{y^{16}}{v^4}-1 \\
 \hat{\chi }_{\text{(11000)}} & v^{10} y^{100}+2 v^6 y^{36}-\frac{y^{36}}{v^6}-\frac{2 y^4}{v^2} \\
 \hat{\chi }_{\text{(12000)}} & v^{12} y^{144}-\frac{y^{64}}{v^8} \\
 \hat{\chi }_{\text{(13000)}} & v^{14} y^{196}-\frac{y^{100}}{v^{10}} \\
 \hat{\chi }_{\text{(30000)}} & v^8 y^{64}-\frac{y^{16}}{v^4} \\
\end{array}

\end{array}
$
}

\end{center}

We note that $\mathcal F_\kappa(\xi_{(00)})=-\xi_{(01)}$ and $\mathcal{F}_\kappa(\xi_{(10)})=\xi_{(10)}$.

\section{${(F_4)}_4$}
${(F_4)}_4$ was checked to order $M=9,m=19$. 
$\kappa=5$, so $y=q^{-\frac{1}{20}}$.

\begin{center}
     
\scalebox{0.55}{$
\begin{array}{c|c|c|c}
 \xi _0 & \xi _1 & \xi _2 & \xi _3 \\\hline
 
\begin{array}{cc}
 \hat{\chi }_{\text{(0000)}} & \frac{y^9}{v^3}-v^7 y^{49} \\
 \hat{\chi }_{\text{(0002)}} & v^9 y^{81}-\frac{y}{v} \\
 \hat{\chi }_{\text{(0004)}} & \frac{y^9}{v^3}-v^7 y^{49} \\
 \hat{\chi }_{\text{(0100)}} & v^{11} y^{121}+v^9 y^{81}-v y-\frac{y}{v} \\
 \hat{\chi }_{\text{(1000)}} & \frac{y^{25}}{v^5}-v^5 y^{25} \\
 \hat{\chi }_{\text{(1100)}} & v^{13} y^{169}+\frac{y^{49}}{v^7}-2 v^3 y^9 \\
 \hat{\chi }_{\text{(2000)}} & -v^7 y^{49}+\frac{y^{49}}{v^7}-v^3 y^9+\frac{y^9}{v^3} \\
 \hat{\chi }_{\text{(2100)}} & v^{15} y^{225} \\
 \hat{\chi }_{\text{(3000)}} & v^{11} y^{121}+\frac{y^{81}}{v^9}-2 v y \\
 \hat{\chi }_{\text{(3100)}} & v^{17} y^{289} \\
 \hat{\chi }_{\text{(4000)}} & \frac{y^{121}}{v^{11}}+\frac{y^{81}}{v^9} \\
 \hat{\chi }_{\text{(4100)}} & v^{19} y^{361} \\
 \hat{\chi }_{\text{(5000)}} & v^{17} y^{289}+\frac{y^{169}}{v^{13}} \\
 \hat{\chi }_{\text{(6000)}} & \frac{y^{225}}{v^{15}} \\
 \hat{\chi }_{\text{(7000)}} & \frac{y^{289}}{v^{17}} \\
\end{array}
 & 
\begin{array}{cc}
 \hat{\chi }_{\text{(0001)}} & v^6 y^{36}-\frac{y^{16}}{v^4} \\
 \hat{\chi }_{\text{(0002)}} & \frac{y^4}{v^2}-v^8 y^{64} \\
 \hat{\chi }_{\text{(0003)}} & v^6 y^{36}-\frac{y^{16}}{v^4} \\
 \hat{\chi }_{\text{(0010)}} & 1-v^{10} y^{100} \\
 \hat{\chi }_{\text{(1001)}} & v^6 y^{36}-\frac{y^{36}}{v^6}+v^4 y^{16}-\frac{y^{16}}{v^4} \\
 \hat{\chi }_{\text{(1010)}} & -v^{12} y^{144}-v^8 y^{64}+v^2 y^4+\frac{y^4}{v^2} \\
 \hat{\chi }_{\text{(2001)}} & -v^{12} y^{144}-\frac{y^{64}}{v^8}+2 v^2 y^4 \\
 \hat{\chi }_{\text{(2010)}} & -v^{14} y^{196}-\frac{y^{36}}{v^6} \\
 \hat{\chi }_{\text{(3001)}} & -\frac{y^{100}}{v^{10}} \\
 \hat{\chi }_{\text{(3010)}} & -v^{16} y^{256}-v^{14} y^{196} \\
 \hat{\chi }_{\text{(4001)}} & -\frac{y^{144}}{v^{12}} \\
 \hat{\chi }_{\text{(4010)}} & -v^{18} y^{324}-\frac{y^{144}}{v^{12}} \\
 \hat{\chi }_{\text{(5001)}} & -\frac{y^{196}}{v^{14}} \\
 \hat{\chi }_{\text{(5010)}} & -v^{20} y^{400} \\
 \hat{\chi }_{\text{(6001)}} & -\frac{y^{256}}{v^{16}} \\
\end{array}
 & 
\begin{array}{cc}
 \hat{\chi }_{\text{(0001)}} & v^9 y^{81}-\frac{y}{v} \\
 \hat{\chi }_{\text{(0002)}} & \frac{y^9}{v^3}-v^7 y^{49} \\
 \hat{\chi }_{\text{(0003)}} & v^9 y^{81}-\frac{y}{v} \\
 \hat{\chi }_{\text{(0010)}} & \frac{y^{25}}{v^5}-v^5 y^{25} \\
 \hat{\chi }_{\text{(1001)}} & v^{11} y^{121}+v^9 y^{81}-v y-\frac{y}{v} \\
 \hat{\chi }_{\text{(1010)}} & -v^7 y^{49}+\frac{y^{49}}{v^7}-v^3 y^9+\frac{y^9}{v^3} \\
 \hat{\chi }_{\text{(2001)}} & v^{13} y^{169}+\frac{y^{49}}{v^7}-2 v^3 y^9 \\
 \hat{\chi }_{\text{(2010)}} & v^{11} y^{121}+\frac{y^{81}}{v^9} \\
 \hat{\chi }_{\text{(3001)}} & v^{15} y^{225} \\
 \hat{\chi }_{\text{(3010)}} & \frac{y^{121}}{v^{11}}+\frac{y^{81}}{v^9} \\
 \hat{\chi }_{\text{(4001)}} & v^{17} y^{289} \\
 \hat{\chi }_{\text{(4010)}} & \frac{y^{169}}{v^{13}} \\
 \hat{\chi }_{\text{(5001)}} & v^{19} y^{361} \\
 \hat{\chi }_{\text{(5010)}} & \frac{y^{225}}{v^{15}} \\
\end{array}
 & 
\begin{array}{cc}
 \hat{\chi }_{\text{(0000)}} & \frac{y^4}{v^2}-v^8 y^{64} \\
 \hat{\chi }_{\text{(0002)}} & v^6 y^{36}-\frac{y^{16}}{v^4} \\
 \hat{\chi }_{\text{(0004)}} & \frac{y^4}{v^2}-v^8 y^{64} \\
 \hat{\chi }_{\text{(0100)}} & v^6 y^{36}-\frac{y^{36}}{v^6}+v^4 y^{16}-\frac{y^{16}}{v^4} \\
 \hat{\chi }_{\text{(1000)}} & 1-v^{10} y^{100} \\
 \hat{\chi }_{\text{(1100)}} & -v^{12} y^{144}-\frac{y^{64}}{v^8}+2 v^2 y^4 \\
 \hat{\chi }_{\text{(2000)}} & -v^{12} y^{144}-v^8 y^{64}+v^2 y^4+\frac{y^4}{v^2} \\
 \hat{\chi }_{\text{(2100)}} & -\frac{y^{100}}{v^{10}} \\
 \hat{\chi }_{\text{(3000)}} & -v^{14} y^{196}-\frac{y^{36}}{v^6} \\
 \hat{\chi }_{\text{(3100)}} & -\frac{y^{144}}{v^{12}} \\
 \hat{\chi }_{\text{(4000)}} & -v^{16} y^{256}-v^{14} y^{196} \\
 \hat{\chi }_{\text{(4100)}} & -\frac{y^{196}}{v^{14}} \\
 \hat{\chi }_{\text{(5000)}} & -v^{18} y^{324}-\frac{y^{144}}{v^{12}} \\
 \hat{\chi }_{\text{(6000)}} & -v^{20} y^{400} \\
\end{array}
 \\
\end{array}
$
}

\end{center}

We note that $\mathcal F_\kappa(\xi_0)=\xi_3$ and $\mathcal{F}_\kappa(\xi_1)=\xi_2$.

\section{$(F_4)_5$}
$(F_4)_5$ was checked to order $M=10,m=20$. A conjecture for the general form is given in \ref{sub:matter-less}. $\kappa=4$, so $y=q^{-\frac{1}{16}}$.
\begin{center}
     
\scalebox{1}{$
\begin{array}{c}
 \xi _0 \\\hline
\begin{array}{cc}
 \hat{\chi }_{\text{(0000)}} & v^8 y^{64}-v^4 y^{16}+\frac{y^{16}}{v^4}-1 \\
 \hat{\chi }_{\text{(1000)}} & v^{10} y^{100}+\frac{y^{36}}{v^6}-2 v^2 y^4 \\
 \hat{\chi }_{\text{(2000)}} & v^{12} y^{144}+\frac{y^{64}}{v^8}-v^4 y^{16}-1 \\
 \hat{\chi }_{\text{(3000)}} & v^{14} y^{196}+\frac{y^{100}}{v^{10}} \\
 \hat{\chi }_{\text{(4000)}} & v^{16} y^{256}+\frac{y^{144}}{v^{12}} \\
 \hat{\chi }_{\text{(5000)}} & v^{18} y^{324}+\frac{y^{196}}{v^{14}} \\
 \hat{\chi }_{\text{(6000)}} & v^{20} y^{400}+\frac{y^{256}}{v^{16}} \\
 \hat{\chi }_{\text{(7000)}} & v^{22} y^{484}+\frac{y^{324}}{v^{18}} \\
\end{array}
\end{array}
$
}

\end{center}

We note $\mathcal F_\kappa(\xi_0)=\xi_0$.

\section{$(E_6)_5$} \label{aa:polynomialsE65}

$(E_6)_5$ was checked to order $M=7,m=16$. $\kappa=7$, so $y=q^{-\frac{1}{28}}$.

\begin{center}
     
\scalebox{0.65}{$
\begin{array}{c|c|c|c}
 \xi _0 & \xi _{-1}+\xi_1 & \xi_{-2}+\xi _2 & \xi _3 \\\hline
 
\begin{array}{cc}
 \hat{\chi }_{\text{(000000)}} & v^{10} y^{100}-2 v^4 y^{16}+\frac{y^{16}}{v^4} \\
 \hat{\chi }_{\text{(000001)}} & v^8 y^{64}-2 v^6 y^{36}+\frac{y^{36}}{v^6} \\
 \hat{\chi }_{\text{(000002)}} & -2 v^8 y^{64}+\frac{y^{64}}{v^8}+v^6 y^{36} \\
 \hat{\chi }_{\text{(000003)}} & \frac{y^{100}}{v^{10}} \\
 \hat{\chi }_{\text{(000004)}} & \frac{y^{144}}{v^{12}} \\
 \hat{\chi }_{\text{(001000)}} & 1-v^{14} y^{196} \\
 \hat{\chi }_{\text{(001001)}} & -v^{16} y^{256} \\
 \hat{\chi }_{\text{(001002)}} & -v^{18} y^{324} \\
 \hat{\chi }_{\text{(100010)}} & -v^{12} y^{144}+2 v^2 y^4-\frac{y^4}{v^2} \\
\end{array}
 & 
\begin{array}{cc}
 \hat{\chi }_{\text{(000010)}} & -v^9 y^{81}+2 v^5 y^{25}-\frac{y^{25}}{v^5} \\
 \hat{\chi }_{\text{(000011)}} & v^7 y^{49}-\frac{y^{49}}{v^7} \\
 \hat{\chi }_{\text{(000012)}} & -\frac{y^{81}}{v^9} \\
 \hat{\chi }_{\text{(000013)}} & -\frac{y^{121}}{v^{11}} \\
 \hat{\chi }_{\text{(000020)}} & v^{11} y^{121}-2 v^3 y^9+\frac{y^9}{v^3} \\
 \hat{\chi }_{\text{(000100)}} & v^{13} y^{169}-2 v y+\frac{y}{v} \\
 \hat{\chi }_{\text{(000101)}} & v^{15} y^{225} \\
 \hat{\chi }_{\text{(000102)}} & v^{17} y^{289} \\
 \hat{\chi }_{\text{(010000)}} & v^{13} y^{169}-2 v y+\frac{y}{v} \\
 \hat{\chi }_{\text{(010001)}} & v^{15} y^{225} \\
 \hat{\chi }_{\text{(010002)}} & v^{17} y^{289} \\
 \hat{\chi }_{\text{(100000)}} & -v^9 y^{81}+2 v^5 y^{25}-\frac{y^{25}}{v^5} \\
 \hat{\chi }_{\text{(100001)}} & v^7 y^{49}-\frac{y^{49}}{v^7} \\
 \hat{\chi }_{\text{(100002)}} & -\frac{y^{81}}{v^9} \\
 \hat{\chi }_{\text{(100003)}} & -\frac{y^{121}}{v^{11}} \\
 \hat{\chi }_{\text{(200000)}} & v^{11} y^{121}-2 v^3 y^9+\frac{y^9}{v^3} \\
\end{array}
 & 
\begin{array}{cc}
 \hat{\chi }_{\text{(000010)}} & -v^{12} y^{144}+2 v^2 y^4-\frac{y^4}{v^2} \\
 \hat{\chi }_{\text{(000011)}} & 1-v^{14} y^{196} \\
 \hat{\chi }_{\text{(000012)}} & -v^{16} y^{256} \\
 \hat{\chi }_{\text{(000013)}} & -v^{18} y^{324} \\
 \hat{\chi }_{\text{(000020)}} & v^{10} y^{100}-2 v^4 y^{16}+\frac{y^{16}}{v^4} \\
 \hat{\chi }_{\text{(000100)}} & v^8 y^{64}-2 v^6 y^{36}+\frac{y^{36}}{v^6} \\
 \hat{\chi }_{\text{(000101)}} & \frac{y^{64}}{v^8} \\
 \hat{\chi }_{\text{(000102)}} & \frac{y^{100}}{v^{10}} \\
 \hat{\chi }_{\text{(010000)}} & v^8 y^{64}-2 v^6 y^{36}+\frac{y^{36}}{v^6} \\
 \hat{\chi }_{\text{(010001)}} & \frac{y^{64}}{v^8} \\
 \hat{\chi }_{\text{(010002)}} & \frac{y^{100}}{v^{10}} \\
 \hat{\chi }_{\text{(100000)}} & -v^{12} y^{144}+2 v^2 y^4-\frac{y^4}{v^2} \\
 \hat{\chi }_{\text{(100001)}} & 1-v^{14} y^{196} \\
 \hat{\chi }_{\text{(100002)}} & -v^{16} y^{256} \\
 \hat{\chi }_{\text{(100003)}} & -v^{18} y^{324} \\
 \hat{\chi }_{\text{(200000)}} & v^{10} y^{100}-2 v^4 y^{16}+\frac{y^{16}}{v^4} \\
\end{array}
 & 
\begin{array}{cc}
 \hat{\chi }_{\text{(000000)}} & v^{11} y^{121}-2 v^3 y^9+\frac{y^9}{v^3} \\
 \hat{\chi }_{\text{(000001)}} & v^{13} y^{169}-2 v y+\frac{y}{v} \\
 \hat{\chi }_{\text{(000002)}} & v^{15} y^{225} \\
 \hat{\chi }_{\text{(000003)}} & v^{17} y^{289} \\
 \hat{\chi }_{\text{(001000)}} & -\frac{y^{49}}{v^7} \\
 \hat{\chi }_{\text{(001001)}} & -\frac{y^{81}}{v^9} \\
 \hat{\chi }_{\text{(100010)}} & -v^9 y^{81}+2 v^5 y^{25}-\frac{y^{25}}{v^5} \\
\end{array}
 \\
\end{array}
$
}

\end{center}

We note that $\mathcal F_\kappa(\xi_0)=\xi_3$ and $\mathcal{F}_\kappa(\xi_1+\xi_{-1})=\xi_{2}+\xi_{-2}$.

As announced in the text, $(E_6)_5$ is a theory for which the Dynkin symmetry akin ambiguity of the $U(1)$ flavor symmetry occurs. The flavor group here is $U(1)_6$. It has 6 integrable representations labeled by $l=-2,-1,0,1,2,3$. We cannot solve for $\xi_{\pm 1},\,\xi_{\pm 2}$ individually due to the $m\mapsto-m$ symmetry relating the characters $\hat \chi^{U(1)}_{\pm l}$. Above, we have displayed the Dynkin symmetric solution of equation \eqref{eq:tosolve} with regard to $E_6$, as the $(E_6)_5$ theory descends from a $E7$ theory with gauge group $E_7$.

\section{$(E_6)_6$}
$(E_6)_6$ was checked to order $M=9,m=18$. A conjecture for the general form is given in \ref{sub:matter-less}. $\kappa=6$, so $y=q^{-\frac{1}{24}}$.
\begin{center}

\scalebox{1}{$
\begin{array}{c}
 \xi _0 \\\hline
\begin{array}{cc}
 \hat{\chi }_{\text{(000000)}} & -v^{11} y^{121}+v^7 y^{49}-2 v^5 y^{25}+\frac{y^{25}}{v^5}+2 v y-\frac{y}{v} \\
 \hat{\chi }_{\text{(000001)}} & -v^{13} y^{169}-2 v^7 y^{49}+\frac{y^{49}}{v^7}+v^5 y^{25}-v y+\frac{2 y}{v} \\
 \hat{\chi }_{\text{(000002)}} & -v^{15} y^{225}-2 v^9 y^{81}+\frac{y^{81}}{v^9}+\frac{2 y^9}{v^3} \\
 \hat{\chi }_{\text{(000003)}} & \frac{y^{121}}{v^{11}}-v^{17} y^{289} \\
 \hat{\chi }_{\text{(000004)}} & \frac{y^{169}}{v^{13}}-v^{19} y^{361} \\
 \hat{\chi }_{\text{(000005)}} & \frac{y^{225}}{v^{15}}-v^{21} y^{441} \\
\end{array}
\end{array}$
}

\end{center}
We note that $\mathcal F_\kappa(\xi_0)=-\xi_0$.

\section{$(E_7)_7$}
$(E_7)_7$ was checked to order $M=9,m=20$. A conjecture for the general form and the action of $\mathcal F_\kappa$ are given in section \ref{subsec:E7_7}. $\kappa=11$, so $y=q^{-\frac{1}{44}}$.
\begin{center}
     
\scalebox{1}{$
\begin{array}{c}
 \xi _0 \\\hline
\begin{array}{cc}
 \hat{\chi }_{\text{(0000000)}} & v^{16} y^{256}-2 v^6 y^{36}+\frac{y^{36}}{v^6} \\
 \hat{\chi }_{\text{(0000001)}} & v^{19} y^{361}-2 v^3 y^9+\frac{y^9}{v^3} \\
 \hat{\chi }_{\text{(0000010)}} & -v^{15} y^{225}+2 v^7 y^{49}-\frac{y^{49}}{v^7} \\
 \hat{\chi }_{\text{(0000100)}} & -v^{18} y^{324}+2 v^4 y^{16}-\frac{y^{16}}{v^4} \\
 \hat{\chi }_{\text{(0001000)}} & v^{17} y^{289}-2 v^5 y^{25}+\frac{y^{25}}{v^5} \\
 \hat{\chi }_{\text{(0100000)}} & -v^{20} y^{400}+2 v^2 y^4-\frac{y^4}{v^2} \\
 \hat{\chi }_{\text{(1000000)}} & v^{14} y^{196}-2 v^8 y^{64}+\frac{y^{64}}{v^8} \\
 \hat{\chi }_{\text{(1000001)}} & v^{21} y^{441} \\
 \hat{\chi }_{\text{(1000010)}} & -v^{13} y^{169}+2 v^9 y^{81}-\frac{y^{81}}{v^9} \\
 \hat{\chi }_{\text{(1100000)}} & -v^{22} y^{484} \\
 \hat{\chi }_{\text{(2000000)}} & v^{12} y^{144}-2 v^{10} y^{100}+\frac{y^{100}}{v^{10}} \\
 \hat{\chi }_{\text{(2000001)}} & v^{23} y^{529} \\
 \hat{\chi }_{\text{(2000010)}} & -\frac{y^{121}}{v^{11}} \\
 \hat{\chi }_{\text{(2100000)}} & -v^{24} y^{576} \\
 \hat{\chi }_{\text{(3000000)}} & \frac{y^{144}}{v^{12}} \\
 \hat{\chi }_{\text{(3000010)}} & -\frac{y^{169}}{v^{13}} \\
 \hat{\chi }_{\text{(4000000)}} & \frac{y^{196}}{v^{14}} \\
\end{array} 
\end{array}$
}

\end{center}

\section{$(E_7)_8$}
$(E_7)_8$ was checked to order $M=10,m=18$. A conjecture for the general form is given in \ref{sub:matter-less}. $\kappa=10$, so $y=q^{-\frac{1}{40}}$.
\begin{center}
     
\scalebox{1}{$
\begin{array}{c}
 \xi _0 \\\hline

\begin{array}{cc}
 \hat{\chi }_{\text{(0000000)}} & -v^{17} y^{289}+v^{13} y^{169}-2 v^7 y^{49}+\frac{y^{49}}{v^7}+2 v^3 y^9-\frac{y^9}{v^3} \\
 \hat{\chi }_{\text{(1000000)}} & -v^{19} y^{361}+v^{11} y^{121}-2 v^9 y^{81}+\frac{y^{81}}{v^9}+2 v y-\frac{y}{v} \\
 \hat{\chi }_{\text{(2000000)}} & \frac{y^{121}}{v^{11}}-v^{21} y^{441} \\
 \hat{\chi }_{\text{(3000000)}} & \frac{y^{169}}{v^{13}}-v^{23} y^{529} \\
 \hat{\chi }_{\text{(4000000)}} & \frac{y^{225}}{v^{15}} \\
\end{array}

\end{array}
$
}

\end{center}

We note that $\mathcal F_\kappa(\xi_0)=-\xi_0$.

\chapter{Higgsing trees} \label{a:higgsing_trees2}
\label{sec:HiggsingTrees}

We can organize the theories considemdtRed in this paper in trees where each line links theories associated by Higgsing. This gives rise to the Higgsing trees we present below, one for each base Hirzebruch surface $\mathbb F_n$ with $n=1,\dots,8,12$. For each theory, we give the gauge group and the flavor group with the level of the corresponding current. We write the rank of the gauge groups beside the class and not as a sub index to lighten notation; for instance $C3_{-1}$ means the gauge group is $C_3$ and the corresponding current is at level $-1$.

Each theory is also assigned a color: In {\color{mdtLime}lime }, we give the theories for which the constants $c$ in the affine ansatz \eqref{eq:affineAnsatzIntro} were computed in \cite{DelZotto:2018tcj}. In {\color{mdtGreen} green}, we give the theories for which we have computed them or have a conjectural form. In {\color{mdtBlue} blue}, we give the theories for which our methods should give complete answers. In {\color{mdtCyan} cyan}, we give the theories for which our methods can be used but the results obtained would still have an ambiguity due to the Dynkin symmetry as explained in section \ref{ss:strategy}. Finally, in {\color{mdtRed} red}, we give the theories for which $h^\vee_G-n\leq 0 $  so the ansatz \eqref{eq:affineAnsatzIntro} cannot be used \footnote{There are just a handful of theories for which this happens and alternative expressions for their elliptic genera where given in \cite{DelZotto:2018tcj}.}, theories for which we do not know the flavor group, or theories for which we do not expect Dynkin symmetry. The structure of the trees is reproduced form \cite{DelZotto:2018tcj}. 

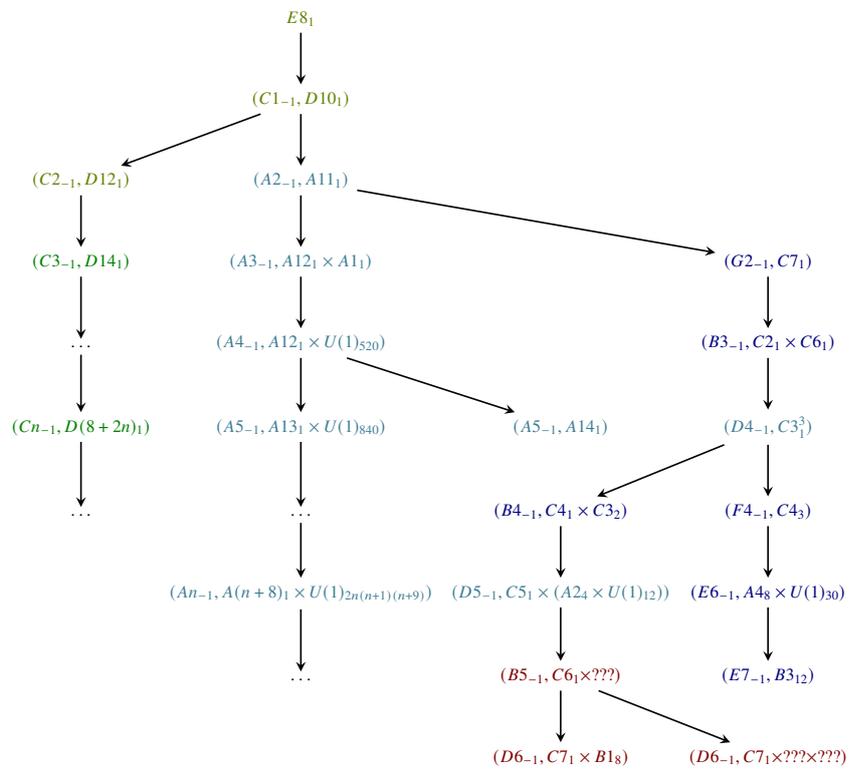
\begin{figure}
    \centering
\[\begin{tikzcd}[scale cd =0.6,column sep = 0.2 ex]
	& \textcolor{mdtLime}{E8_1} \\
	& \textcolor{mdtLime}{(C1_{-1},D10_{1})} \\
	\textcolor{mdtLime}{(C2_{-1},D12_1)} & \textcolor{mdtCyan}{(A2_{-1},A11_1)} \\
	\textcolor{mdtGreen}{(C3_{-1},D14_1)} & \textcolor{mdtCyan}{(A3_{-1},A12_1\times A1_1)} && \textcolor{mdtBlue}{(G2_{-1},C7_1)} \\
	\dots & \textcolor{mdtCyan}{(A4_{-1},A12_1\times U(1)_{520})} && \textcolor{mdtBlue}{(B3_{-1},C2_1\times C6_1)} \\
	\textcolor{mdtGreen}{(Cn_{-1},D(8+2n)_1)} & \textcolor{mdtCyan}{(A5_{-1},A13_1\times U(1)_{840})} & \textcolor{mdtCyan}{(A5_{-1},A14_1)} & \textcolor{mdtCyan}{(D4_{-1},C3_1^3)} \\
	\dots & \dots & \textcolor{mdtBlue}{(B4_{-1},C4_1\times C3_2)} & \textcolor{mdtBlue}{(F4_{-1},C4_3)} \\
	& \textcolor{mdtCyan}{(An_{-1},A(n+8)_1\times U(1)_{2n(n+1)(n+9)})} & \textcolor{mdtCyan}{(D5_{-1},C5_1\times(A2_4\times U(1)_{12}))} & \textcolor{mdtBlue}{(E6_{-1},A4_8\times U(1)_{30})} \\
	& \dots & \textcolor{mdtRed}{(B5_{-1},C6_1\times???)} & \textcolor{mdtBlue}{(E7_{-1},B3_{12})} \\
	&& \textcolor{mdtRed}{(D6_{-1},C7_1\times B1_8)} & \textcolor{mdtRed}{(D6_{-1},C7_1\times???\times???)}
	\arrow[from=1-2, to=2-2]
	\arrow[from=2-2, to=3-1]
	\arrow[from=3-1, to=4-1]
	\arrow[from=4-1, to=5-1]
	\arrow[from=2-2, to=3-2]
	\arrow[from=3-2, to=4-2]
	\arrow[from=4-2, to=5-2]
	\arrow[from=5-1, to=6-1]
	\arrow[from=6-1, to=7-1]
	\arrow[from=5-2, to=6-2]
	\arrow[from=5-2, to=6-3]
	\arrow[from=6-2, to=7-2]
	\arrow[from=7-2, to=8-2]
	\arrow[from=8-2, to=9-2]
	\arrow[from=3-2, to=4-4]
	\arrow[from=4-4, to=5-4]
	\arrow[from=5-4, to=6-4]
	\arrow[from=6-4, to=7-4]
	\arrow[from=6-4, to=7-3]
	\arrow[from=7-3, to=8-3]
	\arrow[from=8-3, to=9-3]
	\arrow[from=9-3, to=10-3]
	\arrow[from=9-3, to=10-4]
	\arrow[from=7-4, to=8-4]
	\arrow[from=8-4, to=9-4]
\end{tikzcd}\]
    \caption{Higgsing tree of $\mathbb F_1$, the E-string. The root is the E-string with its $E_8$ flavor group.}
    \label{fig:F1higgsingtree}
\end{figure}

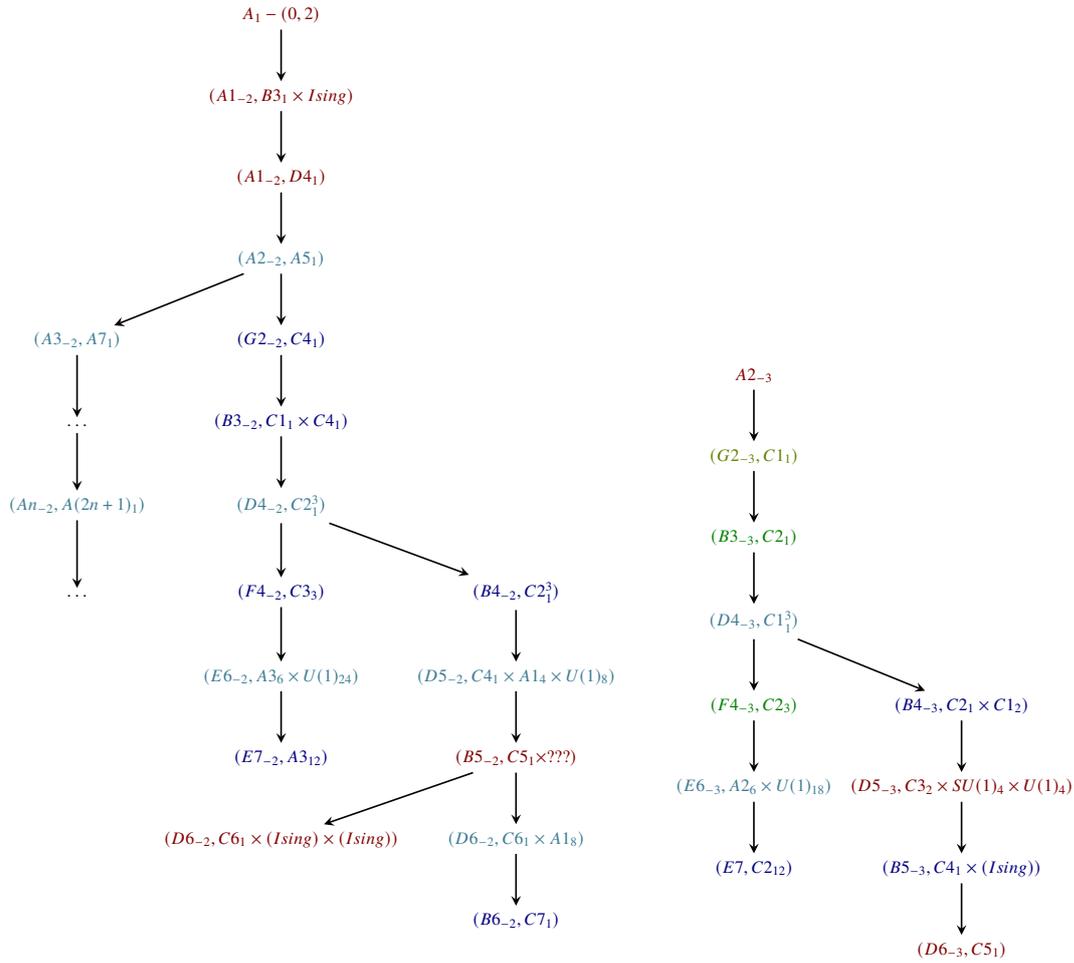
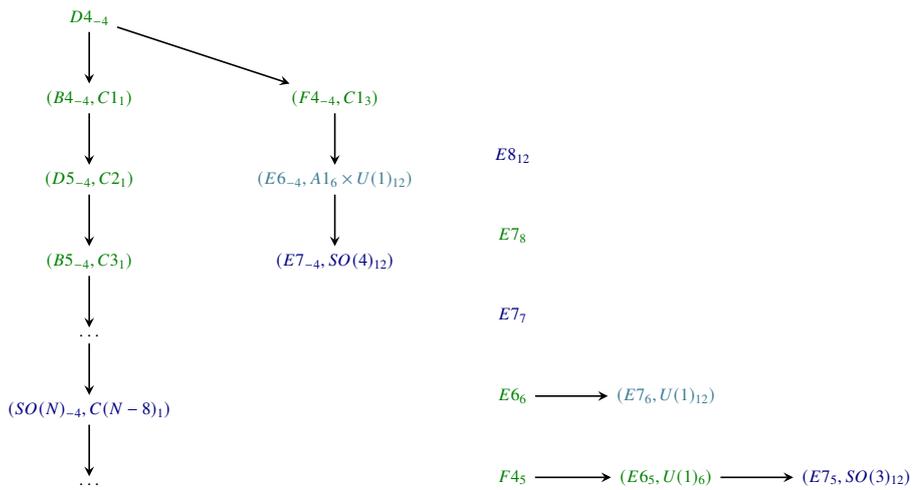
\begin{figure}
    \centering
    \begin{subfigure}[b]{0.58\textwidth}
    \centering
\[\begin{tikzcd}[scale cd =0.6,column sep = 0.2 ex]
&\textcolor{mdtRed}{A_1-(0,2)}\\
	& \textcolor{mdtRed}{(A1_{-2},B3_1\times Ising)} \\
	&  \textcolor{mdtRed}{(A1_{-2},D4_1)} \\
	& \textcolor{mdtCyan}{(A2_{-2},A5_1)} \\
	\textcolor{mdtCyan}{(A3_{-2},A7_{1})} & \textcolor{mdtBlue}{(G2_{-2},C4_1)} \\
	\dots & \textcolor{mdtBlue}{(B3_{-2},C1_1\times C4_1)} \\
	\textcolor{mdtCyan}{(An_{-2},A(2n+1)_1)} & \textcolor{mdtCyan}{(D4_{-2},C2_1^3)} \\
	\dots& \textcolor{mdtBlue}{(F4_{-2},C3_3)} & \textcolor{mdtBlue}{(B4_{-2},C2_1^3)} \\
	& \textcolor{mdtCyan}{(E6_{-2},A3_6\times U(1)_{24})} & \textcolor{mdtCyan}{(D5_{-2},C4_1\times A1_4\times U(1)_8)} \\
	& \textcolor{mdtBlue}{(E7_{-2},A3_{12})} & \textcolor{mdtRed}{(B5_{-2},C5_1\times???)} \\
	& \textcolor{mdtRed}{(D6_{-2},C6_1\times(Ising)\times(Ising))} & \textcolor{mdtCyan}{(D6_{-2},C6_1\times A1_8)} \\
	&& \textcolor{mdtBlue}{(B6_{-2},C7_1)}
    \arrow[from=2-2, to=3-2]
	\arrow[from=3-2, to=4-2]
	\arrow[from=4-2, to=5-1]
	\arrow[from=5-1, to=6-1]
	\arrow[from=6-1, to=7-1]
	\arrow[from=7-1, to=8-1]
	\arrow[from=4-2, to=5-2]
	\arrow[from=5-2, to=6-2]
	\arrow[from=6-2, to=7-2]
	\arrow[from=7-2, to=8-2]
	\arrow[from=8-2, to=9-2]
	\arrow[from=9-2, to=10-2]
	\arrow[from=7-2, to=8-3]
	\arrow[from=8-3, to=9-3]
	\arrow[from=9-3, to=10-3]
	\arrow[from=10-3, to=11-3]
	\arrow[from=10-3, to=11-2]
	\arrow[from=11-3, to=12-3]
	\arrow[from=1-2, to=2-2]
\end{tikzcd}\]
    \caption{Higgsing tree of $\mathbb F_2$. The M-string Higgsing tree. $A_1-(0,2)$ stands for the 6D theory with $(0,2)$ supersymmetry.}
    \label{fig:F2higgsingtree}
    \end{subfigure}
    \hfill
    \begin{subfigure}[b]{0.38\textwidth}
    \centering
    \[\begin{tikzcd}[scale cd =0.6,column sep = 0.2 ex]
	\textcolor{mdtRed}{A2_{-3}} \\
	\textcolor{mdtLime}{(G2_{-3},C1_1)} \\
	\textcolor{mdtGreen}{(B3_{-3},C2_1)} \\
	\textcolor{mdtCyan}{(D4_{-3},C1_1^3)} \\
	\textcolor{mdtGreen}{(F4_{-3},C2_3)} & \textcolor{mdtBlue}{(B4_{-3},C2_1\times C1_2)} \\
	\textcolor{mdtCyan}{(E6_{-3},A2_6\times U(1)_{18})} & \textcolor{mdtRed}{(D5_{-3},C3_2\times SU(1)_4\times U(1)_4)} \\
	\textcolor{mdtBlue}{(E7,C2_{12})} & \textcolor{mdtBlue}{(B5_{-3},C4_1\times (Ising))} \\
	& \textcolor{mdtRed}{(D6_{-3},C5_1)}
	\arrow[from=1-1, to=2-1]
	\arrow[from=2-1, to=3-1]
	\arrow[from=3-1, to=4-1]
	\arrow[from=4-1, to=5-1]
	\arrow[from=5-1, to=6-1]
	\arrow[from=6-1, to=7-1]
	\arrow[from=4-1, to=5-2]
	\arrow[from=5-2, to=6-2]
	\arrow[from=6-2, to=7-2]
	\arrow[from=7-2, to=8-2]
\end{tikzcd}\]
     \caption{Higgsing tree of $\mathbb F_3$.}
    \label{fig:F3higgsingtree}
\end{subfigure}
\hfill
    \begin{subfigure}{0.45\textwidth}
\[\begin{tikzcd}[scale cd =0.6]
	\textcolor{mdtGreen}{D4_{-4}} \\
	\textcolor{mdtGreen}{(B4_{-4},C1_1)} & \textcolor{mdtGreen}{(F4_{-4},C1_3)} \\
	\textcolor{mdtGreen}{(D5_{-4},C2_1)} & \textcolor{mdtCyan}{(E6_{-4},A1_6\times U(1)_{12})} \\
	\textcolor{mdtGreen}{(B5_{-4},C3_1)} & \textcolor{mdtBlue}{(E7_{-4},SO(4)_{12})} \\
	\dots \\
	\textcolor{mdtBlue}{(SO(N)_{-4},C(N-8)_1)} \\
	\dots
	\arrow[from=1-1, to=2-1]
	\arrow[from=2-1, to=3-1]
	\arrow[from=3-1, to=4-1]
	\arrow[from=1-1, to=2-2]
	\arrow[from=2-2, to=3-2]
	\arrow[from=4-1, to=5-1]
	\arrow[from=3-2, to=4-2]
	\arrow[from=5-1, to=6-1]
	\arrow[from=6-1, to=7-1]
\end{tikzcd}\]
     \caption{Higgsing tree of $\mathbb F_4$.}
    \label{fig:F4higgsingtree}
    \end{subfigure}
    \begin{subfigure}{0.45\textwidth}
    \centering
\[\begin{tikzcd}[scale cd =0.6]
	\textcolor{mdtBlue}{E8_{12}} \\
	\textcolor{mdtGreen}{E7_8} \\
	\textcolor{mdtBlue}{E7_7} \\
	\textcolor{mdtGreen}{E6_6} & \textcolor{mdtCyan}{(E7_6,U(1)_{12})} & {} \\
	\textcolor{mdtGreen}{F4_5} & \textcolor{mdtGreen}{(E6_5,U(1)_6)} & \textcolor{mdtBlue}{(E7_5,SO(3)_{12})}
	\arrow[from=4-1, to=4-2]
	\arrow[from=5-2, to=5-3]
	\arrow[from=5-1, to=5-2]
\end{tikzcd}\]
    \caption{Higgsing tree of $\mathbb F_5,\,\mathbb F_6,\,\mathbb F_7,\mathbb F_8,\,\mathbb F_{12},\,$}
    \label{fig:F5-12higgsingtree}
\end{subfigure}
\caption{$\mathbb F_2,\dots,\mathbb F_{12}$ Higgsing trees.}
\label{}
\end{figure}

\newpage

\bibliography{biblio.bib}


\end{document}